\documentclass[twoside]{report} 
\usepackage{epsf,graphicx,axodraw,epsfig}
\usepackage{color}
\usepackage{latexsym}
\usepackage{latexsym,amssymb}
\usepackage{setspace,cite} 
\usepackage{fancyhdr}
\usepackage{a4}

\def\ltsim{\lower3pt\hbox{$\, \buildrel < \over \sim \, $}}
\def\gtsim{\lower3pt\hbox{$\, \buildrel > \over \sim \, $}}
\def\be{\begin{equation}}
\def\ee{\end{equation}}
\def\ba{\begin{eqnarray}}
\def\ea{\end{eqnarray}}
\newcommand{\de}{\partial}  
\def\ga{\mathrel{\raise.3ex\hbox{$>$\kern-.75em\lower1ex\hbox{$\sim$}}}}
\def\la{\mathrel{\raise.3ex\hbox{$<$\kern-.75em\lower1ex\hbox{$\sim$}}}}
\newcommand{\sect}[1]{\section{#1}\setcounter{equation}{0}}
\def\thesection{\arabic{section}}

\def\bo{ { \sqcup\llap{ $\sqcap$} } }
\begin{document}
\newpage

\pagenumbering{arabic}
\setcounter{page}{1} \pagestyle{fancy}
\renewcommand{\chaptermark}[1]{\markboth{\chaptername%
\ \thechapter:\,\ #1}{}}
\renewcommand{\sectionmark}[1]{\markright{\thesection\,\ #1}}



\newpage
\thispagestyle{empty}


\vspace*{1cm}
\begin{center}
{\LARGE\bf Multi-Scale Physics from Multi-Braneworlds\\}
\vspace{2cm}
{ \Large 
{\bf Stavros Mouslopoulos} \\

\vspace{0.5cm}
$\cdot$ Wadham College 1998-2001 $\cdot$\\
$\cdot$ Linacre College 2001-2002 $\cdot$\\
\vspace{0.5cm}
$\cdot$ Department of Physics $\cdot$ 
Theoretical Physics $\cdot$\\ 
$\cdot$ University of Oxford $\cdot$}
\vspace{1cm}
\end{center}

\begin{center}
{\Large\font\oxcrest=oxcrest40
\oxcrest\char'01}
\end{center}
\vspace{7cm}
\begin{center}
{\large  
Thesis submitted for the Degree of Doctor of Philosophy
in the\\
 University of Oxford\\
\vspace{0.3cm}
$\cdot$ Trinity 2002 $\cdot$}
\end{center}
\singlespacing


\begin{abstract}

This Thesis presents a study of higher dimensional brane-world models with
non-factorizable geometry. In the picture of brane-world, Standard
Model fields are assumed to be localized or confined on a lower
dimensional topological defect
(brane) in the higher dimensional space (bulk). 
When the  space is curved, due to the
presence of an   energy density distribution, the non-trivial
geometry can induce localization of gravity across the
extra dimension. This
implies that, in particular constructions, gravity can be 
localized on the brane. The localization of gravity leads to the
realization that, if extra dimensions exists, they need not be
compact. 
  It is shown that in the context of multi-brane world constructions
with localized gravity the phenomenon of multi-localization is
possible. When the latter scenario is realized, the KK spectrum
contains special ultralight and localized KK state(s). Existence of
such states give the possibility that gravitational interactions as we 
realize them are the net effect of the massless graviton and the
special KK state(s). Models that reproduce Newtonian gravity at
intermediate distances even in the absence of massless graviton are
also discussed.
It shown that the massless limit of the  propagator of massive graviton in curved spacetime ($AdS$ or $dS$)
is smooth  in contrast to the case that the
spacetime is flat (vDVZ discontinuity). The latter suggests that in
the presence of local a curvature (\textit{e.g.} curvature induced by
the source) the discontinuity in the graviton propagator disappears 
avoiding the phenomenological difficulties of models with massive gravitons. 
The possibility of generating small neutrino masses through sterile
bulk neutrino in the context of models with non-factorizable geometry
is presented. Additional phenomena related with multi-brane
configurations are discussed.
It is shown that the phenomenon of multi-localization in the context
of multi-brane worlds can also be realized for fields of all
spins. The form of the five dimensional mass terms of the fields is
critical for their localization properties and in the case of the
Abelian gauge field, its localization is possible only for specific
form of mass term.

\vspace*{5cm}

\newpage

\pagenumbering{roman}

\chapter*{} 
\thispagestyle{empty} 
~
\vspace{4cm} 

\begin{flushright}
{\Large {\textit{
Dedicated to Ian I. Kogan.}}}
\end{flushright}

\newpage

\vspace*{8cm}

\pagenumbering{roman}

\begin{center}
\begin{quote}
\it

\dots I leave Sisyphus at the foot of the mountain! One always finds one's burden again. But Sisyphus
   teaches the higher fidelity that negates the gods and raises rocks. He too concludes that all is well.
   This universe henceforth without a master seems to him neither sterile nor futile. Each atom of that
   stone, each mineral flake of that night filled mountain, in itself forms a world. The struggle itself
   toward the heights is enough to fill a man's heart. One must imagine Sisyphus happy.

\end{quote}
\end{center}
\hfill{\small Albert Camus,  ~``The myth of Sisyphus'' 
  ( \textit{c.f.}
Appendix \ref{myth})}

\end{abstract}

\doublespacing

\newpage

\begin{itemize}

\item {\bf{\large Note: }} 

This Thesis was done under the supervision of Professor Graham Ross.
Chapter 1 of this Thesis contains background information only and the
motivation for this work. All subsequent chapters contain original work.

\item {\bf{\large Publications: }} \\


\subitem $\bigstar$ I.~I.~Kogan, S.~Mouslopoulos, A.~Papazoglou, G.~G.~Ross and J.~Santiago,
``A three three-brane universe: New phenomenology for the new  millennium?,''
Nucl.\ Phys.\ B {\bf 584} (2000) 313
[hep-ph/9912552].

\subitem  $\bigstar$ S.~Mouslopoulos and A.~Papazoglou,
``'+-+' brane model phenomenology,''
JHEP {\bf 0011} (2000) 018
[hep-ph/0003207].

\subitem $\bigstar$  I.~I.~Kogan, S.~Mouslopoulos, A.~Papazoglou and G.~G.~Ross,
``Multi-brane worlds and modification of gravity at large scales,''
Nucl.\ Phys.\ B {\bf 595} (2001) 225
[hep-th/0006030].

\subitem $\bigstar$ I.~I.~Kogan, S.~Mouslopoulos and A.~Papazoglou,
``The m $\to$ 0 limit for massive graviton in dS(4) and AdS(4): How to  circumvent the van Dam-Veltman-Zakharov discontinuity,''
Phys.\ Lett.\ B {\bf 503} (2001) 173
[hep-th/0011138].

\subitem $\bigstar$ I.~I.~Kogan, S.~Mouslopoulos and A.~Papazoglou,
``A new bigravity model with exclusively positive branes,''
Phys.\ Lett.\ B {\bf 501} (2001) 140
[hep-th/0011141].

\subitem $\bigstar$ S.~Mouslopoulos,
``Bulk fermions in multi-brane worlds,''
JHEP {\bf 0105} (2001) 038
[hep-th/0103184].

\subitem $\bigstar$ I.~I.~Kogan, S.~Mouslopoulos, A.~Papazoglou and L.~Pilo,
``Radion in multibrane world,''
hep-th/0105255.

\subitem $\bigstar$ 
I.~I.~Kogan, S.~Mouslopoulos, A.~Papazoglou and G.~G.~Ross,
``Multigravity in six dimensions: Generating bounces with flat positive  tension branes,''
hep-th/0107086.

\subitem $\bigstar$
I.~I.~Kogan, S.~Mouslopoulos, A.~Papazoglou and G.~G.~Ross,
``Multi-localization in multi-brane worlds,''
hep-ph/0107307.


\end{itemize}

{\tableofcontents}



\chapter*{Acknowledgments} 
\addcontentsline{toc}{chapter} 
		 {\protect\numberline{Acknowledgments\hspace{-96pt}}} 
The typing of these acknowledgments completes the writing 
of this Thesis. This was by far the hardest part to write given that it took almost four years. This Thesis was completed after changing two Continents, two Universities, two Colleges,  three Offices and over fifteen Temporary Residencies. Well, then let's get on with them:

Working with Graham Ross, was a  great pleasure and  privilege.
I have certainly gained from his knowledge and experience and been 
inspired by him. I hope I will still have the opportunity to continue
 working with him. I particularly thank him for suggesting me to 
spend my fourth year in  Berkeley, California. 

I am also indebted to my second  supervisor, collaborator 
Ian Kogan for  stimulating discussions, his
continuous encouragement and his limitless enthusiasm. 
I wish I had  spent more time discussing with him.

I would like to thank all the people with whom I have had
rewarding collaborations and discussions.  I would like to thank Antonios Papazoglou for collaborating
the past years and for our numerous joint publications. I am also
grateful to Luigi Pilo and Jos\'e Santiago for our common work and for very
fruitful discussions.

Furthermore, I
want to thank all  friends and  colleagues from Theoretical Physics sub-department, especially  my  officemates in the  offices 1.12  and 6.1. 
Mario and Sandra Santos, Shinsuke Kawai, Martin Depken, Anna Durrans, Liliana Velasco-Sevilla, Peter Austing, Peter Richardson, Nuno Reis, Alejandro Ibarra, Bayram Tekin, Alex Nichols,  David Skinner, Martin Schvellinger, Ed Horn, Andrea Jimenez Dalmaroni , Ramon Toldra, Pedro C. Ferreira, Francesc Ferrer and friends from   Wadham and Linacre Colleges:
David Latimer, Guido Sanguinetti , Matteo Semplice, Paolo Matteucci, Ernesto and Laura Dal Bo. Also people from Lawrence Berkeley National Laboratory and University of California at Berkeley: Lawrence Hall, Yashunori Nomura, Zackaria Chacko, Aaron Pierce,  Daniel Larson, Roni and Tami Harnik,  Andrea Pasqua, Steven Oliver, Bianca Cerchiai,  
and especially Take Okui and Yuko Hori.

In respect of funding, I would like first to thank my family for financial support in the first year of my studies and the 
Hellenic State Scholarship Foundation (IKY) for the support during the three following years.
I am also grateful to the sub-Department of Theoretical Physics and to
Wadham College for the travel grands that made it possible for me to
participate to various schools, conferences and workshops during my studies.

Last but not least, I thank my family for their patience, trust and moral support.

\pagestyle{fancy}

\newpage

\addtolength{\headheight}{3pt}
\fancyhead{}
\fancyhead[LE]{\sl\leftmark}
\fancyhead[LO,RE]{\rm\thepage}
\fancyhead[RO]{\sl\rightmark}
\fancyfoot[C,L,E]{}
\pagenumbering{arabic}

\singlespacing  

\chapter{Introduction}
\label{intro}

\section{Introduction}

The idea that our world may have more that three spatial dimensions is
rather old \cite{Kaluza:1921tu,Klein:1926tv}. However, up to the
time of writing this thesis, there is no experimental evidence or indication
for the existence of additional spatial dimensions of any
kind. Nevertheless, there is strong theoretical motivation for considering
spacetimes with more than three spatial dimensions: String theory and
M-theory are theories that try to incorporate quantum gravity in a
consistent way and their  formulation demands for spacetimes
with more that four dimensions. In the context of the previous
theories it seems that a quantum theory of gravity requires
that we live in  ten (String theory) or eleven (M-theory) spacetime dimensions. 

However, apart from the previous theoretical motivation, one is free to ask the
following interesting question: If there are extra dimensions, what is the
phenomenology associated with them ?  This leads to the ``bottom to top''
approach to the physics of extra dimensions: Phenomenological studies
based on simplified (and sometimes over-simplified) field theoretic
(but ``string inspired'') models with extra dimensions. This approach is essential at the
present state since we certainly have not a complete picture of
String theory (the picture is even more obscure for M-theory). Given
the absence of rigid predictions about the
details related to the nature of the extra dimensions of a ``final
theory'', the model building approach can be very instructive
since it can  reveal a whole spectrum of possibilities. 
Obviously the previous has the disadvantage that some (or even most)
of the models considered, may have nothing to do the fundamental theory. 
Another inevitable disadvantage is the appearance of free parameters
in these models something that makes quantitative estimates at best of the 
order-of magnitude and in many cases  not available at all.
Here we should stress that some of the field theoretic models
involving extra dimensions, for example the models with flat extra
dimensions (factorizable geometry), have string theoretic realization
and thus are  theoretically well motivated.
However up to the time of writing this thesis there is no string
theory realization for models with curved extra dimensions
(non-factorizable geometry).

These ``string inspired'' models have recently attracted a lot of
interest since they provide interesting alternative possibilities for
the resolution of longstanding problems of theoretical physics, like:
the gauge hierarchy problem, the cosmological constant problem, the
explanation of the fermions mass hierarchies \textit{etc}. Apart from
these theoretically interesting aspects, these models have also
phenomenological interest since they predict new physics that may be
accessible to future accelerators. Moreover, it turns out that by
exploiting the freedom of the parameter space one can achieve exotic
possibilities where string excitations may be also accessible to
future experiments.    

The material presented in this thesis
could be classified into this category of phenomenological study of
higher dimensional models. 

\section{How many extra dimensions?}

The consistent formulation of String theory (M-theory) demands that
the number of spacetime dimensions is $D=10$ ($D=11$) which means that 
our world has six (seven) extra spatial dimensions. However, from the model
building point of view, the number of extra dimensions is a free
parameter. In particular in the present Thesis we are going to examine 
the phenomenology of models with one or two extra dimensions. The
justification for this is that: 1) Although by considering more
dimensions the phenomenology associated with them will be richer, its
basic characteristics can become apparent in the simplest models with
one or two extra dimensions. 2) Different dimensions can ``open up'' at
different energy scales. If one assumes that for some of the internal
dimensions the compactification radius is much larger than for the
rest, then as experiments probe energy scales corresponding to this
compactification scale, only physics associated with these dimensions
will be accessible while the rest will remain frozen
\footnote{Actually this argument applies literally only in the case of 
flat extra dimensions where the compactification scale sets the scale
of the new physics. Things are different in the case of
non-factrorizable geometry - but similar arguments apply in that case.}. This implies 
that the models with small number of extra dimensions may be
phenomenologically relevant for a range of energy scales.

\section{Hiding Extra Dimensions}

Independently of the number of the extra dimensions that may exist,
experiments confirm that our world, at least up to now, can
be described by four dimensional physics. Thus,  
an important issue in multi-dimensional theories is the mechanism by
which extra dimensions are ``hidden'', so that the spacetime is
effectively four-dimensional at least in the regions that have been probed by
experiments. 
In order to be able to distinguish the five dimensional effects from
the four dimensional, and    since our perception is attached to four dimensions, it is
convenient to give through a well defined procedure, four dimensional 
interpretation of the higher dimensional physics. This is done by the
dimensional reduction procedure, where the original higher dimensional 
physics can described by an effective four dimensional action with the 
price of increasing the number of fields (the number of degrees of
freedom must remain the same): From the four dimensional point of view 
each five dimensional field is
described by an infinite tower of non-interacting fields (Kaluza-Klein (KK) states) with the
same quantum numbers but different masses (degeneracy possible). The zero
mode of this tower (when exists) is used to recover the four
dimensional theory (the existence of the zero mode does not
necessarily imply the recovery of the 4-d theory - further assumptions
usually are needed). The rest of the massive
fields encode the information about the higher dimensional physics and 
their relevance to the four dimensional effective theory decreases the 
more heavy they become. 

Let us briefly discuss how the dimensional reduction is implemented.
One starts with a $4+n$ dimensional
Lagrangian (where $n$ is the number of extra dimensions) that
describes the dynamics of a free field\footnote{Spinor indices suppressed.}
$\Phi(X^{M})=\Phi(x^{\mu},y^{i})$
(with $X^{M}=(x^{\mu},y^{i})$,  $\mu=0$,$1$,$2$,$3$ 
and $i=1$,...,n) propagating in the $4+n$-dimensional 
spacetime. Exploiting the linearity of the equations of motion we can
decompose the higher dimensional field  into normal modes:
\be
\Phi(x^{\mu},y^{i})=\sum_{n} \Phi_{n}(x^{\mu},y^{i})
\ee
Exploiting
the fact that the equations of motion are  partial differential
equations  that can be solved by the method of separation of
variables, every mode of the previous decomposition can be separated
to a part that depends only on the four coordinates of the four
dimensional spacetime 
and a part that depends only on the extra dimensions \textit{i.e.}:
\be
\Phi_{n}(x^{\mu},y^{i})=\Phi_{n}(x^{\mu})\Psi_{n}(y^{i})
\ee
The wavefunctions $\Psi_{n}(y^{i})$ give the localization properties
of the $n$-th state along the extra dimensions (note that $n$ can
describe more than one quantum numbers). 
Our intention is
to ``integrate out'' the extra dimensions ending up with a four dimensional 
Lagrangian that encodes in its form all the details of the higher 
dimensional physics. It can be shown, for particles of all spins,
that the higher dimensional theory can be described by an effective four
dimensional one, in which the effect of the extra dimensions is taken
in account in the mass spectrum and the wavefunctions of the KK states:
\be
S_{5}=\int d^{4}x \prod_{i} dy^{i} L(\Phi(X^{M})) \sim \sum_{n} \int d^{4}x L^{m_{n}}_{n}(\Phi_{n}(x^{\mu}))
\ee
The above equivalence relation holds
provided that the wavefunctions that describe the localization
properties obey a second order partial differential equation 
\be
\hat{O}(y^{i}) \Psi_{n}(y^{i})=m_{n}^{2} \Psi_{n}(y^{i})
\ee
where $m_{n}$ the masses of the KK states and $\hat{O}(y^{i})$ is a hermitian operator the form of which depends
on the geometry and the spin of the particle under consideration. 
The hermicity of the operator $\hat{O}(y)$ ensures  orthonormality
relations for the wavefunctions $\Psi_{n}(y^{i})$  of the form:
\be
\int ( \prod_{i} dy^{i} ) W(y^{i}) \Psi_{m}(y^{i})^{*} \Psi_{n}(y^{i}) = \delta_{mn}
\ee
where $W(y^{i})$ is the appropriate weight.
For the case where $n=1$, it can be shown that the previous second order differential equation
can be brought (after an appropriate change of coordinates, $y
\rightarrow z$, and a
redefinition of the wavefunction $f \rightarrow \hat{f}$ ) to a Schr\"{o}ndiger like form:
 \begin{equation} 
\left\{ -\frac{1}{2}\partial _{z}^{2}+V(z)\right\} \hat{f}^{(n)}(z)=\frac{%
m_{n}^{2}}{2}\hat{f}^{(n)}(z) 
\end{equation} 
where now all the information about the five dimensional physics is
encoded in the form of the potential $V(z)$. Similarly for the cases with $n>1$
following the previous steps we are led to a multi-dimensional Schr\"{o}ndiger
equation.

Summarizing, by implementing the dimensional reduction procedure we can 
describe every higher dimensional field with an infinite number of
four dimensional fields (KK states) with the same quantum numbers. The 
masses and localization properties of these states are found by
solving a second  order Schr\"{o}ndiger differential equation,
where all the information about the five dimensional theory is
contained in the form of the potential.

\section{The Role of Geometry}

The geometry of the higher dimensional space turns out to be of
particular importance in the model building. The topology of the extra 
dimensions (when $n>1$) is also phenomenologically relevant
(\textit{e.g.} see Ref.\cite{Dienes:2001wu}) but its implications is 
generally not as radical in the models that we will consider (we will
mostly work with $n=1$). Let us consider the $n=1$ case for simplicity.
The information related to the geometry can be generally read off the
infinitesimal length element:
\be
ds^2=G_{M N}(X^{K}) dX^{M} dX^{N}=G_{\mu \nu}(X^{K}) dx^{\mu} dx^{\nu}+G_{55}(X^{K}) dy^{2}
\ee
If we demand that the four dimensional spacetime after the reduction
to be Poincare invariant, the most general ansatz has the following
form:
\be
ds^2=A(y) \eta_{\mu \nu} dx^{\mu} dx^{\nu}+B(y) dy^{2}
\label{intr3}
\ee 
(we can always set $B(y)=1$ since it can be absorbed with a
redefinition of variables)
We can classify all five dimensional  models, according to their geometry, into two main categories:

\subsection{Factorizable Geometry}

In this case the extra dimension is considered to be
homogeneous in the sense that there are no preferred places along the 
extra dimension. The spacetime in this case is described be the five 
dimensional flat metric:
\be
ds^2=\eta_{\mu \nu} dx^{\mu} dx^{\nu} + dy^{2}
\ee
with $i=1,2...,n$. 
This is possible only when there is no energy density distribution (or
when it can be neglected). Following the steps of the previous
paragraph we find that the KK decomposition of the
fields is simply a Fourier expansion (in this case we have that
$\hat{O}(y^{i})=\frac{d^{2}}{d y^{2}}$ and the potential of the
Schr\"{o}ndiger equation is trivially zero):
\be
\Phi(x,y)=\sum \Phi_{n}(x,y) =\sum_{n} \Phi_{n}(x) e^{i m_{n} y}
\ee
From the latter we find that in this case the zero mode and all KK states spread along
the extra dimensions with equal probability - since the physically
relevant quantity: $\Phi^{*}_{n}(x,y)\Phi_{n}(x,y)$ is $y$-independent. Thus is this case, there is {\bf{no localization}} of the
states that propagate in the extra dimension. This implies that the
extra dimension must be necessarily compact. The mass spectrum can be 
easily found by imposing the periodicity condition on the wavefunctions:  
$\Phi_{n}(x,y + 2 \pi R) = \Phi_{n}(x,y)$.
The mass spectrum of the KK
states is evenly spaced with mass splitting of the order of
$\frac{1}{R}$ and for this minimal model is given by: $m_{n}=\frac{2 \pi n}{R}$.

Until recently, the main emphasis was
put on KK theories , of this type. In this picture, it is the
compactness of extra dimensions that ensures that the spacetime is
effectively four-dimensional at distances exceeding the
compactification scale (size of the extra dimension). Hence, the size
of extra dimension must be microscopic; a ``common wisdom'' was that
the size was roughly of the order of the Planck scale  (although
compactifications at the electroweak scale were also considered)
\footnote{In the case that only gravity propagates in the bulk, the
size of the extra dimensions can be as large as $1mm$.}.

However, the case of factorizable geometry is not the only
possibility. On the contrary, in more realistic cases  the  presence of energy density distribution 
curves the spacetime and the vacuum is generally non-trivial. This
lead us to spacetimes with  non-factorizable geometry.

\subsection{Non-Factorizable Geometry}

In this case the extra dimensions are not
homogeneous in the sense that different points in the extra dimension
are generally not equivalent. The most general ansatz for these geometries in 
five dimensions with the requirement of Poincare invariance is:
\be
ds^2=A(y) \eta_{\mu \nu} dx^{\mu} dx^{\nu} + dy^2
\ee
where $A(y)$ is a non-trivial function of the extra dimension.
The physical meaning of such a factor is that different points along
the extra dimension have different length scales (something that also
implies that they have different energy scales - something that, as we 
will see, can lead to an elegant explanation of the gauge hierarchy
problem when $A(y)$ is a rapidly changing function.).
In this case, the corresponding  Schr\"{o}ndiger equation has  in
general non-trivial potential something that implies that the form of the wavefunction of the KK states 
is generally non-trivial. For
example, in the case of a massless scalar  field in five dimensions it has the form:
\be
-\frac{d}{dy} \left( A^{4}(y) \frac{d\Psi_{n}(y)}{dy}\right)= m_{n}^{2}
A^{2}(y) \Psi_{n}(y)
\ee
Similar differential equations hold for the wavefunctions of fields of 
all spins.
By solving these differential equations we determine the
wavefunctions $\Psi_{n}(y)$ and the mass spectrum $m_{n}$ of the KK
states. The wavefunction $\Psi_{n}(y)$ gives us information about the
localization properties of the states. The zero modes usually
correspond to the bound states of the quantum mechanical potential
whereas the KK states appear as higher excitations. 
By appropriate choice of the background
geometry one {\bf{ can  achieve localization}} of the zero mode of the field in certain region
of the extra dimension \footnote{Note that  the gauge field 
cannot be localized  by the  background geometry alone - for more details 
see Chapter \ref{multiloc}}. 

As we will see in the following Chapters the non-trivial vacuum
structure is essential for the localization of the spin 2
particle. However, for the fields of spin $<2$, localization can be
achieved also through interactions with other fields.
The localization of the fields  gives the possibility of having 
{\bf{extra dimension with infinite length}} without conflict with the
phenomenology. In this case the existence
of a normalizable zero mode is guaranteed by the requirement of having finite 
compactification volume \footnote{However we will consider cases where 
4-d physics is reproduced even the absence of zero mode - through
special properties of the KK states.(see Chapter \ref{RS})}.

\section{The Brane World Picture}

Recently, the emphasis has shifted towards the ``brane
world'' picture which assumes that ordinary matter (with possible
exceptions of gravitons and other hypothetical particles which
interact very weakly with matter) is trapped to a three-dimensional
sub-manifold  ({\bf{brane}})  embedded in a fundamental multi-dimensional
space ({\bf{bulk}}). In the brane world scenario, extra dimensions may be large, and
even infinite. 

The brane world picture is also attractive due to the fact that lower
dimensional manifolds, $p$-branes are inherent in
string/M-theory. Some kinds of $p$-branes are capable of carrying
matter fields; for example, D-branes have gauge fields residing on
them. Hence, the general idea of brane world appears naturally in
M-theory and indeed, realistic brane-world models based on M-theory
have been proposed \cite{Horava:1996qa,Lukas:1999yy}. 

In the field theoretic approach of these models that we consider
the confinement of the states on the lower dimensional sub-manifold is 
an \textit{ad-hoc} assumption. However, in their context, fields can be localized on the
lower dimensional sub-manifold either through interactions with other 
fields or from the non-trivial geometry (see Chapters \ref{RS}, \ref{multiloc}).   
As a result in the context of field theory the braneworld 
scenario is realized through the localization (and not confinement) of 
the fields.

\section{Which fields feel the extra dimensions?}

In the simplest
formulation of these models no bulk states are assumed to exist and
thus only gravity propagates in the extra dimensions. Nevertheless ``bulk'' (transverse to
the 3-brane space)  physics turns out to be very interesting giving
alternative possible explanations to other puzzles of particle
physics. For example, as we will find in Chapter \ref{neutrino}, by assuming the existence of a  neutral under
Standard Model (SM) spin
$\frac{1}{2}$ fermion in the bulk one can explain  the smallness
of the neutrino masses without invoking the seesaw mechanism and the neutrino
oscillations either in the context of models of large extra dimensions
or in the context of localized gravity models \cite{Dienes:1999sb,Arkani-Hamed:1998vp,Dvali:1999cn,Mohapatra:1999zd,Barbieri:2000mg,Lukas:2000wn,Lukas:2000rg,Cosme:2000ib,Grossman:2000ra,Mouslopoulos:2001uc}.  
In the context of string and M-theory, bulk fermions arise as
superpartners of gravitational moduli, such as, those setting the radii of
internal spaces. Given this origin, the existence of bulk fermions is
unavoidable in any supersymmetric string compactification
and represents a quite generic feature of string
theory\footnote{However, note that brane-world models with non
factorizable geometry have not yet been shown to have string
realizations. For string realizations of models with large extra
dimensions see Ref.\cite{Antoniadis:1998ig}}. This constitutes the most likely origin of such particles
within a fundamental theory and, at the same time, provides the basis
to study brane-world neutrino physics.

However, it is not necessary to confine the SM fields on
the brane. Assuming that also the SM fields can propagate in 
the bulk new interesting possibilities can arise. For example one can
attempt to explain  
the pattern of the SM fermion mass hierarchy by localizing the SM
fermions in different places in the bulk \cite{Arkani-Hamed:2000dc,Mirabelli:2000ks,Dvali:2000ha,delAguila:2000kb}.
The above  gives the motivation for considering the phenomenology 
associated with spin $0$, $\frac{1}{2}$,$1$ fields  propagating in the 
extra dimension(s). 
Furthermore if one additionally wants to 
explore the supersymmetric version of the above models, it is also
necessary to study the phenomenology of spin $\frac{3}{2}$ field.

\section{New Phenomena}

 
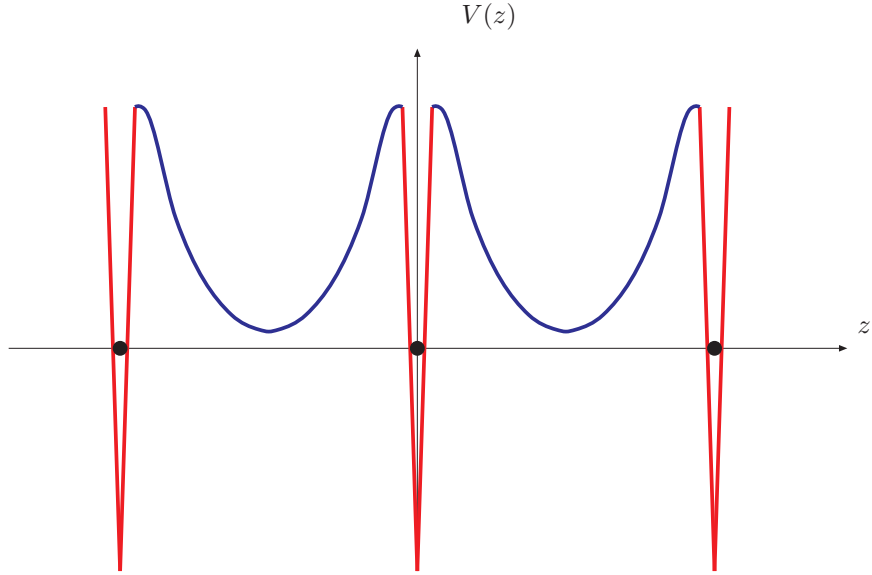
\begin{figure}[t] 
\begin{center} 
\SetScale{0.7}  
\begin{picture}(200,150)(0,50)

\LongArrow(140,0)(140,280)
\LongArrow(-80,120)(370,120)

\SetWidth{2}
\SetColor{Red}
\Line(140,0)(148,250)
\Line(140,0)(132,250)
\SetColor{Black}
\Vertex(140,120){4}


\SetColor{Red}
\Line(-12,250)(-20,0)
\Line(-28,250)(-20,0)
\Line(308,250)(300,0)
\Line(292,250)(300,0)
\SetColor{Black}
\Vertex(-20,120){4}
\Vertex(300,120){4}

\Text(115,210)[l]{$V(z)$}
\Text(270,90)[rb]{$z$}


\SetColor{Blue}
\Curve{(148,250)(153,249)(170,190)(200,138)(215,130)(220,129)(225,130)(240,138)(270,190)(287,249)(292,250)}


\Curve{(-12,250)(-7,249)(10,190)(40,138)(55,130)(60,129)(65,130)(80,138)(110,190)(127,249)(132,250)}


\end{picture} 
\end{center} 
\par 
\vspace*{12mm} 
\caption{The scenario of multi-localization is realized in configurations 
where the corresponding form of potential has potential wells that can 
support bound states. Such a potential is the one that corresponds to the $''++''$ model (see Chapter \ref{5dads}). Positive  tension branes are $\protect\delta 
$-function wells and negative are $\protect\delta$-function barriers.} 
\label{intrfig1}
\end{figure}

\begin{figure}[t]
\begin{center}
\begin{picture}(300,200)(0,50)

\SetWidth{2}
\Line(10,50)(10,250)
\Line(290,50)(290,250)

\SetWidth{0.5}
\Line(10,150)(290,150)

\Text(-10,250)[c]{$''+''$}
\Text(310,250)[c]{$''+''$}


{\SetColor{Green}
\Curve{(10,240)(50,192)(65,181)(80,173)(95,167)(110,162)(130,157)(150,155)}
\Curve{(150,155)(170,157)(190,162)(205,167)(220,173)(235,181)(250,192)(290,240)}
}


{\SetColor{Red}
\DashCurve{(10,240)(50,192)(65,181)(80,173)(95,167)(110,161)(130,155)(150,150)}{3}
\DashCurve{(150,150)(170,145)(190,139)(205,133)(220,127)(235,119)(250,108)(290,60)}{3}
}


{\SetColor{Brown}
\DashCurve{(10,145)(15,149)(20,150)(40,153)(65,158)(140,209)(150,210)}{1}
\DashCurve{(150,210)(160,209)(235,158)(260,153)(280,150)(285,149)(290,145)}{1}
}
\end{picture}
\end{center}

\caption{The  zero mode  (solid line), first (dashed line) and second
(dotted line) KK states wavefunctions in the symmetric $''++''$
model. The wavefunctions of the zero and the first KK mode are localized on the
positive tension branes. Their absolute value differ only in the
central region where they are both suppressed resulting to a very
light first KK state.}
\label{intrfig2}
\end{figure}
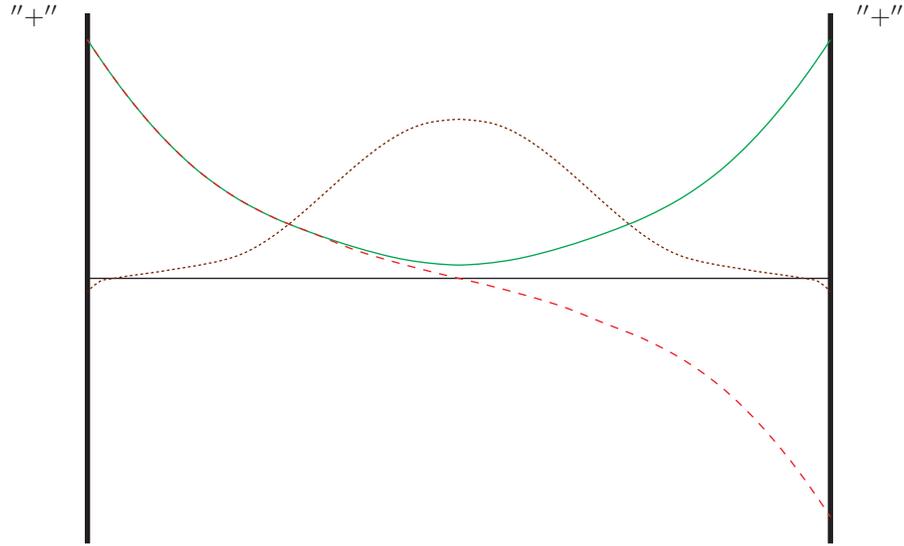

The possibility of localizing  fields in the extra dimensions can lead to new
interesting phenomena: Multi-localization emerges when one considers configurations  such 
that the potential $V(z)$ of the corresponding Schr\"odinger equation 
has at least two  
potential wells\footnote{ \textit{e.g.} in the context of the five dimensional braneworld scenarios, our
world is a three dimensional hypersurface (brane) characterized by
it's tension. In our quantum mechanical description the positive
tension branes correspond to $\delta $-function wells whereas the
negative tension branes to $\delta $-function barriers.}, each of which 
can support a bound state (see Fig.(\ref{intrfig1}) for the $''++''$ case). Models where multi-localization is realized are
interesting since the exhibit non-trivial KK spectrum, with the
appearance of ultralight and localized KK states. This can be
understood as following: 
If we consider the above potential wells separated by an 
infinite distance, then the zero modes are degenerate  and 
massless. However, if the distance between them is finite, due to quantum 
mechanical tunneling the degeneracy is removed and an exponentially small 
mass splitting appears between the states. The rest of levels, which are not 
bound states, exhibit the usual KK spectrum with mass difference 
exponentially larger than the one of the ``bound states'' (see
Fig.(\ref{intrfig2})).
Such an 
example is shown in Fig.(\ref{intrfig2}) where are shown the wavefunctions of the 
graviton in the context of $''++''$ model 
with two positive tension branes at the fixed point boundaries. From it we see that the absolute 
value of these wavefunctions are nearly equal throughout the extra 
dimension, with exception of the central region where the antisymmetric 
wavefunction passes through zero, while the symmetric wavefunction has 
suppressed but non-zero value. The fact that the wavefunctions are 
exponentially small in this central region results in the exponentially 
small mass difference between these states.

The phenomenon of multi-localization is of particular interest since, 
starting from a problem with only one mass scale, 
we are able to create a second scale exponentially 
smaller. Obviously the generation of this hierarchy is due to the tunneling 
effects in our ``quantum mechanical'' problem.

\paragraph{Bigravity - Multigravity}

 The phenomenon of multi-localization and the appearance of
non-trivial KK spectrum can give very interesting applications in the 
gravitational sector of the theory: 

In the simplest case, exploiting the appearance of an ultralight
localized special first KK state we can realize the following
scenario: Gravitational interactions as we realize them are the net
effect of the massless graviton and the massive ultralight KK state. 
In this scenario the large mass gap between  the first KK state and the 
rest of the tower (see Fig.(\ref{intrfig3})) is essential so that Newtonian gravity is recovered
in the intermediate scales. The radical prediction of this scenario is 
that apart from the modifications of gravity at short distances (due
to the ``heavy'' KK states), there will be modifications of gravity at
ultra-large scales due to the fact that the first KK state has non-zero 
mass. In Chapters \ref{RS}, \ref{5dads}, \ref{6D} we present models that exhibit 
such a mass spectrum.

\begin{center}
\begin{figure}
\begin{picture}(350,80)(-100,0)
\SetScale{1.}
\SetOffset(-60,-40)
\LongArrow(0,55)(300,55)

\Text(300,75)[l]{$r$}
\Text(-5,95)[l]{$0$}
\Vertex(0,55){2}

\Text(50,97)[l]{$1 {\rm mm}$}

\Text(230,97)[l]{$10^{26} {\rm cm}$}

\SetScale{1.4}
\C2Text(105,40){Blue}{Red}{Region where Newton's Law}{has been tested}
\SetScale{1.4}
\Vertex(42,39){2}
\Vertex(169,39){2}

\end{picture}
\caption{ } 
\label{intrfig3}
\end{figure}
\end{center}


Of course it is possible to have more than one special KK states -
well separated from the rest of the KK tower contributing to
gravitational interactions - by considering
 configurations with appropriate corresponding potential
(\textit{e.g.} multi-brane configurations). 

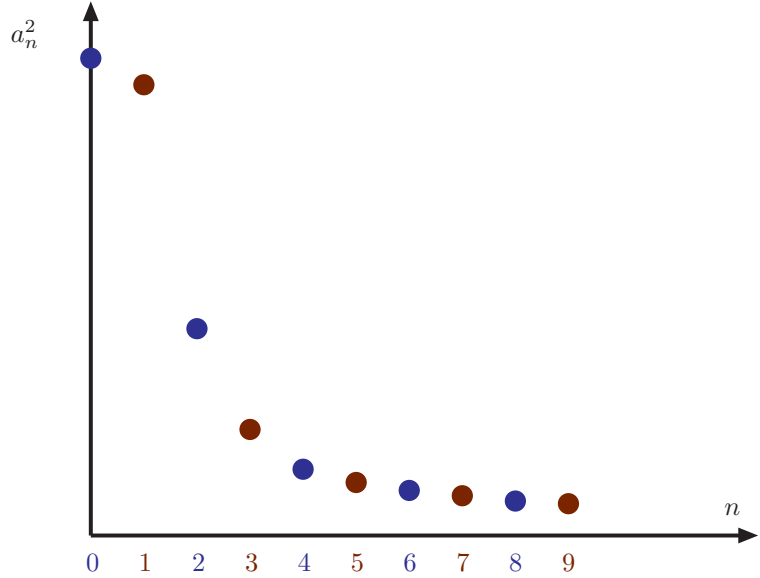
\begin{figure}
\begin{center}
\begin{picture}(300,200)(0,50)
\SetWidth{1.5}

\LongArrow(50,50)(300,50)
\LongArrow(50,50)(50,250)

\SetColor{Blue}
\Vertex(50,230){4}
\Text(51,40)[c]{${\Blue 0}$}
\Vertex(90,128){4}
\Text(91,40)[c]{${\Blue 2}$}
\Vertex(130,75){4}
\Text(131,40)[c]{${\Blue 4}$}
\Vertex(170,67){4}
\Text(171,40)[c]{${\Blue 6}$}
\Vertex(210,63){4}
\Text(211,40)[c]{${\Blue 8}$}

\SetColor{Brown}
\Vertex(70,220){4}
\Text(71,40)[c]{${\Brown 1}$}
\Vertex(110,90){4}
\Text(111,40)[c]{${\Brown 3}$}
\Vertex(150,70){4}
\Text(151,40)[c]{${\Brown 5}$}
\Vertex(190,65){4}
\Text(191,40)[c]{${\Brown 7}$}
\Vertex(230,62){4}
\Text(231,40)[c]{${\Brown 9}$}

\SetColor{Black}
\Text(20,240)[l]{$a_n^2$}
\Text(290,60)[l]{$n$}

\end{picture}
\end{center}
\caption{An suitable behaviour of the coupling, $a(m_{n})$ in the case
of discrete spectrum. 
 In this case the massless graviton exists. However the states with
approximately zero mass contribute to gravity at intermediate
distances due to their significant coupling. The KK states with masses 
that correspond to wavelengths where gravity has been tested have
suppressed coupling. }
\label{intrfig4}
\end{figure}

\begin{figure}[tbp]
\begin{center}
\begin{picture}(400,200)(-50,50)
\SetOffset(50,0)
\SetWidth{1.5}

\LongArrow(-50,80)(250,80)
\LongArrow(-50,80)(-50,250)
\Curve{(-50,230)(-40,229)(-32,224)(-15,150)(40,90)(50,90)(75,90)(100,90)(125,90)(150,90)(175,90)(190,90)(200,90)(225,120)(240,140)}
\DashLine(-45,80)(-45,231){2.5}
\DashLine(-10,80)(-10,130){2.5}
\DashLine(200,80)(200,90){2.5}

\Text(-90,240)[l]{$a(m)^2$}
\Text(-40,70)[c]{$m_{1}$}
\Text(-30,110)[c]{$\Gamma/2$}
\Text(230,70)[l]{$m$}
\Text(200,70)[c]{$m_{0}$}
\Text(-10,70)[c]{$m_{c}$}
\Text(-57,78)[c]{$O$}

\SetWidth{2}
\LongArrow(-50,130)(-10,130)
\LongArrow(-10,130)(-50,130)

\end{picture}
\end{center}
\caption{Again, the  behaviour  of the coupling, $a(m),$ in the case
of continuum spectrum. The region $m>m_{0}$ gives
rise to short distance corrections. The $m_{1}\ll m\ll m_{c}$ region gives
rise to 4D gravity at intermediate distances and 5D gravity at ultra large
distances. For distances $r\gg m_{1}^{-1}$, the zero mode gives the dominant
contribution and thus we return to 4D gravity.}
\label{intrfig5}
\end{figure}
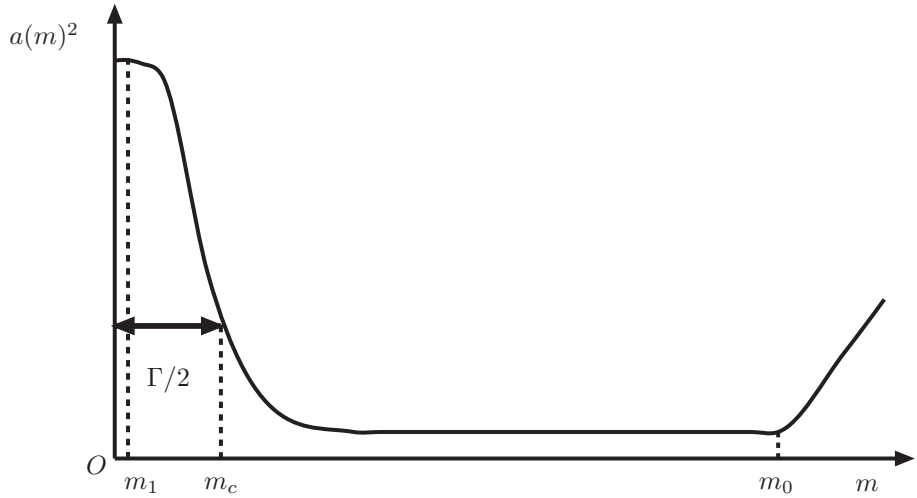

Another way that the KK states can contribute to the gravity at
intermediate distances (or even to reproduce gravity - in the absence
of massless graviton \footnote{This is the case when the
compactification volume is infinite.}) is the following: 
Consider the case that the KK spectrum is dense discrete  or even
continuous . In this case although there is no  mass gap as in the
previous case, the Newtonian gravity can be recovered through the 
special behaviour of the coupling to matter of the KK modes (see
Fig.(\ref{intrfig4}) for the discrete case and Fig.(\ref{intrfig5})
for the continuum). This can
be achieved if the coupling of the KK states (to matter) is
significant  for a band of states with $m_{n} \rightarrow 0$ - states
which will reproduce gravity in the intermediate distances, whereas
it is suppressed for the mass region that corresponds to the
wavelengths that gravity has been tested (see Fig.(\ref{intrfig5})).   
 Heavy states can have significant coupling since they modify gravity
at small distances \footnote{Note that gravity has been test only up
to distances of a fraction of a millimeter}.

Theories of massive gravity in flat spacetime have been considered to
be in conflict with phenomenology due to an apparent discontinuity of
the graviton propagator in the massless limit (vDVZ discontinuity).
In Chapter \ref{mssgr}, however  we show that in the case of $AdS$ and
$dS$ spacetime this discontinuity is absent something that is also
supported by the results of Ref.\cite{Deffayet:2001uk}. Thsese results 
show that theories of massive gravity can be phenomenologically viable 
provided that the mass of the graviton is sufficiently smaller
compared to the characteristic local curvature scale.

\paragraph{Multi-Localization of other fields}

 
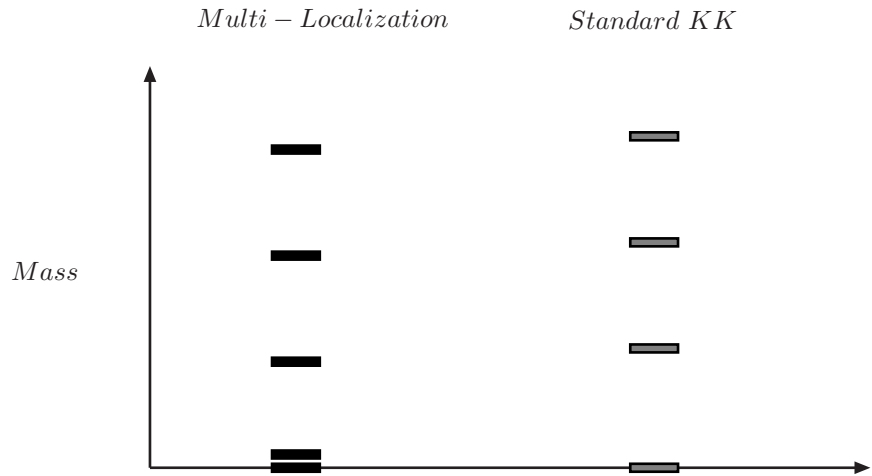
\begin{figure}[t] 
\begin{center} 
\begin{picture}(250,200)(0,70) 
 
\SetWidth{1} 
\LongArrow(-20,100)(-20,250) 
 
\SetWidth{1} 
\LongArrow(-20,100)(250,100) 

\Text(-60,175)[c]{$Mass$} 
\Text(45,270)[c]{$Multi-Localization$} 
\Text(170,270)[c]{$Standard~KK $} 
 
 
 
\GBoxc(35,220)(18,3){0} 
\GBoxc(35,180)(18,3){0} 
\GBoxc(35,140)(18,3){0} 
 
\GBoxc(35,105)(18,3){0} 
 
\GBoxc(35,100)(18,3){0} 
 
 
 

\GBoxc(170,225)(18,3){0.5} 
\GBoxc(170,185)(18,3){0.5} 
\GBoxc(170,145)(18,3){0.5} 
 
 
\GBoxc(170,100)(18,3){0.5} 
 
 
\end{picture} 
\end{center} 
\caption{Comparison of the gravitational spectrum of the $''++''$ or 
$''+-+''$ model with the $''+-''$ 
Randall-Sundrum model.} 
\label{intrfig6}
\end{figure}

As we have mentioned, localization can be
realized by fields of all spins in the context of braneworld models
with  non-factorizable geometry (for some fields in order to achieve
the desired localization, specific mass terms must be added).
Given the latter, in the context of multi-brane models emerges the possibility of multi-localization for all 
the previous fields with appropriate mass terms. When multi-localization is realized the above fields 
apart from the massless zero mode support ultra-light localized KK mode(s). 
An example of the  non-trivial KK spectrum in the context of the
$''++''$ model is given in 
Fig.(\ref{intrfig6}).

In the simplest constructions with two positive branes, that one can
consider, 
there is only one special KK state. However by adding more positive 
tension branes one can achieve more special light states. In the extreme 
example of a infinite sequence of positive branes instead of discrete 
spectrum of KK states we have continuum bands. In the previous case the 
special character of the zeroth band appears as the fact that it is well 
separated from the next.

\section{Summary}

In this introductory Chapter we gave a general outlook of models with
extra spatial dimensions. We gave the general motivation for studying
such models, we discussed their underlying assumptions and classified them
according to their geometry. We presented how the dimensional reduction
is implemented and presented the main characteristics of models with
factorizable and non-factorizable geometry. Finally we presented
briefly the ideas behind the new phenomena that will be analyzed in
the rest of the Thesis such as: Bigravity (or multigravity) and
multi-localization of fields, 
 phenomena that have the same characteristic that 
starting from a problem with only one mass scale,
we are able to create a second scale exponentially 
smaller giving rise to multi-scale physics. The generation of this hierarchy is due to the tunneling 
effects in our ``quantum mechanical'' problem. 
\chapter{Flat Multi-brane Constructions} 
\label{RS}

\section{Introduction}

The importance of the RS construction consists of two main features: 
The first is that in the context of this model gravity is localized and
second is the fact that it provides a geometrical mechanism to
generate the  hierarchy between the Plank and electroweak energy
scales . 
This model has also attracted  a lot of
interest since it belong to the class of brane-world models which
provide an alternative framework within which other problems of particle
physics and cosmology can be addressed.  
In this chapter we will review the prototype model and examine the new 
physics associated with extensions of this model.

\section{Localization of Gravity}

The key feature of the RS model that makes gravity localized, is the
non-trivial background geometry. Thus, let us start building the model based 
on this hint.
For simplicity let us adopt the brane-world picture where all the SM
fields are confined on a three dimensional sub-manifold (brane). Also
for the moment we assume that there are no other, neutral under the SM 
, states in the bulk. Given the previous assumptions, all new physics
will come from the gravitational sector of the theory.  
Our intention is to study the dynamics of gravity  in a  non-trivial background geometry. 
Non-trivial geometry requires some energy density
distribution in order to be created.
The simplest case is to assume a homogeneous distribution of energy
density in the bulk, that is,  the five dimensional bulk is filled with
  energy density \textit{i.e.} cosmological constant 
$\Lambda$.  
 Since gravity
has the characteristic that it creates the background in which the
graviton propagates, we have first to solve the classical Einstein's
equations in order to find the vacuum solution and then perturbe it in
order to study the dynamics.
The action set-up describing five dimensional gravity with a bulk
cosmological constant is:
\be
S=\int d^4 x \int_{-L_1}^{L_1} dy \sqrt{-G} 
\{2 M^3 R - \Lambda \}
\ee 
where $L_{1}$, $-L_{1}$ are the boundaries of our one dimensional manifold (however 
in principle we can have $L_{1}=\infty$), and $M$ the fundamental scale of the five 
dimensional theory. The extra variable $y$  which
parameterizes the extra dimension is thus taking values in the region $[-L_{1},L_{1}]$ 
The variation of action, in respect to the metric leads to the Einstein equations:
\be
R_{MN}-\frac{1}{2}G_{MN}R=-\frac{\Lambda}{4M^3}
G_{MN}
\ee
In the case that $\Lambda=0$, the five dimensional
spacetime is flat and gravity is not localized, according to the
arguments presented in Chapter \ref{intro}. 
An example of this case is  the large extra dimensions - type
models where the zero mode and the KK states that emerge from  the
dimensional reduction procedure are not localized but they spread in
the extra dimension (in this case $L_{1}$ is bounded from above).
However for $\Lambda \ne 0$, we have to solve the Einstein's
equations in order to find the non-trivial vacuum solution.
We are trying to find a solution by making the simple ansatz which has 
the property that the hypersurfaces $y=ct.$ are flat: 
\be
ds^2=e^{-2\sigma(y)}\eta_{\mu\nu}dx^\mu dx^\nu +dy^2
\label{ansatzrs+-+}
\ee
Here the ``warp'' function $\sigma(y)$ is essentially a conformal
factor that rescales the 4D component of the metric. A straightforward
calculation gives us:
\ba
R_{\mu \nu}-\frac{1}{2}G_{\mu \nu} R&=&-3\sigma''e^{-2 \sigma}
\eta_{\mu \nu} + 6 (\sigma ')^{2} e^{-2 \sigma}
\eta_{\mu \nu} \\
R_{55}-\frac{1}{2}G_{55} R&=& 6 (\sigma ')^{2}
\ea
Equating the latter to the energy momentum tensor and assuming that
$\Lambda<0$\footnote{This choice results to $AdS_{5}$ vacuum solution.
 This choice is made so that warp factor $e^{\sigma(y)}$ is a fast
varying function of the $y$ coordinate - giving the possibility of
solving the hierarchy problem, as we will see.} we get the solution:
\be
\sigma(y)= \pm ky
\ee
where $
k=\sqrt{\frac{-\Lambda}{24M^3}}$ is a measure of the curvature of the bulk.
This describes the five dimensional $AdS$ spacetime that the negative
cosmological constant creates. Note that the previous solution is
valid .  The characteristic of the above solution 
is that,  the length scales change exponentially along the extra
dimension. 

Now let us examine the graviton dynamics in the previous background.
This is
determined by considering the (linear) fluctuations of the metric of the
form:
\be
ds^2=\left[e^{-2\sigma(y)}\eta_{\mu\nu} +\frac{2}{M^{3/2}}h_{\mu\nu}
(x,y)\right]dx^\mu dx^\nu +dy^2
\label{perturbrs+-+}
\ee
We expand the field $h_{\mu\nu}(x,y)$ in graviton and KK states plane waves:
\be
h_{\mu\nu}(x,y)=\sum_{n=0}^{\infty}h_{\mu\nu}^{(n)}(x)\Psi^{(n)}(y)
\ee
where
$\left(\partial_\kappa\partial^\kappa-m_n^2\right)h_{\mu\nu}^{(n)}=0$
and fix the gauge as
$\partial^{\alpha}h_{\alpha\beta}^{(n)}=h_{\phantom{-}\alpha}^{(n)\alpha}=0$.
In order the above to be valid the zero mode and KK wavefunctions should obey the
following second order differential equation:
\be
-\frac{1}{2}\frac{d^2 \Psi^{(n)}(y) }{dy^2}+2
(\sigma')^{2} \Psi^{(n)}(y)-\frac{1}{2}e^{2 \sigma} m_{n}^{2} \Psi^{(n)}(y)=0
\ee
This for $m_{0}=0$ (massless graviton) gives 
\be
\Psi^{(0)}(y)=e^{\pm ky}
\ee
From the previous equation we see that the profile of the wavefunction 
is non-trivial. This is an important result: the non-zero energy
distribution ($k \ne 0$) induces non-trivial profile  
to wavefunction of the zero mode (and KK states).
However the zero mode is not localized since the
wavefunction is not normalizable (the zero mode is interpreted as the
4-d graviton and thus its presence in this minimal model is essential
for recovering the 4-d Newton's law).

\subsection{The Single Brane Model (RS2)}

Let us try to modify the previous solution in a way that the graviton
is normalizable. The one solution we considered so far is of the form:
$\Psi^{(0)}(y)=e^{- ky}$. This solution has a good behaviour for $y
\rightarrow \infty$ however in diverges badly for   $y
\rightarrow - \infty$. The other solution, $\Psi^{(0)}(y)=e^{+ ky}$,
has the opposite behaviour. Thus one possibility is to match these two 
different solutions at $y=0$ \textit{i.e.}:
\be
\renewcommand{\arraystretch}{1.5}
\sigma(y) = \left\{\begin{array}{cl} -ky &y\in[0,\infty)\\ 
 ky &y\in(- \infty,0]\\ \end{array} 
\right.
\
\ee
However in this case the function $\sigma'(y)$ is not continuous at
$y=0$. This implies that 
\be
\sigma''(y)=2k \delta (y)
\ee
Note that the latter choice of solution is equivalent to imposing
$Z_{2}$ symmetry (identification $y \rightarrow -y$) 
in the extra dimension around the point $y=0$ (the extra dimension has 
thus the geometry of an orbifold with one fixed point at $y=0$), and choosing that the
graviton has even parity under the reflections $y \rightarrow -y$.
Given that the term $\sigma''(y)$ appears in the Einstein equations,
in order this solution to be consistent we
have to include a brane term in the action. The action in this case
should be:  
\be
S=\int d^4 x \int_{- \infty}^{\infty} dy \sqrt{-G} 
\{-\Lambda + 2 M^3 R\}-\int_{y=0}d^4xV_{0}\sqrt{-\hat{G}^{(0)}}
\ee
where $\hat{G}^{(0)}_{\mu\nu}$ is the induced metric on the brane
and $V_{0}$ its tension. 
The Einstein equations that arise from this
action are:
\be
R_{MN}-\frac{1}{2}G_{MN}R=-\frac{1}{4M^3}
\left(\Lambda G_{MN}+
V_{0}\frac{\sqrt{-\hat{G}^{(0)}}}{\sqrt{-G}}
\hat{G}^{(0)}_{\mu\nu}\delta_M^{\mu}\delta_N^{\nu}\delta(y)\right)
\ee
in this case in order to find a solution, 
the  tension of the brane has to be tuned to $V_0=-\Lambda/k>0$.

The four-dimensional effective theory now follows by considering the
massless fluctuations of the vacuum metric (i.e. $g_{\mu \nu}=e^{-2k|y|}(\eta_{\mu \nu}+h_{\mu \nu}(x))$).  
In order to get the scale of gravitational interactions, we focus on
the curvature term from which we can derive that : 
\begin{equation}
 S_{eff} \supset \int d^4 x \int^{\infty}_{-\infty} d y ~
2 M^3  e^{-2 k |y|}  \sqrt{-g} ~R
\end{equation}
where $R$ denotes the four-dimensional Ricci scalar
 made out of 
$g_{\mu \nu}(x)$. 
 We can 
explicitly perform the $y$ integral to obtain a purely four-dimensional 
action. From this we derive
\begin{equation}
\label{effplanck1}
M_{Pl}^2  =  M^3  \int_{- \infty}^{\infty} dy  e^{-2 k |y|} = 
\frac{M^3}{k}
\end{equation}
The above formula tells us that   the
three mass scales $M_{\rm Pl}$, $M$, $k$ can be taken to be of the same
order. Thus we take $k\sim {\mathcal {O}}(M)$ in order
not to introduce a new hierarchy, with the additional restriction
$k<M$ so that the bulk curvature is small compared to the 5D Planck
scale so that we can trust our solution.

The corresponding differential equation in this case takes the form
Schr\"{o}dinger equation:
\be
\left\{-
\frac{1}{2}\partial_z^2+V(z)\right\}\hat{\Psi}^{(n)}(z)=\frac{m_n^2}{2}\hat{\Psi
}^{(n)}(z)
\ee
with the corresponding potential 
\be
\hspace*{0.5cm} V(z)=\frac{15k^2}{8[g(z)]^2}-
\frac{3k}{2g(z)} \delta(z)
\ee
when new variables and wavefunction in the above equation are defined as:
\be
\renewcommand{\arraystretch}{1.5}
z\equiv\left\{\begin{array}{cl}\frac{e^{ky}-1}{k}&y\in[0,\infty)\\-\frac{e^{-ky}-1}{k}&y\in[-
\infty,0]\end{array}\right.
\
\ee
\be
\hat{\Psi}^{(n)}(z)\equiv \Psi^{(n)}(y)e^{\sigma/2}
\ee
where we have defined the function $g(z)$ as
 $g(z)\equiv k|z|+1$.
In this case the zero mode ($m_{0}=0$) has the form:
\be
\hat{\Psi}^{(0)}=\frac{A}{[g(z)]^{3/2}}= A e^{-3 \sigma(y) /2}
\ee
Given the form of $\sigma(y)$, it is obvious that the above state is normalizable.
The
normalization condition is  
\be
\displaystyle{\int_{- \infty}^{ \infty }
dz\left[\hat{\Psi}^{(0)}(z)\right]^2=1}
\ee
The rest of the spectrum consists of a gapless (starting from
$m=0$) continuum of KK states with wavefunctions:
\be
\hat{\Psi}(z,m)=\sqrt{\frac{g(z)}{k}}\left[A_{1} 
J_2\left(\frac{m_n}{k}g(z)\right)+A_
{2} Y_{2}\left(\frac{m_n}{k}g(z)\right)\right]
\ee
with normalization condition:
\be
\displaystyle{\int_{- \infty}^{ \infty }
dz \hat{\Psi}(z,m) \hat{\Psi}(z,m')  =\delta(m,m')}
\ee

The massless zero mode reproduces the $V(r) \propto \frac{1}{r}$ Newton's Law
potential while the continuum of KK states give small corrections.
A detailed calculation gives:
\begin{equation}
V(r) \sim G_N { m_1 m_2 \over r} +\int_0^{\infty} dm { G_N \over k} 
{ m_1 m_2 e^{- m r} \over r} {m \over k}.
\end{equation}
Note there is a Yukawa exponential suppression in the massive Green's functions
for $m > 1/r$, and the extra power of $m/k$ arises from the suppression of 
continuum wavefunctions at $z = 0$. 
The coupling $G_N/k$ in the second term is nothing but the fundamental
coupling of gravity, $1/M^3$.
Therefore, the potential behaves as
\begin{equation}
V(r) = G_N { m_1 m_2 \over r}\left(1+{1 \over r^2 k^2} \right)
\end{equation}
The latter shows that the theory produces an effective four-dimensional
theory of gravity. The leading term due
to the bound state mode is  the usual Newtonian potential;
the KK modes generate an extremely
suppressed correction term, for $k$ taking the
expected value of order the
fundamental Planck scale and $r$ of the size
tested with gravity.

Summarizing we have found that the set-up consisting of a single
positive tension flat brane embedded in an $AdS_{5}$ bulk
with $Z_{2}$ symmetry imposed can localize gravity in the sense that
the zero mode is peaked on the brane whereas it falls exponentially
away from it. This zero mode reproduces the Newton's potential. 
The continuum KK states on the other hand are suppressed 
near the brane and their presence  results to small corrections in the Newton's potential.

\subsection{The Two Brane Model(RS1)}

We can make the previous one brane model compact by cutting the extra dimension
in symmetric in respect to $y=0$ points (say $y=\pm L_{1}$) and then
identify these endpoints. However, since
 \be
\renewcommand{\arraystretch}{1.5}
\sigma'(y) = \left\{\begin{array}{cl} -k &y\in[0,\infty)\\ 
 k &y\in(- \infty,0]\\ \end{array} 
\right.
\
\ee
the function $\sigma'(y)$ develops a discontinuity at $y=\pm L_{1}$
and thus  $\sigma''(y)$ will give a second $\delta$-function at that
point. Thus we have (for $y \ge 0$)
\be
\sigma''(y)=2k [\delta(y)-\delta(y-L_{1})]
\ee 
Given the latter, in order our solution to be consistent in the 
compact case we must add a second brane term in the action:
\be
S=\int d^4 x \int_{-L_1}^{L_1} dy \sqrt{-G} 
\{-\Lambda + 2 M^3 R\}-\int_{y=0}d^4xV_{0}\sqrt{-\hat{G}^{(1)}}-\int_{y=L_{1}}d^4xV_{1}\sqrt{-\hat{G}^{(1)}}
\ee
in this case in order to find a solution, 
the  tension of the branes has to be tuned to $V_{0}=-V_{1}=-\Lambda/k>0$. 
 Thus in this construction the branes are placed on the orbifold fixed 
points \footnote{ \textit{i.e.} at $y=0$ {\bf{hidden brane}} and $y=L_{1}$ {\bf{visible brane}}} and they have opposite tension. 

Again, the function $\Psi^{(n)}(y)$ will obey a second order differential
equation which after a change of variables reduces to an ordinary
Schr\"{o}dinger equation:
\be
\left\{-
\frac{1}{2}\partial_z^2+V(z)\right\}\hat{\Psi}^{(n)}(z)=\frac{m_n^2}{2}\hat{\Psi
}^{(n)}(z)
\ee
with the corresponding potential 
\be
\hspace*{0.5cm} V(z)=\frac{15k^2}{8[g(z)]^2}-
\frac{3k}{2g(z)}\left[\delta(z)-\delta(z-z_1)-\delta(z+z_1)\right]
\ee
The new variables and wavefunction in the above equation are defined as:
\be
\renewcommand{\arraystretch}{1.5}
z\equiv\left\{\begin{array}{cl}\frac{e^{ky}-1}{k}&y\in[0,L_1]\\-\frac{e^{-ky}-1}{k}&y\in[-
L_1,0]\end{array}\right.
\
\ee
\be
\hat{\Psi}^{(n)}(z)\equiv \Psi^{(n)}(y)e^{\sigma/2}
\ee
and the function $g(z)$ as
 $g(z)\equiv k|z|+1$, where $z_1=z(L_1)$.

This is a quantum mechanical problem with $\delta$-function potentials of
different weight and an extra $1/g^2$ smoothing term (due to the AdS geometry) 
that gives the
potential a double ``volcano'' form. The change of variables has been
chosen so that there are no first derivative terms in the
differential equation. 

This potential is that it always
gives rise to a (massless) zero mode, with wavefunction  given by:
\be
\hat{\Psi}^{(0)}=\frac{A}{[g(z)]^{3/2}}
\ee
The
normalization factor $A$ is determined by the requirement 
$\displaystyle{\int_{- z_{1}}^{ z_{1} }
dz\left[\hat{\Psi}^{(0)}(z)\right]^2=1}$, chosen so that we get the standard 
form 
of the Fierz-Pauli Lagrangian.

For the KK modes the solution is given in terms of Bessel
functions. For $y$ lying in the regions ${\bf A}\equiv\left[0,L_1\right]$, we have:
\be
\hat{\Psi}^{(n)}(z)=\sqrt{\frac{g(z)}{k}}\left[A_{1} 
J_2\left(\frac{m_n}{k}g(z)\right)+A_
{2} Y_{2}\left(\frac{m_n}{k}g(z)\right)\right]
\ee
The wavefunctions are  
 normalized as $\displaystyle{\int_{- z_{1}}^{z_{1}}
dz\left[\hat{\Psi}^{(n)}(z)\right]^2=1}$. 
The boundary conditions result in a
$2\times2$ homogeneous linear system which, in order to have a
non-trivial solution, leads to the vanishing determinant:
\be
\renewcommand{\arraystretch}{1.5}
\left|\begin{array}{cccc}J_1\left(\frac{m}{k}\right)&Y_1\left(\frac{m}{k}\right)
\\\phantom{-
}J_1\left(\frac{m}{k}g(z_1)\right)&\phantom{-
}Y_1\left(\frac{m}{k}g(z_1)\right)\end{array}\right|=0
\ee
(where we have suppressed the subscript $n$ on the masses $m_n$)
This is essentially the mass quantization condition which gives the
spectrum of the KK states.
From the previous condition we can easily workout the mass spectrum for the KK states:
\be
m_{n}=\xi_{n}~k~e^{-k L_{1}}
\ee
(for $n\ge{1}$), where $\xi_{n}$ in the n-th root
of $ J_{1}(x)$.

Following the steps of the previous Section,
we can get get the scale of gravitational interactions:
\begin{equation}
\label{effplanck2}
M_{Pl}^2  =  M^3  \int_{-L_{1}}^{L_{1}} dy  e^{-2 k |y|} = 
\frac{M^3}{k}
[1 - e^{- 2 k L_{1}}].
\end{equation}
The above formula tells us that for large enough $kL_1$  the
three mass scales $M_{\rm Pl}$, $M$, $k$ can be taken to be of the same
order. Thus we take $k\sim {\mathcal {O}}(M)$ in order
not to introduce a new hierarchy, with the additional restriction
$k<M$ so that the bulk curvature is small compared to the 5D Planck
scale so that we can trust our solution.

\subsection{Solving the hierarchy problem in the two brane model}

Let us now how the background vacuum solution of the two brane
configuration can be used in order to solve the gauge hierarchy
problem.
In order to determine the matter field Lagrangian we need to know the 
 coupling of  the 3-brane fields  to the low-energy 
gravitational fields, in particular the metric, $g_{\mu \nu}(x)$. 
 From Eq. (\ref{ansatzrs+-+}) we see that $g_{hid} = 
{g}_{\mu \nu}$. 
This is not the case for the visible sector
 fields;   by Eq. (\ref{ansatzrs+-+}), we have 
$g^{vis}_{\mu \nu} = e^{- 2 k L_{1}} {g}_{\mu \nu}$. By properly 
normalizing the fields we can determine the physical masses. Consider for 
example a fundamental Higgs field,
\begin{equation}
S_{vis} \supset \int d^4 x \sqrt{-g_{vis}} \{ g_{vis}^{\mu \nu} D_{\mu}
 H^{\dagger} D_{\nu} H - \lambda (|H|^2 - v_0^2)^2 \}, 
\end{equation}
which contains one mass parameter $v_0$. Substituting $g^{vis}_{\mu \nu}$ into this action yields 
\begin{equation}
S_{vis} \supset \int d^4 x \sqrt{- {g}} e^{- 4 k L_{1}} 
\{ {g}^{\mu \nu} e^{2 k L_{1}} D_{\mu}
 H^{\dagger} D_{\nu} H - \lambda (|H|^2 - v_0^2)^2 \}, 
\end{equation}
After wave-function rescaling, $H \rightarrow e^{k L_{1}} H$, we 
obtain
\begin{equation}
S_{eff} \supset \int d^4 x \sqrt{- g}  
\{ {g}^{\mu \nu} D_{\mu}
 H^{\dagger} D_{\nu} H - \lambda (|H|^2 - e^{-2 k L_{1}} v_0^2)^2 \}. 
\end{equation}
We see that after this rescaling, the physical mass scales are set 
by the exponentially suppressed scale:
\begin{equation}
\label{weak}
v \equiv e^{- k L_{1}} v_0.
\end{equation}
This result is completely general: any mass parameter 
$m_0$ on the visible 3-brane 
in the fundamental higher-dimensional theory will correspond to a 
physical mass
\begin{equation}
\label{punch}
m \equiv e^{- k L_{1}} m_0
\end{equation}
when measured with the metric $g_{\mu \nu}$, which is the metric 
that appears in the effective Einstein action, since all operators
get rescaled according to their four-dimensional conformal weight.
  If 
$e^{k L_{1}}$ is of order $10^{15}$, this mechanism  produces 
TeV physical mass scales from fundamental mass parameters not far from the 
Planck scale, $10^{18}$ GeV.  Because this geometric factor is an 
exponential, we clearly do not require very large hierarchies among the 
fundamental parameters.

Of course the latter arguments apply not only in the minimal two brane 
mode but in all models with non-factorizable
geometry. Thus, solution to the hierarchy problem can be given in
configurations where $\sigma(L_{Br})<<\sigma(0)$ (where $L_{Br}$ is
the position of the brane in the extra dimension) since all mass scales  
$m$ on the  brane will be related to 
the
mass parameters $m_0$ of the fundamental 5D theory by the conformal (warp) 
factor
\be
m=e^{-\sigma\left(L_{Br}\right)}m_0
\ee


\sect{The $''+-+''$ Model}

Up to now we have considered the two minimal models with one or two
branes placed on the orbifold fixed points. It is interesting to
examine the new physics associated with models with more than two
branes. The next-to-minimal models that we will consider are the
$''+-+''$, $''++-''$, Gregory-Rubakov-Sibiryakov(GRS) and $''+--+''$ models.

\begin{figure}
\begin{center}
\begin{picture}(300,100)(0,50)
\SetWidth{1.5}

\BCirc(150,100){60}
\DashLine(90,100)(210,100){3}

\CCirc(90,100){7}{Black}{Red}
\CCirc(210,100){7}{Black}{Red}

\CCirc(183,148){7}{Black}{Green}
\CCirc(183,52){7}{Black}{Green}

\Text(70,100)[]{${\LARGE{\bf +}}$}
\Text(230,100)[]{${\LARGE{\bf +}}$}
\Text(190,162)[]{$-$}
\Text(190,38)[]{$-$}
\Text(170,140)[]{$L_1$}
\Text(193,112)[]{$L_2$}
\Text(158,60)[]{$-L_1$}

\Text(130,120)[]{$Z_2$}

\LongArrowArc(150,100)(68,4,52)
\LongArrowArcn(150,100)(68,52,4)
\Text(220,135)[l]{$x=k(L_2-L_1)$}

\SetWidth{2}
\LongArrow(150,100)(150,115)
\LongArrow(150,100)(150,85)

\end{picture}
\end{center}
\caption{The $''+-+''$ model with two positive tension branes, $''+''$, on the
orbifold fixed points and a negative,  $''-''$, freely moving in-between.  }
\label{+-+fig}
\end{figure}
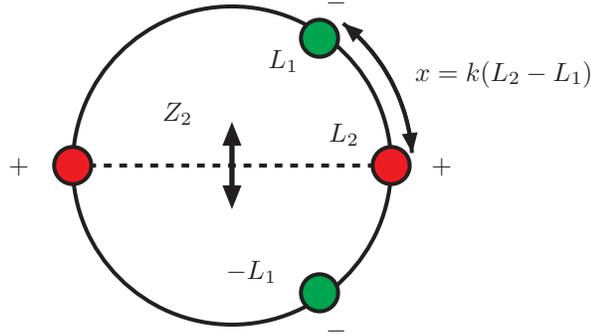

Let us start our discussion with
the $''+-+''$ model which consists of three parallel 3-branes in an $AdS_5$ space with
cosmological constant $\Lambda<0$. The 5-th dimension has the geometry
of an orbifold and the branes are located at
$L_0=0$, $L_1$ and $L_2$ where $L_0$ and $L_2$
are the orbifold  fixed
points (see Fig.(\ref{+-+fig})). Firstly we consider the branes having no  matter on them in
order to find a suitable vacuum solution. The action of this setup is:
\be
S=\int d^4 x \int_{-L_2}^{L_2} dy \sqrt{-G} 
\{-\Lambda + 2 M^3 R\}-\sum_{i}\int_{y=L_i}d^4xV_i\sqrt{-\hat{G}^{(i)}}
\ee
where $\hat{G}^{(i)}_{\mu\nu}$ is the induced metric on the branes
and $V_i$ their tensions. 
 The Einstein equations that arise from this
action are:
\be
R_{MN}-\frac{1}{2}G_{MN}R=-\frac{1}{4M^3}
\left(\Lambda G_{MN}+
\sum_{i}V_i\frac{\sqrt{-\hat{G}^{(i)}}}{\sqrt{-G}}
\hat{G}^{(i)}_{\mu\nu}\delta_M^{\mu}\delta_N^{\nu}\delta(y-L_i)\right)
\ee
 A straightforward
calculation, using the ansatz of eq.(\ref{ansatzrs+-+}) gives us the following differential equations for $\sigma(y)$:
\ba
\left(\sigma '\right)^2&=&k^2\\
\sigma ''&=&\sum_{i}\frac{V_i}{12M^3}\delta(y-L_i)
\ea
where $
k=\sqrt{\frac{-\Lambda}{24M^3}}$. 

The solution of these equations consistent with the orbifold geometry is 
precisely:
\be
\sigma(y)=k\left\{L_1-\left||y|-L_1\right|\right\}
\ee
with the requirement that the brane tensions are tuned to $V_0=-\Lambda/k>0$,
$V_1=\Lambda/k<0$, \mbox{$V_2=-\Lambda/k>0$}. 
If we consider massless fluctuations of this vacuum metric  as in
the previous Section and then integrate over the 5-th dimension, we find
the 4D Planck mass is given by
\be
M_{\rm Pl}^2=\frac{M^3}{k}\left[1-2e^{-2kL_1}+e^{-2k(2L_1-L2)}\right]
\label{Planck3}
\ee
The above formula tells us that for large enough $kL_1$ and  $k\left(2L_1-
L_2\right)$ the
three mass scales $M_{\rm Pl}$, $M$, $k$ can be taken to be of the same
order. Thus we take $k\sim {\mathcal {O}}(M)$ in order
not to introduce a new hierarchy, with the additional restriction
$k<M$ so that the bulk curvature is small compared to the 5D Planck
scale so that we can trust our solution. Furthermore, if we put matter on the 
third brane all the physical masses $m$ on the third brane will be related to 
the
mass parameters $m_0$ of the fundamental 5D theory by the conformal (warp) 
factor
\be
m=e^{-\sigma\left(L_2\right)}m_0=e^{-k\left(2L_1-L_2\right)}m_0
\ee
Thus we can assume that the third brane is our universe and get a solution of 
the Planck hierarchy problem arranging
$e^{-k\left(2L_1-L_2\right)}$ to be of
$\mathcal{O}$$\left(10^{-15}\right)$, \textit{i.e} $2L_1-L_2\approx35k^{-1}$. In this case 
all the parameters of the
model $L_1^{-1}$, $L_2^{-1}$ and k are of the order of Plank scale.

 The KK mass spectrum and wavefunctions are determined 
by considering the (linear) fluctuations of the metric like in eq.(\ref{perturbrs+-+})

Here we have ignored the scalar fluctuations of the metric: the
dilaton and the radion. For an extensive account of the modes see Appendix.
We will return to the discussion of these scalar modes at the end of
this Chapter.  

Following the same steps as in the previous Section we can find that
the function $\Psi^{(n)}(y)$ will obey a second order differential
equation which after a change of variables reduces to an ordinary
Schr\"{o}dinger equation:
\be
\left\{-
\frac{1}{2}\partial_z^2+V(z)\right\}\hat{\Psi}^{(n)}(z)=\frac{m_n^2}{2}\hat{\Psi
}^{(n)}(z)
\ee

\be
{\rm with}\hspace*{0.5cm} V(z)=\frac{15k^2}{8[g(z)]^2}-
\frac{3k}{2g(z)}\left[\delta(z)+\delta(z-z_2)-\delta(z-z_1)-\delta(z+z_1)\right]
\ee


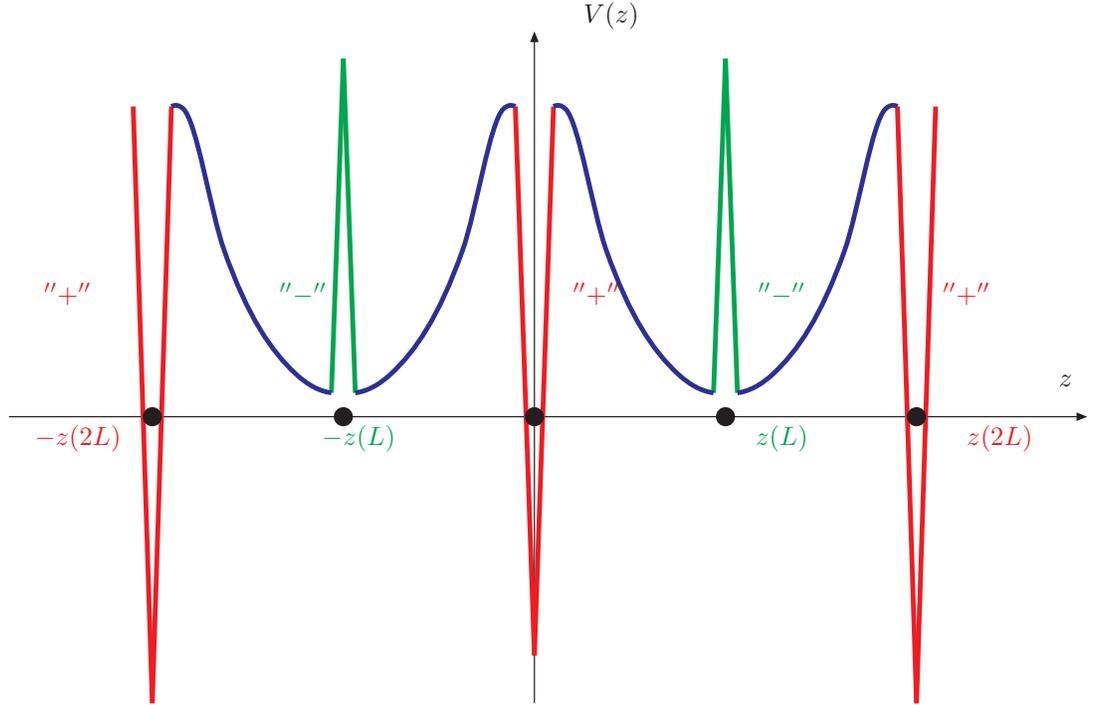
\begin{figure}
\begin{center}
\SetScale{0.9} 
\begin{picture}(200,200)(0,50)
\LongArrow(140,0)(140,280)
\LongArrow(-80,120)(370,120)

\SetWidth{2}
\SetColor{Red}
\Line(140,20)(148,250)
\Line(140,20)(132,250)
\SetColor{Black}
\Vertex(140,120){4}

\SetColor{Green}
\Line(220,270)(225,130)
\Line(220,270)(215,130)
\Line(60,270)(65,130)
\Line(60,270)(55,130)
\Text(220,100)[c]{${\Green {z(L)}}$}
\Text(60,100)[c]{${\Green {-z(L)}}$}
\SetColor{Black}
\Vertex(220,120){4}
\Vertex(60,120){4}

\SetColor{Red}
\Line(-12,250)(-20,0)
\Line(-28,250)(-20,0)
\Line(308,250)(300,0)
\Line(292,250)(300,0)
\Text(290,100)[l]{${\Red {z(2L)}}$}
\Text(-30,100)[r]{${\Red {-z(2L)}}$}
\SetColor{Black}
\Vertex(-20,120){4}
\Vertex(300,120){4}

\Text(145,260)[l]{$V(z)$}
\Text(330,120)[rb]{$z$}

\Text(-50,155)[c]{${\Red {''+''}}$}
\Text(290,155)[c]{${\Red {''+''}}$}
\Text(150,155)[c]{${\Red {''+''}}$}
\Text(220,155)[c]{${\Green {''-''}}$}
\Text(39,155)[c]{${\Green {''-''}}$}

\SetColor{Blue}
\Curve{(148,250)(153,249)(170,190)(200,138)(215,130)}
\Curve{(225,130)(240,138)(270,190)(287,249)(292,250)}

\Curve{(-12,250)(-7,249)(10,190)(40,138)(55,130)}
\Curve{(65,130)(80,138)(110,190)(127,249)(132,250)}

\end{picture}
\end{center}
\vskip 13mm
\caption{The form of the potential $V(z)$ in the case of $''+-+''$ model.}
\end{figure}


The new variables and wavefunction in the above equation are defined as:
\be
\renewcommand{\arraystretch}{1.5}
z\equiv\left\{\begin{array}{cl}\frac{2e^{kL_1}-e^{2kL_1-ky}-
1}{k}&y\in[L_1,L_2]\\\frac{e^{ky}-1}{k}&y\in[0,L_1]\\-\frac{e^{-ky}-1}{k}&y\in[-
L_1,0]\\-\frac{2e^{kL_1}-e^{2kL_1+ky}-1}{k}&y\in[-L_2,-L_1]\end{array}\right.
\
\ee
\be
\hat{\Psi}^{(n)}(z)\equiv \Psi^{(n)}(y)e^{\sigma/2}
\ee
and the function $g(z)$ as $
g(z)\equiv k\left\{z_1-\left||z|-z_1\right|\right\}+1$, where $z_1=z(L_1)$.

This is a quantum mechanical problem with $\delta$-function potentials of
different weight and an extra $1/g^2$ smoothing term (due to the AdS geometry) 
that gives the
potential a double ``volcano'' form. The change of variables has been
chosen so that there are no first derivative terms in the
differential equation. 

This potential  always
gives rise to a (massless) zero mode, with wavefunction:
\be
\hat{\Psi}^{(0)}=\frac{A}{[g(z)]^{3/2}}
\ee
The
normalization factor $A$ is determined by the requirement 
$\displaystyle{\int_{- z_{2}}^{ z_{2} }
dz\left[\hat{\Psi}^{(0)}(z)\right]^2=1}$, chosen so that we get the standard 
form 
of the Fierz-Pauli Lagrangian.

In the specific case where $L_1=L_2/2$
(and with zero hierarchy) the potential and thus the zero mode's
wavefunction is
symmetric with respect to the second brane. When the second brane
moves towards the third one the wavefunction has a minimum on the
second brane but different heights on the other two branes, the difference 
generating the hierarchy between the first and the third brane.
From now on we will focus on the symmetric case since it simplifies
the calculations  without losing the 
interesting characteristics of the model.

For the KK modes the solution is given in terms of Bessel
functions. For $y$ lying in the regions ${\bf A}\equiv\left[0,L_1\right]$ and
${\bf B}\equiv\left[L_1,L_2\right]$, we have:
\be
\hat{\Psi}^{(n)}\left\{\begin{array}{cc}{\bf A}\\{\bf 
B}\end{array}\right\}=\sqrt{\frac{g(z)}{k}}\left[\left\{\begin{array}{cc}A_1\\B_
1\end{array}\right\}J_2\left(\frac{m_n}{k}g(z)\right)+\left\{\begin{array}{cc}A_
2\\B_2\end{array}\right\}Y_2\left(\frac{m_n}{k}g(z)\right)\right]
\ee

The boundary conditions (one for the
continuity of the wavefunction at $z_1$ and three for the
discontinuity of its first derivative at $0$, $z_1$, $z_2$) result in a
$4\times4$ homogeneous linear system which, in order to have a
non-trivial solution, leads to the vanishing determinant:
\be
\renewcommand{\arraystretch}{1.5}
\left|\begin{array}{cccc}J_1\left(\frac{m}{k}\right)&Y_1\left(\frac{m}{k}\right)
&\phantom{-}0&\phantom{-}0\\0&0&\phantom{-
}J_1\left(\frac{m}{k}g(z_2)\right)&\phantom{-
}Y_1\left(\frac{m}{k}g(z_2)\right)\\J_1\left(\frac{m}{k}g(z_1)\right)&Y_1\left(
\frac{m}{k}g(z_1)\right)&\phantom{-}J_1\left(\frac{m}{k}g(z_1)\right)&\phantom{-
}Y_1\left(\frac{m}{k}g(z_1)\right)\\J_2\left(\frac{m}{k}g(z_1)\right)&Y_2\left(
\frac{m}{k}g(z_1)\right)&-J_2\left(\frac{m}{k}g(z_1)\right)&-
Y_2\left(\frac{m}{k}g(z_1)\right)\end{array}\right|=0
\ee
(where we have suppressed the subscript $n$ on the masses $m_n$)
This is essentially the mass quantization condition which gives the
spectrum of the KK states. For each mass we can then determine the wave function 
with normalization $\displaystyle{\int_{- z_{2}}^{ z_{2} }
dz\left[\hat{\Psi}^{(n)}(z)\right]^2=1}$. 
From the form of the potential we can
immediately deduce that there is  a second ``bound'' state, the
first KK state. In the symmetric case, $L_1=L_2/2$, this is simply given by 
reversing the sign of the graviton wave function for $y>L_1$ (it has one zero
at $L_1$). When the second brane moves towards the third this symmetry
is lost and the first KK wave function has a very small value on the
first brane, a large value on the third and a zero very close to the
first brane.

The interaction of the KK states to the SM particles is found as in
Ref. \cite{Han:1999sg} by expanding the minimal
gravitational coupling of the SM Lagrangian $\displaystyle{\int
d^4x\sqrt{-\hat{G}}{\mathcal{L}}\left(\hat{G},SM fields\right)}$ with respect to 
the metric. After the rescaling due to the ``warp'' factor we get:
\ba
{\mathcal{L}}_{int}&=&-\frac{g\left(z_2\right)^{3/2}}{M^{3/2}}\sum_{n\geq
0}
\hat{\Psi}^{(n)}\left(z_2\right)h_{\mu\nu}^{(n)}(x)T_{\mu\nu}\left(x\right)= 
\nonumber
\\&=&-\frac{A}{M^{3/2}}h_{\mu\nu}^{(0)}(x)T_{\mu\nu}\left(x\right)-
\sum_{n>0}\frac{\hat{\Psi}^{(n)}\left(z_2\right)g\left(z_2\right)^{3/2}}{M^{3/2}
}h_{\mu\nu}^{(n)}(x)T_{\mu\nu}\left(x\right)
\ea
with $T_{\mu\nu}$ the energy momentum tensor of the SM
Lagrangian. Thus the coupling suppression of the zero and KK modes to matter is
respectively:
\ba
\frac{1}{c_0}&=&\frac{A}{M^{3/2}}\\
\frac{1}{c_n}&=&\frac{\hat{\Psi}^{(n)}\left(z_2\right)g\left(z_2\right)^{3/2}}{M
^{3/2}}
\ea
For
the zero mode the normalization constant $A$ is
$\frac{M^{3/2}}{M_{\rm Pl}}$ which gives the Newtonian gravitational
coupling suppression $c_0=M_{\rm
Pl}$.

\subsection{The first  and subsequent KK modes: Masses and coupling constants}

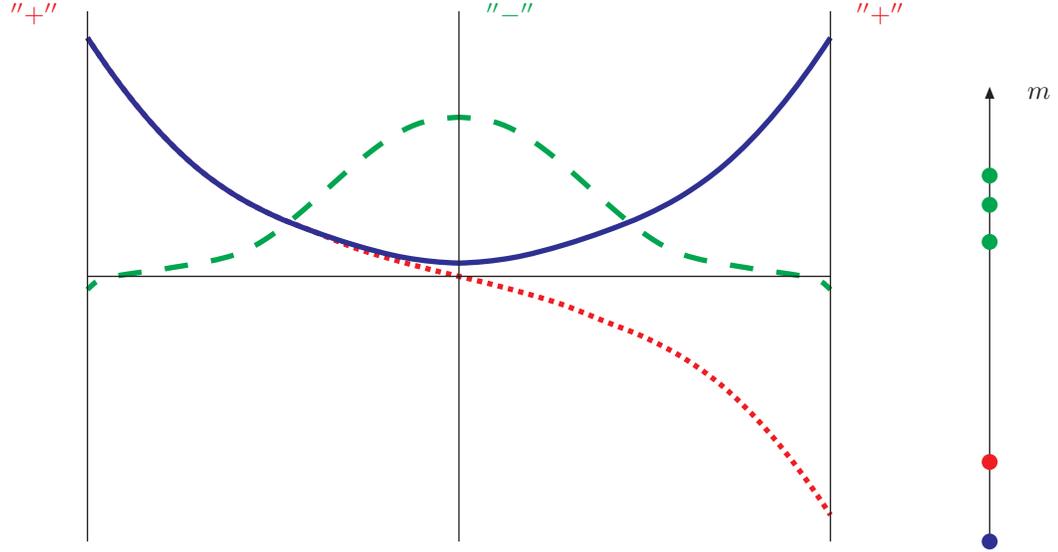
\begin{figure}
\begin{center}
\begin{picture}(300,200)(0,50)

\Text(-10,250)[c]{$\Red{''+''}$}
\Text(310,250)[c]{$\Red{''+''}$}
\Text(170,250)[c]{$\Green{''-''}$}

\SetWidth{2}

\SetColor{Red}
\DashCurve{(10,240)(50,192)(65,181)(80,173)(95,167)(110,161)(130,155)(150,150)}{2}
\DashCurve{(150,150)(170,145)(190,139)(205,133)(220,127)(235,119)(250,108)(290,60)}{2}

\SetColor{Blue}
\Curve{(10,240)(50,192)(65,181)(80,173)(95,167)(110,162)(130,157)(150,155)}
\Curve{(150,155)(170,157)(190,162)(205,167)(220,173)(235,181)(250,192)(290,240)}

\SetColor{Green}
\DashCurve{(10,145)(15,149)(20,150)(40,153)(65,158)(140,209)(150,210)}{8}
\DashCurve{(150,210)(160,209)(235,158)(260,153)(280,150)(285,149)(290,145)}{8}


\SetWidth{.5}
\SetColor{Black}
\Line(10,150)(290,150)
\Line(10,50)(10,250)
\Line(290,50)(290,250)
\Line(150,50)(150,250)

\Text(370,220)[c]{$m$}
\LongArrow(350,50)(350,220)
\SetColor{Green}
\Vertex(350,163){3}
\Vertex(350,177){3}
\Vertex(350,188){3}
\SetColor{Red}
\Vertex(350,80){3}
\SetColor{Blue}
\Vertex(350,50){3}

\end{picture}
\end{center}
\caption{The wavefunctions of the zero mode (solid), first (dotted)
and second KK state (dashed).}

\end{figure}

Let us examine  in more details  the mass spectrum of the $''+-+''$
model. In the case of the symmetric configuration of branes  we have that
for the first KK state:
\be
m_1=2 \sqrt{2} ~k~ e^{-2 x}
\label{mm1}
\ee
and for the rest of the tower:
\be
m_{n+1}= \xi_n ~ k~ e^{-x} ~~~~~~n=1,2,3, \ldots
\label{mm2}
\ee
where $\xi_{2i+1}$ is the $(i+1)$-th root of $J_{1}(x)$ ($i=0,1,2,
\ldots$) and $\xi_{2i}$ is the $i$-th root of $J_{1}(x)$ ($i=1,2,3, \ldots$).
The above approximations become better away from $x=0$
 , $x=0$ and for higher KK levels $n$.
The  mass of the first KK state is  singled out from the rest of the KK tower
as it has an extra exponential suppression that depends on the mass of
the bulk fermion. 
In the case that we have
a hierarchy $w$ (where $w\equiv
\frac{1}{g(z_{2})}=e^{-\sigma(L_{2})}$) the previous mass scales are
multiplied with $w$.

Let us now turn to the coupling behaviour of the states.
In the symmetric configuration, the first KK mode has constant
coupling equal to that of the 4D graviton: 
\begin{equation}
a_{1}=\frac{1}{M_{\ast }}~(=a_{0})~~~~~~~\mathrm{where}~~~M_{\ast }^{2}=%
\frac{2M^{3}}{k}
\label{firstcoupling}
\end{equation}
while the couplings of the rest of the KK tower are exponentially
suppressed: 
\begin{equation}
a_{n+1}=\frac{1}{M_{\ast }}~\frac{e^{-x}}{\sqrt{J_{1}^{2}\left( \frac{%
m_{n}e^{x}}{k}\right) +J_{2}^{2}\left( \frac{m_{n}e^{x}}{k}\right) }}%
~~~~~~n=1,2,3,\ldots  
\label{KKcoupling}
\end{equation}
the latter reveals once more the special character of the first KK
state compared to the rest of the tower: The coupling of the
ultralight KK state is indepented of the separation of the two branes
something that shows that this state is strongly localized on the
positive tension branes.

\subsection{ Bi-Gravity }

Equations (\ref{mm1}) and (\ref{mm2}) show that, for large $x$, the lightest KK mode 
splits off from the remaining tower. This leads to an exotic possibility in 
which the lightest KK mode is the dominant source of Newtonian gravity!

Cavendish experiments and astronomical observations studying the motions of 
distant galaxies have put Newtonian
gravity to test from sub-millimeter distances up to distances that
correspond to $1\%$ of the size of observable 
Universe, searching for violations of the weak equivalence principle
and inverse square law. In the context of the graviton KK modes discussed above 
this constrains $m<10^{-31}{\rm eV}$ or $m>10^{-4}{\rm eV}$. Our exotic scheme corresponds 
to the choice $m_1\approx 10^{-31}{\rm eV}$ and $m_2>10^{-4}{\rm eV}$. In this case, for 
length scales less than $10^{26}{\rm cm}$ gravity is generated by the exchange of {\it 
both} the massless graviton and the first KK mode.

\begin{figure}
\begin{center}
\begin{picture}(320,80)(0,0)
\LongArrow(0,55)(300,55)
\G2Text(175,55){0.6}{Excluded by Observational Data}{and by the
Cavendish Experiments}
\Text(300,65)[l]{$r$}
\Text(-2,65)[l]{$0$}
\Vertex(0,55){2}
\Vertex(45,55){2}
\Vertex(100,55){2}
\Text(25,67)[l]{$10 {\rm \mu m}$}
\Text(40,37)[l]{$m_2^{-1}$}
\Text(73,67)[l]{$1 {\rm mm}$}
\Vertex(250,55){2}
\Text(252,67)[l]{$10^{26} {\rm cm}$}
\Text(252,37)[l]{$m_1^{-1}$}
\LongArrow(45,-10)(45,20)
\LongArrow(252,-10)(252,20)
\Curve{(45,-10)(252,-10)}
\label{Bi-Gravityresults2}
\end{picture}
\end{center}
\caption{Exclusion regions for the Bi-Gravity case and correlation of
the first two KK states}
\label{Bi-Gravity2}
\end{figure}
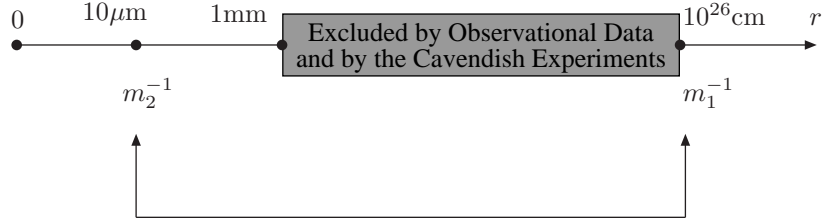

The gravitational potential is computed by the tree level exchange diagrams
of the 4D graviton and KK states which in the Newtonian limit is: 
\begin{equation}
V(r)=-\sum_{n=0}^{N_{\Lambda}}a_n^2\frac{e^{-m_{n}r}}{r}  \label{gravipot2}
\end{equation}
where $a_n$ is the coupling (\ref{firstcoupling}),(\ref{KKcoupling}) and $n=0$ accounts for the
massless graviton. The summation stops at some very high level $N_{\Lambda}$
with mass of the order of the cutoff scale $\sim M$.

In the ``bigravity'' scenario, at distances $r\ll m_{1}^{-1},$ the first KK
state and the 4D graviton contribute equally to the gravitational force, $%
i.e.$ 
\begin{equation}
V_{ld}(r)\approx -\frac{1}{M_{\ast }^{2}}\left( \frac{1}{r}+\frac{e^{-m_{1}r}%
}{r}\right) \approx -\frac{G_{N}}{r}
\end{equation}
where $G_{N}\equiv \frac{2}{M_{\ast }^{2}}$. For distances $r\gtrsim
m_{1}^{-1}$ the Yukawa suppression effectively reduces gravity to half its
strength. Astronomical constraints and the requirement of the observability
of this effect demand that for $k\sim M_{\mathrm{Pl}}$ we should have $x$ in
the region 65-70. Moreover, at distances $r\lesssim m_{2}^{-1}$ the Yukawa
interactions of the remaining KK states are significant and will give rise
to a short distance correction. This can be evaluated by using the
asymptotic expression of the Bessel functions in (\ref{KKcoupling}) since we
are dealing with large $x$ and summing over a very dense spectrum, giving: 
\begin{equation}
V_{sd}(r)=-\frac{G_{N}}{k}\sum_{n=2}^{N_{\Lambda }}\frac{k\pi }{2e^{x}}~%
\frac{m_{n}}{2k}~\frac{e^{-m_{n}r}}{r}  \label{shortpot2}
\end{equation}
At this point we exploit the fact that the spectrum is nearly continuum
above $m_{2}$ and turn the sum to an integral with the first factor in (\ref
{shortpot2}) being the integration measure, {\textit{i.e.}} $\sum \frac{k\pi 
}{2e^{x}}=\sum \Delta m\rightarrow \int dm$ (this follows from eq(\ref
{mm1}) for the asymptotic values of the Bessel roots). Moreover, we can
extend the integration to infinity because, due to the exponential
suppression of the integrand, the integral saturates very quickly and thus
the integration over the region of very large masses is irrelevant. The
resulting potential is now: 
\begin{equation}
V_{sd}(r)=-\frac{G_{N}}{k}\int_{m_{2}}^{\infty }dm~\frac{m}{2k}~\frac{%
e^{-m_{n}r}}{r}
\end{equation}
The integration is easily performed and gives: 
\begin{equation}
V_{sd}(r)\simeq -\frac{G_{N}}{2r}~\frac{1+m_{2}r}{(kr)^{2}}~e^{-m_{2}r}
\end{equation}
We see these short distance corrections are significant only at Planck scale
lengths $\sim k^{-1}$.

\subsection{$''+-+''$ Model Phenomenology}

In this Section we will present a discussion of the phenomenology of the 
KK modes to be expected in high energy colliders, concentrating on the simple and
sensitive to new physics  processes 
$e^+e^-\rightarrow\mu^+\mu^-$ (this analysis is readily generalized to include 
$q\bar{q}$, $gg$ initial and final states) and $e^+e^-\rightarrow\gamma+
\mbox{\textit{missing energy}}$. Since the characteristics of the
phenomenology depend on the parameters of the model ($w$,$k$,$x$) we
explore the
regions of the parameter space that are of special interest (\textit{i.e.} give
hierarchy factor $\mathcal{O}$$\left(10^{15}\right)$ and do not
introduce a new hierarchy between $k$ and $M$ as seen from equation (\ref{Planck3})).  

\subsubsection{ $e^+e^-\rightarrow\mu^+\mu^-$ process}

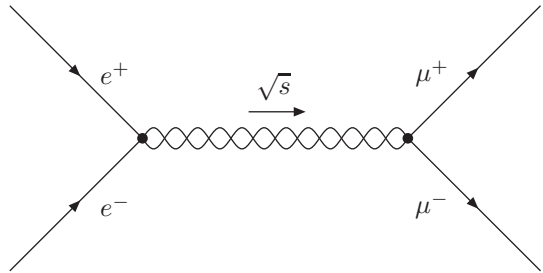
\begin{figure}[b]
\begin{center}
\begin{picture}(300,100)(0,50)
\ArrowLine(50,50)(100,100)
\ArrowLine(50,150)(100,100)
\ArrowLine(200,100)(250,150)
\ArrowLine(200,100)(250,50)
\Vertex(100,100){2}
\Vertex(200,100){2}
\LongArrow(140,110)(160,110)
\Photon(100,100)(200,100){4}{6}
\Photon(100,100)(200,100){-4}{6}
\Text(90,125)[]{$e^+$}
\Text(90,75)[]{$e^-$}
\Text(210,125)[]{$\mu^+$}
\Text(210,75)[]{$\mu^-$}
\Text(150,120)[]{$\sqrt{s}$}
\end{picture}
\caption{$e^+e^-\rightarrow \mu^+ \mu^-$}\label{feyn:diag2}
\end{center}
\end{figure}

Using the Feynman rules of Ref. \cite{Han:1999sg} the contribution of the KK modes to 
$e^+e^-\rightarrow\mu^+\mu^-$ is given by
\be
\sigma\left(e^+e^-\rightarrow\mu^+\mu^-\right)=\frac{s^3}{1280\pi}|D(s)|^2
\ee
where $D(s)$ is the sum over the propagators multiplied by the
appropriate coupling suppressions:
\be
D(s)=\sum_{n>0}\frac{1/c_n^2}{s-m_n^2+i\Gamma_n m_n}
\ee
and $s$ is the center of mass energy of $e^+e^-$.

Note that the bad high energy behaviour (a violation of perturbative unitarity) of this cross section is expected since 
we are working with an effective - low
energy non-renormalizable theory of gravity. We assume our  effective theory is valid up to 
an energy scale
$M_{s}$ (which is ${\mathcal{O}}({\rm TeV})$), which acts as an ultraviolet 
cutoff. The theory that applies  above this scale
is supposed to give a consistent description of quantum gravity. Since this is 
unknown we are only able to determine the contributions of the KK states with 
masses less than this
scale. This means that the summation in the previous formula should
stop at the KK mode with mass near the cutoff.

For the details of the calculation it will be important to know the
decay rates of the KK states. These are given by:
\be
\Gamma_n=\beta\frac{m_n^3}{c_n^2}
\label{gam2}
\ee
where $\beta$ is a dimensionless constant that is between
$\frac{39}{320\pi}\approx0.039$ (in the case that the KK is light
enough, \textit{i.e.} smaller than $0.5 {\rm MeV}$, so that it  decays only to 
massless 
gauge bosons and neutrinos) and $\frac{71}{240\pi}\approx0.094$ (in the case where
the KK is heavy enough that can decay to all SM particles).

If we consider $w$ and $k$ fixed, then when
$x$ is smaller than a certain value $x_0=x_0(w,k)$ we have a widely spaced discrete spectrum (from
the point of view of TeV physics) close to the one of the
RS case with cross section at a KK resonances of the form
$\sigma_{res} \sim s^3/m_ n^8$ (see \cite{Davoudiasl:2000jd}).       
For the discrete 
spectrum there is always a range of values of the $x$ parameter so that the
KK resonances are in the range of energies of
collider experiments. 
In these cases we calculate the excess over the SM contribution which would have been seen either by direct
scanning if the resonance is near the energy at which the experiments
actually run or by means of the process
$e^+e^-\rightarrow\gamma\mu^+\mu^-$ which scans a continuum of energies 
below the center of mass energy of the experiment 
(of course if $k$ 
is raised the KK modes become heavier and there will be a value for which the 
lightest KK mode is above the experimental limits). 

For values of $x$ greater than $x_0(k,w)$ the spacing in the spectrum is
so small that we can safely consider it to be continuous. At this
point we have to note that we consider that the ``continuum'' starts at the point
where the convoluted KK resonances start to overlap. In this case
we substitute in $D(s)$ the sum for $n\geq2$ by an integral over the
mass of the KK excitations, \textit{i.e.}
\be
D(s)_{KK}\approx\frac{1/c_1^2}{s-m_1^2+i\Gamma_1 m_1}+
\frac{1}{\Delta m\; c^2} \int_{m_2}^{M_s}dm\; \frac{1}{s-m^2+i\epsilon}
\ee
where the value of the integral is $\sim i \pi/2\sqrt{s}$ with the principal
value negligible in the region of interest ($\sqrt{s} \ll M_s$) 
and we have considered constant coupling suppression $c$ 
for the modes with $n\geq 2$ (approximation that turns out to be
reasonable as the coupling saturates  quickly as we consider
higher and higher levels).
The first state is singled out because of
its different coupling.

\subsubsection{$e^+e^-\rightarrow \gamma +
\mbox{\textit{missing energy}}$  process}

\begin{figure}[t]
\begin{center}
\begin{picture}(300,100)(0,50)

\ArrowLine(-55,50)(-25,80)
\ArrowLine(-55,150)(-25,120)
\ArrowLine(-25,80)(-25,120)
\Vertex(-25,80){2}
\Vertex(-25,120){2}
\Photon(-25,120)(5,150){4}{6}
\Photon(-25,80)(5,50){4}{6}
\Photon(-25,80)(5,50){-4}{6}
\Text(-50,130)[]{$e^+$}
\Text(-50,70)[]{$e^-$}
\Text(5,130)[]{$\gamma$}
\Text(5,70)[]{KK}

\ArrowLine(45,50)(75,80)
\ArrowLine(45,150)(75,120)
\ArrowLine(75,80)(75,120)
\Vertex(75,80){2}
\Vertex(75,120){2}
\Photon(75,80)(105,50){4}{6}
\Photon(75,120)(105,150){4}{6}
\Photon(75,120)(105,150){-4}{6}
\Text(50,130)[]{$e^+$}
\Text(50,70)[]{$e^-$}
\Text(105,130)[]{KK}
\Text(105,70)[]{$\gamma$}

\ArrowLine(145,70)(175,100)
\ArrowLine(145,130)(175,100)
\Vertex(175,100){2}
\Photon(175,100)(215,100){4}{6}
\Vertex(215,100){2}
\Photon(215,100)(245,70){4}{6}
\Photon(215,100)(245,130){4}{6}
\Photon(215,100)(245,130){-4}{6}
\Text(150,115)[]{$e^+$}
\Text(150,85)[]{$e^-$}
\Text(250,115)[]{KK}
\Text(250,85)[]{$\gamma$}

\ArrowLine(285,70)(315,100)
\ArrowLine(285,130)(315,100)
\Vertex(315,100){2}
\Photon(315,100)(345,70){4}{6}
\Photon(315,100)(345,130){4}{6}
\Photon(315,100)(345,130){-4}{6}
\Text(290,115)[]{$e^+$}
\Text(290,85)[]{$e^-$}
\Text(350,115)[]{KK}
\Text(350,85)[]{$\gamma$}

\end{picture}
\caption{$e^+e^-\rightarrow \gamma$  KK}\label{miss2}
\end{center}
\end{figure}
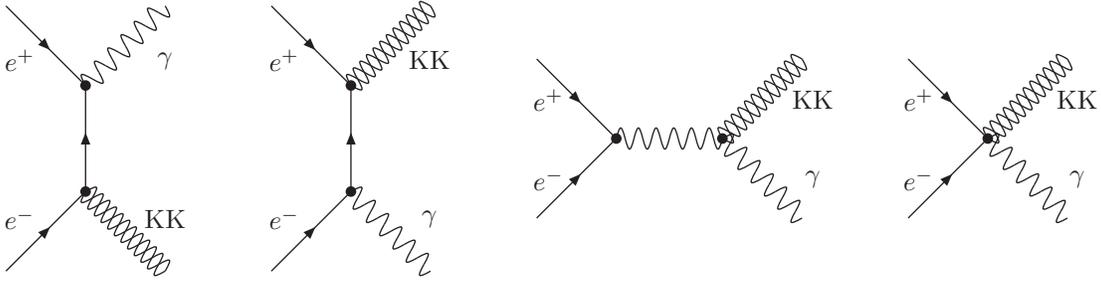

The missing energy processes in the SM (\textit{i.e.}
$e^+e^-\rightarrow \gamma\nu\bar{\nu}$) are well explored and are a
standard way to count the number of neutrino species. In the presence
of the KK modes there is also a possibility that any KK mode produced, if it has large enough
lifetime, escapes from the detector before decaying, thus giving
us an additional missing energy signal. The new diagrams that
contribute to this effect are the ones in the Fig. \ref{miss2}.

The differential cross section of the production of a KK mode plus a photon is
given by Ref. \cite{Giudice:1999ck} and is equal to:
\be
\frac{d\sigma}{dt}(e^+e^-\rightarrow \gamma +  {\rm{KK}})=\frac{\alpha}{16}\sum_{n>0}\frac{1}{c_n^2s}F\left(\frac{t}{s},\frac{m_n^2}{s}\right)
\ee
where $s$, $t$ are the usual Mandelstam variables and the function $F$ is given by:
\ba
F(x,y)=\frac{1}{x(y-1-x)}[&-4x(1-x)(1+2x+2x^2)+y(1+6x+18x^2+16x^3)\nonumber\\
&-6y^2x(1+2x)+y^3(1+4x)\phantom{0}]
\ea

A reasonable size of a detector is of the order of $d=1$m, so we assume that the events of KK production are counted as missing
energy ones if the KK modes survive at least for distance d from the
interaction point (this excludes decays in neutrino pairs which always
give missing energy signal). We can then find a limit on the KK masses that contribute
to the experimental measurement. By a straightforward relativistic
calculation we find that this is the case if:
\be
\Gamma_n< \frac{E_\gamma}{m_nd}=\frac{s-m_n^2}{2\sqrt{s}m_nd}
\ee
From equation \ref{gam2} we see that this can be done if:
\be
m_n<\sqrt{\frac{-c_n^2+c_n\sqrt{c_n^2+8\beta ds^{3/2}}}{4\beta d \sqrt{s}}}
\ee
It turns out that usually only the first KK state mass satisfies this
condition and decays outside the detector. All the other states have
such short lifetimes that decay inside the detector and so are not
counted as missing energy events (again this excludes decays in neutrino pairs). In the regions of the parameter
space where this was not the case, we found that only a very small part of
the KK tower contributed and didn't give any important excess in
comparison with the one from the first state alone. 
Thus, taking only the contribution of the first KK state and imposing
 the kinematic cuts given by the  
experiments on the angular integration, we found the measurable cross section. This cross section
has to be compared to the error of the experimentally measured cross
section because so far the SM predictions coincide with the measured
value.

The most stringent measurement available is the one by OPAL
Collaboration \cite{Abbiendi:1999yu} at $\sqrt{s}=183$GeV. The measured cross section
is $\sigma_{meas}=4.71\pm 0.34$pb so the values of the parameters of
the model that give cross section greater than $0.34$pb are
excluded. Since the main contribution comes from  the first KK state and
because its coupling depends only on the warp factor $w$, we will either
exclude or allow the whole k-x plane for a given $w$. The critical
value of $w$ that the KK production cross section equals to the
experimental error is $w=1.8e^{-35}$. 

It is worth noting that the above cross section is almost constant for
different center of mass  energies $\sqrt{s}$, so ongoing experiments with
smaller errors (provided that they are in accordance with the SM
prediction) will push the bound on $w$ further ahead.

\subsubsection{Cavendish experiments}

A further bound on the parameters of our model can be put from the
Cavendish experiments. The fact that gravity is Newtonian at
least down to millimeter distances implies that the corrections to
gravitational law due to the presence of the KK states must be
negligible for such distances. The gravitational potential is the
Newton law plus a Yukawa potential due to the exchange of the KK
massive particles (in the Newtonian limit):
\be
V(r)=-\frac{1}{M_{\rm Pl}^2}\frac{M_1 
M_2}{r}\left(1+\sum_{n>0}\left(\frac{M_{\rm Pl}}{c_n}\right)^2e^{-m_nr}\right)
\ee

The contribution to the above sum of the second and
higher modes is negligible compared with the one 
of the first KK
state, because they have larger masses and coupling
suppressions. Thus, the
condition for the corrections of the Newton law to be small for
millimeter scale distances is:
\be
x<\tilde{x}=15-\frac{1}{2}{\rm ln}\left(\frac{-{\rm ln}w}{kw}{\rm
GeV}\right)
\label{Cbound2}
\ee

\subsubsection{$k-x$ plots}

\begin{figure}[!h]

\begin{center}
\epsfig{file=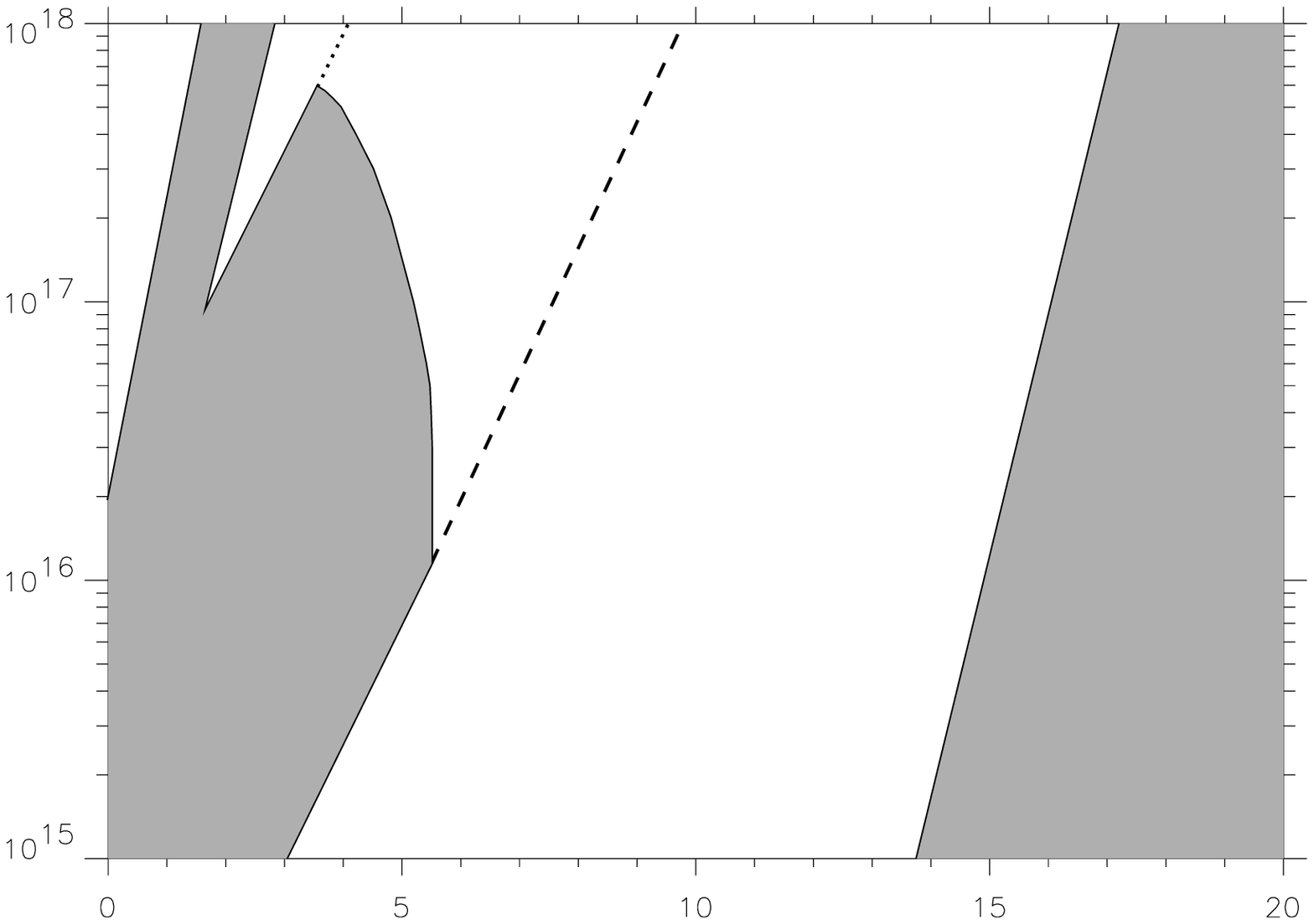,width=12cm}

\begin{picture}(-200,-200)(0,0)
\SetScale{1}

\LongArrow(-110,5)(-70,5)
\rText(-135,5)[l][]{$x$}
\LongArrow(-280,160)(-280,200)
\rText(-315,145)[l][]{$k$ (GeV)}

\rText(-250,230)[l][]{$(1)$}
\rText(-175,190)[l][]{$(2)$}
\rText(-35,145)[l][]{$(3)$}

\end{picture}

\caption{Excluded regions (shaded areas) for $w=4.5e^{-35}$.\label{plot1}}
\end{center}


\begin{center}
\epsfig{file=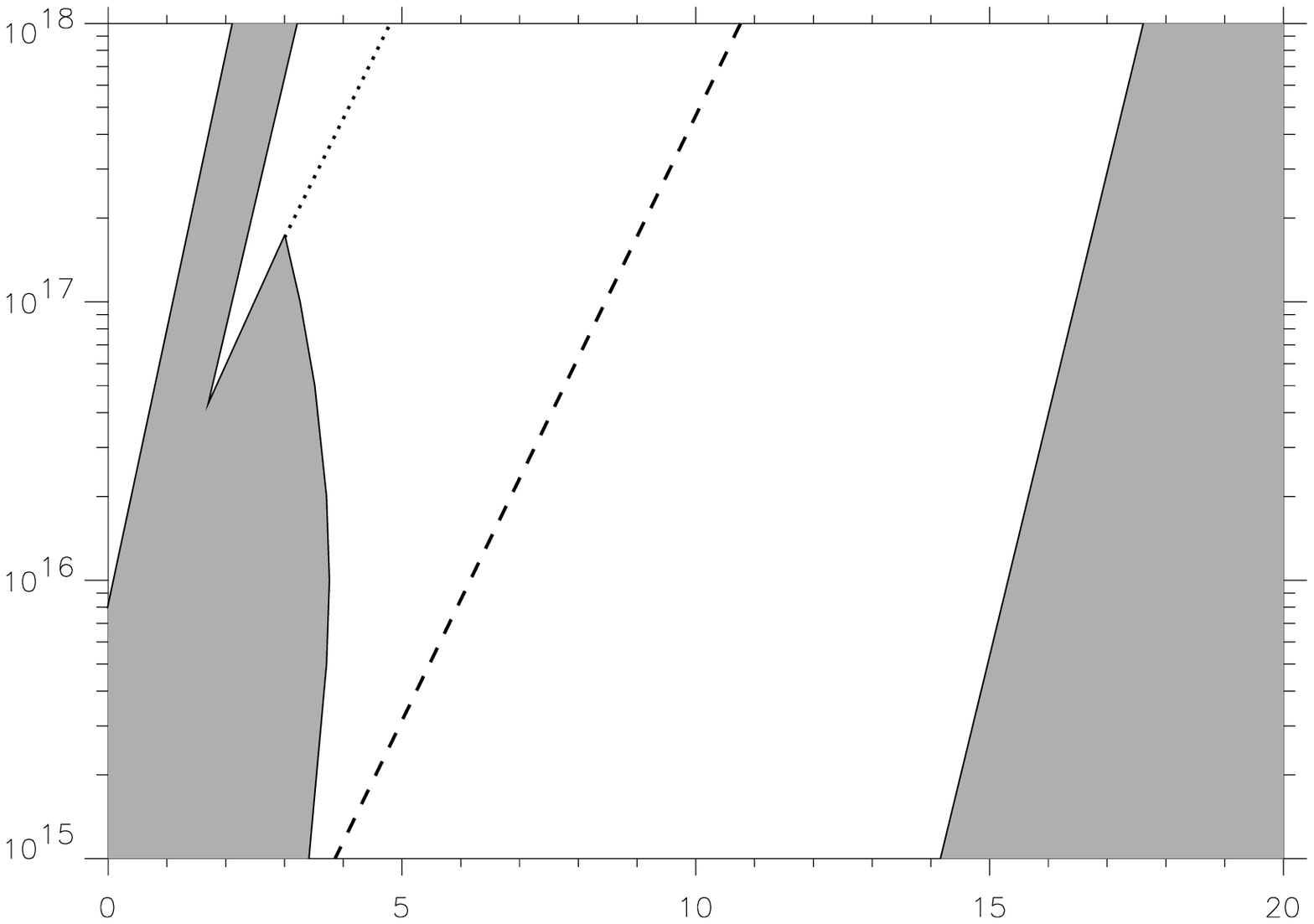,width=12cm}

\begin{picture}(-200,-200)(0,0)
\SetScale{1}

\LongArrow(-110,5)(-70,5)
\rText(-135,5)[l][]{$x$}
\LongArrow(-280,160)(-280,200)
\rText(-315,145)[l][]{$k$ (GeV)}

\rText(-245,220)[l][]{$(1)$}
\rText(-197,150)[l][]{$(2)$}
\rText(-28,145)[l][]{$(3)$}

\end{picture}

\caption{Excluded regions (shaded areas) for $w=10e^{-35}$.\label{plot2}}
\end{center}

\end{figure}

As mentioned above the  range of the parameter space that we
explore is chosen so that it corresponds to the region of physical
interest giving rise to the observed hierarchy between the electroweak
and the Planck scale
i.e. $w\sim 10^{-15}$, $k\sim M_{\rm Pl}$. The allowed  regions
(unshaded areas)
for  $w=4.5e^{-35}$ and $w=10e^{-35}$ are shown in the Figures
~\ref{plot1} and ~\ref{plot2}. The bounds from the previously mentioned
experiments and the form of the diagram will be now explained in detail.

\begin{itemize}
\item {\bf$e^+e^-\rightarrow\mu^+\mu^-$ bounds}
\end{itemize}

As we noted in section 3.1, for relatively small values of $x$ the
spectrum is discrete and as $x$ increases it tends to a continuum (the dashed line shows approximately where
we the spectrum turns from discrete to continuum).
In case of the continuum, for the parameter region that we explore, it
turns out that it does not give any bound since the excess over  the  SM cross
section becomes important only for energies much larger than $200$GeV.
However there are significant bounds coming from the discrete spectrum
region, since generally we
have KK resonances in the experimentally accessible region and  the
convolution of some of them will give significant excess to the SM
background.
The exclusion region coming from $e^+e^-\rightarrow\mu^+\mu^-$, is the
region between the curves (1) and (2).  
The details of the bound depend on the behaviour of the couplings
and the masses. In this case the bounds start when the KK states
have sufficiently large  width and height ({\it i.e.} large  mass and coupling).
This is the reason why curve (2) bends to the left as k increases.
The shape 
of the upper part of the curve (2) comes from the fact that by increasing
$k$ we push the masses of the KK states to larger values so that there
is the possibility that the
first KK state has mass smaller that $20$GeV and at the same time the 
rest of tower is above $200$GeV (the dotted line is where the second KK states is at $200$GeV). The last
region is not experimentally explored at present.
An increase of $w$  decreases all the couplings  and thus this will push
the bound even more to the left.
Decreasing $w$ ({\it e.g.} $w=e^{-35}$), on the contrary will increase the values of the
couplings and there are strict bounds coming both from the discrete and
the continuum.
The $e^+e^-\rightarrow\mu^+\mu^-$ experiments don't give any bound when the first KK
state has mass bigger than $\sim200$GeV, since in this case the KK
resonances cannot be produced from current experiments and the low
energy effects are negligible for the range of parameters that we
examine (this region is represented by the triangle at the upper left
corner of the plot).  As colliders
probe higher center of mass energies the curve (1) will be pushed to
the left and curve (2) to the right.

\begin{itemize}
\item {\bf Missing energy bounds}
\end{itemize}

As we noted in section 3.2, the KK states have generally very
short lifetime. For certain value of $w$, the main contribution to
the cross section comes from the first KK state, since the restriction
on the mass (so that the KK states escape
the detector) means that only a few states contribute, even near
the boundary of Cavendish bounds where the spacing of the tower is
very small.
Decreasing the value of $w$, we increase the coupling of
the first KK state so for the values of $w<1.8e^{-35}$ the contribution
from the first KK state becomes so big that excludes all the region
between $e^+e^-\rightarrow\mu^+\mu^-$ and the Cavendish bound (the region
with the KK tower over $200$GeV always survives).
Variation of the $k$, $x$ parameters in this case does not change
significantly the cross section, because the main contribution comes from
the first KK mode  whose coupling is constant {\it i.e.} independent of $k$,
$x$ and since although its mass, $m_1$ depends on $k$ the cross section is
insensitive to the mass changes because  it is evaluated at energy
$\sqrt{s}=183$GeV where the cross section has saturated.
To summarize, the missing energy bounds either exclude the whole
region between the $e^+e^-\rightarrow\mu^+\mu^-$ and the Cavendish limits or
nothing at all (due to the smallness of the coupling).
Additionally, the missing energy experiments don't give any bound when the first KK
state has mass bigger than $\sim143$GeV, since in this case the emitted
photon has energy smaller than the experimental cuts. Currently
running and forthcoming colliders will push the bound of $w$ to larger
values.

\begin{itemize}
\item {\bf Cavendish bounds}
\end{itemize}

From the discussion in section 3.3 we see that the bound on the parameter
space from Cavendish experiments comes from Eq. (\ref{Cbound2}). The exclusion
region is the one that extends to the right of the line (3) of the
plots.  
  For fixed $k$, $w$  the Cavendish bounds exclude the region $x>\tilde{x}(k,w)$ due to the fact
that the first KK becomes very light (and its coupling remains
constant). Now if we increase $k$, since the masses of the KK are
proportional to it, we will have an exclusion region of
$x>\tilde{x}(k',w)$, with $\tilde{x}(k',w)>\tilde{x}(k,w)$. This explains the form of the
 Cavendish bounds. When we increase the $w$ parameter the whole bound
will move to the right since the couplings of the KK states
decrease. Future Cavendish experiments testing the Newton's law at
smaller distances will push the curve (3) to the left.


\section{The three-brane $''++-''$ Model}

One other extension of the RS model is the $''++-''$ model. In this
case although there are two positive tension branes - no light state
appears since they act as a single positive tension brane resulting to 
a phenomenology similar to the RS one. In this Section
we examine in  detail this configuration.

\begin{figure}[h]
\begin{center}
\begin{picture}(300,125)(0,50)

\SetWidth{1.5}

\SetOffset(0,10)

\BCirc(150,100){60}
\DashLine(90,100)(210,100){3}

\GCirc(90,100){7}{0.9}
\GCirc(183,148){7}{0.9}
\GCirc(183,52){7}{0.9}

\GCirc(210,100){7}{0.2}

\Text(70,100)[]{$+$}
\Text(230,100)[]{$-$}
\Text(190,162)[]{$+$}
\Text(190,38)[]{$+$}
\Text(170,140)[]{$L_1$}
\Text(193,112)[]{$L_2$}
\Text(158,60)[]{$-L_1$}

\Text(130,120)[]{$Z_2$}

\LongArrowArc(150,100)(68,4,52)
\LongArrowArcn(150,100)(68,52,4)
\Text(220,135)[l]{$x=k_2(L_2-L_1)$}

\SetWidth{2}
\LongArrow(150,100)(150,115)
\LongArrow(150,100)(150,85)

\end{picture}
\end{center}
\caption{The brane locations in the three-brane $''++-''$ model. The bulk curvature between the $''+''$ branes is $k_{1}$ and between the $''+''$ and $''-''$ brane
is $k_{2}$.}
\label{++-}
\end{figure}
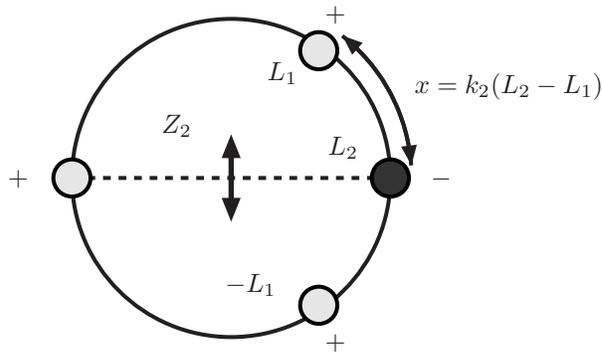

The $''++-''$ model (see Fig.\ref{++-})
consists of three parallel 3-branes in $AdS_{5}$ spacetime with orbifold
topology, two of which are located at the orbifold fixed points $L_{0}=0$
and $L_{2}$ while the third one is moving at distance $L_{1}$ in between. In
order to get 4D flat space with this configuration, it turns out that the $%
AdS_{5}$ space must have different cosmological constants $\Lambda _{1}$ and 
$\Lambda _{2}$ between the first - second and the second - third brane
respectively with $|\Lambda _{2}|>|\Lambda _{1}|$ (see \cite{Hatanaka:1999ac} for
constructions of different bulk cosmological constants). The action of the
above setup (if we neglect the matter contribution on the branes in order to
find a suitable vacuum solution) is given by: 
\begin{eqnarray}
S &=&\int d^{4}x\int_{-L_{2}}^{L_{2}}dy\sqrt{-G}[-\Lambda (y)+2M^{3}R%
]-\sum_{i}\int_{y=y_{i}}d^{4}xV_{i}\sqrt{-\hat{G}^{(i)}}  
\label{action++-} \\
\mathrm{with}~~~\Lambda (y) &=&\left\{ 
\begin{array}{cl}
{\Lambda _{1}} & ,y\in \lbrack 0,L_{1}] \\ 
\Lambda _{2} & ,y\in \lbrack L_{1},L_{2}]
\end{array}
~~~~~~\right. \   \nonumber
\end{eqnarray}
where $\hat{G}_{\mu \nu }^{(i)}$ is the induced metric on the branes, $V_{i}$
are their tensions and $M$ the 5D fundamental scale.Again, we consider 
 the vacuum metric ansatz that respects 4D Poincar\'{e} invariance: 
\begin{equation}
ds^{2}=e^{-2\sigma (y)}\eta _{\mu \nu }dx^{\mu }dx^{\nu }+dy^{2}
\end{equation}

The above ansatz substituted in the Einstein equations requires that $\sigma
(y)$ satisfies the differential equations: 
\begin{equation}
\sigma ^{\prime \prime }=\sum_{i}\frac{V_{i}}{12M^{3}}\delta (y-L_{i})~~~%
\mathrm{and}~~~\left( \sigma ^{\prime }\right) ^{2}=\left\{ 
\begin{array}{cl}
{k_{1}^{2}} & ,y\in \lbrack 0,L_{1}] \\ 
k_{2}^{2} & ,y\in \lbrack L_{1},L_{2}]
\end{array}
\right. \ 
\end{equation}
where $k_{1}=\sqrt{\frac{-\Lambda _{1}}{24M^{3}}}$ and $k_{2}=\sqrt{\frac{%
-\Lambda _{2}}{24M^{3}}}$ are effectively the bulk curvatures in the two
regions between the branes. The solution of these equations for $y>0$ it is
given by: 
\begin{equation}
\sigma (y)=\left\{ 
\begin{array}{cl}
{k_{1}y} & ,y\in \lbrack 0,L_{1}] \\ 
{k_{2}(y-L_{1})+k_{1}L_{1}} & ,y\in \lbrack L_{1},L_{2}]
\end{array}
\right. \ 
\end{equation}
Einstein's equations impose the following conditions on the brane tensions : 
\begin{equation}
V_{0}=24M^{3}k_{1},~~~V_{1}=24M^{3}\frac{k_{2}-k_{1}}{2}%
,~~~V_{2}=-24M^{3}k_{2}
\end{equation}
If we consider massless fluctuations of this vacuum metric and then
integrate over the 5-th dimension, we find the 4D Planck mass is given by: 
\begin{equation}
M_{\mathrm{Pl}}^2={M^3}\left[\frac{1}{k_1}\left(1-e^{-2k_1L_1}\right)+\frac{1%
}{k_2}e^{2(k_2-k_1)L_1}\left(e^{-2k_{2}L_1}-e^{-2k_{2}L_{2}}\right)\right]
\end{equation}

The above formula tells us that for large enough $kL_{1}$ and $kL_{2}$ the
four mass scales $M_{\mathrm{Pl}}$, $M$, $k_{1}$ and $k_{2}$ can be taken to
be of the same order. Thus we take $k_{1},k_{2}\sim {\mathcal{O}}(M)$ in
order not to introduce a new hierarchy, with the additional restriction $%
k_{1}<k_{2}<M$ so that the bulk curvature is small compared to the 5D Planck
scale and we can trust the solution. Furthermore, if we put matter on the
second brane all the physical masses $m$ on it will be related to the mass
parameters $m_{0}$ of the fundamental 5D theory by the conformal ``warp''
factor 
\begin{equation}
m=e^{-\sigma \left( L_{1}\right) }m_{0}=e^{-k_{1}L_{1}}m_{0}
\end{equation}

This allows us to put the SM states on the intermediate $^{\prime \prime
}+^{\prime \prime }$ brane, solving the Planck hierarchy problem by choosing 
$e^{-kL_{1}}$ to be of $\mathcal{O}$$\left( 10^{-15}\right) $, \textit{i.e} $%
L_{1}\approx 35k^{-1}$.

The KK spectrum can be as usual found by considering the linear ``massive''
fluctuations of the metric \footnote{%
Again we will ignored the radion/dilaton modes that could be used to stabilize the
brane positions $L_{1}$ and $L_{2}$. For discussion and possible
stabilization mechanisms see \cite{Goldberger:1999uk}}. 

 The wavefunction $\Psi
^{(n)}(y)$ obeys a second order differential equation and carries all the
information about the effect of the non-factorizable geometry on the
graviton and the KK states. After a change of coordinates and a redefinition
of the wavefunction the problem reduces to the solution of an ordinary
Schr\"{o}dinger equation: 
\begin{equation}
\left\{ -\frac{1}{2}\partial _{z}^{2}+V(z)\right\} \hat{\Psi}^{(n)}(z)=\frac{%
m_{n}^{2}}{2}\hat{\Psi}^{(n)}(z)  
\label{sch++-}
\end{equation}
where the potential $V(z)$ for $z>0$ has the form: 
\begin{eqnarray}
\hspace*{0.5cm}V(z) &=&\frac{15}{8[g(z)]^{2}}\left[ k_{1}^{2}(\theta
(z)-\theta (z-z_{1}))+k_{2}^{2}(\theta (z-z_{1})-\theta (z-z_{2}))\right] 
\nonumber \\
&&-\frac{3}{2g(z)}\left[ k_{1}\delta (z)+\frac{(k_{2}-k_{1})}{2}\delta
(z-z_{1})-k_{2}\delta (z-z_{2})\right]  
\label{potential++-}
\end{eqnarray}

The new coordinates and wavefunction in the above equations are defined by: 
\begin{equation}
\renewcommand{\arraystretch}{1.5}z\equiv \left\{ 
\begin{array}{cl}
\frac{e^{k_{1}y}-1}{k_{1}} & ,y\in \lbrack 0,L_{1}] \\ 
\frac{e^{k_{2}(y-L_{1})+k_{1}L_{1}}}{k_{2}}+\frac{e^{k_{1}L_{1}}-1}{k_{1}}-%
\frac{e^{k_{1}L_{1}}}{k_{2}} & ,y\in \lbrack L_{1},L_{2}]
\end{array}
\right. \ 
\end{equation}
\begin{equation}
\hat{\Psi}^{(n)}(z)\equiv \Psi ^{(n)}(y)e^{\sigma /2}
\end{equation}
with the symmetric $z$ for $y<0$ and the function $g(z)$ as: 
\begin{equation}
g(z)=\left\{ 
\begin{array}{cl}
{k_{1}z+1} & ,z\in \lbrack 0,z_{1}] \\ 
{k_{2}(z-z_{1})+k_{1}z_{1}+1} & ,z\in \lbrack z_{1},z_{2}]
\end{array}
\right. \ 
\end{equation}
where $z_{1}=z(L_{1})$ and $z_{2}=z(L_{2})$. The change of coordinates has
been chosen so that there are no first derivative terms in the differential
equation. Furthermore, in this coordinate system it can be easily seen that
the vacuum metric takes the conformaly flat form: 
\begin{equation}
ds^{2}=\frac{1}{g(z)^{2}}\left( \eta _{\mu \nu }dx^{\mu }dx^{\nu
}+dz^{2}\right)
\end{equation}

The potential (\ref{potential++-}) always gives rise to a massless graviton zero mode
which reflects the fact that Lorentz invariance is preserved in 4D
spacetime. Its wavefunction is given by: 
\begin{equation}
\hat{\Psi}^{(0)}=\frac{A}{[g(z)]^{3/2}}  
\label{zerowave++-}
\end{equation}
with normalization factor $A$ determined by the requirement $\displaystyle{%
\int_{-z_2}^{\phantom{-}z_2} dz\left[\hat{\Psi}^{(0)}(z)\right]^2=1}$,
chosen so that we get the standard form of the Fierz-Pauli Lagrangian (the
same holds for the normalization of all the other states).

For the massive KK modes the solution is given in terms of Bessel functions.
For $y$ lying in the regions $\mathbf{A}\equiv\left[0,L_1\right]$ and $%
\mathbf{B}\equiv\left[L_1,L_2\right]$, we have: 
\begin{equation}
\hat{\Psi}^{(n)}\left\{ 
\begin{array}{cc}
\mathbf{A} &  \\ 
\mathbf{B} & 
\end{array}
\right\}=N_n \renewcommand{\arraystretch}{2} \left\{ 
\begin{array}{cc}
\sqrt{\frac{g(z)}{k_{1}}}\left[\phantom{A_1}Y_2\left(\frac{m_n}{k_{1}}%
g(z)\right)+A_{\phantom{2}}J_2\left(\frac{m_n}{k_{1}}g(z)\right)\right] & 
\\ 
\sqrt{\frac{g(z)}{k_{2}}}\left[B_1Y_2\left(\frac{m_n}{k_{2}}%
g(z)\right)+B_2J_2\left(\frac{m_n}{k_{2}}g(z)\right)\right] & 
\end{array}
\right\}  
\label{wave++-}
\end{equation}

The boundary conditions (one for the continuity of the wavefunction at $%
z_{1} $ and three for the discontinuity of its first derivative at $0$, $%
z_{1}$, $z_{2}$) result in a $4\times 4$ homogeneous linear system which, in
order to have a non-trivial solution, should have a vanishing determinant: 
\begin{equation}
\renewcommand{\arraystretch}{1.5}\left| 
\begin{array}{cccc}
Y_{1}\left( \frac{m}{k_{1}}\right) & J_{1}\left( \frac{m}{k_{1}}\right) & %
\phantom{-}0 & \phantom{-}0 \\ 
0 & 0 & \phantom{---}Y_{1}\left( \frac{m}{k_{2}}g(z_{2})\right) & %
\phantom{---}J_{1}\left( \frac{m}{k_{2}}g(z_{2})\right) \\ 
Y_{1}\left( \frac{m}{k_{1}}g(z_{1})\right) & J_{1}\left( \frac{m}{k_{1}}%
g(z_{1})\right) & -\sqrt{\frac{k_{1}}{k_{2}}}Y_{1}\left( \frac{m}{k_{2}}%
g(z_{1})\right) & -\sqrt{\frac{k_{1}}{k_{2}}}J_{1}\left( \frac{m}{k_{2}}%
g(z_{1})\right) \\ 
Y_{2}\left( \frac{m}{k_{1}}g(z_{1})\right) & J_{2}\left( \frac{m}{k_{1}}%
g(z_{1})\right) & -\sqrt{\frac{k_{1}}{k_{2}}}Y_{2}\left( \frac{m}{k_{2}}%
g(z_{1})\right) & -\sqrt{\frac{k_{1}}{k_{2}}}J_{2}\left( \frac{m}{k_{2}}%
g(z_{1})\right)
\end{array}
\right| =0  
\label{det++-}
\end{equation}
(The subscript $n$ on the masses $m_{n}$ has been suppressed.)

\subsection{Masses and Couplings}

The above quantization condition determines the mass spectrum of the model.
The parameters we have are $k_{1}$, $k_{2}$, $L_{1}$, $L_{2}$ with the
restriction $k_{1}<k_{2}<M$ and ${k_{1}}\sim {k_{2}}$ so that we don't
introduce a new hierarchy. It is more convenient to introduce the parameters 
$x=k_{2}(L_{2}-L_{1})$, $w=e^{-k_{1}L_{1}}$ and work instead with the set $%
k_{1}$, $k_{2}$, $x$, $w$. From now on we will assume that $w\ll 1$ ($w\sim {%
\mathcal{O}}\left( 10^{-15}\right) $ as is required if the model is to
provide an explanation of the hierarchy problem).

For the region $x\gtrsim 1$ it is straightforward to show analytically that
all the masses of the KK tower scale in the same way as $x$ is varied: 
\begin{equation}
m_{n}=\xi _{n}k_{2}we^{-x}  
\label{big++-}
\end{equation}
where $\xi _{n}$ is the $n$-th root of $J_{1}(x)$. This should be compared
with the $"+-+"$ $^{\prime \prime }+-+^{\prime \prime }$model in which $%
m_{1}\propto kwe^{-2x}$ and $m_{n+1}\propto kwe^{-x}$, where $x$ is the
separation between the $^{\prime \prime }-^{\prime \prime }$ and the second $%
^{\prime \prime }+^{\prime \prime }$ brane. This significant difference can
be explained by the fact that in the $^{\prime \prime }++-^{\prime \prime }$%
case the negative tension brane creates a potential barrier between the two
attractive potentials created by the positive tension branes. As a result
the wave function in the region of the $^{\prime \prime }-^{\prime
\prime }$ brane is small due to the tunneling effect. The two attractive potentials
support two bound states, one the graviton and the other the first KK mode.
The mass difference between the two is determined by the wave function in
the neighbourhood of the $^{\prime \prime }-^{\prime \prime }$ brane and is
thus very small. On the other hand the wave function between the two $%
^{\prime \prime }+^{\prime \prime }$ branes in the $^{\prime \prime
}++-^{\prime \prime }$ configuration is not suppressed by the need to tunnel
and hence the mass difference between the zero mode and the first KK mode is
also not suppressed. This has as result the two $^{\prime \prime }+^{\prime
\prime }$ branes behave approximately as one. This becomes even more clear
when $x\gg 1$ where the model resembles the $^{\prime \prime }+-^{\prime
\prime }$ RS construction. Indeed, in this limit the mass spectrum becomes $%
m_{n}=\xi _{n}k_{2}e^{-k_{2}L_{2}}$ which is exactly that of the $^{\prime
\prime }+-^{\prime \prime }$ RS model with orbifold size $L_{2}$ and bulk
curvature $k_{2}$.

In the region $x\lesssim 1$ the relation of eq(\ref{big++-}) breaks down. As
reduce $x$ the second $^{\prime \prime }+^{\prime \prime }$ comes closer and
closer to the $^{\prime \prime }-^{\prime \prime }$ brane and in the limit $%
x\rightarrow 0$ \mbox{({\textit {i.e.}}
$L_{2}=L_{1}$)} the combined brane system behaves as a single $^{\prime
\prime }-^{\prime \prime }$ brane, reducing to the $^{\prime \prime
}+-^{\prime \prime }$ RS model. In this limit the spectrum is given by: 
\begin{equation}
m_{n}=\xi _{n}k_{1}w  
\label{small++-}
\end{equation}
which is just the one of the $^{\prime \prime }+-^{\prime \prime }$ RS
model. In the region $0\leqslant x\lesssim 1$ the mass spectrum interpolates
between the relations (\ref{big++-}) and (\ref{small++-}).

The fact that there is nothing special about the first KK mode is true also
for its coupling on the second $^{\prime \prime }+^{\prime \prime }$ brane.
The interaction of the KK states on the second $^{\prime \prime }+^{\prime
\prime }$ brane is given by: 
\begin{equation}
{\mathcal{L}}_{int}=\sum_{n\geq 0}a_{n}h_{\mu \nu }^{(n)}(x)T_{\mu \nu
}(x)~~,~~\mathrm{with}~~a_{n}=\left[ \frac{g(z_{1})}{M}\right] ^{3/2}\hat{%
\Psi}^{(n)}(z_{1})  
\label{coupl++-}
\end{equation}

In the RS limit ($x=0$) all the states of the KK tower have equal coupling
given by: 
\begin{equation}
a_{n}=\frac{1}{wM_{\mathrm{Pl}}}
\end{equation}

As we increase $x$, the lower a state is in the tower, the more strongly it
couples, {\textit{i.e.}} $a_{1}>a_{2}>a_{3}>\cdots $ (with $a_{1}<(wM_{%
\mathrm{Pl}})^{-1}$) and tends to a constant value for high enough levels.
At some point this behaviour changes, the levels cross and for $x\gtrsim 1$
the situation is reversed and the lower a state is in the tower, the more
weakly it couples, {\textit{i.e.}} $a_{1}<a_{2}<a_{3}<\cdots $. In this
region it is possible to obtain a simple analytical expression for the
couplings: 
\begin{equation}
a_{n}=\frac{8\xi _{n}^{2}}{J_{2}\left( \xi _{n}\right) }\left( \frac{k_{2}}{%
k_{1}}\right) ^{3/2}\frac{1}{wM_{Pl}}e^{-3x}
\end{equation}
Here, we also see that the first KK state scales in exactly the same way as
the remaining states in the tower with respect to $x$, a behaviour quite
different to that in the $"+-+"$ model in which the coupling is $x$%
-independent. Furthermore, the coupling falls as ${e^{-3x}}$, {\textit{i.e.}}
much faster than $e^{-x}$ as one would naively expect. This can be explained
by looking at the origin of the $x$-dependence of the wavefunction. For
increasing $x$ the normalization volume coming from the region between $%
L_{1} $ and $L_{2}$ dominates and the normalization constant in (\ref{wave++-})
scales as $N_{n}\propto e^{-3x}$. This rapid decrease is not compensated by
the increase of the value of the remaining wavefunction (which from (\ref
{wave++-}) is approximately constant). Thus, although the two $^{\prime \prime
}+^{\prime \prime }$ branes in the large $x$ limit behave as one as far as
the mass spectrum is concerned, their separation actually makes the coupling
of the KK modes very different.


\section{The GRS model}

When the compactification volume becomes infinite,
the zero mode fails to be normalizable and thus the theory has no
massless graviton. However gravitational interactions at intermediate
distances can be reproduced if the KK states have specific
behaviour. Such an example is given in this Section through the GRS model.

The GRS model  consists of
one brane
with tension $V_{0} > 0$ and two branes with equal
tensions $V_{1}=-V_{0} /2$
placed at equal distances 
to the right and to the left of the positive tension
brane in the fifth direction.
The two negative tension branes are introduced for simplicity,
to have  $Z_2$ symmetry, $y \to -y$, in analogy to
RS model (hereafter $y$ denotes the fifth coordinate). 
It is assumed that  conventional matter resides on the
positive tension brane, and in what follows we will be
interested in gravitational interactions of this matter.

The physical set-up (for $y>0$) is as follows:
The bulk cosmological constant between the branes, $\Lambda$, is
negative as in the RS model, however, in contrast to that model, is
{\it zero} to the right of the negative tension brane. With
appropriately tuned $\Lambda$,  there exists a
solution to the five-dimensional 
Einstein equations for which both positive and
negative tension branes are at rest at $y=0$ and $y=y_c$ respectively,
$y_c$ being an arbitrary constant
The metric of this solution  for this set-up is
\be
ds^2=a^2(y)\eta_{\mu\nu}dx^{\mu}dx^{\nu}+dy^2
\label{1}
\ee
where
\be
a(y) = \cases{ e^{-ky} & $0<y<y_c$ \cr
e^{-ky_c}\equiv a_- & $y>y_c$\cr}
\label{2}
\ee
Where  $k=\sqrt{\frac{- \Lambda}{24 M^{3}}}$ 
 and $V_{0}=- \frac{\Lambda}{k}$. The four-dimensional hypersurfaces
$y=const.$ are flat, the five-dimensional space-time is flat to the
right of the negative-tension brane and anti-de Sitter between the
branes. The spacetime to the left of the positive tension brane is
of course a mirror image of this set-up.

\begin{figure}
\begin{center}
\begin{picture}(300,100)(0,50)
\SetWidth{1.5}

\Line(-40,100)(340,100)
\DashLine(150,50)(150,150){3}
\CCirc(150,100){7}{Black}{Red}

\CCirc(50,100){7}{Black}{Green}
\CCirc(250,100){7}{Black}{Green}
\LongArrow(160,110)(240,110)
\LongArrow(240,110)(160,110)
\Curve{(200,110)(220,145)(230,150)}
\Text(235,150)[l]{$x=kL$}

\Text(65,80)[]{$-1/2$}
\Text(230,80)[]{$-1/2$}
\Text(140,80)[]{$+$}

\Text(250,120)[c]{$L$}
\Text(50,120)[c]{$-L$}

\Text(170,145)[]{$Z_2$}
\Text(-40,100)[lt]{${\Blue {\underbrace{\phantom{Lonword}}_{{\LARGE{\bf Flat}}}}}$}
\Text(260,100)[lt]{${\Blue {\underbrace{\phantom{Lonword}}_{{\LARGE{\bf Flat}}}}}$}

\SetWidth{2}
\LongArrow(130,130)(170,130)
\LongArrow(170,130)(130,130)

\end{picture}
\end{center}
\caption{The GRS constuction.}
\label{GRS}
\end{figure}
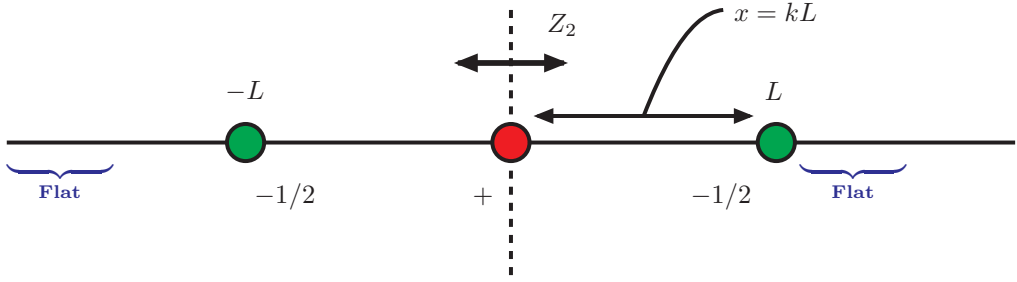

This background has two length scales, $k^{-1}$ and 
$\zeta_c \equiv k^{-1} e^{ky_c}$. We will consider the case of
large enough $z_c$, in which the two scales are 
well separated, $\zeta_c \gg k^{-1}$. We will see that
gravity in this model is effectively four-dimensional
at distances $r$ belonging to the interval
$k^{-1} \ll r \ll \zeta_c (k\zeta_c)^2$, and is five-dimensional
{\it both} at short distances, $r \ll k^{-1}$ (this situation is
exactly the same as in RS model), and at long distances,
$r\gg \zeta_c (k\zeta_c)^2$. In the latter regime of very long
distances the five-dimensional gravitational constant gets effectively
renormalized and no longer coincides with $G_5$.

To find the gravity law experienced by matter residing on
the positive tension brane, let us study
gravitational perturbations about the background
metric (\ref{1}). We will work in the Gaussian Normal (GN) gauge,
$g_{zz}=-1$, $g_{z\mu}=0$.
The linearized theory is described by the metric
\be
ds^2=a^2(y)\eta_{\mu\nu}dx^{\mu}dx^{\nu}+
h_{\mu\nu}(x,y)dx^{\mu}dx^{\nu}+dy^2
\label{3}
\ee
In the 
transverse-traceless gauge, $h^{\mu}_{\mu}=0$,
$h^{\nu}_{\mu,\nu}=0$, and the linearized Einstein equations take
the form for all components of $h_{\mu\nu}$,
\be
\cases{ h'' - 4k^2 h - {1\over a^2} \Box^{(4)}h=0 & $0<y<y_c$\cr
h'' - {1\over a_-^2}\Box^{(4)}h=0 & $y>y_c$\cr}
\label{6*}
\ee
The Israel junction conditions on the branes are
\be
\cases{h'+2kh = 0 & at $y=0$ \cr \left[ h'\right] - 2kh = 0 & at $y=y_c$\cr}
\label{7*}
\ee
where $\left[h'\right]$ is the discontinuity of the
$y$-derivative of the metric perturbation at $z_c$, 
and four-dimensional indices
are omitted. A general perturbation is a
superposition of modes, $h=\psi(y)e^{ip_{\mu}x^{\mu}}$ 
with $p^2=m^2$, where $\psi$ obeys the following set of equations
in the bulk,
\be
\cases{ \psi'' -4 k^2\psi+\frac{m^2}{a^2}\psi = 0 & $0<y<y_c$ \cr
\psi''+\frac{m^2}{a_-^2}\psi=0 & $y > y_c$ \cr}
\label{9}
\ee
with the junction conditions (\ref{7*}) (replacing $h$ by $\psi$).
It is straightforward to check that there are no negative modes,
i.e., normalizable solutions to these equations with $m^2 < 0$.
There are no normalizable solutions with $m^2 \geq 0$ either, 
so the spectrum is continuous, beginning at $m^2 =0$. To write the
modes explicitly, it is convenient to introduce a new coordinate
between the branes, $\zeta=\frac{1}{k}e^{ky}$, in terms of which the
background metric is conformally flat. Then the modes
have the following form,
\be
\psi_m = \cases{ C_m\left[N_1\left(\frac{m}{k}\right)J_2(m\zeta)-
J_1\left(\frac{m}{k}\right)N_2(m\zeta)\right] & $0<y<y_c$\cr
A_m \cos\left(\frac{m}{a_-}(y-y_c)\right)+
B_m \sin\left(\frac{m}{a_-}(y-y_c)\right) & $y>y_c$ \cr}
\label{18}
\ee
where $N$ and $J$ are the Bessel functions.
The constants $A_m, B_m$ and $C_m$ obey two relations due to the
junction conditions at the negative tension brane. Explicitly, 
\ba
A_m&=&C_m\left[N_1\left(\frac{m}{k}\right)J_2(m\zeta_c)
- J_1\left(\frac{m}{k}\right)N_2(m\zeta_c)\right]
\label{AA} \\
B_m&=&C_m \left[N_1\left(\frac{m}{k}\right)J_1(m\zeta_c) -
J_1\left(\frac{m}{k}\right)N_1(m\zeta_c) \right]
\label{BB}
\ea
The remaining overall constant $C_m$ is obtained from
the normalization condition. The latter is determined by the
explicit form of Eq.\ (\ref{9}) and reads
\be
\int~\psi_m^{*}(y) \psi_{m'} (y) \frac{dy}{a^2(y)} = \delta (m-m')
\ee
One makes use of the asymptotic behaviour of $\psi_m$ at
$y \to \infty$ and finds
\be
\frac{\pi}{a_-}(|A_m|^2+|B_m|^2)=1
\ee
which fixes $C_m$  from (\ref{AA}) and (\ref{BB}).

It is instructive to consider two limiting cases. 
At $m\zeta_c\gg 1$ we obtain by making use of the
asymptotics of the Bessel functions,
\be
C_m^2=\frac{m}{2k}\left[J_1^2\left(\frac{m}{k}\right)+
N_1^2\left(\frac{m}{k}\right)\right]^{-1}
\ee
which coincides, as one might expect, with the
normalization factor for the massive modes in RS model. In the opposite case
$m\zeta_c\ll 1$ (notice that this automatically implies
$m/k \ll 1$), the expansion
of the Bessel functions in Eqs.\ (\ref{AA}) and (\ref{BB}) yields
\be
C_{m}^{2}=\frac{\pi}{(k\zeta_c)^3}\left(1+
\frac{4}{(m\zeta_c)^2(k\zeta_c)^4}\right)^{-1}
\label{21}
\ee

It is now straightforward to calculate the static
gravitational potential between two unit masses
placed on the positive-tension brane at a
distance $r$ from each other.  This potential is
generated by the exchange of the massive modes. 
\be
V(r)=G_5\int_0^\infty~dm~\frac{e^{-mr}}{r} ~\psi_m^2 (y=0)
\label{22}
\ee
It is convenient to divide this integral into two parts,
\be
V(r)=G_5\int_0^{\zeta_c^{-1}}~dm ~\frac{e^{-mr}}{r} ~\psi_m^2(0) +
G_5\int_{\zeta_c^{-1}}^\infty ~dm~\frac{e^{-mr}}{r} ~\psi_m^2(0)
\label{23}
\ee
At $r \gg k^{-1}$, the second term in Eq.\ (\ref{23}) 
is small and it is similar to the contribution of
the continuum modes to the gravitational
potential in  RS model. It gives short distance corrections to Newton's law, 
\be
\Delta V_{short}(r) \sim
\frac{G_5}{kr^3} = \frac{G_N}{r}\cdot \frac{1}{k^2r^2}
\label{24}
\ee
where $G_N=G_5k$ is the four-dimensional Newton constant.

Of  greater interest  is the first term in
Eq.\ (\ref{23}) which dominates at $r \gg k^{-1}$.
Substituting the normalization factor (\ref{21}) into this term, we find
\be
V(r)=\frac{G_5}{r}\int_0^{\zeta_c^{-1}}~dm\frac{\pi}{(k\zeta_c)^3}
\left(1+\frac{4}{(m\zeta_c)^2(k\zeta_c)^4}\right)^{-1}
\frac{4k^2}{\pi^2m^2}e^{-mr}
\ee
This integral is always saturated  at 
$m \lesssim r_c^{-1} \ll \zeta_c^{-1}$, where
\be
r_c = \zeta_c (k\zeta_c)^2 \equiv k^{-1} e^{3ky_c}
\label{rrc}
\ee
Therefore, we can extend the integration to infinity and obtain
\ba
V(r) &=& \frac{G_N}{r}\cdot\frac{2}{\pi}\int_0^\infty dx
\frac{e^{-\frac{2r}{r_c}x}}{x^2+1} \label{25} \\
&=& {2G_N\over\pi r} \left [
\mbox{ci} (2r/r_c) \sin (2r/r_c) - \mbox{si} (2r/r_c) \cos (2r/r_c)\right ]
\nonumber
\ea
where $x=mr_c/2$, and $\mbox{ci/si}(t) = -\int_t^\infty {\cos/\sin (u)
\over u}du$ are the sine and cosine integrals. We see that
$V(r)$ behaves in a  peculiar way. At $r\ll r_c$, 
the exponential factor in Eq.\ (\ref{25}) can be set
equal to one and the four-dimensional Newton law is restored,
$V(r)=G_N/r$.
Hence, at intermediate distances,
$ k^{-1} \ll r \ll r_c$, the collection of continuous modes with 
$m \sim r_c^{-1}$ has the same effect as the graviton
bound state in RS model. However,
in the opposite case, $r\gg r_c$, we find
\be
V(r)=\frac{G_Nr_c}{\pi r^2}
\label{5dg}
\ee
which has the form of  ``Newton's law'' of
five-dimensional gravity with a renormalized gravitational constant.

It is clear from Eq.\ (\ref{25}) that at intermediate distances,
$k^{-1} \ll r \ll r_c$, the four-dimensional Newtonian potential
obtains not only short distance corrections, Eq.\ (\ref{24}), 
but also long distance ones, $V(r) = G_N/r + \Delta V_{short}(r)
+ \Delta V_{long}(r)$. The long distance corrections are suppressed by
$r/r_c$, the leading term being
\be
\Delta V_{long}(r)=\frac{G_N}{r}\cdot\frac{r}{r_c}\cdot\frac{4}{\pi}
\left(\ln \frac{2r}{r_c}+{\bf C}-1\right)
\label{26}
\ee
where ${\bf C}$ is the Euler constant. The two types of corrections,
Eqs.\ (\ref{24}) and (\ref{26}), are comparable  at roughly $r\sim
\zeta_c$.
At larger $r$, deviations from the four-dimensional Newton law are
predominantly due to the long-distance effects.

In this scenario, the approximate four-dimensional gravity law is valid over a
finite range of distances. Without strong fine-tuning however,
this range is large, as
required by phenomenology. Indeed, the exponential factor in 
Eq.\ (\ref{rrc}) leads to a very large  $r_c$ even for microscopic
separations, $z_c$, between the branes. As an example, for $k\sim M_{Pl}$
we only require $z_c \sim 50 l_{Pl}$ to have $r_c \sim 10^{28}$ cm,
the present horizon size of the Universe, i.e., with mild assumptions
about $z_c$, the four-dimensional description of gravity is valid from
the Planck to cosmological scales (in this example, long distance
corrections to Newton's gravity law dominate over short distance ones
at $r \lesssim \zeta_c \sim 10^{-13}$ cm).


\section{The four-brane $''+--+''$ Model}

In this section we discuss the four-brane model in order to clarify the
relation between the ``bigravity'' $''+-+''$ model  with the GRS
model. In particular we wish to explore and compare the
modification of gravity at large scales predicted by each model.

In the case of the $''+-+''$  model, in
the limit of very large $x$, gravity results from the net effect of both the
massless graviton and the ultralight first KK state. The modifications of
gravity at very large distances come from the fact that the Yukawa type
suppression of the gravitational potential coming from the KK state turns on
at the Compton wavelength of the state. On the other hand, the GRS model has
a continuous spectrum with no normalizable zero mode. However, the values of
the KK states wavefunctions on the $^{\prime \prime }+^{\prime \prime }$
brane have a ``resonance''-like behaviour \cite{Csaki:2000ei} which give rise to 4D
gravity at distances smaller than the Compton wavelength of its width.
Beyond this scale gravity becomes five-dimensional.

The four-brane GRS configuration can be obtained from the $"+-+"$ model by
``cutting'' the $^{\prime \prime }-^{\prime \prime }$ brane in half, {%
\textit{i.e.}} instead of having one $^{\prime \prime }-^{\prime \prime }$
brane one can take two $^{\prime \prime }-^{\prime \prime }$ branes of half
the tension of the original one ($^{\prime \prime }-1/2^{\prime \prime }$
branes), having flat spacetime between them (see Fig.(\ref{multi}). Finally
if the second $^{\prime \prime }+^{\prime \prime }$ brane is taken to
infinity together with one of the $^{\prime \prime }-1/2^{\prime \prime }$
branes we shall get precisely the GRS picture.

Let us discuss the four-brane $^{\prime \prime }+--+^{\prime \prime }$ model
in more detail. It consists of 5D spacetime with orbifold topology with four
parallel 3-branes located at $L_{0}=0$, $L_{1}$, $L_{2}$ and $L_{3}$, where $%
L_{0}$ and $L_{3}$ are the orbifold fixed points (see Fig.(\ref{multi})). The
bulk cosmological constant $\Lambda $ is negative ({\textit{i.e.}} $AdS_{5}$
spacetime) between the branes with opposite tension and zero ({\textit{i.e.}}
flat spacetime) between the two $^{\prime \prime }-1/2^{\prime \prime }$
branes. The model has four parameters namely $L_{1}$, $L_{2}$ and $L_{3}$
and $\Lambda $. For our present purposes we consider the symmetric
configuration, leaving 3 parameters, $l$, $l_{-}$ and $\Lambda $ where
$l\equiv L_1=L_3-L_2$ and $l_{-}\equiv L_{2}-L_{1}$. In the absence of matter the model is described by
eq(\ref{action++-}) with 
\begin{equation}
\Lambda (y)=\left\{ 
\begin{array}{cl}
{0} & ,y\in \lbrack L_{1},L_{2}] \\ 
{\Lambda}  & ,y\in \lbrack 0,L_{1}]\bigcup [L_{2},L_{3}]
\end{array}
\right.
\end{equation}
By considering the ansatz eq(\ref{ansatzrs+-+}) the ``warp'' function $\sigma (y)$
must satisfy: 
\begin{equation}
\sigma ^{\prime \prime }=\sum_{i}\frac{V_{i}}{12M^{3}}\delta (y-L_{i})~~~%
\mathrm{and}~~~\left( \sigma ^{\prime }\right) ^{2}=\left\{ 
\begin{array}{cl}
{0} & ,y\in \lbrack L_{1},L_{2}] \\ 
k^{2} & ,y\in \lbrack 0,L_{1}] \bigcup [L_{2},L_{3}]
\end{array}
\right. \ 
\end{equation}
where $k=\sqrt{\frac{-\Lambda }{24M^{3}}}$ is a measure of the bulk
curvature and we take $V_{0}=V_{3}=-2V_{1}=-2V_{2} \equiv
V$. The solution for $y>0$ is: 
\begin{equation}
\sigma (y)=\left\{ 
\begin{array}{cl}
{ky} & ,y\in \lbrack 0,L_{1}] \\ 
{kL_{1}} & ,y\in \lbrack L_{1},L_{2}] \\ 
{kL_{1}+k(L_{2}-y)} & ,y\in \lbrack L_{2},L_{3}]
\end{array}
\right. \ 
\end{equation}
Furthermore, 4D Poincare invariance requires the fine tuned relation: 
\begin{equation}
V=-\frac{\Lambda }{k}
\end{equation}

\begin{figure}[t]
\begin{center}
\begin{picture}(300,160)(0,50)

\SetScale{0.9}
\SetOffset(20,40)
\SetWidth{1.5}

\BCirc(150,100){80}
\DashLine(70,100)(230,100){3}

\GCirc(70,100){7.7}{0.9}
\GCirc(230,100){7.7}{0.9}

\GCirc(183,172){4}{0.2}
\GCirc(183,28){4}{0.2}
\GCirc(117,172){4}{0.2}
\GCirc(117,28){4}{0.2}

\LongArrowArc(150,100)(90,298,358)
\LongArrowArcn(150,100)(90,358,298)
\LongArrowArc(150,100)(90,182,242)
\LongArrowArcn(150,100)(90,242,182)

\LongArrowArc(150,100)(90,70,110)
\LongArrowArcn(150,100)(90,110,70)

\Text(45,100)[]{$+$}
\Text(225,100)[]{$+$}
\Text(178,168)[l]{$-1/2$}
\Text(92,168)[r]{$-1/2$}
\Text(165,0)[c]{$-1/2$}
\Text(105,0)[c]{$-1/2$}

\Text(120,110)[]{$Z_2$}

\Text(215,45)[l]{$x=kl$}
\Text(55,45)[r]{$x=kl$}
\Text(135,185)[c]{$x_-=kl_-$}

\Text(100,140)[c]{$L_1$}
\Text(100,40)[c]{$-L_1$}
\Text(170,140)[c]{$L_2$}
\Text(170,40)[c]{$-L_2$}
\Text(195,105)[c]{$L_3$}

\Text(136,149)[c]{$\underbrace{\phantom{abcdefghi}}_{{\Large{\bf FLAT}}}$}
\Text(136,31)[c]{$\overbrace{\phantom{abcdefghi}}^{{\LARGE{\bf FLAT}}}$}

\SetWidth{2}
\LongArrow(150,100)(150,115)
\LongArrow(150,100)(150,85)

\end{picture}
\end{center}
\caption{$^{\prime\prime}+--+^{\prime\prime}$ configuration with scale
equivalent $^{\prime\prime}+^{\prime\prime}$ branes. The distance between
the $^{\prime\prime}+^{\prime\prime}$ and $^{\prime\prime}-1/2^{\prime%
\prime} $ branes is $l=L_1=L_3-L_2$ while the distance between the $%
^{\prime\prime}-1/2^{\prime\prime}$ branes is $l_-=L_2-L_1$. The curvature
of the bulk between the $^{\prime\prime}+^{\prime\prime}$ and $%
^{\prime\prime}-1/2^{\prime\prime}$ branes is $k$.}
\label{multi}
\end{figure}

In order to determine the mass spectrum and the couplings of the KK modes we
consider linear ``massive'' metric fluctuations as in eq(\ref{perturbrs+-+}).
Following the same procedure we find that the function $\hat{\Psi}^{(n)}(z)$
obeys a Schr\"{o}dinger-like equation with potential $V(z)$ of the form: 
\begin{eqnarray}
\hspace*{0.5cm}V(z) &=&\frac{15k^{2}}{8[g(z)]^{2}}\left[ \theta (z)-\theta
(z-z_{1})+\theta (z-z_{2})-\theta (z-z_{3})\right]  \nonumber \\
&&-\frac{3k}{2g(z)}\left[ \delta (z)-\frac{1}{2}\delta (z-z_{1})-\frac{1}{2}%
\delta (z-z_{2})+\delta (z-z_{3})\right]  
\label{poten2}
\end{eqnarray}
The conformal coordinates now are given by: 
\begin{equation}
\renewcommand{\arraystretch}{1.5}z\equiv \left\{ 
\begin{array}{cl}
\frac{e^{ky}-1}{k} & ,y\in \lbrack 0,L_{1}] \\ 
\ (y-l)e^{kl}+\frac{e^{kl}-1}{k} & ,y\in \lbrack L_{1},L_{2}] \\ 
\ -\frac{1}{k}e^{2kl+kl_{-}}e^{-ky}+l_{-}e^{kl}+\frac{2}{k}e^{kl}-\frac{1}{k}
& ,y\in \lbrack L_{2},L_{3}]
\end{array}
\right. \ 
\end{equation}
\begin{equation}
g(z)=\left\{ 
\begin{array}{cl}
{kz+1} & ,z\in \lbrack 0,z_{1}] \\ 
{kz_{1}+1} & ,z\in \lbrack z_{1},z_{2}] \\ 
{k(z_{2}-z)+kz_{1}+1} & ,z\in \lbrack z_{2},z_{3}]
\end{array}
\right. \ 
\end{equation}
where $z_{1}=z(L_{1})$, $z_{2}=z(L_{2})$ and $z_{3}=z(L_{3})$.

The potential (\ref{poten2}) again gives rise to a massless graviton zero mode
whose wavefunction is given by (\ref{zerowave++-}) with the same normalization
convention. Note, however, that in the limit $l_- \rightarrow \infty$ this
mode becomes non-normalizable (GRS case). The solution of the
Schr\"{o}dinger equation for the massive KK modes is: 
\begin{equation}
\hat{\Psi}^{(n)}\left\{ 
\begin{array}{c}
\mathbf{A} \\ 
\mathbf{B} \\ 
\mathbf{C}
\end{array}
\right\}=N_n \renewcommand{\arraystretch}{1.7} \left\{ 
\begin{array}{c}
\sqrt{\frac{g(z)}{k}}\left[\phantom{A_1}Y_{2}\left(\frac{m_n}{k}%
g(z)\right)+A_{\phantom{2}}J_{2}\left(\frac{m_n}{k}g(z)\right)\right] \\ 
~~~~~~~~B_1\cos(m_{n}z)+B_{2}\sin(m_{n}z) \\ 
\sqrt{\frac{g(z)}{k}}\left[C_{1}Y_{2}\left(\frac{m_n}{k}g(z)%
\right)+C_{2}J_{2}\left(\frac{m_n}{k}g(z)\right)\right]
\end{array}
\right\} \   \label{wavemulti}
\end{equation}
where $\mathbf{A}=[0,z_{1}]$, $\mathbf{B}=[z_{1},z_{2}]$, and $\mathbf{C}%
=[z_{2},z_{3}]$. We observe that the solution in the first and third
interval has the same form as in the $^{\prime\prime}+ - +^{\prime\prime}$
model. The new feature is the second region (flat spacetime). The
coefficients that appear in the solution are determined by imposing the
boundary conditions and normalizing the wavefunction.

The boundary conditions (two for the continuity of the wavefunction at $%
z_{1} $, $z_{2}$ and four for the discontinuity of its first derivative at $%
0 $, $z_{1}$, $z_{2}$ and $z_{3}$) result in a $6\times 6$ homogeneous
linear system which, in order to have a non-trivial solution, should have
vanishing determinant. It is readily reduced to a $4\times 4$ set of
equations leading to the quantization condition: 
\begin{equation}
\renewcommand{\arraystretch}{2}{ \footnotesize {\left| 
\begin{array}{cccc}
Y_{2}\left( g_{1}\frac{m}{k}\right) -\frac{Y_{1}\left( \frac{m}{k}\right) }{%
J_{1}\left( \frac{m}{k}\right) }J_{2}\left( g_{1}\frac{m}{k}\right) & -\cos
(mz_{1}) & -\sin (mz_{1}) & 0 \\ 
Y_{1}\left( g_{1}\frac{m}{k}\right) -\frac{Y_{1}\left( \frac{m}{k}\right) }{%
J_{1}\left( \frac{m}{k}\right) }J_{1}\left( g_{1}\frac{m}{k}\right) & %
\phantom{-}\sin (mz_{1}) & -\cos (mz_{1}) & 0 \\ 
0 & -\sin (mz_{2}) & \phantom{-}\cos (mz_{2}) & \phantom{-}Y_{1}\left( g_{2}%
\frac{m}{k}\right) -\frac{Y_{1}\left( g_{3}\frac{m}{k}\right) }{J_{1}\left(
g_{3}\frac{m}{k}\right) }J_{1}\left( g_{2}\frac{m}{k}\right) \\ 
0 & \phantom{-}\cos (mz_{2}) & \phantom{-}\sin (mz_{2}) & -Y_{2}\left( g_{2}%
\frac{m}{k}\right) +\frac{Y_{1}\left( g_{3}\frac{m}{k}\right) }{J_{1}\left(
g_{3}\frac{m}{k}\right) }J_{2}\left( g_{2}\frac{m}{k}\right)
\end{array}
\right| =0}}  \label{det1}
\end{equation}
with $g_{1}=g(z_{1})$, $g_{2}=g(z_{2})$ and $g_{3}=g(z_{3})$. Here we have
suppressed the subscript $n$ on the masses $m_{n}$.

\subsection{The Mass Spectrum}

The above quantization condition provides the mass spectrum of the model. It
is convenient to introduce two dimensionless parameters, $x=kl$ and $%
x_{-}=kl_{-}$ (c.f. Fig.(\ref{multi})) and we work from now on with the set
of parameters $x$, $x_{-}$ and $k$. The mass spectrum depends crucially on the
distance $x_{-}$. We must recover the $''+-+''$ spectrum in the limit $%
x_{-}\rightarrow 0$, and the GRS spectrum in the limit $x_{-}\rightarrow
\infty .$ From the quantization condition, eq(\ref{det1}) it is easy to
verify these features and show how the $^{\prime \prime }+--+^{\prime \prime
}$ spectrum smoothly interpolates between the $''+-+''$ model and the GRS one.
It turns out that the structure of the spectrum has simple $x_{-}$ and $x$
dependence in three separate regions of the parameter space:

\subsubsection{The three-brane \textbf{{$^{\prime \prime }+-+^{\prime \prime
}$ Region}}}

For $x_{-}\lower3pt\hbox{$\, \buildrel < \over \sim \, $}1$ we find that the
mass spectrum is effectively $x_{-}$-independent given by the approximate
form: 
\begin{eqnarray}
m_{1} &=&2\sqrt{2}ke^{-2x} \\
m_{n+1} &=&\xi _{n}ke^{-x}~~~~~~n=1,2,3,\ldots  \label{bimass}
\end{eqnarray}
where $\xi _{2i+1}$ is the $(i+1)$-th root of $J_{1}(x)$ ($i=0,1,2,\ldots $)
and $\xi _{2i}$ is the $i$-th root of $J_{2}(x)$ ($i=1,2,3,\ldots $). As
expected the mass spectrum is identical to the one of the $''+-+''$ model for
the trivial warp factor $w=1$. The first mass is manifestly singled out from
the rest of the KK tower and for large $x$ leads to the possibility of
bigravity.

\subsubsection{The \textbf{Saturation Region}}

For $1\ll x_{-}\ll e^{2x}$ we find a simple dependence on $x_{-}$ given by
the approximate analytic form: 
\begin{eqnarray}
m_{1} &=&2k\frac{e^{-2x}}{\sqrt{x_{-}}} \\
m_{n+1} &=&n\pi k\frac{e^{-x}}{x_{-}}~~~~~~n=1,2,3,\ldots
\end{eqnarray}
As $x_{-}$ increases the first mass decreases less rapidly than the other
levels.

\subsubsection{The \textbf{GRS Region}}

For $x_{-}\gg e^{2x}$ the first mass is no longer special and scales with
respect to {\textit{both}} $x$ and $x_{-}$ in the same way as the remaining
tower: 
\begin{equation}
m_{n}=n\pi k\frac{e^{-x}}{x_{-}}~~~~~~n=1,2,3,\ldots  \label{longmass}
\end{equation}
The mass splittings $\Delta m$ tend to zero as $x_{-}\rightarrow \infty $
and we obtain the GRS continuum of states

The behaviour of the spectrum is illustrated in Figure 3.

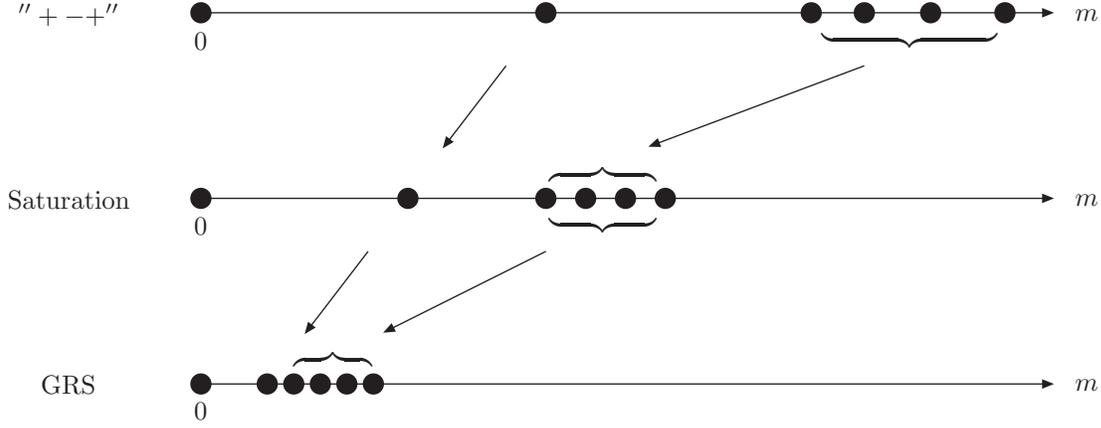
\begin{figure}[h]
\begin{center}
\begin{picture}(300,200)(0,50)

\LongArrow(0,220)(320,220)
\LongArrow(0,150)(320,150)
\LongArrow(0,80)(320,80)

\Vertex(0,220){4}
\Vertex(0,150){4}
\Vertex(0,80){4}

\Vertex(130,220){4}
\Vertex(230,220){4}
\Vertex(250,220){4}
\Vertex(275,220){4}
\Vertex(303,220){4}

\Vertex(78,150){4}
\Vertex(130,150){4}
\Vertex(145,150){4}
\Vertex(160,150){4}
\Vertex(175,150){4}

\Vertex(25,80){4}
\Vertex(35,80){4}
\Vertex(45,80){4}
\Vertex(55,80){4}
\Vertex(65,80){4}

\Text(335,220)[]{$m$}
\Text(335,150)[]{$m$}
\Text(335,80)[]{$m$}

\Text(0,210)[]{$0$}
\Text(0,140)[]{$0$}
\Text(0,70)[]{$0$}

\Text(50,85)[c]{$\overbrace{\phantom{abcdef}}$}
\Text(152,145)[c]{$\underbrace{\phantom{abcdefgh}}$}
\Text(152,155)[c]{$\overbrace{\phantom{abcdefgh}}$}
\Text(268,215)[c]{$\underbrace{\phantom{abcdefghijklm}}$}

\LongArrow(250,200)(170,170)
\LongArrow(130,130)(70,100)

\LongArrow(115,200)(92,170)
\LongArrow(63,130)(40,100)

\Text(-50,220)[c]{$''+-+''$}
\Text(-50,150)[c]{Saturation}
\Text(-50,80)[c]{GRS}

\end{picture}
\end{center}
\caption{The behaviour of the mass of the first five KK states in the three
regions of simple $x$, $x_{-}$ dependence. The first dot at zero stands for
the graviton.}
\label{masses}
\end{figure}

\subsection{Multigravity}

Armed with the details how the spectrum smoothly changes between the $''+-+''$
model ($x_{-}=0)$ and the GRS model ($x_{-}\rightarrow \infty )$, we can now
discuss the possibilities for modifying gravity at large distances. The
couplings of the KK states with matter on the left $^{\prime \prime
}+^{\prime \prime }$ brane are readily calculated by the interaction
Lagrangian (\ref{coupl++-}) with: 
\begin{equation}
a_{n}=\left[ \frac{g(0)}{M}\right] ^{3/2}\hat{\Psi}^{(n)}(0)
\label{couplmulti}
\end{equation}

\subsubsection{\textbf{Bigravity Region}}

In the KPMRS limit, $x_{-}\rightarrow 0,$ the first KK mode has constant
coupling equal to that of the 4D graviton: 
\begin{equation}
a_{1}=\frac{1}{M_{\ast }}~(=a_{0})~~~~~~~\mathrm{where}~~~M_{\ast }^{2}=%
\frac{2M^{3}}{k}
\end{equation}
while the couplings of the rest of the KK tower are exponentially
suppressed: 
\begin{equation}
a_{n+1}=\frac{1}{M_{\ast }}~\frac{e^{-x}}{\sqrt{J_{1}^{2}\left( \frac{%
m_{n}e^{x}}{k}\right) +J_{2}^{2}\left( \frac{m_{n}e^{x}}{k}\right) }}%
~~~~~~n=1,2,3,\ldots  \label{bigcoupl}
\end{equation}

The gravitational potential is computed by the tree level exchange diagrams
of the 4D graviton and KK states which in the Newtonian limit is: 
\begin{equation}
V(r)=-\sum_{n=0}^{N_{\Lambda}}a_n^2\frac{e^{-m_{n}r}}{r}  \label{gravipot}
\end{equation}
where $a_n$ is the coupling (\ref{couplmulti}) and $n=0$ accounts for the
massless graviton. The summation stops at some very high level $N_{\Lambda}$
with mass of the order of the cutoff scale $\sim M$.

In the ``bigravity'' scenario, at distances $r\ll m_{1}^{-1},$ the first KK
state and the 4D graviton contribute equally to the gravitational force, $%
i.e.$ 
\begin{equation}
V_{ld}(r)\approx -\frac{1}{M_{\ast }^{2}}\left( \frac{1}{r}+\frac{e^{-m_{1}r}%
}{r}\right) \approx -\frac{G_{N}}{r}
\end{equation}
where $G_{N}\equiv \frac{2}{M_{\ast }^{2}}$. For distances $r\gtrsim
m_{1}^{-1}$ the Yukawa suppression effectively reduces gravity to half its
strength. Astronomical constraints and the requirement of the observability
of this effect demand that for $k\sim M_{\mathrm{Pl}}$ we should have $x$ in
the region 65-70. Moreover, at distances $r\lesssim m_{2}^{-1}$ the Yukawa
interactions of the remaining KK states are significant and will give rise
to a short distance correction. This can be evaluated by using the
asymptotic expression of the Bessel functions in (\ref{bigcoupl}) since we
are dealing with large $x$ and summing over a very dense spectrum, giving: 
\begin{equation}
V_{sd}(r)=-\frac{G_{N}}{k}\sum_{n=2}^{N_{\Lambda }}\frac{k\pi }{2e^{x}}~%
\frac{m_{n}}{2k}~\frac{e^{-m_{n}r}}{r}  
\label{shortdispot}
\end{equation}
At this point we exploit the fact that the spectrum is nearly continuum
above $m_{2}$ and turn the sum to an integral with the first factor in (\ref
{shortpot}) being the integration measure, {\textit{i.e.}} $\sum \frac{k\pi 
}{2e^{x}}=\sum \Delta m\rightarrow \int dm$ (this follows from eq(\ref
{bimass}) for the asymptotic values of the Bessel roots). Moreover, we can
extend the integration to infinity because, due to the exponential
suppression of the integrand, the integral saturates very quickly and thus
the integration over the region of very large masses is irrelevant. The
resulting potential is now: 
\begin{equation}
V_{sd}(r)=-\frac{G_{N}}{k}\int_{m_{2}}^{\infty }dm~\frac{m}{2k}~\frac{%
e^{-m_{n}r}}{r}
\end{equation}
The integration is easily performed and gives: 
\begin{equation}
V_{sd}(r)\simeq -\frac{G_{N}}{2r}~\frac{1+m_{2}r}{(kr)^{2}}~e^{-m_{2}r}
\end{equation}
We see these short distance corrections are significant only at Planck scale
lengths $\sim k^{-1}$.

\subsubsection{The \textbf{GRS Region}}

In the GRS limit, $x_{-}\gg e^{2x},$ we should reproduce the
``resonance''-like behaviour of the coupling in the GRS model. In the
following we shall see that indeed this is the case and we will calculate
the first order correction to the GRS potential for the case $x_{-}$ is
large but finite.

For the rest of the section we split the wavefunction (\ref{wavemulti}) in
two parts, namely the normalization $N_{n}$ and the unnormalized
wavefunction $\tilde{\Psi}^{(n)}(z)$, {\textit{i.e.}} $\hat{\Psi}%
^{(n)}(z)=N_{n}\tilde{\Psi}^{(n)}(z)$. The former is as usual chosen so that
we get a canonically normalized Pauli-Fierz Lagrangian for the 4D KK modes $%
h_{\mu \nu }^{(n)}$ and is given by: 
\begin{equation}
N_{n}^{2}=\frac{1/2}{2\displaystyle{\int_{\phantom{.}0}^{z_{1}}dz\left[ 
\tilde{\Psi}^{(n)}(z)\right] ^{2}}+\displaystyle{\int_{\phantom{.}%
z_{1}}^{z_{2}}dz\left[ \tilde{\Psi}^{(n)}(z)\right] ^{2}}}
\end{equation}
The value of $\tilde{\Psi}^{(n)}(z)$ on the left $^{\prime \prime }+^{\prime
\prime }$ brane is, for $m_{n}\ll k$: 
\begin{equation}
\tilde{\Psi}_{(n)}^{2}(0)\simeq \frac{16k^{3}}{\pi ^{2}m_{n}^{4}}  \label{un}
\end{equation}

It is convenient to split the gravitational potential given by the relation (%
\ref{gravipot}) into two parts: 
\begin{equation}
V(r)=-\frac{1}{M^{3}}\sum_{n=1}^{N_{x_{-}}-1}\frac{e^{-m_{n}r}}{r}N_{n}^{2}%
\tilde{\Psi}_{(n)}^{2}(0)-\frac{1}{M^{3}}\sum_{n=N_{x_{-}}}^{N_{\Lambda }}%
\frac{e^{-m_{n}r}}{r}N_{n}^{2}\tilde{\Psi}_{(n)}^{2}(0)  \label{gravpot}
\end{equation}
As we shall see this separation is useful because the first $N_{x_{-}}$
states give rise to the long distance gravitational potential $V_{ld}$ while
the remaining ones will only contribute to the short distance corrections $%
V_{sd}$.

\begin{itemize}
\item  \textbf{\ Short Distance Corrections}
\end{itemize}

We first consider the second term. The normalization constant in this region
is computed by considering the asymptotic expansions of the Bessel functions
with argument $\frac{g(z_{1})m_{n}}{k}$. It is easily calculated to be: 
\begin{equation}
N_{n}^{2}=\frac{\pi ^{3}m_{n}^{5}}{32k^{3}g(z_{1})x_{-}}~\left[ \frac{1}{1+%
\frac{2}{x_{-}}}\right]
\end{equation}

If we combine the above normalization with unnormalized wavefunction (\ref
{un}), we find 
\begin{equation}
V_{sd}(r)\simeq -\frac{1}{M^{3}}\sum_{n=N_{x_{-}}}^{N_{\Lambda }}\frac{k\pi 
}{x_{-}e^{x}}~\frac{m_{n}}{2k}~\frac{e^{-m_{n}r}}{r}~\left[ \frac{1}{1+\frac{%
2}{x_{-}}}\right]  
\label{shortpot}
\end{equation}
Since we are taking $x_{-}\gg e^{2x}$, the spectrum tends to continuum, {%
\textit{i.e.}} $N_{n}\rightarrow N(m)$, $\tilde{\Psi}_{(n)}(0)\rightarrow 
\tilde{\Psi}(m)$, and the sum turns to an integral where the first factor in
(\ref{shortpot}) is the integration measure, {\textit{i.e.}} $\sum \frac{%
k\pi }{x_{-}e^{x}}=\sum \Delta m\rightarrow \int dm$ ($c.f.$ eq(\ref
{longmass})). Moreover, as before we can again extend the integration to
infinity. Finally, we expand the fraction involving $x_{-}$ keeping the
first term in the power series to obtain the potential: 
\begin{equation}
V_{sd}(r)\simeq -\frac{1}{M^{3}}\int_{m_{0}}^{\infty }dm~\frac{e^{-mr}}{r}~%
\frac{m}{2k}\left( 1-\frac{2}{x_{-}}\right)
\end{equation}
where $m_{0}=ke^{-x}$. The integral is easily calculated and the potential
reads: 
\begin{equation}
V_{sd}(r)\simeq -\frac{G_{N}}{2r}~\frac{1+m_{0}r}{(kr)^{2}}~(1-\frac{2}{x_{-}%
})~e^{-m_{0}r}
\end{equation}
where we identified $G_{N}\equiv \frac{k}{M^{3}}$ for reasons to be seen
later. The second part of the above potential is the first correction coming
from the fact that $x_{-}$ is finite. Obviously this correction vanishes
when $x_{-}\rightarrow \infty $. Note that the above potential gives
corrections to the Newton's law only at distances comparable to the Planck
length scale.

\begin{itemize}
\item  \textbf{\ Multigravity: 4D and 5D gravity}
\end{itemize}

We turn now to the more interesting first summation in eq(\ref{gravpot}) in
order to show that the coupling indeed has the ``resonance''-like behaviour
for $\Delta m\rightarrow 0$ responsible for 4D Newtonian gravity at
intermediate distances and the 5D gravitational law for cosmological
distances. This summation includes the KK states from the graviton zero mode
up to the $N_{x_{-}}$-th level. The normalization constant in this region is
computed by considering the series expansion of all the Bessel functions
involved. It is easily calculated to be: 
\begin{equation}
N_{n}^{2}\simeq \frac{\pi ^{2}m_{n}^{4}}{4g(z_{1})^{4}x_{-}}~\left[ \frac{1}{%
m_{n}^{2}+\frac{\Gamma ^{2}}{4}+\frac{8k^{2}}{g(z_{1})^{4}x_{-}}}\right]
\end{equation}
where $\Gamma =4ke^{-3x}$. If we combine the above normalization with the
unnormalized wavefunction (\ref{un}), we find that the long distance
gravitational potential is: 
\begin{equation}
V_{ld}(r)=-\frac{1}{M^{3}}\sum_{n=0}^{N_{x_{-}}}\frac{\pi k}{x_{-}e^{x}}~%
\frac{4k^{2}}{\pi g(z_{1})^{3}}~\frac{e^{-m_{n}r}}{r}~\left[ \frac{1}{%
m_{n}^{2}+\frac{\Gamma ^{2}}{4}+\frac{8k^{2}}{g(z_{1})^{4}x_{-}}}\right]
\end{equation}
Again, since we are taking $x_{-}\gg e^{2x},$ the above sum will turn to an
integral with $\sum \Delta m\rightarrow \int dm$. Moreover, we can safely
extend the integration to infinity since the integral saturates very fast
for $m\lesssim \Gamma /4\equiv r_{c}^{-1}\ll ke^{x}$. If we also expand the
fraction in brackets keeping the first term in the power series, we find the
potential: 
\begin{eqnarray}
V_{ld}(r)\simeq &-&\frac{1}{M^{3}}\int_{0}^{\infty }dm~\frac{4k^{2}}{\pi
g(z_{1})^{3}}~\frac{e^{-mr}}{r}~\frac{1}{m^{2}+\frac{\Gamma ^{2}}{4}} 
\nonumber \\
&+&\frac{1}{M^{3}}\int_{0}^{\infty }dm~\frac{32k^{4}}{x_{-}\pi g(z_{1})^{7}}~%
\frac{e^{-mr}}{r}~\frac{1}{(m^{2}+\frac{\Gamma ^{2}}{4})^{2}}
\end{eqnarray}

The first part is the same as in the GRS model potential, whereas the second
one is the first correction that comes from the fact that $x_{-}$ is still
finite though very large. Note that the width of the ``resonance'' scales
like $e^{-3x}$, something that is compatible with the scaling law of the
masses ($m_{n}=n\pi k\frac{e^{-x}}{x_{-}}$), since we are working at the
region where $x_{-}\gg e^{2x}$, {\textit{i.e.}} $m_{n}\ll n\pi ke^{-3x}$.
The above integrals can be easily calculated in two interesting limits.

For $k^{-1}\ll r\ll r_{c}$ the potential is given approximately by : 
\begin{equation}
V_{ld}(r\ll r_{c})\simeq -\frac{G_{N}}{r}~(1-\frac{e^{2x}}{x_{-}})
\end{equation}
where we have identified $G_{N}\equiv \frac{k}{M^{3}}$ to obtain the normal
4D Newtonian potential. Note that since $x_{-}\gg e^{2x}$, the $1/x_{-}$
term is indeed a small correction.

In the other limit, $r\gg r_{c}$, the integrand is only significant for
values of $\ m$ for which the $m^{2}$ term in the denominator of the
``Breit-Wigner'' can be dropped and the potential becomes: 
\begin{equation}
V_{ld}(r\gg r_{c})\simeq -\frac{G_{N}r_{c}}{\pi r^{2}}~(1-\frac{2e^{2x}}{%
x_{-}})
\end{equation}
The fact that Newtonian gravity has been tested close to the present horizon
size require that for $k\sim M_{\mathrm{Pl}}$ we should have $x\gtrsim $
45-50.

Finally we note that if we take the $x_{-}\rightarrow \infty $ we recover
the GRS result 
\begin{equation}
\lim_{x_{-}\rightarrow \infty }V_{+--+}(r,x_{-})=V_{GRS}(r)
\end{equation}

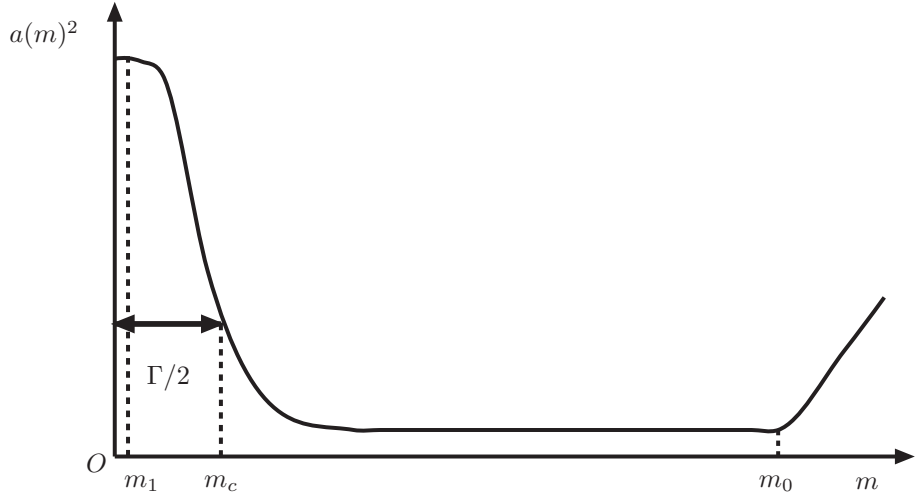
\begin{figure}[tbp]
\begin{center}
\begin{picture}(400,200)(-50,50)
\SetOffset(50,0)
\SetWidth{1.5}

\LongArrow(-50,80)(250,80)
\LongArrow(-50,80)(-50,250)
\Curve{(-50,230)(-40,229)(-32,224)(-15,150)(40,90)(50,90)(75,90)(100,90)(125,90)(150,90)(175,90)(190,90)(200,90)(225,120)(240,140)}
\DashLine(-45,80)(-45,231){2.5}
\DashLine(-10,80)(-10,130){2.5}
\DashLine(200,80)(200,90){2.5}

\Text(-90,240)[l]{$a(m)^2$}
\Text(-40,70)[c]{$m_{1}$}
\Text(-30,110)[c]{$\Gamma/2$}
\Text(230,70)[l]{$m$}
\Text(200,70)[c]{$m_{0}$}
\Text(-10,70)[c]{$m_{c}$}
\Text(-57,78)[c]{$O$}

\SetWidth{2}
\LongArrow(-50,130)(-10,130)
\LongArrow(-10,130)(-50,130)

\end{picture}
\end{center}
\caption{The behaviour of the coupling, $a(m),$ in the limit of $x_{-}\gg
e^{2x}$. Three regions of interest are indicated. The region $m>m_{0}$ gives
rise to short distance corrections. The $m_{1}\ll m\ll m_{c}$ region gives
rise to 4D gravity at intermediate distances and 5D gravity at ultra large
distances. For distances $r\gg m_{1}^{-1}$, the zero mode gives the dominant
contribution and thus we return to 4D gravity.}
\label{coupling}
\end{figure}

\begin{itemize}
\item  \textbf{\ Back to 4D gravity}
\end{itemize}

As we have just seen, probing larger distances than $r_{c},$ the 4D
gravitational potential changes to a 5D one. This is the most significant
characteristic of the GRS model. In the case that $x_{-}$ is large compared
to $e^{2x}$ but still finite, there is another distinct region of interest,
namely $r\gg m_{1}^{-1}$. This follows from the fact that, in this limit,
the spectrum is still discrete. For distances larger than of the order of
the corresponding wavelength of the first KK mode, the contribution to
gravity from the KK tower is suppressed and thus the zero mode gives the
dominant contribution, leading to the 4D Newtonian potential again. In this
case the strength of the gravitational interaction is a small fraction of
the strength of the intermediate 4D gravity. More precisely, the
contribution of the massless graviton is $\frac{1}{x_{-}}$ suppressed and
thus vanishes when $x_{-}\rightarrow \infty $, something that is expected
since in this limit there is no nomalizable zero mode. The gravitational
potential in this case is: 
\begin{equation}
V_{4D}(r)=-\frac{1}{M^{3}}~\frac{1}{r}~N_{0}^{2}\tilde{\Psi}_{(0)}^{2}(0)=-%
\frac{G_{N}}{r}~\frac{e^{2x}}{x_{-}}
\end{equation}
Obviously this 4D region disappears at the limit $x_{-}\rightarrow \infty $
since the spectrum becomes continuum and thus the 5D gravity ``window''
extents to infinity. We should note finally that for the values of $x$ that
we consider here this final modification of gravity occurs at distances well
above the present horizon.

\section{vDVZ discontinuity, negative tension branes and ghosts}

All the  constructions that we have considered up to now  have two important defects. The first
one \cite{Dvali:2001xg}
is that the extra polarizations of the massive KK states do not
decouple in the limit of vanishing mass, the famous van Dam
- Veltman - Zakharov \cite{vanDam:1970vg}
discontinuity, which makes the model  disagree with the experimental
measurements of the bending of
the light by the sun. The second 
one \cite{Dvali:2000km,Pilo:2000et,Kogan:2001qx} is that the moduli (radions) associated with the perturbations of the
$''-''$ branes are necessarily physical ghost fields, therefore
unacceptable. The latter problem is connected to the violation of the
weaker energy condition \cite{Freedman:1999gp,Witten:2000zk} on $''-''$ branes sandwiched between $''+''$
 branes. In this Section we discuss these issues.

\subsection{Graviton propagator in flat spacetime -  The vDVZ discontinuity }

The celebrated  van
Dam - Veltman - Zakharov discontinuity is evident from the different form
of the propagators that correspond to the massive and massless graviton.
In  more detail, the form of the massless graviton propagator in flat
spacetime (in momentum space) has the form: 
\be
G^{\mu\nu;\alpha\beta}=\frac{1}{2}~ \frac{\left(\eta^{\mu\alpha}\eta^{\nu\beta} +
\eta^{\eta\alpha}\eta^{\mu\beta}\right) - \eta^{\mu\nu}\eta^{\alpha\beta} 
  }{p^2 }+ \cdots
\ee
where we have omitted terms that do not contribute when contracted
with a conserved $T_{\mu \nu}$.

On the other hand, the 
four-dimensional  massive graviton propagator (in momentum space) has the form:\be
G^{\mu\nu;\alpha\beta}=\frac{1}{2} ~ \frac{\left(\eta^{\mu\alpha}\eta^{\nu\beta} +
\eta^{\nu\alpha}\eta^{\mu\beta}\right) - \frac{2}{3}~ \eta^{\mu\nu}\eta^{\alpha\beta} 
 }{p^2 -m^2}+\cdots
\ee
In order to be able to answer the question if the graviton can be
massive we have to examine if the theory of massive graviton can
reproduce, at least, the testable predictions of General Relativity.

\paragraph{Newton's Law} 

One of the  essential requirements of the theory of massive graviton 
is to be able to reproduce the
Newton's Law at least at the intermediate distances (where it  has been tested). 

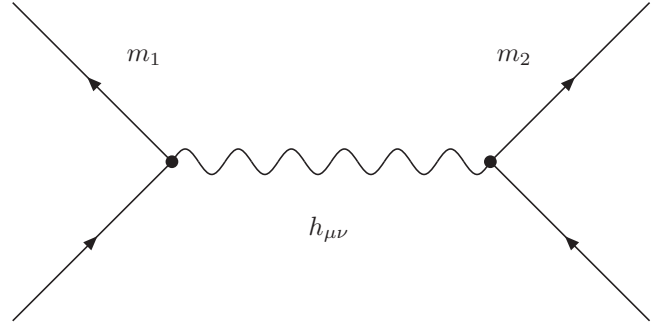
\begin{figure}
\begin{picture}(300,100)(-45,50)
\SetScale{1.2}
\SetOffset(-40,-30)
\ArrowLine(50,50)(100,100)
\ArrowLine(100,100)(50,150)
\ArrowLine(200,100)(250,150)
\ArrowLine(250,50)(200,100)
\Vertex(100,100){2}
\Vertex(200,100){2}
\Photon(100,100)(200,100){4}{6}
\Text(110,160)[]{$m_{1}$}

\Text(250,160)[]{$m_{2}$}

\Text(180,95)[]{$h_{\mu\nu}$}
\end{picture}
\vspace*{8mm} 
\caption{One graviton exchange diagram.} 
\label{mssgrfig1}
\end{figure}

In order to calculate the interaction between two non-relativistic
masses, we consider the non-relativistic limit of one graviton exchange.
The energy momentum tensor of the two masses are:
\ba
T^{(1)}_{\mu \nu}(x)&=& m_{1}u_{\mu}u_{\nu}\delta(\vec{x}-\vec{r_{1}})\nonumber \\ 
T^{(2)}_{\mu \nu}(x')&=& m_{2}u_{\mu}u_{\nu}\delta(\vec{x'}-\vec{r_{2}})
\nonumber
\ea
where $u_{\mu}$ is the four momentum of the particles.
The one graviton exchange with massless graviton gives amplitude:
\be
G_{N}^{0}\int d^4x\int d^4x' T^{1}_{\mu \nu}(x)
G_{0}^{\mu\nu;\alpha\beta}T^{2}_{\alpha \beta}(x') \rightarrow \frac{2}{3}G_{N}^{0}T^{1}_{00}T^{2}_{00}\frac{e^{-i\vec{p}(\vec{x}-\vec{x'})}}{{\vec{p}}^2-i\epsilon }
\ee
On the other hand the one graviton exchange with massive graviton gives:
\be
G_{N}^{m}\int d^4x \int d^4x' T^{1}_{\mu \nu}(x) G_{m}^{\mu\nu;\alpha\beta}T^{2}_{\alpha \beta}(x')\rightarrow\frac{1}{2}G_{N}^{m}T^{1}_{00}T^{2}_{00}\frac{e^{-i\vec{p}(\vec{x}-\vec{x'})}}{{\vec{p}}^2-m^2-i\epsilon }
\ee
The massless graviton gives the usual $\sim 1/r$ potential while the
massive gives $e^{-mr}/r$ which at the $m \rightarrow 0$ limit results 
to the usual $1/r$ behaviour.
However, in order to have identical  limit when $m \rightarrow 0$ the
relationship:
\be
G_{N}^{m}=\frac{3}{4} G_{N}^{0}
\ee
must hold.


\paragraph{Bending of Light} 

The theory of massive graviton must reproduce, apart from the
non-relativistic results of General Relativity, also its relativistic
results: \textit{e.g.} the deflection angle of the light by the Sun.

\vspace*{10mm}

\begin{center}
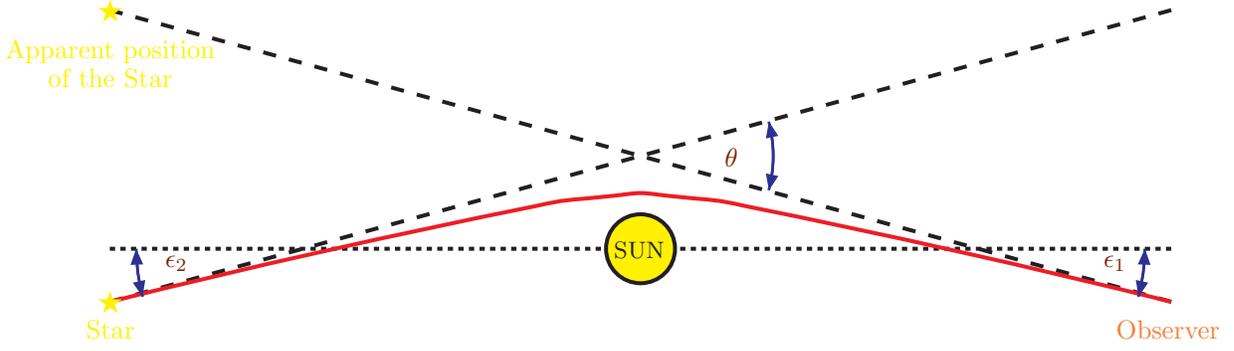
\begin{figure}
\begin{picture}(350,100)(-50,80)
\SetWidth{1.5}

\DashLine(-50,80)(350,190){5}
\DashLine(350,80)(-50,190){5}
\DashLine(-50,100)(350,100){2}

\SetColor{Red}
\Curve{(-50,80)(110,116)(120,118)(130,119)(140,120)(150,121)(160,120)(170,119)(180,118)(190,116)(350,80)}

\CCirc(150,100){13}{Black}{Yellow}

\Text(150,100)[]{\small{SUN}}
\Text(-50,80)[]{\Yellow{$\bigstar$}}
\Text(-50,70)[]{\Yellow{\normalsize{Star}}}
\Text(-50,190)[]{\Yellow{$\bigstar$}}
\Text(-50,175)[]{\Yellow{\normalsize{Apparent position}}}
\Text(-50,165)[]{\Yellow{\normalsize{of the Star}}}
\Text(350,70)[]{\Orange{\normalsize{Observer}}}
\Text(185,135)[]{\Brown{$\theta$}}
\Text(-25,95)[]{\Brown{$\epsilon_{2}$}}
\Text(330,95)[]{\Brown{$\epsilon_{1}$}}

\SetColor{Blue}
\SetWidth{1.}
\LongArrowArc(30,100)(70,180,195)
\LongArrowArc(270,100)(70,-15,0)
\LongArrowArcn(30,100)(70,195,180)
\LongArrowArcn(270,100)(70,0,-15)
\LongArrowArc(150,135)(50,-15,15)
\LongArrowArcn(150,135)(50,15,-15)
\end{picture}
\vspace*{8mm} 
\caption{Deflection of light in the gravitational field of the Sun.} 
\label{mssgrfig2}
\end{figure}
\end{center}
Let us calculate the deflection angle of the light induced by the
gravitational field of a massive body \textit{e.g.} the Sun. In order
to do this we have to calculate the one graviton exchange amplitude
between the source and the light (photon).  
The important difference with the case of non-relativistic sources is that  the electromagnetic field's  energy momentum tensor is traceless {\textit{i.e.}}:
\be
{T_{EM}}^{\mu}_{\mu}=0
\ee
This characteristic of the electromagnetic field will reveal the
different tensor structures of the propagators. In this case  the one
graviton exchange is:
\be
\Black{G^{0,m}_{N}}T_{\mu\nu}(k,q,\epsilon_{\kappa},\epsilon'_{\lambda})
G^{\mu\nu;\alpha\beta}_{0,m}T_{\alpha\beta} \rightarrow \Black{G^{0,m}_{N}}M_{SUN}
T_{00}(k,q,\epsilon_{\kappa},\epsilon'_{\lambda})
G^{00;00}_{0,m}
\ee
Given that 
\be
\theta^{0,m}\propto\Black{G^{0,m}_{N}}
\ee
and since in order to have the usual Newton's Law we should have
\be
G_{N}^{m}=\frac{3}{4} G_{N}^{0}
\ee
we get the following relationship between the deflection angles of the 
massive and massless graviton: 
\be
\theta(m\neq0)\approx\frac{3}{4}\theta(m=0)
\ee
Thus the discontinuity in the massless limit of the propagator of
massive graviton in flat spacetime  has observable phenomenological implications in
standard tests of Einsteinian gravity and particularly in
the  bending of light by the Sun. The latter expression tells us
that if gravity is due to the exchange of a massive spin 2 particle,
then the deflection angle of light would be $25\%$ smaller than if it
corresponds to the  exchange of the massless graviton.  
The fact that the  bending of the light by the sun agrees with 
the prediction of  Einstein's theory  to $1\%$  accuracy,  rules out
the possibility that gravity is due to massive graviton exchange
irrespective  of
how small the mass is\footnote{Of course there remains the possibility 
that a small
fraction of the gravitational interactions are associated with a
massive graviton component in the presence of a dominant massless
graviton component. 
 This can be realized by having an ultralight spin-2 particle with a
very small coupling compared to graviton's one \cite{Kogan:2000wc}.}. 

However, we will see in Chapter \ref{mssgr} in the presence of curvature the discontinuity disappears
allowing for non-zero graviton mass without conflict with the
phenomenology. Moreover even in the case of ``flat spacetime'' the
presence of a source curves the spacetime making the discontinuity to
disappear in  distances smaller that a characteristic scale associated 
with the  scale of the mass of the source.

\subsection{Negative tension branes}

The characteristic of the $''+-+''$ model that gives rise to the Bigravity
scenario is the bounce form of the warp factor. In the case of flat
branes that we have considered up to now it is, it is inevitably
associated with the presence of moving (not on a fixed point) negative 
tension branes. This can be easily understood from the fact that when
we try to match the solution of two positive tension branes, at the
point of the matching the jump of the derivative of the $\sigma(y)$
function has the opposite sign from the one that correspond to the
jump on positive tension branes, giving rise to a negative tension brane.

Similarly, the Multigravity models nessecarily develop moving negative 
tension branes at the points where the five dimensional spacetime
becomes flat.

However, the presence of these objects with negative energy density violate the 
weaker energy condition. This means that models with this bounce
structure cannot be generated dynamically (\textit{e.g.} from a scalar
field) - let us see why:
 In five dimensions with the metric:
\begin{equation}
ds^2=e^{-A(\rho)}\eta_{\mu \nu}dx^{\mu}dx^{\nu}+d\rho^2
\end{equation}
one can readily show that the weaker energy condition requires that:
\begin{equation}
A''\geq 0
\end{equation}
which means that the bounce is linked to moving negative tension
branes and at their position the weaker energy condition is violated.
This violation of the weaker energy condition is associated with 
the presence of scalar field(s) with negative kinetic term (ghost fields).

\subsection{Moduli fields}

In the treatment of gravitational perturbations, in the context of
the previous models, that is presented in this Chapter, we have ignored the presence
of the scalar perturbations.
These excitations describe  
the effect of the fluctuation of the size of the extra dimension and/or  
of the relative positions of the branes. We will distinguish   
these two kinds of modes by calling the former dilaton 
\cite{Charmousis:2000rg} and the  
latter radions \cite{Charmousis:2000rg,Pilo:2000et}.

In the models  considered up to now, we have imposed an orbifold symmetry $Z_2$ acting on the extra dimensional coordinate 
as $y \to -y$. When the topology of the extra dimension 
is $S^1$, the compact case, the $Z_2$ action has two fixed points $y=0, \, L_{1}$ 
and two of the  branes are sitting on fixed points.  As a result of the 
$Z_2$ symmetry the branes in  $y=0, \, L_{1}$ are frozen. Thus in the case 
of the RS1 model there is onle one scalar pertubation associated with
the size of the orbifold.

However in multi-brane constructions (\textit{e.g.} $''+-+''$ model), there are necessarily freely
moving branes,  giving rise to additional scalar perturbations,  
the radion fields, corresponding to the fluctuation of the position of these 
moving branes.

From the detailed calculation that is presented in the Appendix, for
the case of a general three brane configuration, we
have that while the kinetic term of the dilaton field is always
positive, the kinetic term of the radion is negative when the
moving brane is negative. The latter means that the presence of
negative tension moving branes is associated with the presence of
ghost fields in the theory making it theoretically unacceptable.  
Note that the models (with flat 3-branes) that exhibit the bi-gravity
or multi-gravity phenomena have moving negative tension branes.

Radion excitations play an important role in the context of   
multigravity models. As we found in the begining of this  Section, a
 generic problematic feature of multigravity models with  
flat branes is that massive gravitons have extra polarization states which do  
not decouple in the massless limit (  
van Dam - Veltman - Zakharov discontinuity \cite{vanDam:1970vg,Zakharov}).  
However, according to the previous discussion, an equally generic characteristic of these  
models is that they contain moving branes of negative  
tension.  In certain models the radion can help to recover 4D gravity on the  
brane at intermediate distances. Indeed, the role of the radion associated  
with the negative tension brane is  
precisely to cancel the unwanted  massive graviton polarizations and recover  
the correct tensorial structure of the four dimensional graviton propagator  
\cite{Dvali:2000km,Pilo:2000et}, something also seen from the bent 
brane calculations of \cite{Csaki:2000ei,Gregory:2000iu}.  
This happens because the radion in this case is a physical ghost because
it is has a wrong sign kinetic term. This fact of course makes the construction  
problematic because the system is probably quantum mechanically unstable.

Resolutions of the problems mentioned above will be given in Chapters
\ref{mssgr}, \ref{5dads} 
\chapter{Massive Gravity} 
\label{mssgr}

\section{Introduction}

In Chapter \ref{RS} we studied models that suggest that
a part  or all 
of gravitational interactions come from massive gravitons. The massive
gravitons in these models occur as the result of the dimensional reduction
of a theory of gravity in more that four dimensions, something
 well motivated from String Theory. 

In the first kind of models (\textit{e.g.} $''+-+''$ model) apart of the massless graviton, there
exist ultralight Kaluza-Klein (KK) state(s) that lead to  a ``multigravity'' scenario, in the sense that
gravitational interactions  are due to the net effect of the
massless graviton and the ultralight state(s). In this case, the Cavendish
bounds on gravitational interactions are satisfied, since it can be
arranged that the rest of the KK tower is much heavier and contributes well below the
submillimeter region. In this scenario, modifications to gravity at large
scales will appear as we probe distances of the order of the Compton
wavelength of the ultralight KK state(s). The phenomenological signature of
this will be that gravitational interactions will be reduced or almost
switched off (depending on the choice of parameters of the model) at
ultralarge scales.

In the second kind of models (\textit{e.g.} GRS model), there is no
normalizable massless mode and 4D gravity at intermediate scales is
reproduced from a resonance-like behavior \cite{Csaki:2000pp,Dvali:2000rv} of the wavefunctions of the
KK states continuum. In other words 4D gravity in this case is
effectively reproduced from a small band of KK states starting from zero mass. In this
picture modifications of gravity will begin at scales that 
correspond to the width of the resonance that appears in the form of
the coupling of these states to matter. The phenomenological signature
of these modifications will be that the four dimensional Newton's Law
(\textit{i.e.} inverse square)
will  change to a five dimensional one (\textit{i.e.} inverse cube)
at ultralarge distances. In both kind of models, these modifications
 can be confronted with current observations of the CBM
power spectrum or gravitational lensing \cite{Binetruy:2001xv,Uzan:2000mz} and are consistent with the data at present.

However, as we mentioned in Chapter \ref{RS}, these models face phenomenological difficulties in
reproducing certain predictions of General Relativity 
 \cite{Dvali:2000rv}
which are associated to the fact that the extra polarizations of the
massive KK states that contribute to gravity do not
decouple in the limit of vanishing mass, something that is known as the  van Dam
- Veltman - Zakharov \cite{vanDam:1970vg,Zakharov}
discontinuity between massless-massive spin 2 propagator. This  makes
the models of the above two classes to disagree, for example, with the measurement of the bending of
the light by the sun. The other difficulty of these models  is that the moduli (radions) associated with the perturbations of the
$''-''$ branes are necessarily physical ghost fields \cite{Dvali:2000km,Pilo:2000et,Kogan:2001qx}, therefore
unacceptable (for more detailed discussion see Appendix). The latter problem is connected to the violation of the
weaker energy condition \cite{Freedman:1999gp,Witten:2000zk} on $''-''$ branes sandwiched
between $''+''$ branes. 

In the present 	Chapter we will demonstrate that there is actually a way
out of the first problem. The second
problem can be avoided by considering models with only positive
tension branes but this will be addressed in the Chapter \ref{5dads}.
Here we demonstrate that due to an unusual
property of the graviton propagators in $dS_{4}$ or $AdS_{4}$ spacetime, we are able
to circumvent the  van Dam - Veltman - Zakharov no go theorem. In more
detail, it is known that in flat spacetime the $m \rightarrow 0$ limit
of the massive graviton propagator does not give the massless one due
to the non-decoupling of the additional longitudinal components. This
generates  the well known discontinuity between massive and massless states. Considering the massive graviton
propagator in $dS_{4}$ or $AdS_{4}$ spacetime we can show that this result
persists if $m/H \rightarrow \infty$ where $H$ is the  ``Hubble'' parameter,  \textit{i.e.} the
discontinuity is still present in the  $m\rightarrow 0$ limit if it
happens that 
$m/H \rightarrow \infty$. However,
in the case that  $m/H \rightarrow 0$, we will explicitly show that
the $m \rightarrow 0$ limit is smooth. This
 is an important result since it gives us the possibility to circumvent
the van Dam - Veltman - Zakharov no go theorem about the
non-decoupling of the  extra graviton
polarizations. Thus, in the limit that $m/H \rightarrow 0$ 
all the extra polarizations of the graviton decouple, giving an
effective theory with  massless graviton with just two polarization
states.

Here we
have to make an important comment about the relationship of the parameters
$m$, $H$. From a four dimensional point of view these parameters are
independent. However, the models that give the additional KK contributions of
massive gravitons to gravity are higher dimensional models. After the
dimensional reduction, they give us an effective four
dimensional Lagrangian in which in general $m$ and $H$
depend on common parameters (\textit{e.g.} the effective four dimensional
cosmological constant).  The behaviour of $m/H$  is  model dependent
and explicit models  that
satisfy the smoothness requirement as $m/H \rightarrow 0$  can be
found. An interesting example in the context of the $''++''$ model
will be given in Chapter \ref{5dads}.

Furthermore, in such models if we keep $m/H$ finite but small, the
extra graviton polarization states couple more weakly to matter
than the transverse states. This gives us the possibility that we can
have a model with massive gravitons  that do not violate the
observational bounds of \textit{e.g.} the bending of light by the
sun. Moreover, such models predict modifications of gravity at all
scales which could be measured by higher precision observations. However, in such a case 
even though the above can make ``multigravity'' models viable and interesting,  the condition that the
mass of the massive graviton must always scale faster that the ``Hubble'' parameter
 implies that the dramatic long distance effects of modifications of
gravity (\textit{e.g.} reduction of Newton's constant or transition
to 5D law)
will not reveal themselves until super-horizon scales. Thus, in this case the
horizon acts as a curtain that prevents the long distance modifications
of gravity due to the massive KK mode(s) to be observable.

\section{Graviton propagator in $dS_4$ and $AdS_4$ space}

In this section we will present the forms of the massless and massive
graviton propagators in the case of $dS_4$ and $AdS_4$ spacetime with arbitrary
``Hubble'' parameter $H$ and graviton mass $m$. Our purpose is to examine the behaviour of
these propagators in the limit where $H\rightarrow0$ and the limit
where the mass of the graviton tends to zero.
 For simplicity we will do our calculations in Euclidean $dS_4$ or
$AdS_4$ space. We can use for metric the one of the stereographic
projection of the sphere or the hyperboloid\footnote{Note that \cite{D'Hoker:1999jc} and
\cite{Naqvi:1999va} whose results we use in the following have a different
metric convention, but this makes no difference for our calculations.}:
\vspace{0.5cm}
\be
ds^2= \frac{\delta_{\mu\nu}}{\left(1 \mp {H^2 x^2 \over4}\right)^2}dx^{\mu}dx^{\nu}\equiv g^{0}_{\mu\nu}dx^{\mu}dx^{\nu}
\ee
where $x^2=\delta_{\mu\nu}x^{\mu}x^{\nu}$ and the scalar curvature is $R=\mp 12~H^2$. From now on, the upper
sign corresponds to $AdS_4$ space while the lower to $dS_4$ space.
The fundamental invariant in these spaces is the geodesic distance $\mu(x,y)$
between two points $x$ and $y$. For convenience, we will introduce
another invariant $u$ which is related with the geodesic distance 
by the relation $u=\cosh (H \mu)-1$ for $AdS_4$ ~($u\in[0,\infty)$)
and the relation $u=\cos (H \mu)-1$ for $dS_4$ ~($u\in[-2,0]$).  In
the small distance limit $u\sim \pm {\mu^2 H^2 \over 2}$.

This background metric is taken by the the Einstein-Hilbert action:
\be
S=\int d^4x \sqrt{g}\left(2M^2 R-\Lambda\right)
\ee
where the cosmological constant is
$\Lambda=\mp 12~H^2 M^2$ and M the 4D fundamental scale. The spin-2 massless 
graviton field can be obtained by the linear metric
fluctuations $ds^2=\left(g^0_{\mu\nu}+h_{\mu\nu}\right)dx^{\mu}dx^{\nu}$. This
procedure gives us the analog of the Pauli-Fierz graviton action in
curved space:
\ba
\frac{S_0}{2 M^2}= \int d^4x \sqrt{g^0} \left\{-{1\over 4}h \bo^0 h +{1\over
2}h^{\mu \nu}\nabla^0_{\mu}\nabla^0_{\nu}h +{1\over 4}h^{\mu \nu} \bo^0
h_{\mu \nu} - {1\over
2} h^{\mu \nu}\nabla^0_{\nu}\nabla^0_{\kappa}
h^{\kappa}_{\mu}\right.\cr \pm \left. {1\over
2}H^2\left({1\over 2}h^2 +h_{\mu \nu}h^{\mu \nu}\right)\right\} 
\ea
The above action is invariant under the gauge transformation $\delta
h_{\mu\nu}=\nabla^0_{\mu}\xi_{\nu}+\nabla^0_{\nu}\xi_{\mu}$ which
guarantees that the graviton has only two physical degrees of
freedom. This is precisely the definition of masslessness in $dS_4$ or 
$AdS_4$
space (for example see \cite{Buchbinder:2000fy} and references therein). 

The propagator of the above spin-2 massless field can be written in the form:
\be
G^{0}_{\mu\nu;\mu'\nu'}(x,y)=(\partial_{\mu}\partial_{\mu'}u\partial_{\nu}\partial_{\nu'}u+\partial_{\mu}\partial_{\nu'}u\partial_{\nu}\partial_{\mu'}u)G^{0}(u)+g_{\mu\nu}g_{\mu'\nu'}E^{0}(u)+D[\cdots]
\label{mlesspr}
\ee
where $\partial_{\mu}={\partial \over \partial x^{\mu}}$,
$\partial_{\mu'}={\partial \over \partial y^{\mu'}}$. The  last term, denoted $D[\cdots]$, is a total derivative and drops out of the
calculation when integrated with a conserved energy momentum tensor. Thus, all physical information is encoded in the first two terms.

The process of finding the functions $G^{0}$ and $E^{0}$ is quite
complicated and is the result of solving a system of six coupled differential
equations \cite{D'Hoker:1999jc}. We will  present here only the differential equation
that $G^{0}$ satisfies to show the difference between $AdS_4$ and
$dS_4$ space. This equation results from various integrations and has
the general form:
\be
u(u+2)G^{0}(u)''+4(u+1)G^{0}(u)'=C_1 +C_2 u
\label{diff}
\ee
where the constants $C_1$ and $C_2$ are to be fixed by the boundary
conditions. For the case of the $AdS_4$ space \cite{D'Hoker:1999jc}, these
constants were set to zero so that the  $G^{0}$ function vanishes at
the boundary  at infinity ($u \rightarrow \infty$). Using the same condition also for the  $E^{0}$ function, the 
exact form of them was found to be:
\ba
G^{0}(u)&=&\frac{1}{8 \pi^2 H^2}\left[\frac{2(u+1)}{u(u+2)}-\log \frac{u+2}{u}\right]\cr
E^{0}(u)&=&-\frac{ H^2}{8
\pi^2}\left[\frac{2(u+1)}{u(u+2)}+4(u+1)-2(u+1)^2\log \frac{u+2}{u}\right]
\ea

For the case of the $dS_4$ space we iterated the procedure of
Ref.\cite{D'Hoker:1999jc} imposing the condition \cite{Allen:1986wd} that the  $G^{0}$ and $E^{0}$
functions should be non-singular at the antipodal point
($u=-2$). The constants $C_1$ and $C_2$ were kept non-zero and played
a crucial role in finding a consistent solution. It is straightforward 
to find the full expression of these functions, but we only need to know their short distance behaviour. Then with this accuracy the answer is:
\ba
G^{0}(u)&=&-\frac{1}{8 \pi^2 H^2}\left[\frac{1}{u}+\log (-u)\right] +
\cdots \cr
E^{0}(u)&=&\frac{H^2}{8
\pi^2}\left[\frac{1}{u}+2(u+1)^2\log (-u)\right]+\cdots 
\ea

If we define $\Pi^{0}(u)=\frac{1}{H^4}\frac{E^{0}(u)}{G^{0}(u)}$, then 
for short distances ($H^2x^2 \ll 1$) where $u \rightarrow 0$ we get:
\ba
g_{\mu\nu}g_{\mu'\nu'}&\rightarrow & \delta_{\mu\nu} \delta_{\mu'\nu'} 
\cr
\partial_{\mu}\partial_{\nu}u &\rightarrow& \mp H^2 \delta_{\mu \nu} \cr G^0(u) &\rightarrow& {1 \over 4 \pi^2 H^4 \mu^2} \cr
\Pi^{0}(u) &\rightarrow& -1
\ea
and so we recover
the short distance limit of the massless flat Euclidean space propagator:
\be
G^{0}_{\mu\nu;\mu'\nu'}(x,y)=\frac{1}{4 \pi^2
\mu^2}(\delta_{\mu\mu'}\delta_{\nu\nu'}+\delta_{\mu\nu'}\delta_{\nu\mu'}-\delta_{\mu\nu}\delta_{\mu\nu'})+\cdots
\label{mlessprop}
\ee
Of course this is just as expected.

In order to describe a spin-2 massive field it is necessary to add to the above 
action  a Pauli-Fierz mass term:
\be
\frac{S_m}{2 M^2}=\frac{S_0}{2 M^2}- \frac{m^2}{4}\int d^4 x \sqrt{g^0}(h_{\mu \nu}h^{\mu \nu}-h^2)
\ee
By adding this term we immediately lose the gauge invariance
associated with the $dS_4$ or $AdS_4$ symmetry group and the
massive gravitons acquire five degrees of freedom.

The propagator of this massive spin-2 field  can again be 
written in the form:
\be
G^{m}_{\mu\nu;\mu'\nu'}(x,y)=(\partial_{\mu}\partial_{\mu'}u\partial_{\nu}\partial_{\nu'}u+\partial_{\mu}\partial_{\nu'}u\partial_{\nu}\partial_{\mu'}u)G^{m}(u)+g_{\mu\nu}g_{\mu'\nu'}E^{m}(u)+D[\cdots]
\label{mivepr}
\ee
The last term
of the propagator in eq. (\ref{mivepr}), denoted $D[\cdots]$, is again a total derivative and thus
drops out of the calculation when integrated with a conserved $T_{\mu
\nu}$.

At his point we should emphasize that in case of an arbitrary massive
spin-2 field, the absence of
gauge invariance means that there is no guarantee that the field will
couple to a conserved current. However, in the context of a higher
dimensional theory  whose symmetry group is spontaneously broken by
some choice of vacuum metric,  the massive spin-2 graviton KK states 
couple to a conserved $T_{\mu \nu}$. One can understand this by the
following example. Consider the case of the most simple KK
theory, the one with one compact extra dimension. By the time 
we choose a vacuum metric \textit{e.g.} $g^0_{MN}={\rm diag}\left(\eta_{\mu \nu},1\right)$, the higher dimensional
symmetry is broken. If we denote the
graviton fluctuations around
the background metric by $h_{\mu \nu}$, $h_{\mu 5}$ and $h_{55}$, there is still the gauge freedom:
\ba
\delta h_{\mu \nu}&=&\partial_{\mu}\xi_{\nu}+\partial_{\nu}\xi_{\mu}\nonumber\\
\delta h_{\mu 5}&=&\partial_{\mu}\xi_{5}+\partial_{5}\xi_{\mu}\label{gauge}\\
\delta h_{5 5}&=&2\partial_{5}\xi_{5}\nonumber
\ea

If we Fourier decompose these fields, their $n$-th Fourier mode acquires
a mass $m_n \propto n$ with $n=0,1,2,\dots$, but there is mixing
between them. This means for example that $h^{(n)}_{\mu \nu}$ is not a 
massive spin-2 eigenstate \textit{etc.}. However, we can exploit the gauge
transformations (\ref{gauge}) to gauge away the massive $h^{(n)}_{\mu
5}$ and  $h^{(n)}_{55}$ and construct a pure spin-2 field (see for
example \cite{Cho:1992rq} and references therein). For a comprehensive account of KK theories see \cite{Appelquist:1987nr}. The new massive spin-2 field $\rho^{(n)}_{\mu \nu}$ is
invariant under (\ref{gauge}) and so its Lagrangian does not exhibit a
gauge invariance of the form $\delta \rho_{\mu
\nu}=\partial_{\mu}\chi_{\nu}+\partial_{\nu}\chi_{\mu}$. However,
since is originates from a Lagrangian that has the gauge invariance
(\ref{gauge}), it is bound to couple to a conserved $T_{\mu
\nu}$. The argument goes on for more complicated choices of vacuum
metric as for example warped metrics which are recently very popular
in brane-world constructions.

Again the functions $G^{m}$ and $E^{m}$ result from a complicated system of differential
equations \cite{Naqvi:1999va}. In that case, the differential equation
that $G^{m}$ satisfies is:
\be
u(u+2)G^{m}(u)''+4(u+1)G^{m}(u)'\mp \left({m\over H}\right)^2 G^{m}(u)=C_1 +C_2 u
\label{diffm}
\ee
where the constants $C_1$ and $C_2$ are to be fixed by the boundary
conditions. For the case of the $AdS_4$ space \cite{Naqvi:1999va}, these
constants were set to zero so that the  $G^{0}$ function vanishes at
the boundary  at infinity. Imposing 
additionally  the condition of fastest falloff at infinity ($u
\rightarrow \infty$) \cite{Allen:1986wd}, the exact form of the  $G^{m}$ and  $E^{0}$
function was found to be:
\ba
G^{m}(u)&=&\frac{\Gamma(\Delta)\Gamma(\Delta-1)}{16 \pi^2
\Gamma(2\Delta-2)
H^2}\left(\frac{2}{u}\right)^{\Delta}F(\Delta,\Delta-1,2\Delta-2,-{2\over 
u})\cr
E^{m}(u)&=&-\frac{2}{3}~\frac{\Gamma(\Delta-1) H^2}{16
\pi^2 \Gamma(2\Delta
-2)[2+(m/H)^2]}\left(\frac{2}{u}\right)^{\Delta}\times\cr
& \times & \left\{ \renewcommand{\arraystretch}{1.5} \begin{array}{l} \phantom{-} 3[2+(m/H)^2]
\Gamma(\Delta-2) u^2 F(\Delta-1,\Delta-2,2\Delta-2,-{2\over u})
\\-3(u+1)u F(\Delta-1,\Delta-1,2\Delta-2,-{2\over u})\\+[3+(m/H)^2]\Gamma(\Delta)F(\Delta,\Delta-1,2\Delta-2,-{2\over u}) \end{array} \right\}
\ea
where $\Delta={3\over 2}+{1\over 2}\sqrt{9+4(m/H)^2}$. 

For the case of the $dS_4$ space we iterated the procedure of
Ref.\cite{Naqvi:1999va} imposing the condition \cite{Allen:1986wd} that the  $G^{m}$ and $E^{m}$
functions should be non-singular at the antipodal point
($u=-2$) and also finite as $m \rightarrow 0$. Again we kept the
constants $C_1$ and $C_2$  non-zero to obtain a consistent solution. It is straightforward to find the full expression of these functions, but we only need to know their short distance behaviour. Then with this accuracy the answer is:
\ba
G^{m}(u)&=&\frac{\Gamma(\Delta)\Gamma(3-\Delta)}{16 \pi^2
H^2}\left[F(\Delta,3-\Delta,2,{u+2\over 2})-1\right]+\cdots \cr
E^{m}(u)&=&-\frac{2}{3}~\frac{\Gamma(\Delta)\Gamma(3-\Delta) H^2}{4
\pi^2 [2-(m/H)^2]}~\times\cr
& \times & \left\{ \renewcommand{\arraystretch}{1.5} \begin{array}{l} -3[2-(m/H)^2]\left[{2(u+2)\over(\Delta-1)(\Delta-2)}F(\Delta-1,2-\Delta,2,{u+2\over 2})+{u(u+2)\over2}\right]
\\-3(u+1)\left[{2\over(\Delta-1)(\Delta-2)} F(\Delta-1,2-\Delta,1,{u+2\over 2})+(u+1)\right]
\\+[3-(m/H)^2]\left[F(\Delta,3-\Delta,2,{u+2\over 2})-1\right]
\end{array} \right\}+\cdots \phantom{aaa}
\ea
where $\Delta={3\over 2}+{1\over 2}\sqrt{9-4(m/H)^2}$. 

If we define
$\Pi^{m}(u)=\frac{1}{H^4}\frac{E^{m}(u)}{G^{m}(u)}$, then 
for short distances ($H^2x^2 \ll 1$) where $u \rightarrow 0$ we get:
\ba
G^m(u) &\rightarrow& {1 \over 4 \pi^2 H^4 \mu^2} \cr
\Pi^m(u) &\rightarrow& -{2\over 3}~\frac{~~3 \pm \left( {m \over H}
\right)^2}{~~2 \pm \left( {m \over H} \right)^2}
\label{ratio}
\ea

It is interesting to consider two massless flat limits. In the first
one  $m \rightarrow 0$ and $H \rightarrow 0$  while ~$m/H \rightarrow
\infty$. In this case, from
eq. (\ref{ratio}) we see that we
recover the Euclidean propagator for a massive graviton in flat space:
\be
G^{m}_{\mu\nu;\mu'\nu'}(x,y)=\frac{1}{4 \pi^2
\mu^2}(\delta_{\mu\mu'}\delta_{\nu\nu'}+\delta_{\mu\nu'}\delta_{\nu\mu'}-{2\over 
3}\delta_{\mu\nu}\delta_{\mu'\nu'})+\cdots
\ee
This is in agreement with the van Dam - Veltman - Zakharov
theorem. The second limit has $m \rightarrow 0$ and $H \rightarrow 0$
but $m/H \rightarrow 0$. In this case the propagator passes smoothly
to the one of the flat massless
case (\ref{mlessprop}):
\be
G^{0}_{\mu\nu;\mu'\nu'}(x,y)=\frac{1}{4 \pi^2
\mu^2}(\delta_{\mu\mu'}\delta_{\nu\nu'}+\delta_{\mu\nu'}\delta_{\nu\mu'}-\delta_{\mu\nu}\delta_{\mu\nu'})+\cdots
\ee
This is in contrary to the van Dam - Veltman  - 
Zakharov discontinuity in flat
space.

In general, we may consider the limit with $m/H$ finite. Then for
small   $m/H$ the contribution to the $\delta_{\mu\nu}\delta_{\mu'\nu'}$ 
structure is  $ - 1  \pm m^2/6 H^2$.  Since observations agree to $1\%$ 
accuracy with the prediction of Einstein gravitational theory for the
bending of light by the sun, we obtain the limit ${m
\over H} \la 0.1$.

\section{Discussions and conclusions}

In summary, in this Chapter we showed that, by considering physics in
$dS_{4}$ or$AdS_{4}$ spacetime, 
one can circumvent the van Dam - Veltman - Zakharov 
theorem about non-decoupling of  the extra polarization states of a massive
graviton. It was shown that the smoothness of the $m\rightarrow{0}$
limit is ensured if the $H$ (``Hubble'') parameter, associated with
the horizon of $dS_{4}$ or
$AdS_{4}$ space, tends to zero slower than $m$. The above  requirement
can be  realized in various models and an interesting example
will be given in the next Chapter. Furthermore, if we keep $m/H$
finite, we can obtain models where massive
gravitons contribute to 
gravity and still have acceptable and interesting
phenomenology. Gravity will then be modified at all scales with
testable differences from the Einstein theory in future higher
precision observations. However, the dramatic modifications of gravity at large
distances that ``multigravity'' models suggest, will be hidden by the existence of the horizon which will
always be well before the scales that modifications would become
relevant.

 One loop effects in
the massive graviton propagator in $AdS_4$ were discussed  in 
\cite{Dilkes:2001av,Duff:2001zz}. 
Of course, purely four-dimensional theory with massive
graviton is  not  well-defined  and it 
 is certainly true that if the mass term is 
added by hand in purely four-dimensional theory a lot of problems will 
emerge as it was shown in the classical paper of 
\cite{Boulware:1972my}.  If however  the  underlying theory is  a 
higher dimensional one, the graviton(s)
mass terms appear dynamically and this is a different story. 
All quantum corrections must be calculated in a 
higher-dimensional theory, where a larger number of
graviton degrees of freedom is present naturally (a massless
five-dimensional graviton has the same number of  degrees of freedom
as a massive four-dimensional one). 

  Moreover, the smoothness
of the limit $m \rightarrow  0$ 
  is not only a property of the $AdS_4$ space but holds 
for any background where the characteristic curvature invariants are  
non-zero \cite{Kogan:2001qx}. For physical processes taking place 
 in some region a curved space with a characteristic average curvature, 
the effect of graviton mass is controlled by positive powers of the 
ratios  $m^2/R^2$  where $R^2$ is a characteristic curvature invariant
(made
from Riemann and Ricci tensors or scalar curvature).
  A very interesting  argument  supporting  the
conjecture that  there is a smooth limit for phenomenologically 
 observable amplitudes in brane gravity  with ultra-light gravitons 
 is based on a very interesting paper   \cite{Vain}

In that paper it  was shown  that there is a smooth
limit for a   metric  around a spherically symmetric  source with a
mass  $M$   in a theory with
massive graviton  with  mass $m$ for  small (\textit{i.e.} smaller than 
 $m^{-1}(mM/M_P^2)^{1/5}$) distances.
   The discontinuity reveals itself at large distances. The
non-perturbative  solution discussed in \cite{Vain} was found in a 
limited range of distance from
the center and it is still unclear if it can be
smoothly continued to spatial infinity (this problem was stressed in 
 \cite{Boulware:1972my}). Existence of this smooth
continuation depends on the full  nonlinear structure of the
theory. If one adds a mass term by hand the smooth asymptotic at
infinity may not exit.  However, it 
 seems  plausible   that 
 in all cases when modification of gravity at large distances comes
from consistent higher-dimensional models, the global  smooth solution 
can exist because in this case there is 
a unique non-linear structure related to the mass term  which is
dictated by the  underlying 
higher-dimensional theory. In  a  paper \cite{Deffayet:2001uk} an 
example of a 5D cosmological solution  was discussed which contains an
explicit 
interpolation between perturbative and non-perturbative regimes: a 
direct analog
of large and small distances in the Schwarschild case.

\chapter{5D Bigravity from $AdS_{4}$ branes} 
\label{5dads}

\section{Introduction}

The $''+-+''$ model was the prototype  model of the  class of brane
universe models  that suggest that
a part or all of gravitational interactions come from massive
gravitons. In the first case (``bigravity''), gravitational interactions
are the net effect of a massless graviton and a finite number of KK
states that have sufficiently small mass (or/and coupling) so that
there is no conflict with phenomenology. 
In the second case (``multigravity''), in the absence of a massive
graviton, the normal Newtonian gravity at intermediate scales is reproduced by
special properties of the lowest part of the KK tower (with mass close to zero). These models predict that at
sufficiently large scales, which correspond to the Compton wavelength
of the KK states involved, 
modifications to either  the Newtonian coupling constant or  the inverse
square law will appear due to the Yukawa suppression of the KK states
contribution.

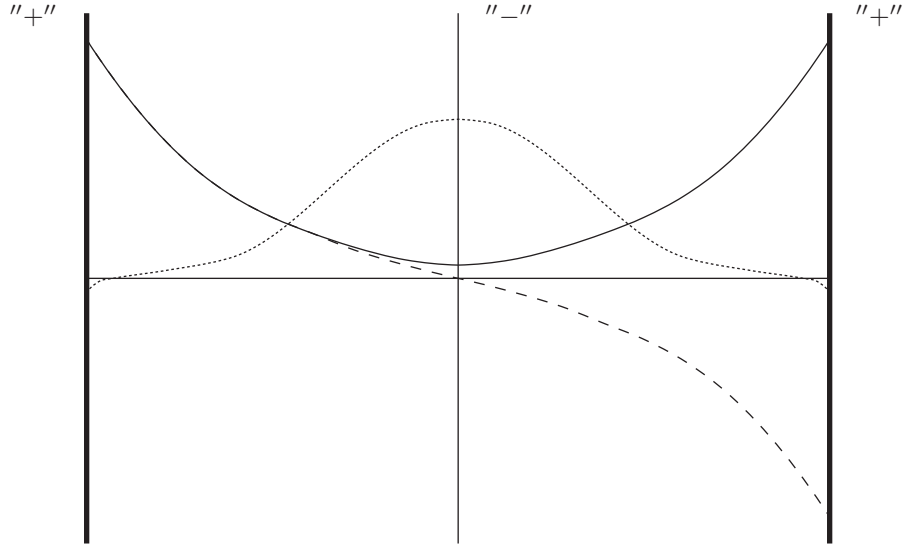
\begin{figure}[t]
\begin{center}
\begin{picture}(300,200)(0,50)

\SetWidth{2}
\Line(10,50)(10,250)
\Line(290,50)(290,250)

\SetWidth{0.5}
\Line(150,50)(150,250)
\Line(10,150)(290,150)

\Text(-10,250)[c]{$''+''$}
\Text(310,250)[c]{$''+''$}
\Text(170,250)[c]{$''-''$}


\Curve{(10,240)(50,192)(65,181)(80,173)(95,167)(110,162)(130,157)(150,155)}
\Curve{(150,155)(170,157)(190,162)(205,167)(220,173)(235,181)(250,192)(290,240)}


\DashCurve{(10,240)(50,192)(65,181)(80,173)(95,167)(110,161)(130,155)(150,150)}{4}
\DashCurve{(150,150)(170,145)(190,139)(205,133)(220,127)(235,119)(250,108)(290,60)}{4}


\DashCurve{(10,145)(15,149)(20,150)(40,153)(65,158)(140,209)(150,210)}{1}
\DashCurve{(150,210)(160,209)(235,158)(260,153)(280,150)(285,149)(290,145)}{1}

\end{picture}
\end{center}

\caption{The graviton (solid line), first (dashed line) and second
(dotted line) KK states wavefunctions in the symmetric $''+-+''$ model. The
wavefunctions are not smooth on the $''-''$ branes. The same pattern
have also the $''++''$ model wavefunctions with the position of the
$''-''$ brane corresponding to the minimum of the warp factor. The
wavefunctions are then smooth.}
\label{wfunct}

\end{figure}

However, the above construction and its variations, have two important defects. The first
one \cite{Dvali:2000rv}
is that the extra polarizations of the massive KK states do not
decouple in the limit of vanishing mass, the famous van Dam
- Veltman - Zakharov \cite{vanDam:1970vg,Zakharov}
discontinuity, which makes the model  disagree with the experimental
measurements of the bending of
the light by the sun. The second 
one \cite{Dvali:2000km,Pilo:2000et,Kogan:2001qx} is that the moduli (radions) associated with the perturbations of the
$''-''$ branes are necessarily physical ghost fields, therefore
unacceptable. The latter problem is connected to the violation of the
weaker energy condition \cite{Freedman:1999gp,Witten:2000zk} on $''-''$ branes sandwiched between $''+''$
 branes.  Two mechanisms for solving these problems
have been suggested. The first mechanism  \cite{Csaki:2000ei,Gregory:2000iu} involves a cancellation of the
additional polarizations by the radion fields, but has necessarily the 
ghost field problem. The second mechanism \cite{Kogan:2000cv,Kogan:2000xc} involves a
cancellation of both the extra massive graviton polarizations and the
radion field by some bulk dynamics which are necessary to stabilize
the system, based on a scenario described in \cite{Kanti:2000nz}. The latter mechanism is however non-local in the extra
dimension and because of this may not be very attractive. 

In the present Chapter, with the help of results of Chapter \ref{mssgr}, we will demonstrate that there is actually a way
out of both these problems. We will use a two brane model with only
$''+''$ branes which was known in
\cite{Kaloper:1999sm,Kanti:2000nz,Kanti:2000rd,Kanti:2000cz} to exhibit a bounce
of the ``warp'' factor, and therefore is bound to have ``bigravity''
by general arguments presented in Chapter \ref{intro}. This can 
be achieved if we consider two $AdS_4$ branes in $AdS_5$ bulk. The
weaker energy condition is satisfied so there is 
no ghost modulus in this setup. Furthermore, as was shown in
Chapter \ref{mssgr}, in $AdS$ space it is possible to  circumvent the
van Dam - Veltman - Zakharov no go theorem about the non-decoupling of 
the massive graviton extra polarization states. The price we pay
is that there is a remnant negative cosmological constant on the
brane. This sets an horizon scale and unfortunately the Compton
wavelength of the  light state of  the system lies exponentially far
from this horizon. This does not change even if we consider a highly asymmetric
version of this model. As a result, this ``bigravity'' model makes
no predictions for observable deviations from Newtonian gravity at
ultra-large distances. In addition, although theoretically there exist
modifications of  General Relativity  at all scales due to the
additional polarization states of the massive graviton, they are 
so highly suppressed that they are not observable.

\sect{The two positive brane model}

The model consists of two  3-branes with tensions $V_{1}$ and $V_{2}$ respectively, in an $AdS_5$ space with
five dimensional cosmological constant $\Lambda<0$. The $5$-th dimension has the geometry
of an orbifold and the branes are located at its fixed points,
\textit{i.e.}  $L_0=0$ and $L_{1}=L$. Due to orbifolding, we can
restrict ourselves to the region $0\leq{z}\leq{L}$, imposing the suitable boundary conditions to our solutions. Firstly, we  find a suitable vacuum solution. The action of this setup is:
\begin{equation}
S=\int d^4 x \int_{-L}^{L} dz \sqrt{-G} 
\{-\Lambda + 2 M^3 R\}+\sum_{i}\int_{z=L_i}d^4xV_i\sqrt{-\hat{G}^{(i)}}
\end{equation}
where $\hat{G}^{(i)}_{\mu\nu}$ is the induced metric on the branes. 
 The Einstein equations that arise from this
action are:
\be
R_{MN}-\frac{1}{2}G_{MN}R=-\frac{1}{4M^3}
\left(\Lambda G_{MN}-
\sum_{i}V_i\frac{\sqrt{-\hat{G}^{(i)}}}{\sqrt{-G}}
\hat{G}^{(i)}_{\mu\nu}\delta_M^{\mu}\delta_N^{\nu}\delta(z-L_i)\right)
\ee
In order to find a specific form of the equations of motion we need to
write a metric ansatz which will take in account the spacetime symmetries of the
3-brane. Since we would like not to restrict our model to flat solutions on the
branes, we should make a choice which will let us interpolate between
the maximally symmetric space-times in four dimensions, \textit{i.e.} de-Sitter,
Minkowski and Anti-de-Sitter. The metric ansatz \cite{Kaloper:1999sm} that accomplishes this is the following:
\be
ds^2=a^{2}(z)(-dt^{2}+e^{2Ht}d\vec{x}^{2}) +b^{2}(z)dz^2
\ee
where $H$ is the ``Hubble'' parameter and is determined in terms of the brane
tension $V_{1}$ and the bulk cosmological constant  $\Lambda$ from
Einstein's equations. The $z$-dependent function $a(z)$ is the ``warp''
factor that is essential for the localization of gravity and also for
producing the hierarchy between the two branes. 
 In the  case of flat brane solution, \textit{i.e.} the
effective cosmological constant on the brane is zero, we have $H=0$. On
the other hand if we demand a de-Sitter solution on the brane,
\textit{i.e.} the effective cosmological constant on the branes is
positive, we have
$H^{2}>0$. In the case of Anti-de-Sitter solution, \textit{i.e.} the effective
cosmological constant on the branes is negative,  we have $H^{2}<0$
and thus $H$ is imaginary. In order to get a physical 
interpretation of the latter case it is necessary to analytically continue the solution by
a coordinate transformation of the form $t=-ix_1'$, $x_1=it'$, $x_2=x_2'$
and $x_3=x_3'$. After this transformation the metric ansatz can be written
in the  
following form:
\be
ds^2=a^{2}(z)(dx_1'^{2}+e^{2Hx_1'}(-dt'^2+dx_2'^2+dx_3'^2)) +b^{2}(z)dz^2
\ee
Furthermore, in order to have a more compact notation for all cases of maximally
symmetric spaces and
simplify our calculations, it is useful to bring the metric ansatz in the
form:
\be
ds^2=\frac{a^2(z)}{(1-\frac{H^{2}x^{2}}{4})^2}\eta_{\mu\nu}dx^{\mu}dx^{\nu} +b^2(z)dz^2
\ee
where $x^2=\eta_{\mu\nu}x^{\mu}x^{\nu}$. It can be shown that the
Ricci scalar for this metric is $R=-12H^2$. Thus  this metric represents all
maximally symmetric spaces: Minkowski for $H^2=0$, Anti-de-Sitter for
$H^2>0$ and de-Sitter for $H^2<0$. From
now on we shall choose the gauge $b(z)=1$ where our coordinate system
is Gaussian Normal. 
A straightforward
calculation of the Einstein's equations gives us the following differential equations for $a(z)$:
\ba
a'^2(z)&=&H^2-\frac{\Lambda}{24M^3}a^2(z)\\
a ''(z)&=&-\sum_{i}\frac{V_i}{12M^3}a(z)\delta(z-L_i)-\frac{\Lambda}{24M^3}a(z)
\ea
By solving the above equations we find that the solution can be written
in the form :
\be
a(z)=\cosh(k|z|)+\frac{V_{1}k}{\Lambda}\sinh(k|z|)
\ee
with
\be
\renewcommand{\arraystretch}{1.5}
|H^2|=\left\{\begin{array}{cl}\frac{k^2}{\Lambda^2}(V_{1}^{2}k^2-\Lambda^2)&,\frac{|\Lambda|}{k}<V_{1}
~~{\rm for}~dS_{4}~{\rm branes}\\
0&,\frac{|\Lambda|}{k}=V_{1}~~{\rm for~flat~branes}\\
\frac{k^2}{\Lambda^2}(\Lambda^2-V_{1}^{2}k^{2})&,\frac{|\Lambda|}{k}>V_{1}~~{\rm for}~AdS_{4}~{\rm branes}\end{array}\right.
\
\ee
where we have normalized $a(0)=1$ and assumed $V_{1}>0$. Also we have
defined $k\equiv{\sqrt{\frac{-\Lambda}{24 M^3}}}$.

Additionally, in order to have this solution, the brane tensions
$V_{1}$, $V_{2}$, the bulk cosmological
constant $|\Lambda|$ and the position of the second brane $L$ must be
related through the equation: 
\be
\tanh(kL)=k|\Lambda|\frac{V_{1}+V_{2}}{|\Lambda|^{2}+k^{2}V_{1}V_{2}}
\label{ten}
\ee
Let us now restrict ourselves to the case of $AdS_{4}$ spacetime  on the two
branes which will turn out to be the most interesting. In this case
the condition $\frac{|\Lambda|}{k}>V_{1}$ must hold. Hence, we can
define $\tanh(kz_{0})\equiv\frac{k V_{1}}{|\Lambda|}$ and
write the solution in the form: 
\be
a(z)=\frac{\cosh(k(z_{0}-|z|))}{\cosh(kz_{0})}
\ee
from which it is clear that the ``warp'' factor has a minimum at
$z=z_{0}$. From this point we can see the role of the
$AdS_{4}$ on the branes, \textit{i.e.} the role of the condition
$\frac{|\Lambda|}{k}>V_{1}$. This condition allows us to have the bounce
form of the ``warp'' factor (\textit{i.e.} a minimum in the ``warp''
factor) allowing the second brane to have
positive tension and give us, as we will see shortly, a
 phenomenology quite similar to the $''+-+''$ model.
This can be easily seen from the eq.(\ref{ten}) which relates the brane
tensions and the distance between the branes. From this equation we indeed
see that by placing the second brane
after the minimum of the ``warp'' factor we can make the tension of
the second brane positive and thus both branes that appear in the model
have positive tension avoiding the problems associated with 
negative tension branes. 
In fact it is clear that the present model mimics the characteristics
of the $''+-+''$ model since what we effectively do is to reproduce the
effect of the presence of a negative tension brane, \textit{i.e.} the bounce form
of the ``warp'' factor, with another mechanism allowing a negative four
dimensional cosmological constant on the brane. Note that in the
limit that ${V_{1}}\rightarrow\frac{|\Lambda|}{k}$ (flat limit) the
minimum of the ``warp'' factor tends to infinity and if we wish to
have a brane at a finite point, it will necessarily have negative
tension.

The relationship between the 4D effective fundamental scale  $M_{*}$\footnote{the factor $2M^{2}_{*}$ multiplies the four dimensional Ricci scalar
in the Lagrangian after dimensionally reducing} and the five dimensional
fundamental scale $M$ can be easily found by dimensional reduction to be:
\be
M_{\rm
*}^2=\frac{M^3}{k\cosh^{2}(kz_{0})}\left[kL+\sinh(kL)\cosh(k(L-2z_{0}))\right]
\label{plank}
\ee
The above formula tells us that for finite $L$ the compactification
volume is finite and thus the zero mode is normalizable. In the case
where we send the second brane to infinity, the compactification
volume becomes infinite which indicates that the zero mode becomes
non-normalizable. Note that $M_{*}$ is not
necessarily equal to $M_{Pl}$ since as will see shortly, at least for
a sector of the parameter space of our model, gravity is the result not
only of the massless graviton but also of an ultralight KK state.

The ``warp'' factor renormalizes the physical scales of the theory as
 in the RS1 model. Thus, all 
mass parameters $m_0$ on a brane placed at the point $z$ are rescaled as
\be
m=a(z)m_{0}
\ee
Hence, assuming some kind of stabilization mechanism which fixes the
positions of the branes, one can choose  a distance between the
two branes such that this rescaling  leads to the creation of a desired mass
hierarchy.

However, since we consider non-flat solutions on the branes, we
have to make sure that the four dimensional effective cosmological
constant does not contradict  present experimental and
observational bounds. Recent  experimental data favour a positive
cosmological constant, nevertheless since
zero cosmological constant is not completely ruled out it can be argued that also
a tiny negative cosmological constant can be acceptable within the experimental
uncertainties. The effective cosmological constant on the two branes
is:
\be
\Lambda_{4d}=-12H^{2}M_{*}^2=-\frac{12}{\cosh^2(kz_{0})}k^{2}M_{*}^{2}
\ee
From the above formula we can see that we can make the cosmological
constant small enough $|\Lambda_{4d}| \la 10^{-120} M_{\rm Pl}^4$ if we
choose large enough $kz_{0}$, \textit{i.e.} $kz_{0}\ga {135}$. This
however will make observable deviations from Newtonian gravity at
ultra-large scales impossible as we will see in the 
next section.

To determine the phenomenology of the model we need to know the KK
spectrum that follows from the dimensional reduction. This is
determined by considering the (linear) fluctuations of the metric
around the vacuum solution that we found above. We can write the
metric perturbation in the form:
\be
ds^2=\left[a(z)^{2}g^{AdS}_{~\mu\nu} +\frac{2}{M^{3/2}}h_{\mu\nu}
(x,z)\right]dx^\mu dx^\nu +dz^2
\ee
where $g^{AdS}_{~\mu\nu}$ is the vacuum solution. Here we have ignored the radion mode that could be used to stabilize
the brane positions $z=L_{0}$ and $z=L_{1}$, assuming some
stabilization mechanism. We
expand the field $h_{\mu\nu}(x,z)$ into graviton and KK plane waves:
\be
h_{\mu\nu}(x,z)=\sum_{n=0}^{\infty}h_{\mu\nu}^{(n)}(x)\Psi^{(n)}(z)
\ee
where we demand
$\left(\nabla_\kappa\nabla^\kappa +2 H^2-m_n^2\right)h_{\mu\nu}^{(n)}=0$
and additionally 
\mbox{$\nabla^{\alpha}h_{\alpha\beta}^{(n)}=h_{\phantom{-}\alpha}^{(n)\alpha}=0$}.
The function $\Psi^{(n)}(z)$ will obey a second order differential
equation which after a change of variables and a redefinition of the wavefunction reduces to an ordinary
Schr\"{o}dinger-type equation:
\be
\left\{-
\frac{1}{2}\partial_w^2+V(w)\right\}\hat{\Psi}^{(n)}(w)=\frac{m_n^2}{2}\hat{\Psi
}^{(n)}(w)
\ee
where the potential is given by:
\ba
V(w)=&-&
\frac{9\tilde{k}^{2}}{8}~+~\frac{15\tilde{k}^2}{8}\frac{1}{\cos^{2}\left(\tilde{k}(|w|-w_{0})\right)}\cr      &-&\frac{3k}{2}\left[ \tanh(kz_{0})\delta(w)+\frac{\sinh(k(L-z_{0}))\cosh(k(L-z_{0}))}{\cosh^{2}(kz_{0})}
\delta(w-w_{1})\right] 
\ea
with $\tilde{k}$  defined as
$\tilde{k}\equiv{\frac{k}{\cosh(kz_{0})}}$. The new variables  and the redefinition of the wavefunction are
related with the old ones by:
\be
w\equiv {\rm sgn}(z)\frac{2}{\tilde{k}}\left[\arctan\left(\tanh(\frac{k(|z|-z_{0})}{2})\right)+\arctan\left(\tanh(\frac{kz_{0}}{2})\right)\right]
\
\ee
\be
\hat{\Psi}^{(n)}(w)\equiv\frac{1}{\sqrt{a(z)}}\Psi^{(n)}(z)
\ee
Thus in terms of the new coordinates, the branes are  at $w_{L_{0}}=0$
 and $w_{L}$, with the minimum of the potential  at $w_{0}={2 \over \tilde{k}}\arctan\left(\tanh(\frac{kz_{0}}{2})\right)$. Also note
that with this transformation the point $z=\infty$ is mapped to the
finite point $w_{\infty}={2 \over \tilde{k}}\left[{\pi \over 4} + \arctan\left(\tanh(\frac{kz_{0}}{2})\right)\right]$.

From now on we restrict ourselves to the symmetric configuration of the two
branes with respect to the minimum  $w_{0}$ (\textit{i.e.} the first
brane at 0 and the
second at  $2w_{0}$ ), since the important characteristics of the model
appear independently of the details of the configuration.
Thus, the model has been reduced to a ``quantum mechanical problem''
with $\delta$-function potentials wells of
the same weight and an extra smoothing term in-between (due to the AdS
geometry). This  gives the potential a double ``volcano'' form. 

An interesting characteristic of this potential is that it always (for
the compact cases \textit{i.e.} $w_{L}<w_{\infty}$)
gives rise to a normalizable massless zero mode, something that is
expected since the volume of the extra dimension is finite.  The zero
mode wavefunction is given by:
\be
\hat{\Psi}^{(0)}(w)=\frac{A}{[\cos(\tilde{k}(w_{0}-|w|))]^{3/2}}
\ee
where the normalization factor $A$ is determined by the requirement 
$\displaystyle{\int_{-w_{L}}^{w_{L}}
dw\left[\hat{\Psi}^{(0)}(w)\right]^2=1}$, chosen so that we get the standard 
form 
of the Fierz-Pauli Lagrangian.

The form of the zero mode resembles the one  of the zero mode of the $''+-+''$
model, \textit{i.e.} it has a bounce form with the turning at $w_{0}$ 
(see figure 1). In the
case of the $''+-+''$ the cause for this was the presence of the
$''-''$ brane. In the present model it turns out that by considering
$AdS$ spacetime on the branes we get the same effect.

In the case that we send the second brane to infinity
(\textit{i.e.} $w\rightarrow {w_{\infty}}$) the zero mode fails to be normalizable
due to singularity of the wavefunction exactly at that point. This can
be also seen from eq.(\ref{plank}) which implies that at this limit $M_{*}$
becomes infinite (\textit{i.e.} the coupling of the zero mode becomes zero). Thus in this limit the
model has no zero mode and all gravitational interactions must be
produced by the ultralight first KK mode\footnote{The spectrum in this
case was discussed by L. Randall and A. Karch at \cite{Karch:2001ct}.}. 

Considering the Schr\"{o}dinger equation for $m\ne0$ we can determine
the wavefunctions of the KK tower. It turns out that the differential
equation can be brought to a hypergeometric form, and hence  the
general solution  is given in terms two hypergeometric functions: 
\be
\renewcommand{\arraystretch}{1.5}
\begin{array}{c}\hat{\Psi}^{(n)}=\cos^{5/2}(\tilde{k}(|w|-w_{0}))\left[C_{1}~F(\tilde{a}_{n},\tilde{b}_{n},\frac{1}{2};\sin^{2}(\tilde{k}(|w|-w_{0})))~~~~~~~~\right.
\\ \left.  ~~~~~~~~~~~~~~~~~~~+C_{2}~|\sin(\tilde{k}(|w|-w_{0}))|~F(\tilde{a}_{n}+\frac{1}{2},\tilde{b}_{n}+\frac{1}{2},\frac{3}{2};\sin^{2}(\tilde{k}(|w|-w_{0})))\right]
\end{array}
\ee
where
\ba
\tilde{a}_{n}=\frac{5}{4}+\frac{1}{2}\sqrt{\left(\frac{m_{n}}{\tilde{k}}\right)^2+\frac{9}{4}}
\cr \tilde{b}_{n}=\frac{5}{4}-\frac{1}{2}\sqrt{\left(\frac{m_{n}}{\tilde{k}}\right)^2+\frac{9}{4}}
\ea
The boundary conditions (\textit{i.e.} the jump of the wave function at the points
$w=0$, $w_{L}$) result in a
$2\times2$ homogeneous linear system which, in order to have a
non-trivial solution, leads to the vanishing determinant. In the
symmetric configuration which we consider, this procedure can be
simplified by considering even and odd functions with respect to the
minimum of the potential $w_{0}$.

In more detail, the odd eigenfunctions obeying
the  b.c. $\hat{\Psi}^{(n)}(w_{0})=0$  will  have $C_1=0$ and thus the form:
\be
\hat{\Psi}^{(n)}=C_{2}\cos^{5/2}(\tilde{k}(|w|-w_{0}))|\sin(\tilde{k}(|w|-w_{0}))|~F(\tilde{a}_{n}+\frac{1}{2},\tilde{b}_{n}+\frac{1}{2},\frac{3}{2};\sin^{2}(\tilde{k}(|w|-w_{0})))
\ee
On the other hand, the even eigenfunctions obeying
the b.c. $\hat{\Psi}^{(n)}~'(w_{0})=0$  will have $C_2=0$ and thus the form:
\be
\hat{\Psi}^{(n)}=C_{1}\cos^{5/2}(\tilde{k}(|w|-w_{0}))F(\tilde{a}_{n},\tilde{b}_{n},\frac{1}{2};\sin^{2}(\tilde{k}(|w|-w_{0})))
\ee
The remaining boundary condition is given by:
\be
\hat{\Psi}^{(n)}~'(0)+\frac{3k}{2}\tanh(kz_{0})\hat{\Psi}^{(n)}(0)=0
\ee
and determines the mass spectrum of the KK states. From this
quantization condition we get that the KK spectrum has a special first
mode similar to the one of the $"+-+"$ model. For $kz_0 \ga 5$  the  mass 
of the first mode is given by the approximate relation:
\be
m_1=4\sqrt{3}~k~e^{-2kz_{0}}
\label{m1}
\ee
In contrast, the masses of the next levels, if we  put together the
results for even and odd wavefunctions, are given by the formula:
\be
m_{n+1}=2\sqrt{n(n+3)}~k~e^{-kz_{0}}
\label{mr}
\ee
with $n=1,2,...$.

We note that the first KK state has a different scaling law with respect
to the position of the minimum of the ``warp'' factor compared
to the rest of the KK tower, since it scales as $ e^{-2kz_{0}}$ while
the rest of the tower scales as $e^{-kz_{0}}$. Thus the first
KK state is generally much lighter than the rest of the tower. It is clear that
this mass spectrum resembles the one of the $''+-+''$ model. The deeper
reason for this is again the common form of the ``warp'' factor. In both
cases the ``warp'' factor has a minimum due to its ``bounce''
form. The graviton wave function  follows
the form of the ``warp'' factor, \textit{i.e.} it is symmetric with respect to $w_{0}$, while
 the wavefunction of the first KK state is antisymmetric in
respect to $w_{0}$ (see figure 1). The absolute values of the two wavefunctions  are almost identical in all
regions except near $w_{0}$ (the symmetric is nonzero while the
antisymmetric is zero at $w_{0}$). The graviton
wavefunction is suppressed by the factor
$\frac{1}{cosh^{2}(kz_{0})}$ at $w_{0}$
which brings it's value close to zero for reasonable values of
$kz_{0}$. Thus, the mass
difference which is determined by the wavefunction near  $w_{0}$ is expected  to be generally very small, a fact which formally appears as
the extra suppression factor $e^{-kz_{0}}$ in the formula of $m_{1}$
in comparison with the rest of the KK tower. 

In the case that we consider an asymmetric brane configuration,
for example
$w_{L}>2w_{0}$ the spectrum is effectively independent of the position
of the second brane $w_{L}$ (considering $kz_{0}\ga 5$). Thus, even in the case that we place the
second brane at $w_{\infty}$, \textit{i.e.} the point which corresponds to infinity in the
$z$-coordinates, the spectrum is given essentially by the same
formulas. In the case that the second brane is placed at
$w_{0}<w_{L}<2w_{0}$,  some dependence on the position of the second
brane (\textit{i.e.} dependence on the scale hierarchy between the branes) is
present. Nevertheless, the main characteristics of the spectrum remain
the same, \textit{i.e.} the first KK state is special and always much lighter
than the others. In conclusion, the key parameter which determines the
spectrum is the position of the minimum of the ``warp'' factor.

Returning to our wavefunction solutions, we should note that for each eigenfunction the normalization constants $C_{1}$ and $C_{2}$
can be determined by the normalization condition 
$\displaystyle{\int_{-w_{L}}^{w_{L}}dw\left[\hat{\Psi}^{(n)}(w)\right]^2=1}$
 which is such that we get the standard form of the Fierz-Pauli
Lagrangian for the KK states. 
Knowing the normalization of the wavefunctions, it is straightforward
to calculate
the strength of the interaction of the KK states with the SM
fields confined on the brane\footnote{In the symmetric configuration it does not make any
difference which brane is our universe.}. This can be calculated  by expanding the minimal
gravitational coupling of the SM Lagrangian $\displaystyle{\int
d^4x\sqrt{-\hat{G}}{\mathcal{L}}\left(\hat{G},SM fields\right)}$ with respect to 
the metric. In this way we get:
\ba
{\mathcal{L}}_{int}&=&-\frac{1}{M^{3/2}}\sum_{n\geq
0}
\hat{\Psi}^{(n)}\left(w_{\rm brane}\right)h_{\mu\nu}^{(n)}(x)T_{\mu\nu}\left(x\right)= 
\nonumber
\\&=&-\frac{A}{M^{3/2}}h_{\mu\nu}^{(0)}(x)T_{\mu\nu}\left(x\right)-
\sum_{n>0}\frac{\hat{\Psi}^{(n)}\left(w_{\rm brane}\right)}{M^{3/2}
}h_{\mu\nu}^{(n)}(x)T_{\mu\nu}\left(x\right)
\ea
with $T_{\mu\nu}$ the energy momentum tensor of the SM
Lagrangian. Thus, the coupling of the zero and the first KK
mode to matter are respectively:
\ba
\frac{1}{c_0}&=&\frac{A}{M^{3/2}}=\frac{1}{M_{*}} \label{c1} \\
\frac{1}{c_1}&=&\frac{\hat{\Psi}^{(n)}\left(w_{\rm brane}\right)}{M
^{3/2}}\simeq\frac{1}{M_{*}} \label{c2}
\ea
where $A$ is the zero mode normalization constant which turns out
to be $\frac{M^{3/2}}{M_{\rm *}}$. We should also note that the
couplings of the rest of the KK states are much smaller and scale as $e^{-kz_{0}}$.

Exploiting the different mass scaling of the first KK relative to the
rest we can ask whether it is possible to realize a ``bigravity''
scenario similar to that in $"+-+"$ model. 
 In that model by appropriately choosing the position of the minimum of the ``warp''
factor, it was possible to make the first KK state have mass such that the
corresponding wavelength is  of the order of the cosmological scale that gravity
has been tested and at the same time have the rest of the KK tower wavelengths
below 1mm  (so that there is no conflict with Cavendish bounds). In
this scenario the gravitational interactions are due to the net effect of the massless graviton and the first ultralight KK
state. From eq.(\ref{c1}), (\ref{c2}) it can be understood that in the symmetric
configuration the massless graviton and the special KK state
contribute by the same amount to the gravitational interactions.
In other words:
\be
\frac{1}{M_{Pl}^2}=\frac{1}{M_{*}^2}+\frac{1}{M_{*}^2}~~~ \Rightarrow~~~ M_{Pl}=\frac{M_{*}}{\sqrt{2}}
\ee

 In the present model, the fact that the effective four dimensional
cosmological constant should be set very close to zero, requires that
the  ``warp'' factor is constrained by  $kz_{0}\geq{135}$ and thus,
in this case, the spectrum of the KK states will be very dense (tending
to continuum) bringing more states close to zero mass. The KK states
that have masses which correspond to wavelengths larger than $1mm$ have
sufficiently small coupling so that there is no conflict with 
phenomenology (the situation is exactly similar to the RS2 case
where the coupling of the  KK states is proportional to their mass and thus
it is decreasing for lighter KK states).  The fact that the spectrum
tends to a continuum shadows
the special role of the first KK state. Moreover, it is interesting to note that
at the limit where the minimum of the ``warp'' factor is sent to infinity ($w_{\infty}$)
 the special behaviour of the first KK persists and does not catch the
other levels (by changing its scaling law) as
was the case in \cite{Kogan:2000xc}. This means that the limit  $w\rightarrow
w_{\infty}$ will indeed
be identical to \textit{two}  RS2, but on the other hand it is interesting to note
that what we call graviton in the RS2 limit is actually the
combination of a massless graviton
\textit{and} the ``massless'' limit of the special first KK
state. This ``massless'' limit exists as we will see in the next
section and  ensures that locality is respected by the
model, since physics on the brane does not get affected from the
boundary condition at infinity.

\section{Discussion and conclusions}

The fact that we have a ultralight graviton in our spectrum is at
first sight worrying because it is well known that in the flat space
the tensor structure of the propagators of the massless and of the massive
graviton are different \cite{vanDam:1970vg,Zakharov} and that there is no smooth limit between them
when $m \rightarrow 0$. The bending of the light by the sun agrees with 
the prediction of the Einstein theory to $1\%$  accuracy. This
is sufficient to rule out any scenario which a significant component
of gravity is due to a  massive graviton, however
light its mass could be. However, as was shown in Chapter \ref{mssgr}, the situation in $AdS$ space is
quite different. There it was shown
that if we could arrange  ${m_1 \over H} \la 0.1$ there is no
discrepancy with standard tests of Einsteinian gravity as for example
the bending of the light by the sun.

In the particular model we have at hand, it is ${m_1 \over H} \sim
e^{-k z_0}$ so we can easily accommodate the above bound. Then, the
Euclidean propagator (in configuration space) of the massive KK states   for relatively large $z_0$ will be given by:
\be
G^{m}_{\mu\nu;\mu'\nu'}(x,y)=\frac{1}{4 \pi^2
\mu^2}(\delta_{\mu\mu'}\delta_{\nu\nu'}+\delta_{\mu\nu'}\delta_{\nu\mu'}-\left(
1-{1\over 6}e^{-2 kz_0}\right)\delta_{\mu\nu}\delta_{\mu'\nu'})
\ee
where $\mu$ is the geodesic distance between two points. In the above, 
we have omitted terms that do not contribute when integrated with 
a conserved $T_{\mu \nu}$. For $kz_0 \ga 2.3$ there in no problem with the bending of
light. However, if our aim is to see modifications of gravity at
ultra-large distances, this is impossible because the Compton wavelength of our ultralight
graviton will be $e^{kz_0}$ times bigger than the horizon $H^{-1}$ of the
$AdS_4$ space on our brane due to equations (\ref{m1}), (\ref{mr}). The ``Hubble''
parameter follows $m_2$ rather than $m_1$.
 
What happens if one takes  an asymmetric version of this model 
where $L>2z_0$ is that the spectrum  does not get significantly modified so we
are effectively in the same situation. In the 
case where $z_0<L<2z_0$ the spectrum will behave like 
the one of the $''+-+''$ model. Then for $\omega \ll 1$ we will have $m_{1} \sim \omega e^{-2kx} M
\sim e^{-2kx} M_{\rm Pl}$, $H \sim e^{-kz_{0}} M \sim e^{-kx} M_{\rm
Pl}$ where $M \approx  M_{\rm Pl}/\omega$ is the fundamental scale, $\omega \approx e^{-(2z_0-L)}$ is
the ``warp'' factor and
$x=L-z_0$. Again, the  ultralight graviton is hiding well beyond the
$AdS$ horizon. However, the coupling of the remaining of the
KK tower to matter will be different than the symmetric case and one may have different  corrections to  Newton's law  on the left and right branes.

In summary, in this paper we presented a  ``bigravity'' model with two
$AdS_4$ branes in $AdS_5$ bulk which has a lot of similarities with
the $''+-+''$  model. The fact that we have no $''-''$
branes removes the ghost state problem and furthermore, due to some
amazing property of the $AdS$ space we are able to circumvent the van
Dam - Veltman - Zakharov discontinuity of the graviton
propagator. This  makes the model compatible with the predictions of
General Relativity in the small graviton mass limit since  the extra degrees of freedom of the massive
graviton practically decouple. However, the presence of the AdS horizon prevents the
modifications of gravity at large distances to become
observable.

\chapter{Bigravity in six dimensions} 
\label{6D}

\section{Introduction}  

In the previous Chapters we have presented models that give the
interesting possibility of modifications of gravity at large (cosmological) distances
in the context of bigravity (or multigravity) scenario.  However, none of the
presented models can be considered phenomenologically viable: In the
case of the $''+-+''$ model, the presence of the moving brane that
violates the weaker energy condition and its
assocciation to the appearence of a ghost scalar field (radion), makes
this model unacceptable. In the case of the $''++''$ model, the
presence of $AdS_{4}$ spacetime on the branes allows us to avoid the
negative tension brane without losing the main characteristics of the 
$''+-+''$ model  that give rise to an ultra-light KK state. However, the presence 
of a remnant negative cosmological constant on the brane, contrary to
experimental indications for the presence of a positive one, make the model phenomenologically
disfavoured. Furhtermore, the modifications of gravity at large
distances  predicted by this model are hidden behind the $AdS$ horizon. 

In the light of the  results of  \cite{Vain,Deffayet:2001uk}, for
having a phenomenologically viable model, it is
sufficient to look for bounce type solution without negative tension
branes, even when the branes are flat. The previous, although is
impossible in five dimensions,  can be achieved in the case that we
consider brane configurations in six dimensions. This is basically
due to the fact  that in
this case the internal space is non-trivial giving us the freedom to
achieve the desired configurations.

\section{One brane models in six dimensions}

At first we will discuss the one brane solutions in six dimensions to
get an insight of the more complicated multigravity
configurations. In the following section we will review the
Gherghetta-Shaposhnikov single brane model and then we will  generalize it for the case of
two branes.

\subsection{The minimal one brane model}

The simplest one brane model consists of a four brane embedded
in six dimensional $AdS$ space. One of the lognitudinal dimensions of
the four brane is compactified to a Planck lenght radius $R$ while the 
transverse to the four brane extra dimension is infinite. Let us see
how this model can be constructed.

The most general spherically symmetric ansatz that one can write down
in six dimensions and which preserves four dimensional Poincar\'e invariance  is:
\begin{equation}
ds^2=\sigma(\rho)\eta_{\mu \nu}dx^{\mu}dx^{\nu}+d\rho^2+\gamma(\rho)d\theta^2
\end{equation}
where $\theta$ is the compactified dimension with range $[0,2\pi]$ and
$\rho$ is the infinite radial dimension. Note the difference with the
five dimensional case: Even in the spherically symmetric case the
metric ansatz in general is described by two functions: apart from the 
familiar warp factor $\sigma(\rho)$, there is the function
$\gamma(\rho)$ which is assocciated with the geometry of the internal
space (in the five dimensional case this function was removed by a
simple coordinate transformation).

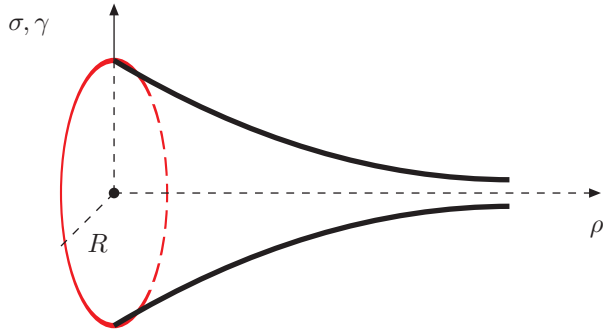
\begin{figure}[b]
\vskip10mm
\begin{center}
\begin{picture}(200,100)(0,50)

\SetWidth{2}
\SetColor{Red}
\Oval(0,100)(50,20)(0)
\SetColor{Black}
\Curve{(0,150)(150,105)(300,150)}
\Curve{(0,50)(150,95)(300,50)}
\Vertex(0,100){2}
\Text(-40,160)[lb]{$\sigma,\gamma$}

\CBox(150,50)(300,150){White}{White}
\SetWidth{.5}
\DashLine(0,100)(180,100){3}
\LongArrow(180,100)(182,100)

\LongArrow(0,150)(0,170)
\DashLine(0,100)(0,150){3}
\DashLine(0,100)(-20,80){3}

\SetWidth{3}
\SetColor{White}
\Line(10,135)(20,135)
\Line(10,125)(20,125)
\Line(10,115)(20,115)
\Line(10,105)(25,105)
\Line(10,95)(25,95)
\Line(10,85)(20,85)
\Line(10,75)(20,75)
\Line(10,65)(20,65)

\SetWidth{2}
\SetColor{Black}
\Text(180,90)[lt]{$\rho$}
\Text(-10,85)[lt]{$R$}

\end{picture}
\end{center}

\caption{The minimal one four-brane model warp functions $\sigma(\rho)$, $\gamma(\rho)$.}
\end{figure}

The Einstein equations for the above metric ansatz is
\footnote{Ignoring the three-branes located on the
conical singularities}:
\begin{equation}
R_{MN}-\frac{1}{2}G_{MN}R={1 \over 4M^4}(T^{(B)}_{MN}+T^{(br)}_{MN})
\end{equation}
where $T^{(B)}_{MN}$ is the bulk energy momentum tensor,
$T^{(br)}_{MN}$ the one of the various four branes.

In the absence of four dimensional cosmological constant we have that 
the $(\theta,\theta)$ component of the above equation is:
\begin{equation}
2{\sigma'' \over \sigma}+{1 \over 2}\left({\sigma' \over 
\sigma}\right)^2=-{\Lambda_{\theta} \over 4M^4}-{V^i_{\theta} \over
4M^4}\delta(\rho-\rho_i)
\label{ttD}
\end{equation}

The $(\rho,\rho)$ component is:
\begin{equation}
{3 \over 2}\left({\sigma' \over \sigma}\right)^2+{\sigma'\gamma' \over 
\sigma \gamma}=-{\Lambda_{\rho} \over 4M^4}
\label{rrD}
\end{equation}

Finally the $(\mu,\nu)$ component is:
\begin{equation}
{3 \over 2}{\sigma'' \over \sigma}+{3 \over 4}{\sigma'\gamma' \over
\sigma \gamma}-{1 \over 4}\left({\gamma' \over \gamma}\right)^2+{1
\over 2}{\gamma'' \over \gamma}=-{\Lambda_0 \over 4M^4}-{V^i_0 \over 4M^4}\delta(\rho-\rho_i)
\label{mnD}
\end{equation}
Note that we have allowed for inhomogeneous cosmological constant and
brane tensions.

It is straightforward to solve the Einstein equations for  bulk
energy momentum tensor $T^{(B)~N}_{~~~M}=-\Lambda \delta_M^N$ (here we 
have chosen homogeneous bulk cosmological constant:
$\Lambda_{\theta}=\Lambda_{\rho}=\Lambda_{0}$ ) with the four  brane contribution:
\begin{equation}
T^{(br)~N}_{~~~M}=-\delta(\rho)\left(\begin{array}{ccc}V
~\delta_{\mu}^{\nu}&~&~\\~&0&~\\~&~&V \end{array}\right)
\end{equation}
From the above equations we find that the solution for the two warp factors is:\begin{equation}
\sigma(\rho)=e^{-k \rho} ~~~,~~~ \gamma(\rho)=R^2 e^{-k \rho}
\end{equation}
with $k^2=-{\Lambda \over 10M^4}$, where the arbitrary integrations constant $R$ is just the radius of
the four brane. The Einstein equations impose the usual fine tuning
between the bulk cosmological constant and the tension of the four brane:
\begin{equation}
V=-{8\Lambda \over 5k}
\end{equation}
Let us note at this point that in Ref.\cite{Gherghetta:2000qi} this fine tuning was absent
because a smooth local defect was considered instead of a four
brane. However, the fine tuning emerges between the different
components of the defect energy momentum tensor. The physics of the
four brane idealization and the one of the defect model is the same.

The four dimensional Kaluza-Klein decomposition can be carried out as
usual by considering the following graviton perturbations (we ignore
the scalar and vector modes):
\begin{equation}
ds^2=\sigma(\rho)\left[\eta_{\mu \nu}+h_{\mu \nu}(\rho,\theta)\right]dx^{\mu}dx^{\nu}+d\rho^2+\gamma(\rho)d\theta^2
\end{equation}
We expand the graviton perturbations in a complete set of radial
eigenfunctions and Fourier angular modes:  
\begin{equation}
h_{\mu \nu}(\rho,\theta)=\sum_{n,l} \phi_n(\rho)e^{il\theta} h^{(n,l)}_{\mu \nu}(x)
\end{equation}
The differential equation for the radial wavefunctions $\phi$ is:
\begin{equation}
\phi''-{5 \over 2}k\phi'+\left(m^2-{l^2 \over R^2}\right)e^{k\rho}\phi=0
\end{equation}
with normalization  $\int_0^\infty d\rho \sigma \sqrt{\gamma} \phi_m
\phi_n=\delta_{mn}$. We can convert this equation to a two dimensional 
Schr\"{o}dinger-like equation by the following redefitions:
\begin{equation}
z={2 \over k}\left(e^{{k \over 2}\rho}-1\right)~~~,~~~\hat{\Psi}=\sigma^{3/4}\phi
\end{equation}
so that 
\begin{equation}
-{1 \over 2 \sqrt{\gamma}} \de_z\left(\sqrt{\gamma}\de_z\hat{\Psi}\right)+V_{eff}\hat{\Psi}={m^2 \over 2}\hat{\Psi}~~~,~~~V_{eff}(z)=\frac{21k^2}{32\left({kz \over
2}+1\right)^2}+{l^2 \over 2R^2}-{3k \over 4}\delta(z)
\end{equation}

\begin{figure}[t]
\vskip10mm
\begin{center}
\begin{picture}(200,100)(0,50)

\SetWidth{2}
\SetColor{Black}
\Curve{(-50,110)(150,5)(350,110)}
\SetColor{Red}
\Curve{(-50,130)(150,25)(350,130)}
\SetColor{Green}
\Curve{(-50,150)(150,45)(350,150)}
\CBox(150,5)(350,150){White}{White}
\SetColor{Black}
\Curve{(149,5)(200,3)(230,2)}
\SetColor{Red}
\Curve{(149,25)(200,23)(230,22)}
\SetColor{Green}
\Curve{(149,45)(200,43)(230,42)}
\SetColor{Black}
\Vertex(-50,0){2}

\SetWidth{.5}
\LongArrow(-50,0)(250,0)
\LongArrow(-50,0)(-50,170)

\SetWidth{2}
\SetColor{Black}
\Text(250,20)[lt]{$\rho$}
\Text(-40,160)[lb]{$V_{eff}$}
\Text(70,100)[lt]{$l=0$}
\Text(70,120)[lt]{${\Red{l=1}}$}
\Text(70,140)[lt]{${\Green{l=2}}$}

\end{picture}
\end{center}

\vskip15mm
\caption{The form of the effective potential for different angular
quantum numbers $l$.}
\end{figure}
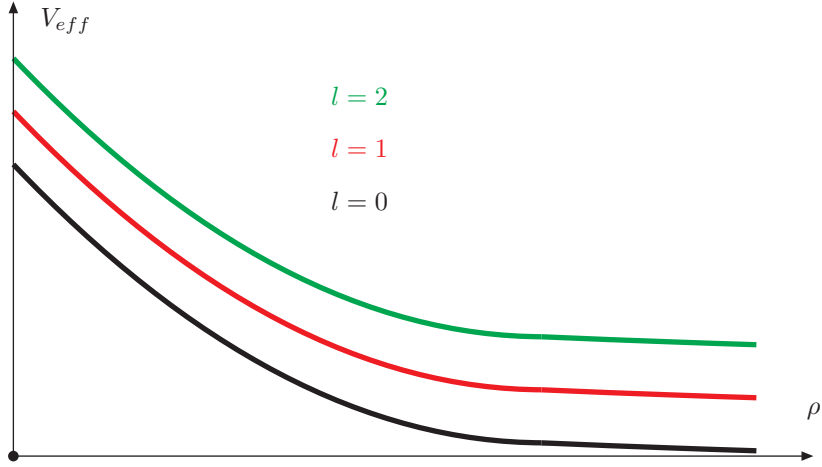

From the form of the potential we can easily deduce that the angular
excitations spectrum will consist of continua starting from a gap of
the order ${l^2 \over R^2}$ and thus can be safely ignored. For the s-wave ($l=0$) there is a normalizable zero mode which is a
constant in the $\rho$ coordinates, \textit{i.e.} $\phi_0=ct.$ The
KK tower for the s-waves will again form a continuum but this time
gapless and their wavefunctions are given by:
\begin{equation}
\phi_m=N_m e^{{5 \over 4}k \rho}\left[J_{3/4}\left({2m \over k}\right)Y_{5/2}\left({2m \over k}e^{{k \over 2} \rho}\right)-Y_{3/4}\left({2m \over k}\right)J_{5/2}\left({2m \over k}e^{{k \over 2} \rho}\right)\right]
\end{equation}

The correction to the Newton's law due to the s-modes can be easily calculated and one
finds that it is more suppressed than the five dimensional RS2 case:
\begin{equation}
\Delta V=-{1 \over {\mathcal{O}}(M_{Pl}^5)}{1 \over r^4}
\end{equation}

\section{Bigravity in six dimensions}

The flat one four-brane models considered in the previous sections have the
characteristic that gravity is localized on the brane in the same way
as in the five dimensional analogue (RS2). In this section we will
show how we can construct realistic multi-localization scenarios for
gravity by consistently pasting two one brane solutions. We will consider a
two four-brane bigravity model which additionally contains in general
a three-brane associated to the existence of a conical
singularity. This two brane configuration can be realized by allowing for an inhomogeneous four brane 
tension.

\subsection{The conifold model}

We are interested in a two brane generalization of the minimal one
brane model. In order to achieve this,  we do not impose
any constraints for the four-brane tension: 
\begin{equation}
T^{(br)~N}_{~~~M}=-\delta(\rho)\left(\begin{array}{ccc}V_0~
\delta_{\mu}^{\nu}&~&~\\~&0&~\\~&~&V_{\theta}\end{array}\right)
\end{equation}
while we still demand for a homogeneous bulk energy momentum tension
of the form:
\begin{equation}
T^{(B)~N}_{~~~M}=-\Lambda \delta_M^N
\end{equation}
The Einstein equations for $\sigma(\rho)$ and $\gamma(\rho)$ in this
case  give the following solutions for the warp factors:
\begin{equation}
\sigma(\rho)=\cosh^{4/5}\left[{5 \over 4}k (\rho-\rho_0) \right] ~~~,~~~\gamma(\rho)=R^2\frac{\cosh^{6/5}\left[{5 \over 4}k \rho_0 \right]}{\sinh^2\left[{5 \over 4}k \rho_0 \right]}\frac{\sinh^2\left[{5 \over 4}k  (\rho-\rho_0) \right]}{\cosh^{6/5}\left[{5 \over 4}k  (\rho-\rho_0) \right]}
\end{equation}
with $k^2=-{\Lambda \over 10M^4}$, where we have normalized as $\sigma(0)=1$ and $\gamma(0)=R^2$. From the above relations it is obvious  that both $\sigma(\rho)$ and
$\gamma(\rho)$ have a bounce form. However, we note that
$\gamma(\rho_{0})=0$, that is,  $\gamma(\rho)$ is vanishing at the
minimum of the warp factor $\sigma(\rho)$. This is a general
characteristic of the solutions even when the branes are non-flat. From eq.(\ref{ttD}) and eq.(\ref{rrD}) we can easily find that
$\gamma(\rho)=C\frac{(\sigma'(\rho))^2}{\sigma(\rho)}$ (where C is an
integration constant) which implies
that whenever we have a bounce in the warp factor the function $\gamma(\rho)$
will develop a zero.

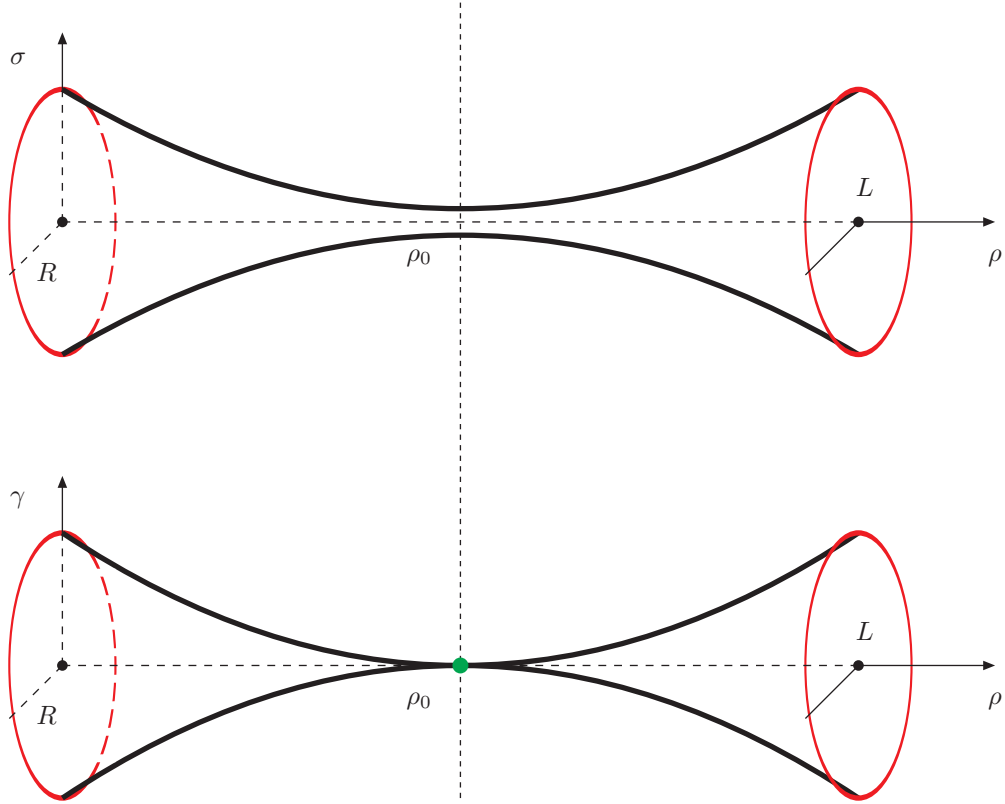
\begin{figure}[t]
\vskip10mm
\begin{center}
\begin{picture}(200,100)(0,50)

\SetWidth{2}
\SetColor{Red}
\Oval(-50,100)(50,20)(0)
\SetColor{Black}
\Curve{(-50,150)(100,105)(250,150)}
\Curve{(-50,50)(100,95)(250,50)}
\Vertex(-50,100){2}
\Vertex(250,100){2}

\SetColor{Red}
\Oval(250,100)(50,20)(0)

\SetColor{Black}
\SetWidth{.5}
\LongArrow(-50,150)(-50,170)
\DashLine(-50,100)(-50,150){3}
\DashLine(-50,100)(-70,80){3}
\Line(250,100)(230,80)

\SetColor{Black}
\SetWidth{.5}
\DashLine(-50,100)(250,100){3}
\LongArrow(250,100)(300,100)

\SetWidth{3}
\SetColor{White}
\Line(-40,135)(-30,135)
\Line(-40,125)(-30,125)
\Line(-40,115)(-30,115)
\Line(-40,105)(-25,105)
\Line(-40,95)(-25,95)
\Line(-40,85)(-30,85)
\Line(-40,75)(-30,75)
\Line(-40,65)(-30,65)

\SetWidth{2}
\SetColor{Black}
\Text(300,90)[lt]{$\rho$}
\Text(80,90)[lt]{$\rho_0$}
\Text(250,110)[lb]{$L$}
\Text(-60,85)[lt]{$R$}
\Text(-70,160)[lb]{$\sigma$}

\end{picture}
\end{center}

\vskip5mm

\vskip15mm
\begin{center}
\begin{picture}(200,100)(0,50)

\SetWidth{2}
\SetColor{Red}
\Oval(-50,100)(50,20)(0)
\SetColor{Black}
\Curve{(-50,150)(100,100)(250,150)}
\Curve{(-50,50)(100,100)(250,50)}
\Vertex(-50,100){2}
\Vertex(250,100){2}

\SetColor{Red}
\Oval(250,100)(50,20)(0)

\SetColor{Black}
\SetWidth{.5}
\LongArrow(-50,150)(-50,170)
\DashLine(-50,100)(-50,150){3}
\DashLine(-50,100)(-70,80){3}
\Line(250,100)(230,80)

\SetColor{Black}
\SetWidth{.5}
\DashLine(-50,100)(250,100){3}
\LongArrow(250,100)(300,100)
\SetColor{Black}
\SetWidth{.5}
\DashLine(100,50)(100,350){2}

\SetWidth{2}
\SetColor{Green}
\Vertex(100,100){3}

\SetWidth{3}
\SetColor{White}
\Line(-40,135)(-30,135)
\Line(-40,125)(-30,125)
\Line(-40,115)(-30,115)
\Line(-40,105)(-25,105)
\Line(-40,95)(-25,95)
\Line(-40,85)(-30,85)
\Line(-40,75)(-30,75)
\Line(-40,65)(-30,65)

\SetWidth{2}
\SetColor{Black}
\Text(300,90)[lt]{$\rho$}
\Text(80,90)[lt]{$\rho_0$}
\Text(250,110)[lb]{$L$}
\Text(-60,85)[lt]{$R$}
\Text(-70,160)[lb]{$\gamma$}

\end{picture}
\end{center}

\caption{On the top, the form of $\sigma(\rho)$ as a function of
$\rho$. $\sigma(\rho)$ has a bounce form with the minimum at
$\rho_{0}$. At the bottom, the corresponding $\gamma(\rho)$ function 
 which also has a similar bounce form. However $\gamma(\rho)$ vanishes 
at the point that corresponds to the minimum of the warp factor (at
$\rho=\rho_{0}$). This point in general corresponds to a conical singularity.}

\end{figure}

In order to examine the nature of this singularity we 
examine the form of the metric at the vicinity of the point
$\rho=\rho_{0}$. Taking in account that in this limit we have 
$\sigma(\rho) \rightarrow 1$ and $\gamma(\rho) \rightarrow \beta^2 (\rho-\rho_{0})^2$
the metric becomes:
\begin{equation}
ds^2=\eta_{\mu \nu}dx^{\mu}dx^{\nu}+d\rho^2 + \beta^2 (\rho-\rho_{0})^2 d\theta^2
\end{equation}
where $\beta^2 \equiv \frac{25 k^2 R^2}{16} \frac{\cosh^{6/5}(\frac{5}{4}k\rho_{0})}{\sinh^{2}(\frac{5}{4}k\rho_{0})}$
From the form of the metric it is clear  that for general values of
the $\beta$ parameter there will be a conical singularity with a
corresponding 
deficit angle $\delta=2\pi(1-\beta)$. The exhistence of this conifold singularity is
connected with the presence of a 3-brane at
$\rho=\rho_{0}$. In order
to find the tension one has to carefully examine the Einstein tensor
at the vicinity of that point. For this reason we write the metric for 
the internal manifold in the conformally flat form:
\begin{equation}
ds^2=\eta_{\mu \nu}dx^{\mu}dx^{\nu}+f(r)(dr^2 + r^2 d\theta^2)
\end{equation}
with $f(r)=r^{2(\beta-1)}$ and $\rho-\rho_{0}=\beta^{-1}r^{\beta}$
In these coordinates it is easy to see how the three brane appears on
the conifold point. The Einstein tensor can be calculated for $\rho
\rightarrow \rho_{0}$:
\begin{equation}
R_{MN}-\frac{1}{2}G_{MN}R=\left(\begin{array}{ccc}\frac{\nabla^{2}\log(f(r))}{2f(r)}~
\eta_{\mu \nu}&~&~\\~&0&~\\~&~&0\end{array}\right)
\end{equation}
where $\nabla^{2}$ is the flat two dimensional Laplacian. Now given that $\nabla^{2}\log(r)=2 \pi \delta^{(2)}({\bf{r}})$ and by comparing with:
\begin{equation}
R_{MN}-\frac{1}{2}G_{MN}R=-\frac{V_3}{4M^{4}}\frac{\sqrt{-\hat{G}}}{\sqrt{-G}}\hat{G}_{\mu\nu}\delta^{\mu}_{M}\delta^{\nu}_{N}\delta(r)
\end{equation}
where $\delta^{(2)}({\bf{r}})={\delta(r) \over 2\pi r}$, we find that the tension of the 3-brane is:
\begin{equation}
V_{3}=4(1-\beta)M^{4}={2M^4 \over \pi}\delta
\end{equation}
Thus, if there is angle deficit $\delta>0$ ($\beta<1$) the tension of
the brane is  positive, whereas if there is angle excess $\delta<0$
($\beta>1$) the tension of the brane is negative. At the critical
value $\beta=0$ there is no conical singularity at all and we have a
situation where two locally flat spaces touch each other at one point.

For the previous solution to be consistent the brane tensions of the
four branes for the symmetric configuration must be tuned as 
\begin{equation}
V_{\theta}=-{8\Lambda \over 5k}\tanh\left[{5 \over 2}k
\rho_0\right]~~~,~~~V_0={3 \over 8}V_{\theta}+{8\Lambda^2 \over 5k^2}{1 \over V_{\theta}}
\end{equation}

Thus, the above contruction consists of two positive tension four-branes placed at the end of the compact space and an intermediate three 
brane due to the conifold singularity with tension depending on the
parameters of the model. In the limit $\rho_0 \rightarrow \infty$ we
correctly obtain two identical minimal one four-models for the case
where $\delta=2\pi$ ($\beta=0$).

The differential equation for the radial wavefunction $\phi$ of 
the graviton excitations reads:
\begin{equation} 
\phi''+2\left({\sigma' \over \sigma}+{\gamma' \over
4\gamma}\right)\phi'+\left({m^2 \over \sigma}-{l^2  \over \gamma}\right)\phi=0
\label{diffD}
\end{equation}
normalized as $\int_0^\infty d\rho \sigma \sqrt{\gamma} \phi_m
\phi_n=\delta_{mn}$.

There is an obvious normalizable zero mode with $\phi_0=ct.$ and a
tower of discrete KK states. From a general separability and locality
argument we immediately see that there should be an ultralight KK
state at least for the cases of $\delta>0$ ($\beta<1$) where the two
disconnected one four-brane limit is well behaved.

We can transform this differential equation to a two
dimensional Schr\"{o}dinger-like one by the coordinate change ${dz
\over d\rho}=\sigma^{-1/2}\equiv g(z)$ and the usual wavefunction
redefinition. Then the effective potential reads:
\begin{equation}
V_{eff}={15 \over 8}\left({\de_z g \over g}\right)^2-{3 \over 4}{\de_z^2 g \over
g}-{3 \over 8}{\de_z g \de_z \gamma \over g \gamma}+{l^2 \over 2\gamma 
g^2}
\label{potD}
\end{equation}

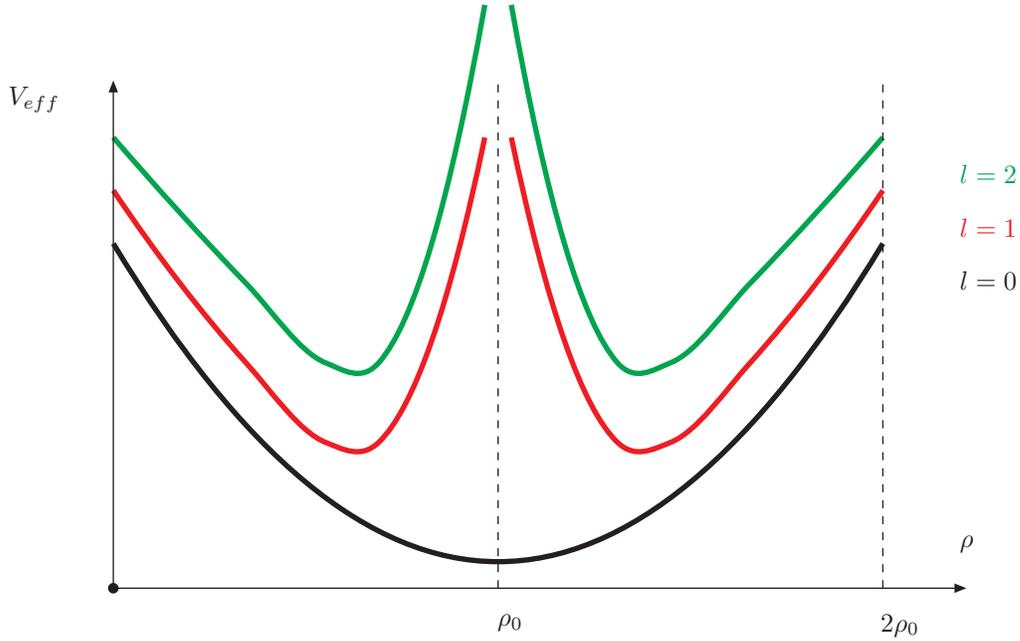
\begin{figure}[t]
\vskip20mm
\begin{center}
\begin{picture}(200,100)(0,50)

\SetWidth{2}
\SetColor{Black}
\Curve{(-70,110)(75,-10)(220,110)}
\SetColor{Red}
\Curve{(-70,130)(-20,65)(10,35)(30,35)(70,150)}
\Curve{(80,150)(120,35)(140,35)(170,65)(220,130)}
\SetColor{Green}
\Curve{(-70,150)(-20,95)(10,65)(30,65)(70,200)}
\Curve{(80,200)(120,65)(140,65)(170,95)(220,150)}
\SetColor{Black}
\Vertex(-70,-20){2}

\SetWidth{.5}
\LongArrow(-70,-20)(250,-20)
\LongArrow(-70,-20)(-70,170)
\DashLine(220,-20)(220,170){3}
\DashLine(75,-20)(75,170){3}

\SetWidth{2}
\SetColor{Black}
\Text(250,0)[lt]{$\rho$}
\Text(-110,160)[lb]{$V_{eff}$}
\Text(250,100)[lt]{$l=0$}
\Text(250,120)[lt]{${\Red{l=1}}$}
\Text(250,140)[lt]{${\Green{l=2}}$}
\Text(75,-30)[lt]{$\rho_0$}
\Text(220,-30)[lt]{$2\rho_0$}

\end{picture}
\end{center}

\vskip25mm
\caption{The form of the effective potential for different angular
quantum numbers $l$.}
\end{figure}

We cannot write an explicit analytic formula for of the above
potential in the $z$-coordinates because the coordinate transformation 
is not analytically invertible. However, since the transformation is
monotonic, we can easily sketch the form of the potential by
calculating it in the $\rho$ coordinates. From this procedure we see
that the potential for the s-wave is finite at the conical singularity 
$\rho_0$ but has a divergence for all the angular excitations with
$l\neq0$. This means that only the s-wave excitations with communicate
the two parts of the  conifold and the other excitations will be
confined in the two semicones.

Finally, one could move the second brane to infinity and obtain a six
dimensional analogue of the localy localized model. In that case gravity on
the four-brane at $\rho=0$ will be mediated by only the ultralight
state since the  graviton zero mode will not be normalizable.

\section{Discussion and conclusions}

In this Chapter we have constructed, a brane theory
which can lead to multigravity models and their associated
modifications of gravity at large distances {\textit{without}}
introducing moving negative tension branes. The constructions are made
possible by going to a six dimensional theory. The way that the no-go theorem for the presence of a bounce in the
warp factor in five dimensions without
negative tension branes  is evaded in six dimensions is
obvious. In five dimensions with the metric:
\begin{equation}
ds^2=e^{-A(\rho)}\eta_{\mu \nu}dx^{\mu}dx^{\nu}+d\rho^2
\end{equation}
one can readily show that the weaker energy condition requires that:
\begin{equation}
A''\geq 0
\end{equation}
which means that the bounce is linked to moving negative tension
branes and at their position the weaker energy condition is violated.
In the case that the branes were $AdS$ one could have a bounce without
having negative tension branes and still satisfy the weaker energy
condition because the above relation is modified to \cite{Karch:2001ct}:
\begin{equation}
A''\geq -2H^2 e^{A}
\end{equation}
However, such models do not lead to modifications of gravity at large
distances and moreover the remnant negative cosmological constant is
in conflict with current observations.

In the six dimensional case with metric:
\begin{equation}
ds^2=e^{-A(\rho)}\eta_{\mu \nu}dx^{\mu}dx^{\nu}+d\rho^2+e^{-B(\rho)}d\theta^2
\end{equation}
from the weaker energy condition one finds two inequalities which can be cast into the following relation:
\begin{equation}
{1 \over 6} (B')^2 -{1 \over 3}B''-{1 \over 6}A'B'\leq A'' \leq -{1
\over 2} (B')^2 +B''-{3 \over 2}A'B'+ 2(A')^2
\end{equation}
This shows that the bounce can be obtained with only positive tension 
branes without violating the weaker energy condition and guarantees
that our constructions will evade the ghost field problem encountered in the five
dimensional case.

However, apart from the construction with the singular model, one 
can construct six dimensional models with the same properties (bounce
form of the warp factor) without  conical singularity. This is
possible if one allows for also a inhomogeneous bulk cosmological
constant. In order this to be realized some additional fine tunings
between the components of the bulk cosmological constant are required.
The solution for $\sigma(\rho)$ in this case is identical as in the
singular model but $\gamma(\rho)$ is non-vanishing in all the region
between the branes. The latter implies that the model is free of
conical singularities. The form of the potential of the corresponding
two dimensional Schr\"odinger equation again ensures that this model
has a light state making the bigravity scenario possible (for more
details see Ref.\cite{Kogan:2001yr}).

\chapter{Fermions in Multi-brane worlds} 
\label{neutrino}

\section{Introduction}

The study of bulk fermion  fields, although not something new \cite{Jackiw:1976fn,Rubakov:1983bb}, turns out to be of particular interest 
in the context of brane-world scenarios both in the case of models
with large
extra dimensions \cite{Arkani-Hamed:1998rs,Arkani-Hamed:1999nn,Antoniadis:1998ig} (factorizable geometry) and in models of localized
gravity \cite{Gogberashvili:1998vx,Randall:1999ee,Randall:1999vf} (non factorizable geometry) since they can provide possible
new ways to explain the smallness of the neutrino masses, neutrino
oscillations and the pattern of
fermion mass hierarchy. 

In the context of string and M-theory, bulk fermions arise as
superpartners of gravitational moduli, such as, those setting the radii of
internal spaces. Given this origin, the existence of bulk fermions is
unavoidable in any supersymmetric string compactification
and represents a quite generic feature of string
theory\footnote{However, note that brane-world models with non
factorizable geometry have not yet been shown to have string
realizations. For string realizations of models with large extra
dimensions see Ref.\cite{Antoniadis:1998ig}}. This constitutes the most likely origin of such particles
within a fundamental theory and, at the same time, provides the basis
to study brane-world neutrino physics.

In the traditional approach the small neutrino masses are a result of
the see-saw mechanism, in which a large right-handed Majorana mass
$M_{R}$ suppresses the eigenvalues of the neutrino mass matrix
leading to the light neutrino mass $m_{\nu}\sim \frac{m_{fermion}^{2}}{M_{R}}$. The neutrino
mixing explanations of the atmospheric and solar neutrino anomalies
require that $M_{R}$ to be a superheavy mass scale $> 10^{10}$ GeV.     

In the case of large extra dimensions \cite{Arkani-Hamed:1998rs,Arkani-Hamed:1999nn,Antoniadis:1998ig}, despite  the absence
of  a high scale like $M_{R}$ (since in such models the fundamental scale can be
as low as 1 TeV), small neutrino masses \cite{Dienes:1999sb,Arkani-Hamed:1998vp} (Dirac or Majorana) can arise
from an intrinsically higher-dimensional mechanism. The idea is that any fermionic state
that propagates in the bulk, being a Standard Model (SM) singlet can
be identified with
a sterile neutrino which through it's coupling to the SM
left-handed neutrino can generate small neutrino mass. In
the case of factorizable geometry, the smallness of the induced masses  is due to the fact that
the  coupling is suppressed by the large volume of
the internal bulk manifold.
In other words, the interaction probability between the bulk fermion
zero mode, the Higgs
and Lepton doublet fields (which are confined to a brane) is small
because of the large volume of bulk compared to the thin wall where the SM
states are confined, resulting  a highly
suppressed coupling. In the context of these models one can attempt to
explain the atmospheric and solar neutrino anomalies (see e.g. \cite{Dienes:1999sb,Dvali:1999cn,Barbieri:2000mg,Lukas:2000wn,Lukas:2000rg,Cosme:2000ib}).

\begin{figure}[t]
\begin{center}
\begin{picture}(300,200)(0,50)

\SetWidth{2}
\Line(10,50)(10,250)
\Line(290,50)(290,250)

\SetWidth{0.5}
\Line(150,50)(150,250)
\Line(10,150)(290,150)

\Text(-10,250)[c]{$''+''$}
\Text(310,250)[c]{$''+''$}
\Text(170,250)[c]{$''-''$}


{\SetColor{Green}
\Curve{(10,240)(50,192)(65,181)(80,173)(95,167)(110,162)(130,157)(150,155)}
\Curve{(150,155)(170,157)(190,162)(205,167)(220,173)(235,181)(250,192)(290,240)}
}


{\SetColor{Red}
\DashCurve{(10,240)(50,192)(65,181)(80,173)(95,167)(110,161)(130,155)(150,150)}{3}
\DashCurve{(150,150)(170,145)(190,139)(205,133)(220,127)(235,119)(250,108)(290,60)}{3}
}


{\SetColor{Brown}
\DashCurve{(10,145)(15,149)(20,150)(40,153)(65,158)(140,209)(150,210)}{1}
\DashCurve{(150,210)(160,209)(235,158)(260,153)(280,150)(285,149)(290,145)}{1}
}
\end{picture}
\end{center}

\caption{The right-handed fermion zero mode  (solid line), first (dashed line) and second
(dotted line) KK states wavefunctions in the symmetric $''+-+''$
model. The same pattern can occur
for the corresponding wavefunctions in the $''++''$ model. The
wavefunctions of the zero and the first KK mode are localized on the
positive tension branes. Their absolute value differ only in the
central region where they are both suppressed resulting to a very
light first KK state.}

\end{figure}
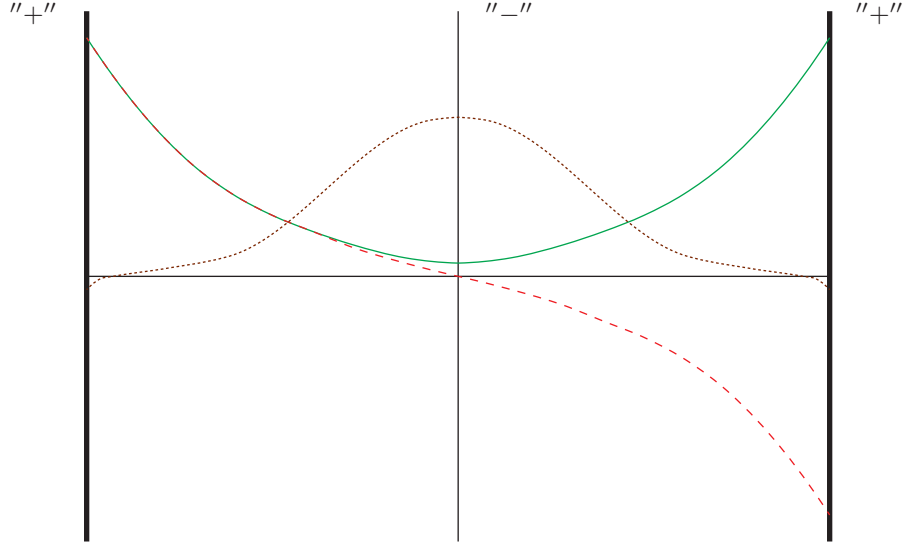

In the context of brane world models with localized gravity \cite{Gogberashvili:1998vx,Randall:1999ee,Randall:1999vf} (non
factorizable geometry) small neutrino masses can again be achieved,
without invoking the see-saw mechanism. In more detail, in this case  the
mechanism generating the small coupling between the Lepton doublet and
the Higgs which live on the brane and the  right-handed sterile neutrino zero mode is not due to the
compactification volume (which is now small)  but due to the fact that
the sterile neutrino wavefunction can be localized [18-30]
on a distant brane.
One may thus arrange that the overlap between this mode and the SM
brane is sufficiently small. 
In this case the $AdS_{5}$ geometry localizes the fermion zero mode on
negative tension branes.  
Localization can occur on positive branes if a mass term of
the appropriate form is added to counterbalance the effect of
the $AdS_{5}$ geometry, by applying the ideas presented
in Ref.\cite{Jackiw:1976fn,Rubakov:1983bb}. Such a mass term appears
naturally in the case that, the branes arise
as the limiting cases of domain walls that are created
from a five-dimensional scalar field with an nontrivial ground state
(kink or multi-kink) \cite{Kehagias:2000au,Kehagias:2000dg} \footnote{However, negative
tension branes cannot be generated by this mechanism (\textit{i.e.}
using a scalar field) if the branes are 
flat. However, we will use the $''+-+''$  as a toy
models since, its phenomenology  related with the bulk fermion
resembles the one corresponding to the $''++''$ model, which  is
possible to be contructed by a scalar field that will naturally couple 
to the fermion field giving it the desired localization. For more
details see Appendix}. In this case the scalar field  naturally
couples to the bulk fermion through an non trivial ``mass'' term which 
can naturally induce localization to the bulk modes on positive
tension branes.

In this Chapter we analyze the localization of a bulk fermion  the
mass spectrum and the coupling between SM neutrino and bulk states in
the context of multi-brane worlds (e.g. see Ref.\cite{Hatanaka:1999ac}). We discuss in detail the conditions
and the options for the localization in relation to the form of the
bulk mass term.
We also discuss  
the possibility of generating small neutrino masses in the context of
$''++-''$, $''+-+''$\footnote{We consider the $''+-+''$ configuration
as a toy-model ignoring the phenomenological difficulties associated
with the  presence of a negative tension brane \cite{Pilo:2000et} since it's
phenomenology is very similar to the $''++''$ which includes only
positive tension branes.}, $''++''$ models. We determine for which
regions of the parameter space lead to a solution of the
hierarchy problem and generation of small neutrino masses.
The study of $''+-+''$ and
$''++''$ models reveals the possibility of  an ultralight KK state of
the bulk fermion  analogous to the  KK graviton in the gravitational sector 
\cite{Kogan:2000wc,Mouslopoulos:2000er,Kogan:2000xc,Kogan:2001vb}.
This is due to the fact that, in this region the wavefunction of the
right-handed bulk fermion states obeys a similar equation to that of
the graviton. 
The  above fermion state, when exists, imposes even more severe
constraints on the parameter space of these
models. 



\section{General Framework}

Following the framework introduced in Ref.\cite{Grossman:2000ra}, we
consider a spinor $\Psi$  in a five dimensional $AdS_{5}$ space-time, where the
extra dimension is  compact and has the geometry of an orbifold
$S^{1}/Z_{2}$. The  $AdS_{5}$ 
background geometry is described by \footnote{We will assume that the
background metric is not modified by the presence of the bulk fermion,
that is, we will neglect the back-reaction on the metric from the bulk
fields.}:
\be
ds^2=e^{-2\sigma(y)}\eta_{\mu\nu}dx^\mu dx^\nu - dy^2
\label{neutr1}
\ee
where the warp factor $\sigma(y)$  depends on the details of the model considered. For the moment we assume that
we have a model with a number of positive and negative tension flat
branes (the sum of the brane tensions should be zero if one wants 
flat four dimensional space on the branes) and that this function is known ( it can be found by
looking the system gravitationally). 

The action for a Dirac Spinor of a  mass $m$  in such a background is given
by \footnote{We do not include a Majorana mass term, ${\Psi^{T}}C\Psi$,
which is forbidden if the bulk fermion has a conserved lepton number.}:
\be
S=\int d^4 x \int dy \sqrt{G} 
\{  E^{A}_{\alpha}\left[
\frac{i}{2} \bar{\Psi} \gamma^{\alpha} \left( \overrightarrow{{\partial}_{A}}-\overleftarrow{\partial_{A}} \right)
\Psi + \frac{\omega_{bcA}}{8} \bar{\Psi} \{
\gamma^{\alpha},\sigma^{bc} \} 
\Psi \right] - m \Phi \bar{\Psi}\Psi \}
\label{neutr2}
\ee
where 
$G=det(G_{AB})=e^{-8\sigma(y)}$.
The four dimensional representation of the Dirac matrices in
five-dimensional flat space is chosen to be:~$\gamma^{\alpha}=(\gamma^{\mu},i \gamma_{5})$.
The inverse vielbein is given by
$E^{A}_{\alpha}=diag(e^{\sigma(y)},e^{\sigma(y)},e^{\sigma(y)},e^{\sigma(y)},1)$.
Due to the fact that the vielbein is symmetric, the contraction of $\omega_{bcA}$
\footnote{ $\omega_{\mu ab}=\frac{1}{2}(\partial_{\mu}e_{b
\nu}-\partial_{\nu} e_{b \mu}){e_{a}}^{\nu}-\frac{1}{2}(\partial_{\mu}e_{a
\nu}-\partial_{\nu} e_{a \mu}){e_{b}}^{\nu}
-\frac{1}{2}{e_{a}}^{\rho}{e_{b}}^{\sigma}(\partial_{\rho}e_{c
\sigma}-\partial_{\sigma}e_{c \rho}){e^{c}}_{\mu}$, where
${e_{\mu}}^{\nu}$ is the vielbein. } 
with the corresponding term in the action gives vanishing contribution.
The mass term is assumed to be generated by a Yukawa coupling with a
scalar field $\Phi$ which has a nontrivial stable vacuum $\Phi(y)$.

\subsection{The mass term}

A few comments on presence of the mass term are in order. 
The  motivation for introducing a mass term  of this form comes
from the need to  localize  fermion zero modes in the
extra dimension. This is discussed in more details in the next
paragraph.
Note that 
the ``localization'' of the wavefunction of a state does not
necessarily reflect the actual localization of the state, since in
one has to take in account the nontrivial geometry of the extra
dimension - something that is done when we calculate physical
quantities. Also when we note that a state is localized on a brane
we mean that this holds irrespectively if the space is compact or not
(thus the state should be normalizable even in the non-compact case). 
 In order to have a localized
state (zero mode) on a positive brane it is necessary to have an
appropriate bulk mass term.
This is because \cite{Bajc:2000mh,Chang:2000nh} the background $AdS_{5}$ geometry itself
has the opposite effect favouring localization on negative tension branes. 
As we will discuss, this leads to  a critical mass
$m_{cr}$ below which  the localization is still on the
negative brane (if we consider a configuration of a positive and one
negative brane), for $m=m_{cr}$ the is no localization and for
$m>m_{cr}$ the zero mode is localized on the positive tension brane
\footnote{One may ask why one should localize the fermion zero mode on
a positive tension brane and not on a negative ? In the case of RS
model it is obvious that if we demand to solve the hierarchy problem
and in the same time to create small neutrino masses through this
mechanism, one should localize the fermion zero mode on the positive
brane. In the case of
multi-brane models this is not a necessity and thus, in principle,
another  possibility (of course in this case the mass  term - if
needed - should have
different form (e.g. for $m \rightarrow -m$)).}.

Let us return to the specific form  of the mass term and it's generality
since this will become important in later discussions.
Since the mass term is a key element for the localization let us, for
a moment assume that it has the form $m\Phi(y) \bar{\Psi}\Psi$ where
$\Phi(y)$ is the vev of a scalar field which has a nontrivial stable vacuum (its
vev does not depend on  the remaining  spatial dimensions). 
Now if we do a simple calculation, in the case of flat extra dimension
(the general arguments will apply also in our background eq.(\ref{neutr1}), taking
in account also the effects of the $AdS_{5}$ geometry), we find that the above configuration
implies that the zero mode  satisfies a Schr\"ondiger equation with
potential of the form $V(y)=\Phi^{2}(y) -\Phi'(y)$. In
order to localize one state the profile of $\Phi(y)$ should be such
that it creates a  potential well. The way to do this is to assume that the
ground state of the scalar field has a kink or a multi-kink profile \cite{Jackiw:1976fn,Rubakov:1983bb,Kehagias:2000au}. Although
 the details of these profiles depend on the form of the
potential of the scalar field $\Phi$,  if we demand strong
localization of the states, the kink profiles tend to $\theta$-
functions. This implies that the function $\Phi$ in eq.(\ref{neutr2}) can be
considered  as an arbitrary combination of $\theta$-functions
(compatible with the symmetries of the action).   
However, as shown in Ref.\cite{Kehagias:2000au} the same field  $\Phi$
can be used in order to create the branes themselves. This restricts
the possible form of the mass term. Note that negative tension branes
cannot be generated by this mechanism and thus the $''+-+''$ model cannot
considered to be generated from such a scalar field. However if one allows for
$AdS_{4}$ spacetime on the branes, then no negative tension branes
occur (thus the model becomes $''++''$) and it can be shown that the configuration can be generated from 
such a scalar field. It turns out that the phenomenology
accossiated with the bulk fermion in the $''+-+''$ model is similar
to the one of the $''++''$ model if we assume that
the mass term  has a (multi-) kink form, which can be
parametrized as
$\Phi(y)=\frac{\sigma'(y)}{k}$  up to
a sign (this naive parametrization works also in the case of non-flat
branes although the relationship between the scalar and the warp
factor is non-linear).\footnote{Note thought, that if we choose the
opposite sign for the mass term (i.e. $m \rightarrow -m$) the
localization of the fermion zero mode will always be on the negative
tension branes. Taking in account the fact that the latter occurs also,
in a region of the parameter space, in the case that our mass term
choice is the one that appears in eq.(\ref{neutr2}),  we will not consider this
possibility separately since we can easily, as we will see, generalize our results
for $m<0$ (the presence of such a mass term sharpens the localization
of the states on negative branes).}
 The previous argument also supports the $\theta$-function
form of the mass term (and not for example a $\tanh(y)$ profile) since we
assume that the branes are infinitely thin. Note thought that  in 
the $''++''$ model  due to the $AdS_{4}$
geometry on the branes the, $\sigma'(y)$ function does not have just a
$\theta$-function form  but it also involves  kink profiles.

 If one assumes that
the mass is generated by coupling to a scalar field, different from the
one that creates the branes, it can have any form allowed by the
dynamics and the symmetries of the action. For example it can take
 the form: $-m \left( \theta(y)-\theta(-y)) \right)
$. In this case, it  will tend to induce localization  on the brane
siting at the origin of the orbifold. Nevertheless  this will not be
satisfactory option in multi-brane models where the desired
$M_{Pl}/M_{EW}$ hierarchy is not
generated between the two first branes (e.g. $''+-+''$ or even in the
$''++''$ model)
since in these cases it  will not generally be possible to 
simultaneously solve the hierarchy problem and
generate small neutrino masses (it would work though in the $''++-''$
model - or it could work in the cases where the background induced
localization dominates but in that case the mass term would be
unnecessary anyway). In any case since we are interested in the most
economic, in terms of parameters and fields, models we will not
consider these possibilities.

The geometry we consider has a $Z_{2}$ symmetry ( $y \rightarrow -y$ ). 
Under this the fermion parity is defined as: $\Psi(-y)=\gamma_{5}\Psi(y)$
(i.e. $\Psi_{L}(-y)=-\Psi_{L}(y)$, $\Psi_{R}(-y)=\Psi_{R}(y)$ ) and
 changes the sign of a Lagrangian mass term of the form: $m\bar{\Psi}\Psi=m(\bar{\Psi}_{L}\Psi_{R}+\bar{\Psi}_{R}\Psi_{L})$.
The full mass term  however is invariant under the $Z_{2}$ since the function
$\sigma'(y)$ is also odd under the reflections $y \rightarrow -y$ . 
With this definition of
parity  one of the wavefunctions will be
symmetric and the other antisymmetric with respect to the center of the orbiford
. Note that this implies that the odd wavefunction will be zero
at the orbifold fixed points (i.e. zero coupling to fields confined to
that points).  
Since we would like in what follows to use the
right-handed component in order to give mass to SM neutrinos, which
could be  confined on a brane at an orbifold fixed point,  we choose the
right-handed wavefunction to be even (i.e. non-vanishing coupling)
and the left-handed to be odd.

\subsection{The KK decomposition}

It is convenient to write the action in terms of the fields: $\Psi_{R}$
and $\Psi_{L}$ where
$\Psi_{R,L}=\frac{1}{2}(1\pm \gamma_{5})\Psi$ and
$\Psi=\Psi_{R}+\Psi_{L}$.
The action becomes:

\ba
S=\int d^4 x \int dy  
\{  e^{-3\sigma}\left( \bar{\Psi}_{L}i \gamma^{\mu} \partial_{\mu} \Psi_{L} +
\bar{\Psi}_{R}i \gamma^{\mu} \partial_{\mu} \Psi_{R} \right) -  e^{-4\sigma} m
\left(\frac{\sigma'(y)}{k}\right) \left( \bar{\Psi}_{L}\Psi_{R} +
\bar{\Psi}_{R}\Psi_{L} \right) \nonumber \\ -\frac{1}{2}\left[ \bar{\Psi}_{L} (e^{-4\sigma}\partial_{y}+\partial_{y}e^{-4\sigma} ) \Psi_{R} -
\bar{\Psi}_{R}(e^{-4\sigma}\partial_{y}+\partial_{y}e^{-4\sigma} ) \Psi_{L}  \right] 
\ea
writing   $\Psi_{R}$ and
$\Psi_{L}$ in the form:
\be
\Psi_{R,L}(x,y)=\sum_{n}\psi^{R,L}_{n}(x)e^{2\sigma(y)}f_{n}^{R,L}(y)
\ee
the action can be brought in the form
\be
S=\sum_{n} \int d^4 x \{\bar{\psi}_{n}(x) i \gamma^{\mu} \partial_{\mu} \psi_{n}(x) -
m_{n}\bar{\psi}_{n}(x) \psi_{n}(x) \}
\ee
provided the wavefunctions obey the following equations
\ba
\left( -\partial_{y}
+m\frac{\sigma'(y)}{k}\right)f^{L}_{n}(y)=m_{n}e^{\sigma(y)}f^{R}_{n}(y)
\nonumber \\
\left( \partial_{y} +m\frac{\sigma'(y)}{k}\right)f^{R}_{n}(y)=m_{n}e^{\sigma(y)}f^{L}_{n}(y)
\ea
and  the orthogonality relations (taking  account of the $Z_{2}$ symmetry):
\be
\int_{-L}^{L} dy  e^{\sigma(y)} {f^{L}}^{*}_{m}(y) f^{L}_{n}(y)=\int_{-L}^{L} dy e^{\sigma(y)} {f^{R}}^{*}_{m}(y) f^{R}_{n}(y)=\delta_{mn}
\ee
where we assume that the length of the orbifold is $2L$.

We solve the above system of coupled differential equations by
substituting $ f^{L}_{n}(y)$ from the second in the first equation. Thus we end up
with a second order differential equation, which can always be brought
to a Schr\"odinger form by a convenient coordinate transformation from y
to z coordinates  related through
$\frac{dz}{dy}=e^{\sigma(y)}$, the coordinate transformation  chosen to
 eliminate the terms involving first derivatives. 
Thus we end up with the differential equation of the form:

\be
\left\{-
\frac{1}{2}\partial_z^2+V_{R}(z)\right\}\hat{f}^{R}_{n}(z)=\frac{m_n^2}{2}\hat{f}^{R}
_{n}(z)
\label{neutr8}
\ee

\be
{\rm with}\hspace*{0.5cm} V_{R}(z)=\frac{\nu(\nu+1)(\sigma'(y))^{2}}{2[g(z)]^2}-
\frac{\nu}{2[g(z)]^2}\sigma''(y)
\label{neutr9}
\ee
Here $\hat{f}^{R}_{n}(z)=f^{R}_{n}(y)$ and we have defined $\nu\equiv\frac{m}{k}$ and
$g(z)\equiv e^{\sigma(y)}$. The left handed wavefunctions are given by
\footnote{Note that it can be shown that the left-handed component
obeys also a similar Schr\"odinger equation with $V_{L}(z)=\frac{\nu(\nu-1)k^{2}}{2[g(z)]^2}+
\frac{\nu}{2[g(z)]^2}\sigma''(y)$ which is the same as $V_{R}$ for
$\nu \rightarrow -\nu$. }:

\be
f^{L}_{n}(y)= \frac{e^{-\sigma(y)}}{m_{n}} \left( \partial_{y} +m\frac{\sigma'(y)}{k}\right)f^{R}_{n}(y)
\label{neutr10}
\ee                   
The form of eq.(\ref{neutr8})  and (\ref{neutr9}) are
 exactly the same as that satisfied by the graviton
when $\nu=\frac{3}{2}$ \cite{Kogan:2000wc}.
For any $\nu$, we note that before orbifolding  the system supports
two zero modes (left-handed and right-handed) \cite{Grossman:2000ra}. 
  However, the orbifold
compactification leaves only a chiral right-handed zero mode. 

We  note that since the bulk fermion mass, $m$, is a
parameter that appears in the original five dimensional Lagrangian
its ``natural'' value  is of the order of the five dimensional Planck
scale $M_{5}$. Now since we assume that  $k<M_{5}$ (in order to trust our
perturbative analysis when we consider the configuration
gravitationally) it is clear that the
``physical'' value of $\nu$ is $\nu>1$. However, we will always
comment on the behaviour of our results out of this region (even for
negative values). 
  
We are particularly interested in the coupling of the bulk spinor  to
the SM neutrinos since  this is the way that the neutrino masses will
be generated. In order to avoid weak scale neutrino masses and lepton
number violating interactions we assign lepton number $L=1$ to the
bulk fermion state and thus the only gauge invariant coupling is of
the form

\be
S_{Y}=-\int d^{4} x \sqrt{- g_{Br}}\{Y_{5}\bar{L_{0}}(x)\widetilde{H_{0}}(x)\Psi_{R}(x,L_{Br})+h.c.\} 
\ee
where $H_{0}$ is the SM Higgs field, $L_{0}$ is the SM lepton doublet, $\widetilde{H_{0}}=i\sigma_{2}H_{0}^{*}$, $g^{Br}_{\mu\nu}$ is the
induced metric on the brane and
$g_{Br}=det(g^{Br}_{\mu\nu})$. The Yukawa parameter $Y_{5}$ has mass dimension
$-\frac{1}{2}$ and thus since it appears a parameter in the five
dimensional action, its ``natural'' value is  $Y_{5} \sim
\frac{1}{\sqrt{M_{5}}} \sim \frac{1}{\sqrt{k}}$.

 To obtain canonical normalization for the kinetic terms of the SM
neutrino we perform the following field rescalings $H_{0}\rightarrow
e^{\sigma(L_{Br})} H$,~$L_{0}\rightarrow
e^{3 \sigma(L_{Br})/2} L$ , where $L_{Br}$ is the position of the
brane that SM is confined. This gives 

\be
S_{Y}=- \sum_{n\ge0}\int d^{4} x \{y_{n}\bar{L}(x)\widetilde{H}(x)\psi_{n}^{R}(x)+h.c.\} 
\ee
where 
\be
y_{n} \equiv e^{\sigma(L_{Br})/2}~Y_{5}~f^{R}_{n}(L)=(g(z_{Br}))^{1/2}~Y_{5}~{\hat{f}^{R}}_{n}(z_{L})
\label{neutr13}
\ee
From the above interaction terms we can read off the mass matrix
$\mathbf{M}$ that appears in the Lagrangian as  $\bar{\psi}^{\nu}_{L}\mathbf{M}\psi^{\nu}_{R}+ h.c.$
where we have defined
$\psi^{\nu}_{L}=(\nu_{L},\psi_{1}^{L},\ldots,\psi_{n}^{L})$
and
$\psi^{\nu}_{R}=({\psi_{0}}^{R},\psi_{1}^{R},\ldots,\psi_{n}^{R})$.
The mass matrix for the above class of models has the
following form
\begin{displaymath}
\mathbf{M}=
\left( \begin{array}{cccc}
\upsilon y_{0} & \upsilon y_{1} &  \ldots &  \upsilon y_{n} \\
0 & m_{1} &  \ldots &  0 \\
\vdots & 0 &  \ddots &  0 \\
0 & 0 &  \ldots &  m_{n} 
\end{array} \right)
\end{displaymath}


\section{Neutrinos in RS model}

For completeness of our analysis, we first briefly review the case of bulk
fermion spinors in the RS model \cite{Grossman:2000ra}.
This model consists of one positive (hidden) and one negative  tension
brane (where
the SM fields are confined)
placed on the fixed points ($y=0$, $L_{1}$) of a $S^{1}/Z_{2}$
orbifold (for details see \cite{Randall:1999ee}). In this case
the background geometry is described by eq.(\ref{neutr1})  where  $\sigma(y)=k|y|$.
The convenient choice of variable, for the reasons
described in the previous section, is:

\be
z\equiv\frac{e^{ky}-1}{k} ~~~y\in[0,L_1]
\ee
Since in this model we have $(\sigma'(y))^2=k^2$ and $\sigma''(y)=
2kg(z) \left[\delta(z)-\delta(z-z_1)\right]$, the potential appearing
in eq.(\ref{neutr8}) of the Schr\"odinger equation that the wavefunction of the
right-handed bulk fermion is (for $z\ge 0$):
\be
{\rm}\hspace*{0.5cm} V_{R}(z)=\frac{\nu(\nu+1)k^{2}}{2[g(z)]^2}-
\frac{\nu}{2g(z)} 2k \left[\delta(z)-\delta(z-z_1)\right]
\ee
Here we have defined
$g(z)\equiv kz+1$ and $z_1\equiv z(L_1)$.

This potential always
gives rise to a (massless) zero mode.  It is given by
\be
\hat{f}_{0}^{R}(z)=\frac{A}{[g(z)]^{\nu}}
\ee

From the above expression it seems that the zero mode is always
localized on the positive tension brane for all values of
$\nu$. Nevertheless, by taking the second brane to infinity, we find that the zero
mode is normalizable  in the case that $\nu> \frac{1}{2}$ and that it
fails to be normalizable  when $0 \le \nu \leq \frac{1}{2}$.
The above, as we mentioned, shows that only when $\nu> \frac{1}{2}$
the zero mode is localized on the first brane. For $\nu= \frac{1}{2}$
there is no localization and for $ 0 \le \nu < \frac{1}{2}$ it is localized
on the negative brane. Another way to see the above is to find  the coupling of the KK states to mater
of a ``test'' brane as a function of the distance from the first
(hidden) brane. From eq.(\ref{neutr13}) we can find that in the case of $\nu > \frac{1}{2}$ the
coupling decreases as we go away from the first brane, on the other
hand it is constant when $\nu=\frac{1}{2}$ (no localization), and increases when
$0 \le \nu< \frac{1}{2}$ (localization on the second brane).
In any case as we previously mentioned the ``natural'' value for $\nu$
can be considered to be greater than unity (having already restricted
ourselves in the region $\nu > 0$) and thus we will assume in
the following discussions that the right-handed zero mode is always
localized on the hidden positive tension brane and we will  briefly discuss
the rest possibilities. Note that  all the following expressions for the
masses and the coupling are valid under the assumption that  $\nu >
\frac{1}{2}$, as the  results for the rest of the parameter space are different.
In this case  the normalization constant is $A\simeq \sqrt{k
(\nu-\frac{1}{2})}$.

Apart from the zero mode we have to consider the left and right-handed
KK modes which correspond to solutions for $m_{n}>0$. The solutions
for the right-handed wavefunctions in
this case are given in terms of Bessel
functions \footnote{Note that in the case that $\nu=N+\frac{1}{2}$, where $N$ is an
integer, the two linearly independent solutions are: $J_{N+1}$ and
$Y_{N+1}$. Although for our calculations we have assumed that
$ \nu \neq N +\frac{1}{2}$, all the results for the mass spectrum and the
couplings are valid also in the special cases when  $\nu=N+\frac{1}{2}$.}:

\be
{\hat{f}}^{R}_{n}(z)=\sqrt{\frac{g(z)}{k}}\left[A J_{\nu+\frac{1}{2}}\left(\frac{m_n}{k}g(z)\right)+B
J_{- \nu -\frac{1}{2}}\left(\frac{m_n}{k}g(z)\right)\right]
\ee
These solutions must obey the following  boundary conditions:
\ba
{{\hat f}^{R}_{n}}~'({0}^{+})+\frac{k \nu}{g(0)} {\hat f }^{R}_{n}(0)=0 \nonumber \\
{{\hat f}^{R}_{n}}~'({z_{1}}^{-})+\frac{k \nu}{g(z_{1})} {\hat f }^{R}_{n}(z_{1})=0
\ea

The wave functions of the left-handed KK states can be easily
extracted from eq.(\ref{neutr10}).
The  boundary conditions give a 2~x~2 system for $A$,$B$ which, in order
to have a nontrivial solution, should  have vanishing determinant. This
gives the quantization of the spectrum.
For $\nu> \frac{1}{2}$ the quantization condition  can be approximated by a simpler one:
$J_{\nu-\frac{1}{2}}\left(\frac{m_{n}g(z_{1})}{k}\right)=0$
This implies that the KK spectrum of the bulk state is:

\be
m_{n}=\xi_{n}~k~e^{-k L_{1}}
\ee
(for $n\ge{1}$), where $\xi_{n}$ in the n-th root
of $ J_{\nu-\frac{1}{2}}(x)$ . This means that if one is interested in solving the
hierarchy in the context of this model, i.e. $w \equiv e^{-k L_{1}}\sim 10^{-15}$
the mass of the first  bulk spinor KK state will be of the order of
1 TeV and the spacing between the tower will be of the same order.
To, summarize the spectrum in this case consists of a chiral
right-handed massless zero mode and a tower of Dirac KK states with
masses that start from 1 TeV (if a solution of the hierarchy is required)
with $\sim$1 TeV spacing.
The other important point for the phenomenology is the coupling of the
bulk spinors to the SM neutrino.
It is easy, using eq.(\ref{neutr13}), to find that the zero mode couples as
\be
\upsilon ~ y_{0}=\upsilon ~ Y_{5} ~ \sqrt{k (\nu-\frac{1}{2})}~~ \left(
\frac{1}{g(z_{1})} \right)^{\nu-\frac{1}{2}} \simeq \upsilon ~ \sqrt{\nu-\frac{1}{2}}~~ w^{\nu-\frac{1}{2}}
\ee
since, the hierarchy factor is defined as $w \equiv
\frac{1}{g(z_{1})}$ and, as mentioned in the previous section,  $Y_{5} \sim
\frac{1}{\sqrt{k}}$, $\upsilon \sim 10^2$ GeV and $\nu >
\frac{1}{2}$. 

In a similar fashion we can find  the couplings of the SM neutrino to
bulk KK states. In this particular model it turns out that this
coupling does not depend on the fermion mass or the size of the orbifold  and thus
it is a constant. By a simple calculation we find  that 

\be
\upsilon y_{n}~\simeq \sqrt{2} \upsilon ~ Y_{5} \sqrt{k} \simeq
\sqrt{2} ~ \upsilon 
\ee

Thus from the above we  see that the KK tower couples to SM neutrino with a TeV
strength.
In order to find the mass eigenstates and the mixing between the SM
neutrino and the sterile bulk modes one has to diagonalize the matrix
$\mathbf{M M^{\dag}}$ (actually one finds the squares of the mass
eigenvalues). By performing the above diagonalization,
 choosing  $ e^{-k L_{1}}\sim 10^{-15}$  it turns out that the
mass of the neutrino will be of the order
$m_{\nu} \sim 10^{2} ~(10^{-15})^{\nu-\frac{1}{2}}$  (e.g. for $\nu=\frac{3}{2}$, $m_{\nu}
\sim 10^{-4}$ eV), and the masses of the bulk states are of the order of
1TeV with a 1TeV spacing. From the last calculations it appears that one
can easily create a small neutrino mass and at the same time arrange
for  the desired mass hierarchy when $\nu > \frac{1}{2}$. 
 Apart from creating  small masses, one has to  check that the mixing between the SM neutrino and
the KK tower is  small enough  so that there is no conflict with phenomenology. It was shown in
Ref.\cite{Grossman:2000ra} that this can be done for this model
without fine-tuning. Note that the parameter space: $\nu \le
\frac{1}{2}$ (including negative values) is not of interest in the
present discussion \footnote{This could be of particular interest if
one uses the above mechanism to localize SM fermions on the negative
tension brane and in the same time solving the hierarchy problem
(e.g. see Ref.\cite{delAguila:2000kb}).} since it would be impossible to solve the hierarchy
problem and in the same time to assign small masses to neutrinos.


\section{Neutrinos in $''++-''$ model}

Since we are interested in studying the characteristics of bulk
fermion modes in multi-brane configurations we add to the $''+-''$ RS
model another positive tension brane
 where now SM fields will be confined. Thus we  end up with two different
configurations: the $''++-''$ model which will be the subject of this
section
and the  $''+-+''$ model which will be the subject of the next section.

The $''++-''$ model consists of two positive  and one negative tension
brane. The first positive brane is placed on the origin of the orbifold
at $y=0$ the second (where the SM fields are confined), which is freely moving, is place at $y=L_{1}$ and
the negative brane is placed at the second fixed point of the orbifold
at $y=L_{2}$.

In the present model the convenient choice of  variables  is defined as:
\be
\renewcommand{\arraystretch}{1.5}
z\equiv\left\{\begin{array}{cl}\frac{2e^{k_{1}L_1}-1}
{k_{1}}&y\in[0,L_1]\\\frac{e^{k_{2}(y-L_1)+k_{1}L{1}}}{k_{2}}+\frac{e^{k_{1}L_{1}}-1}
{k_{1}}-\frac{e^{k_{1}L_1}}
{k_{2}}&y\in[L_1,L_2]\end{array}\right.
\
\ee
Note the presence of two bulk curvatures, namely $k_{1}$ and $k_{2}$ 
in this model, which is the price that we have to pay in order to place
two positive branes next to each other  ($k_{1}<k_{2}$ but with $k_{1}
\sim k_{2}$ so that we don't introduce another hierarchy. For details see Ref.\cite{Kogan:2000xc}).
In terms of the new variables we can find that the potential
$V_{R}(z)$ of the Schr\"odinger equation that corresponds to the present
model has the form (for $z \ge 0$):

\ba
\hspace*{0.5cm} V_{R}(z)=&\frac{\nu (\nu+1)}{2[g(z)]^2}(k_{1}^2(\theta(z)-\theta(z-z_{1}))+k_{2}^2(\theta(z-z_{1})-\theta(z-z_{2})))\nonumber\\&-
\frac{\nu}{2g(z)} 2\left[k_{1}\delta(z)+\frac{(k_{2}-k_{1})}{2}\delta(z-z_1)-k_{2}\delta(z-z_2)\right]
\ea
since $\sigma''(y)=2 g(z) \left[k_{1}\delta(z)+\frac{(k_{2}-k_{1})}{2}\delta(z-z_1)-k_{2}\delta(z-z_2)\right]$
and  $(\sigma'(y))^{2}=k_{1}^2$ for $y\in[0,L_{1}]$ and 
 $(\sigma'(y))^{2}=k_{2}^2$ for $y\in[L_{1},L_{2}]$. 
The function $g(z)$ is defined as: 
\be
\renewcommand{\arraystretch}{1.5}
g(z)=\left\{\begin{array}{cl}{k_{1}z+1}&z\in[0,z_{1}]\\{k_{2}(z-z_{1})+k_{1}z_{1}+1}&z\in[z_{1},z_{2}]\end{array}\right.
\
\ee
where $z_{0}=0$, $z_1=z(L_1)$ and $z_{2}=z(L_{2})$ are the positions of the
branes in terms of the new variables.

This potential always
gives rise to a (massless) zero mode whose wavefunction  is given by
\be
\hat{f}_{0}^{R}(z)=\frac{A}{[g(z)]^{\nu}}
\ee

The discussion of the previous section  about the state localization applies
in this model as well . For $\nu > \frac{1}{2}$ the zero mode is
localized on the first brane. For the case  $\nu = \frac{1}{2}$
 there is no localization again. For  $\nu < \frac{1}{2}$ it is localized on
the negative tension brane, as expected. 
In the case  $\nu > \frac{1}{2}$  we find that the normalization
factor of the zero mode is
$A\simeq\sqrt{k_{1}(\nu-\frac{1}{2})}$ which is the same as in the
case of RS for $k=k_{1}$ (not surprisingly since it is strongly localized on the
first brane).   

The wavefunctions for the right-handed  KK modes are given in terms of Bessel
functions. For $y$ lying in the regions ${\bf A}\equiv\left[0,L_1\right]$ and
${\bf B}\equiv\left[L_1,L_2\right]$, we have:

\be
\hat{\Psi}^{(n)}\left\{\begin{array}{cc}{\bf A}\\{\bf 
B}\end{array}\right\}=\left\{\begin{array}{cc}\sqrt{\frac{g(z)}{k_{1}}}\left[A_{1}J_{\nu+\frac{1}{2}}\left(\frac{m_n}{k_{1}}g(z)\right)+B_{1}J_{-\nu-\frac{1}{2}}\left(\frac{m_n}{k_{1}}g(z)\right)\right]\\\sqrt{\frac{g(z)}{k_{2}}}\left[A_
{2}J_{\nu+\frac{1}{2}}\left(\frac{m_n}{k_{2}}g(z)\right)+B_{2}J_{-\nu-\frac{1}{2}}\left(\frac{m_n}{k_{2}}g(z)\right)\right]\end{array}\right\}
\ee

with boundary conditions:

\ba
{{\hat{f}}^{R}_{n}}~'({0}^{+})+\frac{k_{1} \nu}{g(0)}{\hat{f}}^{R}_{n}(0)=0
\nonumber \\
{{\hat f}^{R}_{n}}({z_{1}}^{+})-{\hat f}^{R}_{n}({z_{1}}^{-})=0 \nonumber \\
{{\hat f}^{R}_{n}}~'({z_{1}}^{+})-{{\hat f}^{R}_{n}}~'({z_{1}}^{-})-\frac{2 
\nu}{g(z_{1})}\left( \frac{k_{2}-k_{1}}{2} \right) {\hat f}^{R}_{n}(z_{1})=0 \nonumber \\
{{\hat f}^{R}_{n}}~'({z_{2}}^{-})+\frac{k_{2} \nu}{g(z_{2})}{\hat f}^{R}_{n}(z_{2})=0
\ea

The above boundary conditions result  to a 4~x~4 homogeneous system for
$A_{1}$, $B_{1}$, $A_{2}$ and $B_{2}$ which, in order to have a nontrivial
solution, should have a vanishing determinant. This imposes a quantization condition from
which we are able to extract the mass spectrum of the bulk spinor. The
spectrum consists, apart from the chiral (right-handed) zero mode
(massless) which was mentioned earlier, of
a tower of Dirac KK modes.

In this case in order to provide a solution to the hierarchy problem we
have to arrange the distance between the first two branes so that 
we create the desired hierarchy $w$. In the
present model we have an additional parameter which is the distance
between the second and the third brane $x\equiv k_{2}(L_{2}-L_{1})$
For the region where $x\gtsim 1$ we can find analytically that all the masses the KK tower (including the
first's) scale the same way as we vary the length of the orbifold
$L_{2}$ : 

\be
m_{n}=\zeta_{n}~wk_{2}~e^{-k_{2}L_{2}}
\label{neutr28}
\ee
 where $\zeta_{n}$ is the n-th root of $J_{\nu-\frac{1}{2}}(x)=0$.

In the region $x< 1$ the previous relation for the mass
spectrum breaks down. This is expected since for $x=0$
($L_{2}=L_{1}$)  the $''++-''$ model becomes $''+-''$ (RS) and
the quantization condition becomes approximately
$J_{\nu-\frac{1}{2}}\left(\frac{m}{k_{1}}g(z_1)\right)=0$, which is
identical to the RS condition. So for
$0\leq x\leq 1$ the quantization condition (and thus the mass spectrum) interpolates between
the previous two relations.

Let us now turn to the coupling between the SM neutrino which lives on
the second positive brane with the bulk right-handed zero mode and the
rest of the KK tower. We can easily derive that zero mode couples in
that same way  as in the RS case (the normalization of the zero mode
is approximately the same):
\be
 \upsilon ~ y_{0}=\upsilon ~ Y_{5} ~ \sqrt{k_{1}(\nu-\frac{1}{2})}~~
\left(\frac{1}{(g(z_{1}))}\right)^{\nu-\frac{1}{2}} \simeq \upsilon ~  \sqrt{\nu-\frac{1}{2}}~~
w^{\nu-\frac{1}{2}}
\ee
where, as we previously mentioned,   $Y_{5} \sim
\frac{1}{\sqrt{k_{1}}}$,~ $\upsilon \sim 10^{2}$ GeV, $\nu >
\frac{1}{2}$ and $w \equiv \frac{1}{g(z_{1})}$.
On the other hand the coupling of the SM neutrino to bulk KK states is
given by:
\be
\upsilon ~ y_{n}\sim \upsilon ~ \sqrt{\nu} 
\left(\frac{k_{2}}{k_{1}}\right)^{3/2} \frac{8\zeta_{n}^2}{J_{\nu+\frac{1}{2}}\left(\zeta_{n}\right)}~e^{-3x}
\label{neutr30}
\ee
where $\zeta_{n}$ is the n-th root of $J_{\nu-\frac{1}{2}}(x)=0$.

All approximations become better away from  $\nu = \frac{1}{2}$, $x=0$,
and for higher KK levels.
Note the strong  suppression in the coupling scaling law. This rapid
decrease, which is distinct among the models that we will consider,
also appears in the coupling (to matter) behaviour of the graviton KK
states and  a detailed explanation can be found in Ref.\cite{Kogan:2000xc}.  

Thus from the above we conclude that for $\nu > \frac{1}{2}$ the
phenomenology of this model resembles, in the general characteristics,
the one of RS. Of course in the present model there is an extra
parameter, $x$, which controls the details of the masses and couplings
of the KK states. Since the zero mode coupling is independent of $x$
the general arguments of the previous section about creating small
neutrino masses apply here as well, at least for small $x$ . By
increasing $x$ we make the KK tower lighter, as we see from eq.(\ref{neutr28}), but
 we avoid large mixings between the SM neutrino and the left-handed
bulk states due to the fact that the coupling between the SM neutrino
and the right-handed bulk states drops much faster according with eq.(\ref{neutr30}).

Note that  in the case  $\nu < \frac{1}{2}$ (negative values included) the zero mode 
will be localized on the negative brane and thus one could arrange the
parameter $x$ so that the exponential suppression of the bulk
fermion's zero mode coupling  on the
second brane is such that gives small neutrino masses. Thus in this
case it seems that we are able to solve the hierarchy problem by
localizing the graviton on the first positive brane and in the same
time create small neutrino masses by localizing the bulk fermion zero
mode on the negative brane. Nevertheless, one should make sure that no
large mixings are induced in this case.


\section{Neutrinos in $''+-+''$ model}

We now turn to examine bulk spinors in the $''+-+''$ model, which was analyzed
in detail in Ref.\cite{Kogan:2000wc,Mouslopoulos:2000er}. The model consists of two positive tension
branes placed at the orbifold fixed points and a third, negative brane
which is freely moving in-between. SM field are considered to be
confined on the second positive brane.
Of course the
presence of a moving negative brane is problematic  since it gives
rise to a radion field with negative kinetic term (ghost state)
\cite{Dvali:2000km,Pilo:2000et} in the gravitational sector.
 Nevertheless  we are interested in the  general characteristics of
this model . The interesting feature of this
model is the bounce form of the warp factor which gives rise to an
ultralight graviton KK state as described in Ref.\cite{Kogan:2000wc}. It was shown in Ref.\cite{Kogan:2001vb}
that exactly this feature, of a bounce in the warp factor, can be
reproduced even in the absence of negative branes in the $''++''$
model where this is done by sacrificing the flatness of the branes
(the spacetime on the branes in this case is $AdS_{4}$). 
Thus we will handle the $''+-+''$ as a toy model since in this
case there can be simple analytical calculations of the coupling etc. The
general characteristics will persist in the $''++''$ case.

We are interested to see if this configuration as well as  an ultralight graviton supports an ultralight spinor
field. In order to see this we should check the form of the potential
of the Schr\"odinger equation that the right-handed component obeys. 
We can easily find that the  potential is (for $z \ge 0$):
\be
{\rm }\hspace*{0.5cm} V_{R}(z)=\frac{\nu(\nu+1) k^2}{2[g(z)]^2}-
\frac{\nu}{2g(z)} 2k \left[\delta(z)+\delta(z-z_2)-\delta(z-z_1)\right]
\ee
since $(\sigma'(y))^2=k^2$ and $\sigma''(y)= 2k g(z) \left[\delta(z)+\delta(z-z_2)-\delta(z-z_1)\right]$.
The convenient choice of variables in this case is:
\be
\renewcommand{\arraystretch}{1.5}
z\equiv\left\{\begin{array}{cl}\frac{2e^{kL_1}-e^{2kL_1-ky}-
1}{k}&y\in[L_1,L_2]\\\frac{e^{ky}-1}{k}&y\in[0,L_1]\end{array}\right.
\
\ee
and the function $g(z)$ is defined as $
g(z)\equiv k\left\{z_1-\left||z|-z_1\right|\right\}+1$, where $z_1=z(L_1)$.

As in the previous cases,  the above  potential always supports a (massless) zero mode
with wavefunction of the form:
\be
\hat{f}_{0}^{R}(z)=\frac{A}{[g(z)]^{\nu}}
\ee
In this case the different localization behaviour as a function of
$\nu$ is the following: For $\nu > \frac{1}{2}$ the zero mode is
localized on the positive branes (thus fails
to be normalizable when we send the right positive brane to infinity
but is normalizable when we send both negative and positive to
infinity). For the case $\nu < \frac{1}{2}$  the localization of
the zero mode is on the negative tension brane, as expected. 
In the case  $\nu > \frac{1}{2}$  and for strong hierarchy $w$ we find that the normalization
factor of the zero mode is
$A\simeq\sqrt{k_{1}(\nu-\frac{1}{2})}$
~ (Note that in the case of ``weak'' hierarchy one should be careful
with the assumptions on which the approximations are based on
e.g. for $w=1$ the result must be divided $\sqrt{2}$).   
For the KK modes the solution is given in terms of Bessel
functions. For $y$ lying in the regions ${\bf A}\equiv\left[0,L_1\right]$ and
${\bf B}\equiv\left[L_1,L_2\right]$, we have:
\be
{\hat{f}}^{R}_{n}\left\{\begin{array}{cc}{\bf A}\\{\bf 
B}\end{array}\right\}=\sqrt{\frac{g(z)}{k}}\left[\left\{\begin{array}{cc}A_1\\B_
1\end{array}\right\}J_{\frac{1}{2}+\nu}\left(\frac{m_n}{k}g(z)\right)+\left\{\begin{array}{cc}A_
2\\B_2\end{array}\right\}J_{-\frac{1}{2}-\nu}\left(\frac{m_n}{k}g(z)\right)\right]
\ee
with boundary conditions:
\ba
{{{\hat{f}}^{R}}_{n}}~'({0}^{+})+\frac{k \nu}{g(0)}{{\hat{f}}^{R}}_{n}(0)=0
\nonumber \\
{{{\hat f}^{R}}_{n}}({z_{1}}^{+})-{{{\hat f}^{R}}_{n}}({z_{1}}^{-})=0 \nonumber \\
{{{\hat f}^{R}}_{n}}~'({z_{1}}^{+})-{{{\hat f}^{R}}_{n}}~'({z_{1}}^{-})-\frac{2 k
\nu}{g(z_{1})}{{\hat f}}^{R}_{n}(z_{1})=0 \nonumber \\
{{{\hat f}^{R}}_{n}}~'({z_{2}}^{-})-\frac{k \nu}{g(z_{2})}{\hat f}^{R}_{n}(z_{2})=0
\ea

The  boundary conditions give a 4~x~4 linear homogeneous system for
$A_{1}$, $B_{1}$, $A_{2}$ and $B_{2}$, which, in order to have a nontrivial
solution should have  vanishing determinant. This imposes a quantization condition from
which we are able to extract the mass spectrum of the bulk spinor. The
spectrum consists, apart from the chiral (right-handed) zero mode
(massless) which was mentioned earlier, by
a tower of Dirac KK modes. Nevertheless due to the fact that there are
two positive tension branes present in the model there are now two
``bound'' states in a similar fashion with
Ref.\cite{Kogan:2000wc,Kogan:2001vb} (for $\nu > \frac{1}{2}$). One is the the right-handed zero mode which is
massless and it is localized on the positive brane placed at the
origin of the orbifold and the second is the ultralight right-handed
first KK state which is localized on the second positive brane placed at
the other orbifold fixed point. This can be seen by examining the mass
spectrum and the coupling behaviour of the first KK state in
comparison with the rest of the tower.
  
Firstly let us examine the mass spectrum. In the case that we have
a hierarchy $w$ (where $w\equiv \frac{1}{g(z_{2})}=e^{-\sigma(L_{2})}$) we can find appropriate analytical expressions
for the mass spectrum.

For the first KK state
\be
m_1=\sqrt{4 {\nu}^2 -1 } ~kw~ e^{-(\nu+\frac{1}{2}) x}
\ee
and for the rest of the tower
\be
m_{n+1}= \xi_n ~ kw~ e^{-x} ~~~~~~n=1,2,3, \ldots
\ee
where $\xi_{2i+1}$ is the $(i+1)$-th root of $J_{\nu-\frac{1}{2}}(x)$ ($i=0,1,2,
\ldots$) and $\xi_{2i}$ is the $i$-th root of $J_{\nu+\frac{1}{2}}(x)$ ($i=1,2,3, \ldots$).
The above approximations become better away from the $\nu=\frac{1}{2}$
 , $x=0$ and for higher KK levels $n$.
The first mass is manifestly singled out from the rest of the KK tower
as it has an extra exponential suppression that depends on the mass of
the bulk fermion. By contrast the rest of the KK tower has only a
very small dependence on the mass of the bulk fermion thought the
root of the Bessel function $\xi_{n}=\xi_{n}(\nu)$ which turns out to
be just a linear dependence in $\nu$. Note there is a difference between the
graviton ultralight state (discussed in \cite{Kogan:2000wc,Mouslopoulos:2000er}) and this spinor state: In the case of
gravity the unltralight KK state the mass  scales as a function of $x$
was  $ e^{-2x}$, on the other hand the scaling law in the
case of the ultralight spinor is of the form  $ e^{-(\nu+\frac{1}{2})x}$.
 From the above it seems that the latter can be done much lighter that
the graviton first KK state for a given $x$ by increasing the parameter $\nu$. This
is easy to understand since the role of the mass term, with the kink
or multi-kink profile, is to localize the wavefunction
$\hat{f}(z)$. By increasing the parameter $\nu$ all we do is to force
the  absolute value of the wavefunction of the first KK state and the
massless right-handed zero
mode to become increasingly similar to each other: For example, in the symmetric configuration, the difference between the
 zero mode and the first KK state wavefunctions comes from
the central region of the $''+-+''$ configuration, where the first KK state
wavefunction is zero (since it is antisymmetric) thought the
zero mode's is very small due to the exponential suppression of the wavefunction, but non zero.
 By increasing $\nu$ we force  the value of the zero mode wavefunction at the
middle point
to get closer to zero and thus to resemble even more the
first KK state,
something that appears in the mass spectrum as the  fact that the mass of
the first KK state is approaching to zero. On the other hand the mass
eigenvalues that correspond to the rest
of the tower of KK states will increase linearly their mass by increasing the $\nu$ parameter
since those are not bound states (the first mode has also such a linear
dependence in $\nu$ but it is negligible compared with the exponential
suppression associated with $\nu$ ).

Now let us turn to the behaviour of the coupling of the zero mode and
the KK states to matter living on the third (positive) brane.
As in the previous cases the right-handed zero mode couples to SM left-handed neutrino as
\be
\upsilon ~ y_{0}=\upsilon ~ Y_{5} ~ \sqrt{k (\nu-\frac{1}{2})}~~ {\left(
\frac{1}{g(z_{2})} \right) }^{\nu-\frac{1}{2}} \simeq \upsilon  ~ \sqrt{\nu-\frac{1}{2}}~~
{ w^{\nu-\frac{1}{2}}}
\ee
since $Y_{5} \sim \frac{1}{\sqrt{k}}$. 
From the above relationship we
see that the coupling of the zero mode to SM neutrino will 
generally be suppressed by the hierarchy factor to some power, the power depending on
the bulk fermion mass.  This way one may  readily obtain a very small
coupling.
 The coupling of the zero mode is independent of
$x$. This  is another way to see the localization of this mode
on the first brane (the normalization of the wavefunction is
effectively independent of $x$). Since this model supports a
second ``bound state'' (first KK state) which is localized on the
second brane, we expect something similar to occur in the coupling
behaviour of this state. Indeed, similarly to the graviton case \cite{Kogan:2000wc,Mouslopoulos:2000er}, we can
show that the coupling of this state to the SM neutrino for fixed $w$
is constant, i.e. independent of the $x$ parameter. Taking in account
the result of the graviton KK state $a_{1}=\frac{1}{wM_{Pl}}$ and by
comparing the graviton-matter and spinor matter coupling we can easily
see that the coupling of this special mode will  be of the order of
the electroweak scale:

\be
\upsilon ~y_{1}~\simeq \sqrt{\nu-\frac{1}{2}} ~ \upsilon 
\ee

Let us now consider the coupling of the rest right-handed KK states to
the SM neutrino. We  find that

\be
\upsilon ~y_{n}~\simeq \sqrt{\nu-\frac{1}{2}} ~ \upsilon ~ e^{-x} 
\ee
for $n=0,1,2...$.
From the above relationship we see that the rest of KK states will
generally have exponentially suppressed coupling compared to the first
special state.

The appearance of this special first ultralight and generally strongly
coupled KK state, as in the graviton
case, is going to have radical implication to  the phenomenology of
the model. Let us  consider the
following example suppose that $\nu=\frac{3}{2}$ and that we also require a hierarchy of
the order: $w \sim 10^{-15}$. In this case
the zero mode's coupling is  $ \upsilon y_{0} \sim  \upsilon
10^{-15} \simeq 10^{-4}$ eV a result independent of the $x$
parameter. On the other hand one can check that the rest of KK tower
will have masses $m_{n}\simeq 10^{3}~e^{-x}$ GeV (for $n=2,3...$) with coupling
$ \upsilon y_{n} \sim  \upsilon e^{-x}$. Up to this point the
phenomenology associated with this model is similar to the  RS case
i.e. tiny coupling of the right-handed and generally heavy KK states
with relatively strong coupling. However, taking in account the
special KK state, we have the possibility of obtaining  a much lighter state
with large coupling (effectively independent of how light this state
is). Having a light sterile state whose right-handed mode has strong
coupling to the SM neutrino is potentially dangerous. In such a case
we find that the dominant
contribution to the mass eigenstate  of the lightest mode
(neutrino) $\nu_{phys}$ will come from the left-handed component of
this special sterile mode and not the weak eigenstate $\nu$.
Of course something like this is not
acceptable since there are strict constrains for the mixing of SM
neutrino to sterile states. Since the mass spectrum depends
exponentially on the the $x$ parameter which determines the distance
between the branes, the above argument impose strong constraints on
it's  possible values. 

Finally, in the case that $\nu < \frac{1}{2}$ the bulk right-handed zero
mode is localized on the negative tension brane. In this case a new
possibility arises: By localizing the graviton wavefunction on the
first brane we can explain the SM gauge hierarchy (by setting $w$ to
the desired value ) and by localizing the bulk fermion zero mode on
the negative tension brane to induce small neutrino mass (for
appropriate value of the $x$ parameter) for the SM
neutrino which is confined on the right positive tension brane.  
 Note that for $\nu < \frac{1}{2}$ there is no special bulk spinor KK state
and thus there is no immediate danger of inducing large
neutrino mixing from such a state. However, the presence of the
ultralight graviton KK state is restricting our parameter space as following: 
 In order to solve the gauge hierarchy problem (assuming the SM on the
third brane) we have to fix the one
parameter of the model: $w\sim 10^{-15}$. Since the fermion zero mode is localized
on the intermediate negative brane we have to arrange the distance
between this and the third brane (for given $\nu$), $x$, so that the 
coupling is sufficiently  suppressed in order to give reasonable
neutrino masses. This implies that 
$e^{- |\nu-\frac{1}{2}| x} \sim 10^{-13}$,   if one considers the mass of
the neutrino of the order of $10^{-1}$ eV. From the bounds derived in \cite{Mouslopoulos:2000er}
 we find, for $k\sim10^{17}$ GeV, that in order the ultralight
graviton KK state not to induce modifications of gravity at distances
where Cavendish experiments take place and not to give visible
resonances to  $e^{+}e^{-}\rightarrow \mu^{+} \mu^{-}$ processes, we should have
$4.5<x<15$ or $x<1$. The latter implies certain restrictions to the
values of $\nu$: $1.5<-\nu<6.2$ or $-\nu>29.4$. In the above regions
it is possible to simultaneously create the gauge hierarchy and small
neutrino masses consistently, with the mechanism described earlier.


\section{Neutrinos in $''++''$ model}

As we mentioned in the previous section the $''++''$ model mimics the
interesting characteristics of the $''+-+''$ model without having any
negative tension brane. Thus since the warp factor has a bounce form
this model also supports an ultralight graviton as  was shown in
Ref.\cite{Kogan:2001vb,Karch:2001ct,Miemiec:2000eq,Schwartz:2001ip}. According to the previous discussion, we should expect
that the model will support a ultralight sterile neutrino as well. 
This can be easily shown again by considering the form of the
potential of the differential equation that the right-handed component
is obeying, which will again turn out to be
of the same form as the graviton. 
 
Now it turns out that for the construction of  such a $''++''$
configuration, it is essential to have $AdS_{4}$ geometry on both
branes (for details see Ref.\cite{Kogan:2001vb}). Thus in this case the background geometry is described by:

\be
ds^2=\frac{e^{-2\sigma(y)}}{(1-\frac{H^{2}x^{2}}{4})^2}\eta_{\mu\nu}dx^{\mu}dx^{\nu}
- dy^2
\ee
where the corresponding inverse vielbein is given by 
\be
{E}_{\alpha}^{A}=diag(e^{\sigma(y)}{(1-\frac{H^{2}x^{2}}{4})},e^{\sigma(y)}{(1-\frac{H^{2}x^{2}}{4})},e^{\sigma(y)}{(1-\frac{H^{2}x^{2}}{4})},e^{\sigma(y)}{(1-\frac{H^{2}x^{2}}{4})},1).
\ee

Since now the brane is no longer flat the previous calculations for
the action will be slightly modified. We briefly discuss these modifications.
Following the same steps of the flat case, we write $\Psi=\Psi_{R}+\Psi_{L}$ where
$\Psi_{R,L}=\frac{1}{2}(1\pm \gamma_{5})\Psi$. Since the connection part of the Lagrangian again doesn't give any
contribution (since the vielbein is again symmetric) the  action becomes:

\ba
S=\int \sqrt{\hat{G}} ~d^4x ~ dy  
\{  e^{-3\sigma} {{\hat{E}}}^{A}_{a}\left( \bar{\Psi}_{L}i\gamma^{a}\partial_{A} \Psi_{L} +
\bar{\Psi}_{R}i \gamma^{a} \partial_{A} \Psi_{R} \right) -  e^{-4\sigma} m
\frac{\sigma'(y)}{k} \left( \bar{\Psi}_{L}\Psi_{R} +
\bar{\Psi}_{R}\Psi_{L} \right) \nonumber \\ -\frac{1}{2}\left[ \bar{\Psi}_{L} (e^{-4\sigma}\partial_{y}+\partial_{y}e^{-4\sigma} ) \Psi_{R} -
\bar{\Psi}_{R}(e^{-4\sigma}\partial_{y}+\partial_{y}e^{-4\sigma} ) \Psi_{L}  \right] \}
\ea
where
$\hat{E}_{\alpha}^{A}={(1-\frac{H^{2}x^{2}}{4})}\delta_{\alpha}^{A}$
with ($a$,$A$=0,1,2,3) is the induced vielbein and $\hat{G}$ the
determinant of the induced metric.
For convenience and in order to be able to use results
of Ref.\cite{Kogan:2001vb}, we set $A(y)\equiv e^{-\sigma(y)}$ where

\be
A(y)=\frac{\cosh(k(y_{0}-|y|))}{\cosh(ky_{0})}
\ee
is the equivalent ``warp'' factor in this case, which is found by
considering the configuration gravitationally.
Note that the profile of the scalar field that is assumed to generate 
the above $''++''$ configuration will not have the exact form
$\sigma'(y)=- \frac{A'(y)}{A(y)}$ but, as it is shown in the appendix, 
it can be approximated by this. This is done in order to be able to
have some analytic results for the wavefunctions and the mass spectrum.
From the above relation it is clear that the ``warp'' factor has the
desired bounce form, with a minimum at $y_{0}$. 
The position of the minimum, something that it is going to be important
for the phenomenology of the model, is  defined from the relationship
$\tanh(ky_{0})\equiv\frac{k V_{1}}{|\Lambda|}$, where $V_{1}$ is the
tension of the first brane and $\Lambda$ is the five dimensional
cosmological constant. As we mentioned in the
introduction, the profile of the mass term, $\frac{A'(y)}{A(y)}$, in
the present model is not a simple combination of
$\theta$-functions. In particular there are two $\theta$-function
profiles near the orbifold fixed points  which give rise to the
positive branes, but there is also an intermediate kink profile of the
form $-\tanh(k(y-y_{0}))$ which is associated with the presence of the bounce
(this could give rise to a $''-''$ brane as a limit , resulting to the familiar $''+-+''$
configuration). Note that even though there is no brane at the position
of the minimum of the ``warp'' factor the kink profile is expected to
act in the same way, and thus induce localization of the fermion zero
mode in specific regions of the parameter space, exactly as in the $''+-+''$ model.

As in the flat-brane case, we can decompose the left-handed and
right-handed fermion fields into KK states with nontrivial profile
wavefunctions $f^{L}_{n},f^{R}_{n}$ (in respect to the fifth dimension)  in order to be able to bring the
Lagrangian into the form 

\be
S=\sum_{n} \int d^4x \sqrt{-\hat{g}}
\{\hat{E}_{\alpha}^{A}\bar{\psi}_{n}(x) i \gamma^{\alpha} \partial_{A} \psi_{n}(x) -
m_{n}\bar{\psi}_{n}(x) \psi_{n}(x) \}
\ee

where
the wavefunctions $f^{L}_{n}(y)$, $f^{R}_{n}(y)$  should obey the following equations

\ba
\left( -\partial_{y}
+\frac{m}{k}\frac{A'(y)}{A(y)}\right)f^{L}_{n}(y)=m_{n} A^{-1}(y) f^{R}_{n}(y)
\nonumber \\
\left( \partial_{y} +\frac{m}{k}\frac{A'(y)}{A(y)}\right)f^{R}_{n}(y)=m_{n}A^{-1}(y)f^{L}_{n}(y)
\ea

with the following orthogonality relations:

\be
\int_{-L}^{L} dy A^{-1}(y) {f^{L}_{m}}^{*}(y) f^{L}_{n}(y)=\int_{-L}^{L} dy A^{-1}(y) {f^{R}_{m}}^{*}(y) f^{R}_{n}(y)=\delta_{mn}
\ee

Again we solve the above system of differential equations by finding
the second order differential equation that it implies for the
right-handed component of the spinor.
It is always possible to make the coordinate transformation from y
coordinates to z coordinates  related through:
$\frac{dz}{dy}=A^{-1}(y)$
and bring the differential equations in the familiar form:

\be
\left\{-
\frac{1}{2}{\partial_z}^2+V_{R}(z)\right\}{\hat{f}^{R}}_{n}(z)=\frac{m_n^2}{2}
{\hat{f}^{R}}_{n}(z)
\ee
where we have defined $\hat{f}^{R}_{n}(z)=f^{R}_{n}(y)$
\ba
{\rm with}\hspace*{0.5cm} V_{R}(z)&=&\frac{\nu}{2}A(y)A''(y)+
\frac{{\nu}^{2}}{2}(A'(y))^{2} \nonumber \\
&=&-\frac{{\nu}^2\tilde{k}^{2}}{2}~+~\frac{\nu(\nu+1)\tilde{k}^2}{2}\frac{1}{\cos^{2}\left(\tilde{k}(|z|-z_{0})\right)}\cr      &-&k\nu\left[ \tanh(ky_{0})\delta(z)+\frac{\sinh(k(L-y_{0}))\cosh(k(L-y_{0}))}{\cosh^{2}(ky_{0})}
\delta(z-z_{1})\right] 
\ea
with $\tilde{k}$  defined as
$\tilde{k}\equiv{\frac{k}{\cosh(ky_{0})}}$.
The new variable $z$ is  related to the old one $y$ through the relationship:
\be
z\equiv {\rm sgn}(y)\frac{2}{\tilde{k}}\left[\arctan\left(\tanh(\frac{k(|y|-y_{0})}{2})\right)+\arctan\left(\tanh(\frac{ky_{0}}{2})\right)\right]
\
\ee
Thus in terms of the new coordinates, the branes are  placed at $z_{1}=0$
 and $z_{L}$, with the minimum of the potential  at $z_{0}={2 \over \tilde{k}}\arctan\left(\tanh(\frac{ky_{0}}{2})\right)$. Also note
that with this transformation the point $y=\infty$ is mapped to the
finite point $z_{\infty}={2 \over \tilde{k}}\left[{\pi \over 4} +
 \arctan\left(\tanh(\frac{ky_{0}}{2})\right)\right]$.

We can now proceed to the solution of the above equations for the
right-handed components, while the left-handed wavefunctions can be
easily evaluated using eq.(\ref{neutr10}) (taking in account the definition $A
\equiv e^{- \sigma(y)}$ and the change of variables). The zero
mode wavefunction is given by:
\be
\hat{f}_{0}^{R}(z)=\frac{C}{[\cos(\tilde{k}(z_{0}-|z|))]^{\nu}}
\ee
where $C$ is the normalization factor. If we send one of the two
branes to infinity (i.e. $z_{1} \rightarrow z_{\infty}$) and at the
same time keep $z_{0}$ fixed we find that the zero mode is
normalizable only in the cases where $\nu < \frac{1}{2}$. In the other
cases ($\nu \ge \frac{1}{2}$) the wavefunction fails to be
normalizable due to the fact that
it is too singular at $z_{\infty}$. Note though that in the case where
$\nu < \frac{1}{2}$ the first KK will not be special (i.e. will have
almost the same behaviour as the rest of the KK tower).
We also note that again  the zero mode is chiral i.e. there is no solution for
$f_{n}^{L}$ when $m_{0}=0$ that can satisfy the boundary conditions
(antisymmetric wavefunction). From the above we see that the
localization behaviour of zero mode in the $''++''$ model is the same 
as in the $''+-+''$ model. Note that for $\nu < \frac{1}{2}$ the zero
mode will be localized near $y_{0}$  despite the absence of any brane
at that point. This is because, as we mentioned, the $-
\frac{A'(y)}{A(y)}$ factor, which can be considered as the vacuum
expectation value of a scalar field, has a kink profile in the
neighbourhood of $y_{0}$ which induces the localization (this kink
becomes the negative tension brane in the flat brane limit). 
By considering cases with $m_{n}\neq0$, we find the wavefunctions  for the KK tower :

\be
\renewcommand{\arraystretch}{1.5}
\begin{array}{c}{\hat{f}^{R}}_{n}(z)=\cos^{\nu+1}(\tilde{k}(|z|-z_{0}))\left[C_{1}~F(\tilde{a}_{n},\tilde{b}_{n},\frac{1}{2};\sin^{2}(\tilde{k}(|z|-z_{0})))~~~~~~~~\right.
\\ \left.  ~~~~~~~~~~~~~~~~~~~+C_{2}~|\sin(\tilde{k}(|z|-z_{0}))|~F(\tilde{a}_{n}+\frac{1}{2},\tilde{b}_{n}+\frac{1}{2},\frac{3}{2};\sin^{2}(\tilde{k}(|z|-z_{0})))\right]
\end{array}
\ee
where
\ba
\tilde{a}_{n}=\frac{\nu+1}{2}+\frac{1}{2}\sqrt{\left(\frac{m_{n}}{\tilde{k}}\right)^2+{\nu}^2}
\cr
\tilde{b}_{n}=\frac{\nu+1}{2}-\frac{1}{2}\sqrt{\left(\frac{m_{n}}{\tilde{k}}\right)^2+{\nu}^2
}
\ea
The  boundary conditions are given by:
\ba
{\hat{f}^{R}}_{n}~'({0}^{+})+k \nu \tanh(k y_{0}){\hat{f}^{R}}_{n}(0)=0
\nonumber \\
{\hat{f}}^{R}_{n}~'({z_{L}}^{-})-k \nu \frac{\sinh(k(L-y_{0}))}{\cosh(k y_{0})}{\hat{f}^{R}}_{n}(z_{L})=0
\ea
the above conditions determine the mass spectrum of the KK states. 
By studying the mass spectrum of the KK states it turns out that
it  has a special first
mode similar to the one of the $''+-+''$ model as expected. 
For example, for the symmetric configuration ($w=1$), by approximation we can analytically find the following expressions
for the mass of this special state :

\be
m_1=2\sqrt{4 {\nu}^2+3}~k~\left(e^{-ky_{0}}\right)^{\nu+\frac{1}{2}}
\ee
In contrast, the masses of the next levels are given by the formulae:
For odd states
\be
m_{n}=2  \sqrt{(n+1)(n+1+\nu)}~k ~e^{-ky_{0}}
\ee
with $n=0,1,2,...$, and for even states
\be
m_{n}= 2 \sqrt{( n + \frac{3}{2} )( n + \frac{3}{2}+ \nu)} ~k~ e^{-ky_{0}}
\ee
with $n=0,1,2,...$.

Again the  mass of the first KK state is manifestly singled out from the rest of the KK tower
as it has an extra exponential suppression that depends on the mass of
the bulk fermion. The above characteristics persist in the more
physically interesting  asymmetric
case ($w<<1$). In this case we find $m_{1} \sim 
m_{1}^{0}(\nu)~kw~(e^{-kx})^{\nu+\frac{1}{2}}$ and $m_{n} \sim m_{n}^{0}(\nu)~  kw~ e^{-kx}$ where
$x=L_{1}-y_{0}$ is the distance of the second brane from the minimum
of the warp factor, and where $m_{1}^{0} \sim 1$, $m_{n}^{0}$ has a
linear dependence on
$n$ (for higher levels )  and  a $\sim \sqrt{\nu}$ dependence  . 
  The first KK state is
localized on the second positive brane and, as in the case of
 $''+-+''$  model, its coupling to SM neutrinos
remains constant if we keep the hierarchy parameter $w$ fixed.
The phenomenology of this model will be similar to the
$''+-+''$ model and thus we do not consider it
separately.


\section{ Bigravity and Bulk spinors }

\begin{figure}[t]
\begin{center}
\begin{picture}(300,200)(0,70)

\SetWidth{1}
\LongArrow(-20,100)(-20,250)
\LongArrow(160,100)(160,250)

\SetWidth{1}
\LongArrow(-20,100)(120,100)
\LongArrow(160,100)(300,100)

\Text(-60,175)[c]{$Mass$}
\Text(45,270)[c]{$''+-+''~Model$}
\Text(225,270)[c]{$RS~Model$}
\Text(15,90)[c]{G}
\Text(70,90)[c]{S}
\Text(195,90)[c]{G}
\Text(250,90)[c]{S}



\GBoxc(15,220)(18,3){0}
\GBoxc(15,180)(18,3){0}
\GBoxc(15,140)(18,3){0}

\GBoxc(15,105)(18,3){0}

\GBoxc(15,100)(18,3){0}




\GBoxc(70,225)(18,3){0.5}
\GBoxc(70,185)(18,3){0.5}
\GBoxc(70,145)(18,3){0.5}

\GBoxc(70,107)(18,3){0.5}

\GBoxc(70,100)(18,3){0.5}




\GBoxc(195,220)(18,3){0}
\GBoxc(195,180)(18,3){0}
\GBoxc(195,140)(18,3){0}

\GBoxc(195,100)(18,3){0}




\GBoxc(250,225)(18,3){0.5}
\GBoxc(250,185)(18,3){0.5}
\GBoxc(250,145)(18,3){0.5}

\GBoxc(250,100)(18,3){0.5}


\end{picture}
\end{center}

\caption{On the left, the mass spectrum of the graviton (G) (first column)
and bulk spinor (S) (second column)
KK states in the $''+-+''$-Bigravity model. On the right, for comparison, the
corresponding spectrum for the case of RS model. The figures are not in
scale and the details of the bulk spinor mass spectrum depend on the
additional parameter $\nu$ and thus in the above figure we have
assumed $\nu \simeq \frac{3}{2}$ (for higher value of $\nu$ we could have the fermion
first KK state to be lighter that the first graviton KK state).}

\end{figure}
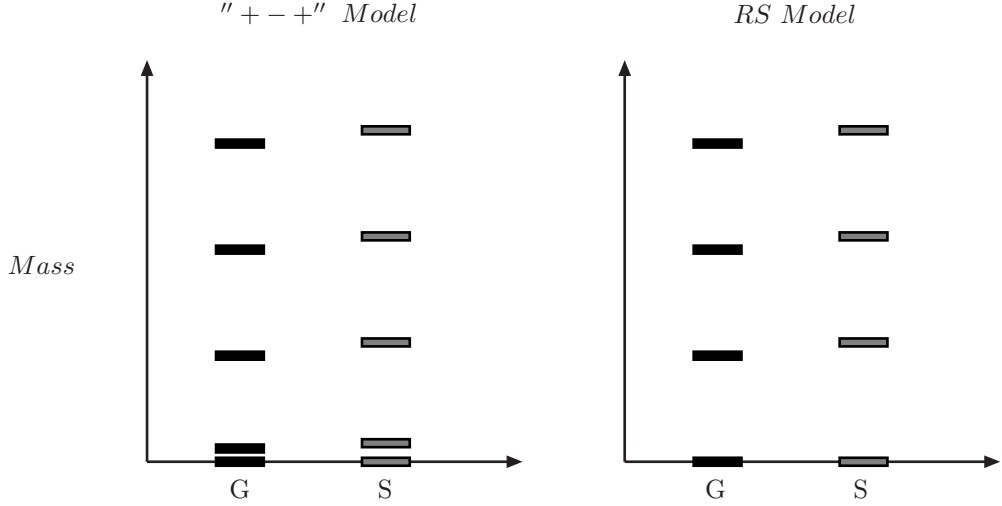

 Bigravity \cite{Kogan:2000wc,Mouslopoulos:2000er,Kogan:2001vb}
(multigravity \cite{Gregory:2000jc,Kogan:2000cv,Kogan:2000xc}) is the possibility
that gravitational interactions do not exclusively come from a massless graviton
, but instead they can be the net effect of a massless graviton
and one or more KK states (continuum of KK states) or even a single  massive
KK mode \cite{Kogan:2001vb,Karch:2001ct} without conflict with General
Relativity predictions \cite{Kogan:2001uy,Porrati:2001cp}.   
This is based on the different
scaling laws of the mass between the first and the rest of KK states. 
Since the first graviton KK state has mass with an additional exponential
suppression we can realize
the scenario that the first KK state is so light that its wavelength
is of the order of the observable universe and thus any observable
effect of its non-vanishing mass to be out of the experimental reach, and
in the same time the rest of the KK tower has masses above the scale
that Cavendish experiments have tested Newtonian gravity at small distances.

In the two previous sections we have shown that in the case of models,
where Bigravity can be realized, there is also an ultralight KK that
corresponds to the bulk spinor assuming the existence of the mass term for the
bulk spinor  that appears in eq.(\ref{neutr2}) (with $\nu > \frac{1}{2}$) . The subject of this section is to
investigate if the two above possibilities are compatible: Can we have
a Bigravity scenario and a consistent neutrino phenomenology? For the
sake of simplicity the discussion below will be concentrated to the
$''+-+''$ model but, analogous arguments
should apply to the case of $''++''$ model. 

Let us review briefly the Bigravity scenario.
The graviton ultralight first KK state has mass
$m^{(G)}_{1}=2wk~e^{-2x}$ and coupling $a^{(G)}_{1}=\frac{1}{wM_{Pl}}$. The rest
of the KK tower has masses :
\be
m_{n+1}^{(G)}= \xi_n ~ kw~ e^{-x} ~~~~~~n=1,2,3, \ldots
\ee
where $\xi_{2i+1}$ is the $(i+1)$-th root of $J_{1}(x)$ ($i=0,1,2,
\ldots$) and $\xi_{2i}$ is the $i$-th root of $J_{2}(x)$ ($i=1,2,3,
\ldots$). The couplings of these states scale as $a_{n}
\propto e^{-x}$.
In order to achieve the Bigravity scenario, as we mentioned in the
beginning of this section, we have the following  constrains on the
range of masses of the KK states: $m^{(G)}_{1}<10^{-31}{\rm eV}$
or $m^{(G)}_{2}>10^{-4}{\rm eV}$ (where Planck suppression is considered for
the ``continuum '' of states above $10^{-4}$eV). Our exotic scheme corresponds 
to the choice $m^{(G)}_1\approx 10^{-31}{\rm eV}$ and $m^{(G)}_2>10^{-4}{\rm eV}$. In this case, for 
length scales less than $10^{26}{\rm cm}$ gravity is generated by the exchange of {\it 
both} the massless graviton and the first KK mode. This implies, (taking into 
account the different coupling
suppressions of the massless graviton and the first KK state) that
the gravitational coupling as we measure it is related to the
parameters of our model by:
\be
\frac{1}{M_{\rm
Pl}^2}=\frac{1}{M_{5}^2}\left(1+\frac{1}{w^2}\right)\approx
\frac{1}{(w M_{5})^2} \Rightarrow M_{\rm Pl}\approx w M_{5}
\ee
We see that the mass scale on our brane, $w M_{5}$, is now the Planck scale so, 
although the ``warp'' factor, $w$, may still be small (i.e. the fundamental 
scale  $M_{5} >>M_{Planck}$), we do not  solve  the Planck hierarchy
problem. Using the equations for the mass spectrum  and assuming as before that $k\approx M_{5}$, 
we find that $m_{1}=2~kw~e^{-2x}\approx M_{Planck}e^{-2x}$. For $m_1=10^{-31}{\rm eV}$ we 
have $m_{2}\approx 10^{-2}{\rm eV}$. This comfortably satisfies the
bound $m>10^{-4}{\rm eV}$. 

Now let us see what this implies for the neutrino physics.
Let us first consider the case where $\nu > \frac{1}{2}$.
In this case, as mentioned above, there will be also an ultralight bulk
fermion KK mode. By forcing the graviton first mode to have  a
tiny mass we also force the first spinor KK state to become very light
 (even in the best case where $\nu \rightarrow \frac{1}{2}$ this state will
have mass of the order of a fraction of eV. For larger values of $\nu$
the state becomes
even lighter) since they are related through: $m_{1} \sim
e^{(\frac{3}{2}-\nu)x}~ m^{G}_{1}$. This  is unacceptable since this mode has a constant coupling of the
order of the weak scale, $\upsilon$, which will induce large mixing of
the neutrino with the left-handed component of this state.
However note that in the limit $\nu \rightarrow \frac{1}{2}$ the
special first fermion KK mode will become a normal one losing it's localization
and thus the above may not apply. Unfortunately, this is not the case as
in this limit  the fermion
zero mode is delocalized (for $\nu = \frac{1}{2}$ the coupling is
constant across the extra dimension) giving no possibility of inducing
small neutrino masses (since the compactification volume is very
small)\footnote{Trying to use this window of the  parameter space 
seems like a fine-tuning though since
one needs to delocalize the fermions first KK state enough in order to
have small coupling to SM neutrino and on the other hand to prevent
the delocalization of the fermion zero mode with the same mechanism
($\nu \rightarrow \frac{1}{2}$).}.
 
 Despite the  severe constraints in the above scenario, due to the
presence of the ultralight bulk fermion KK state,
by no means the Bigravity scenario is excluded in  the case of
$\nu>\frac{1}{2}$ since one can always consider the possibility of
placing the SM on  a  brane (with tiny tension so that
the background is not altered) between the negative and the second
positive brane, so that the coupling of the bulk fermion first KK
state to SM neutrinos  is sufficiently small while  a part of
gravitational interactions will still be generated from the ultralight
graviton first KK state.

Let us turn now to the case $0 \le \nu < \frac{1}{2}$. In this case there is
no special bulk spinor KK state and thus the above arguments do not
apply. In this case  the bulk fermion zero mode is
localized on the negative tension brane. Nevertheless, in order for the
Bigravity scenario to be possible we should have $x \simeq 60$. If we
now try to generate neutrino masses with the  mechanism described in
the first section we will find that the coupling of the zero mode to
SM neutrino will be by far too small to provide  consistent results.
Thus in this case although Bigravity is realized we cannot generate
neutrino masses consistently. 

The case of  $\nu=\frac{1}{2}$ is of no interest since in this case
there is no localization (the fermion zero mode has constant coupling
across the extra dimension). The case where $\nu < 0$ resembles the $0
\le \nu < \frac{1}{2}$ case, with the only difference that the
localization on negative branes will be even sharpner, making the
situation even worse.


\section{Discussion and conclusions}

We have studied bulk fermion fields in various multi-brane  models with localized
gravity. The chiral zero mode that these models support can be
identified as a right-handed sterile neutrino. In this case small
neutrino Dirac masses can naturally appear due
to an analogous (to graviton) localization of the bulk fermion zero
mode wavefunction without
invoking a see-saw mechanism.  For models in which the localization
of the fermion zero mode is induced by the same scalar field that
forms the
branes  the localization behaviour of this mode can resemble
the graviton's at least in a region of  parameter space.
The latter implies that the  $''++''$ model can support,
in addition to  the ultralight graviton KK state, an
ultralight localized and strongly coupled  bulk fermion KK mode.
This fermion state, when exists, imposes even more severe
constrains on the parameter space of  $''++''$
models. In the case that one requires the Bigravity be realized
the light fermion KK mode can induce too large mixing between the neutrino
and the KK tower and thus it restricts even more the allowed parameter
space of the relevant models.
 
As a general remark  we see that the appearance of
multi-localization in the multli-Brane world
picture  and its relation to the existence of ultralight states in
the KK spectrum is not a characteristic of the graviton only, but can
also occur in spin $\frac{1}{2}$ fields. Moreover in the next Chapter
will be shown that also spin  $0$, $1$ and $\frac{3}{2}$ bulk states with
appropriate bulk mass terms can exhibit multi-localalization.

\chapter{Multi-Localization} 
\label{multiloc}
 
\section{Introduction} 
 
In 
the simplest formulation of the braneworld models no bulk matter states are assumed 
to exist and thus only gravity propagates in the extra dimensions. 
Nevertheless ``bulk'' (\textit{i.e.} transverse to 3-brane space dimensions) 
physics turns out to be very interesting giving alternative explanations to 
other puzzles of particle physics. For example, as we found in Chapter 
\ref{neutrino}, by assuming the existence of 
a Standard Model (SM) neutral spin $\frac{1}{2}$ fermion in the bulk one can 
explain the smallness of the neutrino masses without invoking the seesaw 
mechanism (also see Refs.\cite{Dienes:1999sb,Arkani-Hamed:1998vp,Dvali:1999cn,Mohapatra:1999zd,Barbieri:2000mg,Lukas:2000wn,Lukas:2000rg,Cosme:2000ib,Grossman:2000ra}). However, it is not necessary to confine the SM fields to the brane. 
Assuming that the SM fields can propagate in the bulk interesting new 
possibilities arise. For example one can attempt to explain the pattern of 
the SM fermion mass hierarchy by localizing the SM fermions at different 
places in the bulk \cite{Arkani-Hamed:2000dc,Mirabelli:2000ks,Dvali:2000ha,delAguila:2000kb}. These 
considerations give the motivation for considering the phenomenology 
associated with spin $0$, $\frac{1}{2}$ and $1$ fields propagating in extra 
dimension(s). Since our discussions will be limited to models with localized 
gravity and in particular to Randall-Sundrum (RS) type constructions, we 
will be interested in the phenomenology of fields that live in a slice of $ 
AdS_{5}$ spacetime. If one also wants to explore the supersymmetric version 
of the above models, it is also necessary to study the phenomenology of spin  
$\frac{3}{2}$ field on the same background geometry. 
 
In Chapter \ref{RS} we have seen  that in the context of RS type models  the graviton  is 
localized on positive tension branes and suppressed on the negative
ones. 
In Ref.\cite{Goldberger:1999wh} it was also shown 
that the $AdS_{5}$ background geometry of these models can localize the zero 
mode of a massless scalar field on positive tension branes (see also
\cite{Mintchev:2001mf}). Moreover, according to the results of
Chapter \ref{neutrino}, the same background localizes 
the spin $\frac{1}{2}$ fermions on negative tension branes . The same localization behaviour holds for spin $\frac{3}{2}$ fermions \cite{Bajc:2000mh,Oda:2000kh,Oda:2000wa}. However, the $AdS_{5}$ background 
geometry cannot localize massless Abelian gauge fields
\cite{Pomarol:2000ad,Davoudiasl:2000tf,Davoudiasl:2001wi}\footnote{However 
 one can circumvent this ``no-go theorem'' for photon localization on
the brane by considering graviphotons originating from odd-dimensional 
self-duality equations rather than Maxwell equations - see Ref.\cite{Duff:2001jk}}. 
 
The addition of mass terms modifies the localization properties. For 
example, it was shown in Chapter \ref{neutrino} 
that the addition of an appropriate mass term in the action of a spin $\frac{1}{2}$ field can result in the localization of the zero mode that resembles 
that of the graviton (the magnitude of the mass term in this case controls 
the extent of localization). In the present Chapter we show that the same can 
occur in the case of spin $0$, $\frac{3}{2}$ fields with appropriate mass 
terms. Furthermore we show that by adding a mass term of a particular form 
in the action of a massless Abelian gauge field we can achieve the desired 
localization of the massless zero mode which can be made to resemble that of 
the graviton. The mass term in this case must necessarily consist of a five 
dimensional bulk mass part and a boundary part. 
 
From the above it is clear that particles of all spins, with appropriate 
bulk mass terms, can exhibit zero mode localization on positive tension 
branes, just as for the graviton. In the context of multi-brane models with 
localized gravity the above implies a further interesting possibility: the 
phenomenon of multi-localization in models that contain at least two 
positive tension branes. Multi-localization, as we will see, is closely 
related to the appearance of light, localized strongly coupled
Kaluza-Klein (KK) states 
(their coupling to matter can be even larger than the coupling of the zero 
mode). Thus the mass spectrum of multi-localized fields is distinct from the 
mass spectrum of singly localized fields, resulting in the possibility of 
new phenomenological signals. Anomalously light states may also arise in 
theories without multi-localization (\textit{i.e.} even in configurations 
with one positive brane) in models with twisted boundary conditions. 
 
The appearance of light KK states can be of particular phenomenological 
interest. For example in  Chapter \ref{RS}  we examined the case of
the graviton, where multi-localization and 
thus the appearance of light KK graviton excitations, gives rise to the 
exciting possibility of Bi-gravity  (Multi-gravity) where part (or even all) ofgravitational interactions can come from massive spin $2$ particle(s) (KK 
state(s)). In this case the large mass gap between the anomalously light KK state(s) 
and the rest of the tower is critical in order to avoid modifications of 
Newton's law at intermediate distances. 
 
In Chapter \ref{neutrino}, we found that anomalously light spin $\frac{1}{2}$ KK states can also arise when a bulk 
fermion is multi-localized. The non-trivial 
structure of the KK spectrum in this case, with the characteristic mass gap 
between the light state(s) and the rest of the tower, can be used for 
example to construct models with a small number of active or sterile 
neutrinos involved in the oscillation (the rest will decouple since they 
will be heavy).

The organization of this Chapter is as follows: In the next Section we review 
the general framework and discuss the general idea of multi-localization in 
the context of multi-brane models with localized gravity. In Section 3 we study the 
multi-localization properties of a bulk scalar field. In Section 4 we review 
the situation of a bulk fermion field. In Section 5 we study in detail the 
possibilities of localization and multi-localization of an Abelian gauge 
field. In Section 6 present the possibility of multi-localization in the 
case of a gravitino and finally in Section 7 we review the same phenomena 
for the case of graviton. In Section 
8 we discuss how multi-localization is realized in the context of 
supersymmetric versions of the previous models. The overall implications and 
conclusions are presented in Section 9. 
 
\section{General Framework - The idea of Multi-Localization} 
 
The original formulation of multi-localization of gravity, presented
in Chapter \ref{RS} , was obtained in 
five dimensions for the case that there is more than one positive tension 
brane. If the warp factor has a ``bounce'', in the sense that it has a 
minimum (or minima) between the positive tension branes, the massless 
graviton appears as a bound state of the attractive potentials associated 
with the positive tension branes with its wave function peaked around them. 
Moreover in this case there are graviton excitations corresponding to 
additional bound states with wave functions also peaked around the positive 
tension branes. They are anomalously light compared to the usual Kaluza 
Klein tower of graviton excitations. The reason for this is that the 
magnitude of their wavefunction closely approximates that of the massless 
mode, differing significantly only near the position of the bounce where the 
wave function is exponentially small. The mass they obtain comes from this 
region and as a result is exponentially suppressed relative to the usual KK 
excitations. 
 
The first models of this type (\textit{e.g.} $''+-+''$) involved negative tension branes sandwiched 
between the positive tension branes. That this is 
necessary in the case of flat branes with vanishing cosmological constant in 
four dimensions is easy to see because, for a single flat brane, the minimum 
of the warp factor is at infinity and thus any construction with another 
positive brane at a finite distance will have a discontinuity in the 
derivative of the warp factor at the point of matching of the solutions and 
thus at that point a negative tension brane will emerge. 
 
However  free negative tension brane(s) 
violate the weaker energy condition and lead to  ghost radion
field(s). Nevertheless, in Chapter \ref{5dads} we have shown that this 
can be avoided through the use of
$AdS_{4}$ branes \cite{Kogan:2001vb} (see \cite{Gorsky:2000rz} for a
different possibility involving an external four-form field). In this case the minimum of a single 
brane is at finite distance and thus one can match the solution for two 
positive branes without introducing a negative tension brane. However a 
drawback of this approach is that the four dimensional cosmological constant 
is negative in conflict with the current indications for a positive 
cosmological constant. 
 
However, as it was shown  in Chapter \ref{6D}, it is possible to obtain the ``bounce'' and the 
related multilocalisation for zero cosmological constant in four dimensions 
without the need for negative tension branes \cite{Kogan:2001yr}. As
we found in Chapter  \ref{6D},
 this is possible even in cylindrically symmetric models if one allows 
for non-homogeneous brane tensions and/or bulk cosmological constant. In 
this case the structure of the effective four dimensional theory is very 
similar to the five dimensional case for the modes which do not depend on 
the new angular co-ordinate. In particular one finds a massless graviton and 
anomalously light massive modes with wavefunctions peaked around the 
positive tension branes. 
 
In this Chapter we are interested in whether spin $0,\frac{1}{2},1$,
and $\frac{3}{2}$ fields can similarly show the phenomena of
multi-localisation we found for the graviton in suitable curved backgrounds. We will show that 
the curved background can also induce localisation for the case of spin $0, 
\frac{1}{2}$ and $\frac{3}{2}$ fields but not for a massless vector field. 
However, even in flat spacetime in higher dimensions, it is also possible to 
induce localisation by introducing mass of a very specific form for these 
fields. We will show that this effect can localise all fields with spin $ 
\leq \frac{3}{2},$ including the graviton. 
 
For simplicity we will work mainly with the five dimensional 
compactification. In the context of the discussion of the multi-localisation 
of fields of spin $\leq \frac{3}{2},$ what is important is the nature of the 
curved background as determined by the warp factor. Thus the general 
features are applicable to all models of multi-localisation with 
the same warp factor profile. However the interpretation of the mass terms 
needed to achieve multi-localisation differs. In the case of models with 
negative tension branes the masses correspond to a combination of a constant 
bulk mass together with a brane term corresponding to the coupling to 
boundary sources. In the case of models without negative tension branes the 
mass must have a non-trivial profile in the bulk. In certain cases this 
profile may be guaranteed by supersymmetry. 
 
In the five dimensional models considered here, the fifth dimension $y$ is 
compactified on an orbifold, $S^{1}/Z_{2}$ of radius $R$, with $-L \le y \le 
L $. The five dimensional spacetime is a slice of $AdS_{5}$ which is 
described by\footnote{ 
We will assume that the background metric is not modified by the presence of 
the bulk fields, that is, we will neglect the back-reaction on the metric 
from their presence.}:  
\begin{equation} 
ds^{2}=e^{-2 \sigma(y)} \eta_{\mu \nu}dx^{\mu}dx^{\nu}+dy^2 
\label{multi1}
\end{equation} 
where the warp factor $\sigma(y)$ depends on the details of the model 
considered. 
 
Since we are interested in the phenomenology of fields propagating in the 
above slice of $AdS_{5}$ our goal is to determine the mass spectrum and 
their coupling to matter. Starting from a five dimensional Lagrangian, in 
order to give a four dimensional interpretation to the five dimensional 
fields, one has to implement the dimensional reduction. This procedure 
includes the representation of the five dimensional fields $\Phi (x,y)$ in 
terms of the KK tower of states:  
\begin{equation} 
\Phi (x,y)=\sum_{n=0}^{\infty }\Phi ^{(n)}(x)f^{(n)}(y) 
\label{multi2}
\end{equation} 
where $f^{(n)}(y)$ is a complete orthonormal basis spanning the compact 
dimension. The idea behind this KK decomposition is to find an equivalent 4D 
description of the five dimensional physics associated with the field of 
interest, through an infinite number of KK states with mass spectrum and 
couplings that encode all the information about the five dimensions. The 
function $f^{(n)}(y)$ describes the localization of the wavefuntion of the 
n-th KK mode in the extra dimension. It can be shown that $f^{(n)}(y)$ obeys 
a second order differential equation which, after a convenient change of 
variables and/or a redefinition\footnote{The form of the redefinitions depend on the spin of the field.} of the 
wavefunction, reduces to an ordinary Schr\"{o}dinger equation:  
\begin{equation} 
\left\{ -\frac{1}{2}\partial _{z}^{2}+V(z)\right\} \hat{f}^{(n)}(z)=\frac{ 
m_{n}^{2}}{2}\hat{f}^{(n)}(z) 
\end{equation} 
The mass spectrum and the wavefunctions (and thus the couplings) are 
determined by solving the above differential equation. Obviously all the 
information about the five dimensional physics is contained in the form of 
the potential $V(z)$. For example in the case of the graviton the positive 
tension branes correspond to attractive $\delta $-function potential wells 
whereas negative tension branes to $\delta $-function barriers. The form of 
the potential between the branes is determined by the $AdS_{5}$ background. 
 
\subsection{Multi-Localization and light KK states} 
 
 
\begin{figure}[t] 
\begin{center} 
\SetScale{0.7}  
\begin{picture}(200,150)(0,50) 
\LongArrow(140,0)(140,280) 
\LongArrow(-80,120)(370,120) 
 
\SetWidth{1} 
 
\SetColor{Green} 
\Line(140,20)(148,250) 
\Line(140,20)(132,250) 
\SetColor{Black} 
 
\SetColor{Red} 
\Line(240,270)(245,130) 
\Line(240,270)(235,130) 
\Line(40,270)(45,130) 
\Line(40,270)(35,130) 
\SetColor{Black} 
 
\SetColor{Green} 
\Line(-15,140)(-20,0) 
\Line(-25,140)(-20,0) 
\Line(305,140)(300,0) 
\Line(295,140)(300,0) 
\SetColor{Black} 
 
\SetColor{Black} 
 
 
\Text(100,200)[l]{$V(z)$} 
\Text(260,70)[rb]{$z$} 
 

\SetColor{Black} 
\Curve{(148,250)(153,249)(170,190)(200,138)(220,131)(235,130)} 
\Curve{(245,130)(260,131)(295,140)}

\Curve{(45,130)(60,131)(80,138)(110,190)(127,249)(132,250)} 
\Curve{(-15,140)(20,131)(35,130)}

\end{picture} 
\end{center} 
\par 
\vspace*{8mm} 
\caption{The scenario of multi-localization is realized in configurations 
where the corresponding form of potential has potential wells that can 
support bound states. Such a potential is the one that corresponds to the $%
^{\prime\prime}+-+^{\prime\prime}$ model. Positive branes are $\protect\delta 
$-function wells and negative are $\protect\delta$-function barriers.} 

\label{multifig1}
\end{figure} 
 
Multi-localization emerges when one considers configuration of branes such 
that the corresponding potential $V(z)$ has at least two ($\delta $-function)%
\footnote{In the infinitely thin brane limit that we consider, the wells associated 
with positive branes are $\delta $-functions.} potential wells, each of which 
can support a bound state (see Fig.(\ref{multifig1}) for the $^{\prime \prime }+-+^{\prime 
\prime }$ case). If we consider the above potential wells separated by an 
infinite distance, then the zero modes are degenerate  and 
massless. However, if the distance between them is finite, due to quantum 
mechanical tunneling the degeneracy is removed and an exponentially small 
mass splitting appears between the states. The rest of levels, which are not 
bound states, exhibit the usual KK spectrum with mass difference 
exponentially larger than the one of the ``bound states'' (see Fig.(\ref{multifig2})). The 
above becomes clearer if one examines the form of the wavefunctions. In the 
finite distance configuration the wavefunction of the zero mode is the 
symmetric combination ($\hat{f}_{0}=\frac{\hat{f}_{0}^{1}+\hat{f}_{0}^{2}}{%
\sqrt{2}}$) of the wavefunctions of the zero modes of the two wells whereas 
the wavefunction of the first KK state is the antisymmetric combination ($%
\hat{f}_{0}=\frac{\hat{f}_{0}^{1}-\hat{f}_{0}^{2}}{\sqrt{2}}$). Such an 
example is shown in Fig.(\ref{multifig3}) where are shown the wavefunctions of the 
graviton in the context of $^{\prime \prime }+-+^{\prime \prime }$ model 
with two positive tension branes at the fixed point boundaries and a 
negative tension brane at the mid-point. From it we see that the absolute 
value of these wavefunctions are nearly equal throughout the extra 
dimension, with exception of the central region where the antisymmetric 
wavefunction passes through zero, while the symmetric wavefunction has 
suppressed but non-zero value. The fact that the wavefunctions are 
exponentially small in this central region results in the exponentially 
small mass difference between these states. 
 
The phenomenon of multi-localization is of particular interest since, 
starting from a problem with only one mass scale (the inverse radius of 
compactification), we are able to create a second scale exponentially 
smaller. Obviously the generation of this hierarchy is due to the tunneling 
effects in our ``quantum mechanical'' problem. 
 
 
\begin{figure}[t] 
\begin{center} 
\begin{picture}(250,200)(0,70) 
 
\SetWidth{1} 
\LongArrow(-20,100)(-20,250) 
 
\SetWidth{1} 
\LongArrow(-20,100)(250,100) 

\Text(-60,175)[c]{$Mass$} 
\Text(45,270)[c]{$''+-+''~Model$} 
\Text(170,270)[c]{$''+-''~RS~Model$} 
 
 
 
\GBoxc(35,220)(18,3){0} 
\GBoxc(35,180)(18,3){0} 
\GBoxc(35,140)(18,3){0} 
 
\GBoxc(35,105)(18,3){0} 
 
\GBoxc(35,100)(18,3){0} 
 
 
 

\GBoxc(170,225)(18,3){0.5} 
\GBoxc(170,185)(18,3){0.5} 
\GBoxc(170,145)(18,3){0.5} 
 
 
\GBoxc(170,100)(18,3){0.5} 
 
 
\end{picture} 
\end{center} 
\caption{Comparison of the gravitational spectrum of the $^{\prime \prime 
}+-+^{\prime \prime }$ model with the $^{\prime \prime \prime }+-^{\prime }$ 
Randall-Sundrum model.} 

\label{multifig2}

\end{figure} 
 

\begin{figure}
\begin{center}

\SetScale{0.5}
\begin{picture}(180,100)(0,50)

\SetWidth{2}
\Line(10,50)(10,250)
\Line(290,50)(290,250)

\SetWidth{0.5}
\Line(150,50)(150,250)
\Line(10,150)(290,150)



{\SetColor{Green}
\Curve{(10,240)(50,192)(65,181)(80,173)(95,167)(110,162)(130,157)(150,155)}
\Curve{(150,155)(170,157)(190,162)(205,167)(220,173)(235,181)(250,192)(290,240)}
}


{\SetColor{Red}
\DashCurve{(10,240)(50,192)(65,181)(80,173)(95,167)(110,161)(130,155)(150,150)}{3}
\DashCurve{(150,150)(170,145)(190,139)(205,133)(220,127)(235,119)(250,108)(290,60)}{3}
}


{\SetColor{Brown}
\DashCurve{(10,145)(15,149)(20,150)(40,153)(65,158)(140,209)(150,210)}{1}
\DashCurve{(150,210)(160,209)(235,158)(260,153)(280,150)(285,149)(290,145)}{1}
}
\end{picture}

\end{center}

\vspace*{3mm} 

\caption{The wavefunctions of the three first  modes in  the $''+-+''$ 
model. The
zero mode (solid line), first (dashed line )  and second (dotted line)
KK states. Note that the absolute value of the wavefunctions of the
zero mode and the first KK state almost coincide except for the
central region of the configuration where they are both suppressed.
}
\label{multifig3}


\vspace*{3mm}


\begin{center}

\SetScale{0.5}

\begin{picture}(180,100)(0,50)

\SetWidth{2}
\Line(-30,50)(-30,250)
\Line(330,50)(330,250)

\SetWidth{2.5}
\LongArrow(-70,150)(-110,150)
\LongArrow(370,150)(410,150)

\SetWidth{0.5}
\Line(150,50)(150,250)
\Line(-30,150)(330,150)



{\SetColor{Green}
\Curve{(-30,240)(10,192)(25,181)(40,173)(55,167)(70,162)(90,157)(110,155)(130,155)(150,155)(170,155)(190,155)}
\Curve{(190,155)(210,157)(230,162)(245,167)(260,173)(275,181)(290,192)(330,240)}
}


{\SetColor{Red}
\DashCurve{(-30,240)(10,192)(25,181)(40,173)(55,167)(70,161)(90,155)(110,155)(130,152)(150,150)}{3}
\DashCurve{(150,150)(170,147)(190,145)(210,145)(230,139)(245,133)(260,127)(275,119)(290,108)(330,60)}{3}
}


{\SetColor{Brown}
\DashCurve{(-30,145)(-10,145)(0,147)(15,149)(20,150)(40,153)(65,158)(140,200)(150,201)}{1}
\DashCurve{(150,201)(160,200)(235,158)(260,153)(280,150)(285,149)(290,147)(300,147)(310,145)(330,145)}{1}
}
\end{picture}

\end{center}

\vspace*{3mm} 

\caption{The effect of stretching the configuration by moving the two wells further
apart. Note that the wavefunction of the zero mode and the first KK
state remain localized but the remaining of the modes,  not being
bound states, will stretch along the extra dimension.}
\label{multifig4}



\begin{center}

\SetScale{0.5}

\begin{picture}(180,100)(0,50)

\SetWidth{2}
\Line(-120,50)(-120,250)
\Line(420,50)(420,250)

\SetWidth{0.5}
\Line(-120,150)(130,150)
\Line(170,150)(420,150)

\SetWidth{2.5}
\LongArrow(-140,150)(-180,150)
\LongArrow(440,150)(480,150)

\SetWidth{0.5}

\Text(76,75)[c]{$\dots$}


{\SetColor{Green}
\Curve{(-120,240)(-80,192)(-65,181)(-50,173)(-35,167)(-20,162)(0,157)(40,155)(60,155)(80,155)}
\Curve{(200,155)(250,155)(300,157)(320,162)(335,167)(350,173)(365,181)(380,192)(420,240)}
}


{\SetColor{Red}
\DashCurve{(-120,240)(-80,192)(-65,181)(-50,173)(-35,167)(-20,162)(0,157)(40,155)(60,155)(80,155)}{3}
\DashCurve{(200,145)(250,145)(300,143)(320,138)(335,133)(350,127)(365,119)(380,108)(420,60)}{3}
}


{\SetColor{Brown}
\DashCurve{(-120,145)(-80,155)(-65,155)(-50,155)(-35,155)(-20,155)(0,155)(40,155)(60,155)(80,155)}{1}
\DashCurve{(200,155)(250,155)(300,155)(320,155)(335,155)(350,155)(365,155)(380,155)(420,145)}{1}
}
\end{picture}

\end{center}

\vspace*{3mm} 

\caption{ The case when the distance between the branes become infinite. The zero
mode (which still exists if the compactification volume is finite) and
the first KK state become degenerate. The wavefunction of the second KK
state spreads along the extra dimension.}

\label{multifig5}

\end{figure}

 
\subsection{Locality - light KK states and separability} 
 
Summarizing, if in a single brane configuration (with infinite extra 
dimension) the zero mode of a field is localized on the brane, then in a 
multi-brane world scenario it will be multi-localized and as a consequence 
light KK states will appear in its spectrum. The latter is assured from the 
following locality argument: In the infinite separation limit of the 
multi-brane configuration physics on each brane should depend on local 
quantities and not on physics at infinity. The latter assures the smoothness 
of the limit of infinite separation of branes in the sense that at the end 
of the process the configuration will consists of identical and independent 
single brane configurations. As a result the above locality argument also 
ensures the appearance of light states which at the above limit will become 
the zero modes of the one brane configurations. 
 
\section{Multi-Localization of spin 0 field} 
 
\subsection{$^{\prime\prime}+-+^{\prime\prime}$ Model} 
 
Let us now explore whether multi-localization of spin $0$ fields can be 
realized in the context of the RS type of models. We start our discussion 
from the simplest case of a real scalar field propagating in a five 
dimensional curved background described by the metric of eq.(\ref{multi1}) where the 
function $\sigma (y)$ is the one that corresponds to the $^{\prime \prime 
}+-+^{\prime \prime }$ configuration. The action for a massive bulk scalar 
field in this case is:  
\begin{equation} 
S=\frac{1}{2}\int d^{4}x\int dy\sqrt{G}\left( G^{AB}\partial _{A}\Phi 
\partial _{B}\Phi +m_{\Phi }^{2}\Phi ^{2}\right)  
\end{equation} 
where $G=det(G_{AB})=e^{-8\sigma (y)}$. Under the $Z_{2}$ symmetry $m_{\Phi 
}^{2}$ should be even. We take $m_{\Phi }^{2}$ of the form:  
\begin{equation} 
m_{\Phi }^{2}=C+\Sigma _{i}D_{i}\delta (y-y_{i}) 
\end{equation} 
where $D_{i}=\pm 1$ for a positive or negative brane respectively. The first 
term corresponds to a constant five dimensional bulk mass and the second to 
the coupling of the scalar field to boundary sources \footnote{%
Here we have assumed that the magnitude of the boundary mass term 
contribution is the same for all branes. This is needed in order to have a 
zero mode.}. The mass can be rewritten in the form  
\begin{equation} 
m_{\Phi }^{2}=\alpha (\sigma ^{\prime }(y))^{2}+\beta \sigma ^{\prime \prime 
}(y) 
\label{multi6}
\end{equation} 
Taking in account the form of the vacuum, the above action can be written as  
\begin{equation} 
S=\frac{1}{2}\int d^{4}x\int dy\left( e^{-2\sigma (y)}\eta ^{\mu \nu 
}\partial _{\mu }\Phi \partial _{\nu }\Phi -\Phi \partial _{5}(e^{-4\sigma 
(y)}\partial _{5}\Phi )+m_{\Phi }^{2}e^{-4\sigma (y)}\Phi ^{2}\right)  
\end{equation} 
In order to give a four dimensional interpretation to this action we go 
through the dimensional reduction procedure. Thus we decompose the five 
dimensional field into KK modes  
\begin{equation} 
\Phi (x,y)=\sum_{n}\phi _{n}(x)f_{n}(y) 
\end{equation} 
Using this decomposition, the above action can be brought in the form  
\begin{equation} 
S=\frac{1}{2}\sum_{n}\int d^{4}x\{\eta ^{\mu \nu }\partial _{\mu }\phi 
_{n}(x)\partial _{\nu }\phi _{n}(x)+m_{n}^{2}\phi _{n}^{2}(x)\} 
\end{equation} 
provided the KK wavefunctions obey the following second order differential 
equation  
\begin{equation} 
-\frac{d}{dy}\left( e^{-4\sigma (y)}\frac{df_{n}(y)}{dy}\right) +m_{\Phi 
}^{2}e^{-4\sigma (y)}f_{n}(y)=m_{n}^{2}e^{-2\sigma (y)}f_{n}(y) 
\end{equation} 
with the following orthogonality relations (taking in account the $Z_{2}$ 
symmetry):  
\begin{equation} 
\int_{-L}^{L}dye^{-2\sigma (y)}{f}_{m}^{\ast }(y)f_{n}(y)=\delta _{mn} 
\label{multi11}
\end{equation} 
where we assume that the length of the orbifold is $2L$. 
 
The linear second order differential equation can always be brought to a 
Schr\"ondiger form by a redefinition of the wavefunction and by a convenient 
coordinate transformation from y to z coordinates related by : $\frac{dz}{dy}%
=e^{\sigma(y)}$. The coordinate transformation is chosen to eliminate the 
terms involving first derivatives. Thus we end up with the differential 
equation of the form:  
\begin{equation} 
\left\{- \frac{1}{2}\partial_z^2+V(z)\right\}\hat{f}_{n}(z)=\frac{m_n^2}{2}%
\hat{f} _{n}(z) 
\end{equation} 
where the potential is given by  
\begin{equation} 
\hspace*{0.5cm} V(z)=\frac{\frac{15}{4}(\sigma^{\prime}(y))^{2}+m_{\Phi}^{2}%
}{2[g(z)]^2}- \frac{\frac{3}{2}}{2[g(z)]^2}\sigma^{\prime\prime}(y) 
\end{equation} 
where $g(z)\equiv e^{\sigma(y)}$ and we have made a redefinition of the 
wavefunction:  
\begin{equation} 
\hat{f}_{n}(z)=e^{-\frac{3}{2} \sigma(y)} f_{n}(y) 
\end{equation} 
 
Note that for $m_{\phi }=0$ the above Schr\"{o}endiger equation is identical 
to that of the graviton. This implies that the mass spectrum of a massless 
scalar field of even parity is identical to the graviton's and thus supports 
an ultralight KK state(s). Addition of a bulk mass term ($\beta =0$), 
results in the disappearance of the zero mode from the spectrum (the 
ultralight state also is lost). Nevertheless, by considering a mass term of 
the more general form (with $\alpha \neq 0$ and $\beta \neq 0$ ), which has 
the characteristic that it changes both terms of the potential of eq.(\ref{multi15}), 
we can not only recover the zero mode but in addition have ultralight KK 
state(s). In this case the corresponding potential will be  
\begin{equation} 
\hspace*{0.5cm}V(z)=\frac{\left( \frac{15}{4}+\alpha \right) (\sigma 
^{\prime }(y))^{2}}{2[g(z)]^{2}}-\frac{\left( \frac{3}{2}-\beta \right) }{%
2[g(z)]^{2}}\sigma ^{\prime \prime }(y) 
\label{multi15}
\end{equation} 
This is of the general form given in Appendix A. A massless mode exists if $%
\alpha =\beta ^{2}-4\beta $ in which case the wavefunction is ${\hat{f}}%
(z)\propto e^{(\beta -3/2)\sigma (y)}$. From equation (\ref{multi11}) we see that $%
f(y)e^{-\sigma (y)}\propto e^{(\beta -1)\sigma (y)}$ is the appropriately 
normalised wavefunction in the interval $[-L,L]$. This is localised on the 
positive tension brane for $\beta >1$ and on the negative tension brane for $%
\beta <1$. When the condition for the zero mode is satisfied we find the 
mass of the first ultralight KK state to be given by.  
\begin{equation} 
m_{1}\approx \sqrt{4{\nu }^{2}-1}~kw~e^{-(\nu +\frac{1}{2})x} 
\end{equation} 
where $\nu =\frac{3}{2}-\beta $ and for the rest of the KK tower  
\begin{equation} 
m_{n+1}\approx \xi _{n}~kw~e^{-x}~~~~~~n=1,2,3,\ldots  
\end{equation} 
where $\xi _{2i+1}$ is the $(i+1)$-th root of $J_{\nu -\frac{1}{2}}(x)$ ($%
i=0,1,2,\ldots $) and $\xi _{2i}$ is the $i$-th root of $J_{\nu +\frac{1}{2}%
}(x)$ ($i=1,2,3,\ldots $). The above approximations become better away from 
the $\nu =\frac{1}{2}$ , $x=0$ and for higher KK levels, $n$. The first mass 
is singled out from the rest of the KK tower as it has an extra exponential 
suppression that depends on the mass of the bulk fermion. In contrast, the 
rest of the KK tower has only a very small dependence on the mass of the 
bulk fermion through the root of the Bessel function $\xi _{n}=\xi _{n}(\nu ) 
$ which turns out to be just a linear dependence on $\nu $. 
 
\subsection{$''++''$ model} 
 
We now consider a model which exhibits Bi-gravity but does not require 
negative tension branes. It is built using two positive 
tension branes and leads to $AdS$ in four dimensions. A discussion of this 
model appears in Appendix \ref{multilocapp}.  
 
The discussion of the localisation of spin 0 fields in the $^{\prime \prime 
}++^{\prime \prime }$ case follows similar lines to that of the $^{\prime 
\prime }+-+^{\prime \prime }$ case. In the absence of any mass term for the 
scalar the potential has the form given in eq.(\ref{multi80}) with $\nu =\frac{3}{2}$. 
Thus again the spectrum of the scalar KK tower is identical to that of the 
graviton. However the structure changes on the addition of a five 
dimensional mass term for the scalar. If one adds a constant bulk mass term 
there is no longer a zero mode and the ultralight state is also lost. In 
this case, however, it is not possible to recover the zero mode and light 
states by adding a boundary term corresponding to coupling to boundary 
sources. The reason is that a boundary term is no longer equivalent to a 
term proportional to $\sigma ^{\prime \prime }$
(\textit{c.f.}~Appendix \ref{multilocapp} ). As a 
result, up to the constant bulk term we have added, the bulk potential still 
has the form of eq.(\ref{multi80}) with $\nu =\frac{3}{2}$ and, due to the constant 
bulk mass term, there is no zero mode. We see that the multi-localisation by 
a constant bulk plus brane mass term was special to the case with negative 
tension branes. If one is to achieve the same in the case without negative 
tension branes it is necessary to add a mass term of the form given in 
eq.(\ref{multi6}) which cannot now be interpreted as a five dimensional bulk mass term 
plus a coupling of the scalar field to boundary sources. Note that if one  
\textit{\ does} choose a scalar mass term of the form given in eq.(\ref{multi6}) the 
remainder of the discussion applies to the $^{\prime \prime }++^{\prime 
\prime }$ case too and one can generate the multilocalised scalar field 
configurations discussed above. 
 
Of course the question is whether such a scalar mass term can be justified. 
As we will discuss in Section 9 supergravity can generate such mass terms in 
some, but not all, cases. For the remainder it seems unlikely as $\sigma 
^{\prime }$ and $\sigma ^{\prime \prime }$ are related to the metric and the 
underlying geometry of the compactification and it is difficult to see why a 
mass term of the form given in eq.(\ref{multi6}) should arise. One possible explanation 
may follow if one can realise the ideas of reference \cite{Kehagias:2000au}. 
In this case the geometry of compactification is driven by scalar field 
vacua with kink profiles along the extra dimension. Perhaps the coupling to these scalar fields will induce a mass term of 
the form given in eq.(\ref{multi6}). 
 
 
\section{Multi-Localization of spin $\frac{1}{2}$ field} 
 
As has been shown in Chapter \ref{neutrino} multi-localization can 
appear also to spin $\frac{1}{2}$ fields with appropriate mass terms. Here 
for completeness we briefly review this case. The $AdS_{5}$ background 
geometry localizes the chiral zero mode on negative tension branes. However 
the addition of a mass term \cite{Rubakov:1983bb,Jackiw:1976fn} can alter 
the localization properties of the fermion so that it is localized on 
positive tension branes. The starting point again will be the action for a 
spin $\frac{1}{2}$ particle in the curved five dimensional background of 
eq.(\ref{multi2}):  
\begin{equation} 
S=\int d^4 x \int dy \sqrt{G} \{ E^{A}_{\alpha}\left[ \frac{i}{2} \bar{\Psi} 
\gamma^{\alpha} \left( \overrightarrow{{\partial}_{A}}-\overleftarrow{%
\partial_{A}} \right) \Psi + \frac{\omega_{bcA}}{8} \bar{\Psi} \{ 
\gamma^{\alpha},\sigma^{bc} \} \Psi \right] - m(y) \bar{\Psi}\Psi \} 
\end{equation} 
where $G=det(G_{AB})=e^{-8\sigma(y)}$. Given the convention of of eq.(\ref{multi2}) we 
adopt the ``mostly hermitian'' representation of the Dirac matrices. The 
four dimensional representation of the Dirac matrices is chosen to be $%
\gamma^{a}=( \gamma^{\mu},~~\gamma^{5} )$ with $(\gamma^{0})^{2}=-1,~ 
(\gamma^{i})^{2}=1,~ (\gamma^{5})^{2}=1$. We define $\Gamma^{M}=E^{M}_{a}%
\gamma^{a}$ and thus we have $\{\gamma^{a},\gamma^{b}\}=2 \eta^{ab}$ and $%
\{\Gamma^{a},\Gamma^{b}\}=2 g^{ab}(y)$, where $\eta^{ab}=diag(-1,1,1,1,1)$. 
The vielbein is given by  
\begin{equation} 
E^{A}_{\alpha}=diag(e^{\sigma(y)},e^{\sigma(y)},e^{\sigma(y)},e^{%
\sigma(y)},1) 
\end{equation} 
 
Here we 
choose the mass term to have a (multi-) kink profile $m(y)=\frac{\sigma 
^{\prime}(y)}{k}$. It is convenient to write the action in terms of the 
fields: $\Psi_{R}$ and $\Psi_{L}$ where $\Psi_{R,L}=\frac{1}{2}(1\pm 
\gamma_{5})\Psi$ and $\Psi=\Psi_{R}+\Psi_{L}$. The action becomes:  
\begin{eqnarray} 
S=\int d^4 x \int dy \{ e^{-3\sigma}\left( \bar{\Psi}_{L}i \gamma^{\mu} 
\partial_{\mu} \Psi_{L} + \bar{\Psi}_{R}i \gamma^{\mu} \partial_{\mu} 
\Psi_{R} \right) - e^{-4\sigma} m \left(\frac{\sigma^{\prime}(y)}{k}\right) 
\left( \bar{\Psi}_{L}\Psi_{R} + \bar{\Psi}_{R}\Psi_{L} \right)  \nonumber \\ 
-\frac{1}{2}\left[ \bar{\Psi}_{L} 
(e^{-4\sigma}\partial_{y}+\partial_{y}e^{-4\sigma} ) \Psi_{R} - \bar{\Psi}%
_{R}(e^{-4\sigma}\partial_{y}+\partial_{y}e^{-4\sigma} ) \Psi_{L} \right] 
\end{eqnarray} 
writing $\Psi_{R}$ and $\Psi_{L}$ in the form:  
\begin{equation} 
\Psi_{R,L}(x,y)=\sum_{n}\psi^{R,L}_{n}(x)e^{2\sigma(y)}f_{n}^{R,L}(y) 
\end{equation} 
the action can be brought in the form  
\begin{equation} 
S=\sum_{n} \int d^4 x \{\bar{\psi}_{n}(x) i \gamma^{\mu} \partial_{\mu} 
\psi_{n}(x) - m_{n}\bar{\psi}_{n}(x) \psi_{n}(x) \} 
\end{equation} 
provided the wavefunctions obey the following equations  
\begin{eqnarray} 
\left( -\partial_{y} +m\frac{\sigma^{\prime}(y)}{k}%
\right)f^{L}_{n}(y)=m_{n}e^{\sigma(y)}f^{R}_{n}(y)  \nonumber \\ 
\left( \partial_{y} +m\frac{\sigma^{\prime}(y)}{k}%
\right)f^{R}_{n}(y)=m_{n}e^{\sigma(y)}f^{L}_{n}(y) 
\end{eqnarray} 
and the orthogonality relations (taking account of the $Z_{2}$ symmetry):  
\begin{equation} 
\int_{-L}^{L} dy e^{\sigma(y)} {f^{L}}^{*}_{m}(y) f^{L}_{n}(y)=\int_{-L}^{L} 
dy e^{\sigma(y)} {f^{R}}^{*}_{m}(y) f^{R}_{n}(y)=\delta_{mn} 
\label{multi24}
\end{equation} 
where we assume that the length of the orbifold is $2L$. 
 
We solve the above system of coupled differential equations by substituting $%
f^{L}_{n}(y)$ from the second in the first equation. Thus we end up with a 
second order differential equation, which can always be brought to a 
Schr\"ondiger form by a convenient coordinate transformation from y to z 
coordinates related through $\frac{dz}{dy}=e^{\sigma(y)}$. This gives the 
differential equation of the form:  
\begin{equation} 
\left\{- \frac{1}{2}\partial_z^2+V_{R}(z)\right\}\hat{f}^{R}_{n}(z)=\frac{%
m_n^2}{2}\hat{f}^{R} _{n}(z) 
\end{equation} 
with potential  
\begin{equation} 
\hspace*{0.5cm} V_{R}(z)=\frac{\nu(\nu+1)(\sigma^{\prime}(y))^{2}}{2[g(z)]^2}%
- \frac{\nu}{2[g(z)]^2}\sigma^{\prime\prime}(y) 
\label{multi26}
\end{equation} 
Here $\hat{f}^{R}_{n}(z)=f^{R}_{n}(y)$ and we have defined $\nu\equiv\frac{m%
}{k}$ and $g(z)\equiv e^{\sigma(y)}$. The left handed wavefunctions are 
given by \footnote{%
Note that it can be shown that the left-handed component obeys also a 
similar Schr\"odinger equation with $V_{L}(z)=\frac{\nu(\nu-1)k^{2}}{%
2[g(z)]^2}+ \frac{\nu}{2[g(z)]^2}\sigma^{\prime\prime}(y)$ which is given by  
$V_{R}$ with $\nu \rightarrow -\nu$.}:  
\begin{equation} 
f^{L}_{n}(y)= \frac{e^{-\sigma(y)}}{m_{n}} \left( \partial_{y} +m\frac{%
\sigma^{\prime}(y)}{k}\right)f^{R}_{n}(y) 
\end{equation} 
For $\nu=\frac{3}{2}$ that the form of eq.(\ref{multi26}) is  exactly the same as that 
satisfied by the graviton. The solution has the form ${\hat{f}}%
_{n}^{R}(z)\propto e^{-\nu \sigma (y)}. $ From equation (\ref{multi24}) we see that $%
f^{R}(y)e^{\sigma (y)/2}\propto e^{(1/2-\nu )\sigma (y)}$ is the 
appropriately normalised wavefunction. 
 
There are three regions of localization: For $\nu<\frac{1}{2}$ the zero mode 
is localized on negative tension branes, for $\nu=\frac{1}{2}$ it is not 
localized and for $\nu>\frac{1}{2}$ it is localized on positive tension 
branes. The study of the spectrum of the above differential equation 
provides the spectrum (for $\nu>\frac{1}{2}$): For the first KK state we 
find (for the symmetric configuration)  
\begin{equation} 
m_1=\sqrt{4 {\nu}^2 -1 } ~kw~ e^{-(\nu+\frac{1}{2}) x} 
\end{equation} 
and for the rest of the tower  
\begin{equation} 
m_{n+1}= \xi_n ~ kw~ e^{-x} ~~~~~~n=1,2,3, \ldots 
\end{equation} 
where $\xi_{2i+1}$ is the $(i+1)$-th root of $J_{\nu-\frac{1}{2}}(x)$ ($%
i=0,1,2, \ldots$) and $\xi_{2i}$ is the $i$-th root of $J_{\nu+\frac{1}{2}%
}(x)$ ($i=1,2,3, \ldots$). The above approximations become better away from 
the $\nu=\frac{1}{2}$, $x=0$ and for higher KK levels $n$. The first mass is 
manifestly singled out from the rest of the KK tower as it has an extra 
exponential suppression that depends on the mass of the bulk fermion. By 
contrast the rest of the KK tower has only a very small dependence on the 
mass of the bulk fermion thought the root of the Bessel function $ 
\xi_{n}=\xi_{n}(\nu)$ which turns out to be just a linear dependence in $\nu$%
. The special nature of the first KK state appears not only in the 
characteristics of the mass spectrum but also in its coupling behaviour. As 
it was shown in Chapter \ref{neutrino} the coupling to matter of the 
right-handed component of the first KK state is approximately constant 
(independent of the separation of the positive branes). 
 
It is instructive to examine the localization behaviour of the modes as the 
separation between the two $\delta $-function potential wells increases. In 
the following we assume that $\nu >\frac{1}{2}$ so that multi-localization 
is realized. In the case of infinite separation we know that each potential 
well supports a single chiral massless zero mode and that the rest of the 
massive modes come with Dirac mass terms. This raises an interesting 
question: How from the original configuration with finite size which has 
only one chiral mode do we end up with a configuration that has two chiral 
modes ? The answer to this question is found by examining the localization 
properties of the first special KK mode. From Figs (\ref{multifig3},\ref{multifig4},\ref{multifig5}) we see that the 
right-handed component of the first KK state (dashed line) is localized on 
the positive tension brane whereas the left-handed component (dotted line) 
is localized in the central region of the configuration. As we increase the 
distance separating the two potential wells the right-handed component 
remains localized on the positive tension branes whereas the left-handed 
state starts to spread along the extra dimension. We see that the second 
chiral mode that appears in the infinite separation limit is the 
right-handed component of the first KK state. This is possible since in that 
limit the left-handed component decouples (since it spreads along the 
infinite extra dimension). In this limit, the chiral zero mode that each 
potential well supports can be considered as the combination of the zero 
mode of the initial configuration and the massless limit of the first KK 
state. 
 
Our discussion of fermion localisation applies also to models of the ``$++"$ 
type without negative tension branes. The major difference is that the 
fermion mass term are no-longer constant in the bulk. As in the case of the 
scalars it remains to be seen whether such mass terms actually arise in 
models in which the geometry is determined by non-trivial scalar field 
configurations. 
 
 
\section{Localization and Multi-Localization of spin $1$ field} 
 
We now turn to the study of an Abelian gauge field. In the context of string 
theory it is natural to have gauge fields living in their world-volume of 
D-branes (these gauge fields emerge from open strings ending on the 
D-branes). However, in the case of a domain wall it turns out that it is 
difficult to localize gauge bosons in a satisfactory way. The problem has 
been addressed by several authors Ref.\cite{Dvali:1997xe,Dvali:2001rx,Kehagias:2000au,Shaposhnikov:2001nz,Tachibana:2001xq}. In this section we argue that the localization of gauge boson fields is 
indeed technically possible for particular forms of its five dimensional 
mass term (a similar mass term has been  considered by \cite{Ghoroku:2001zu}). Our starting point is the Lagrangian for an Abelian gauge boson 
in five dimensions:  
\begin{equation} 
S=\int d^{4}x\int dy\sqrt{G}\left[ -\frac{1}{4}G^{MK}~G^{NL}~F_{MN}~F_{KL}-%
\frac{1}{2}\alpha (\sigma ^{\prime }(y))^{2}A_{M}A^{M}-\frac{1}{2}\beta 
\sigma ^{\prime \prime }(y)A_{\mu }A^{\mu }\right]  
\end{equation} 
where $F_{MN}=\partial _{M}A_{N}-\partial _{N}A_{M}$. Again we have assumed 
a mass term allowed by the symmetries of the action of the form:  
\begin{equation} 
m^{2}=\alpha (\sigma ^{\prime }(y))^{2}+\beta \sigma ^{\prime \prime }(y) 
\end{equation} 
Of course it is important to be able to generate the above mass term in a 
gauge invariant way. This can be readily done through the inclusion in the 
Lagrangian of the term:  
\begin{equation} 
\left( \alpha (\sigma ^{\prime }(y))^{2}+\beta \sigma ^{\prime \prime 
}(y)\right) \left((D^{M}\phi )^{\ast }(D_{M}\phi )-V(\phi )\right) 
\end{equation} 
Here we have added a five dimensional charged Higgs field, $\phi $. If the 
potential $V(\phi )$ triggers a vacuum expectation value for $\phi $, it 
will spontaneously break gauge invariance both in the bulk and on the brane 
and generate a vector mass term of the required form. The resulting action 
(in the gauge $A_{5}=0$)  is  
\begin{equation} 
S=\int d^{4}x\int dy\sqrt{\hat{G}}\left[ -\frac{1}{4}\hat{G}^{\mu \kappa }~%
\hat{G}^{\nu \lambda }~F_{\mu \nu }~F_{\kappa \lambda }-\frac{1}{2}%
e^{-2\sigma (y)}(\partial _{5}A_{\nu })(\partial _{5}A_{\lambda })\hat{G}%
^{\nu \lambda }-\frac{1}{2}m^{2}A_{\mu }A^{\mu }\right]  
\end{equation} 
where:  
\begin{equation} 
m^{2}=\alpha (\sigma ^{\prime }(y))^{2}+\beta \sigma ^{\prime \prime }(y) 
\end{equation} 
Performing the KK decomposition  
\begin{equation} 
A^{\mu }(x,y)=\sum_{n}A_{n}^{\mu }(x)f_{n}(y) 
\end{equation} 
this can be brought in the familiar action form for massive spin 1 particles 
propagating in flat space-time  
\begin{equation} 
S=\sum_{n}\int d^{4}x\left[ -\frac{1}{4}\eta _{\mu \kappa }~\eta _{\nu 
\lambda }~F_{n}^{\mu \nu }~F_{n}^{\kappa \lambda }-\frac{1}{2}%
m_{n}^{2}A_{n}^{\mu }A_{n}^{\nu }\right]  
\end{equation} 
provided that $f_{n}(y)$ satisfies the following second order differential 
equation  
\begin{equation} 
-\frac{d}{dy}\left( e^{-2\sigma (y)}\frac{df_{n}(y)}{dy}\right) 
+m^{2}e^{-2\sigma (y)}f_{n}(y)=m_{n}^{2}f_{n}(y) 
\end{equation} 
with the following orthogonality relations (taking in account the $Z_{2}$ 
symmetry):  
\begin{equation} 
\int_{-L}^{L}dy{f}_{m}^{\ast }(y)f_{n}(y)=\delta _{mn} 
\label{multi38}
\end{equation} 
where we assume that the length of the orbifold is $2L$. As before this can 
be brought to a Schr\"{o}dinger form by a redefinition of the wavefunction 
and by a convenient coordinate transformation from y to z coordinates 
related through: $\frac{dz}{dy}=e^{\sigma (y)}$. Thus we end up with the 
differential equation of the form:  
\begin{equation} 
\left\{ -\frac{1}{2}\partial _{z}^{2}+V(z)\right\} \hat{f}_{n}(z)=\frac{%
m_{n}^{2}}{2}\hat{f}_{n}(z) 
\end{equation} 
where  
\begin{equation} 
\hspace*{0.5cm}V(z)=\frac{\frac{3}{4}(\sigma ^{\prime }(y))^{2}+m^{2}}{%
2[g(z)]^{2}}-\frac{\frac{1}{2}}{2[g(z)]^{2}}\sigma ^{\prime \prime }(y) 
\end{equation} 
where  
\begin{equation} 
\hat{f}_{n}(z)=e^{-\frac{1}{2}\sigma (y)}f_{n}(y) 
\label{multi41}
\end{equation} 
and we have defined for convenience $g(z)\equiv e^{\sigma (y)}$. Let us now 
examine the localization properties of the gauge boson modes. For $m=0$ 
there exists a zero mode with wavefunction:  
\begin{equation} 
\hat{f}_{n}(z)=Ce^{-\frac{1}{2}\sigma (y)}=\frac{C}{\sqrt{g(z)}} 
\end{equation} 
where C is a normalization constant. From eq.(\ref{multi38}) it is clear that in this 
case the appropriately normalized wavefunction $f(y)$ is constant along the 
extra dimension and thus that the gauge boson is delocalized. For $m\neq 0$ 
with $\alpha \neq 0$ and $\beta =0$ the zero mode becomes massive. We can 
recover the zero mode by allowing for the possibility of $\beta \neq 0$. In 
this case the potential can be written as  
\begin{equation} 
\hspace*{0.5cm}V(z)=\frac{\left( \alpha +\frac{3}{4}\right) (\sigma ^{\prime 
}(y))^{2}}{2[g(z)]^{2}}-\frac{\left( \frac{1}{2}-\beta \right) }{2[g(z)]^{2}}%
\sigma ^{\prime \prime }(y) 
\end{equation} 
This is of the general form given in Appendix \ref{multilocapp}. A massless mode exists if $%
\alpha =\beta ^{2}-2\beta $ in which case the wavefunction ${\hat{f}}%
(z)\propto e^{(\beta -1/2)\sigma (y)}$. From equation (\ref{multi41}) we see that $%
f(y)e^{-\sigma (y)}\propto e^{\beta \sigma (y)}$ is the appropriately 
normalised wavefunction in the interval $[-L,L]$. This is localised on the 
positive tension brane for $\beta >0$ and on the negative tension brane for $%
\beta <0$. When the condition for the zero mode is satisfied we find the 
mass of the first ultralight KK state to be given by (for the symmetric 
configuration):  
\begin{equation} 
m_{1}=\sqrt{4{\nu }^{2}-1}~kw~e^{-(\nu +\frac{1}{2})x} 
\end{equation} 
where $\nu =\frac{1}{2}-\beta $ and for the rest of the tower  
\begin{equation} 
m_{n+1}=\xi _{n}~kw~e^{-x}~~~~~~n=1,2,3,\ldots  
\end{equation} 
where $\xi _{2i+1}$ is the $(i+1)$-th root of $J_{\nu -\frac{1}{2}}(x)$ ($%
i=0,1,2,\ldots $) and $\xi _{2i}$ is the $i$-th root of $J_{\nu +\frac{1}{2}%
}(x)$ ($i=1,2,3,\ldots $). Again, the first KK is singled out from the rest 
of the KK tower as it has an extra exponential suppression that depends on 
the mass parameter $\nu $. In contrast the rest of the KK tower has only a 
very small dependence on the $\nu $ parameter thought the root of the Bessel 
function $\xi _{n}=\xi _{n}(\nu )$ which turns out to be just a linear 
dependence in $\nu $. 
 
Once again our discussion applies unchanged to the case without negative 
tension branes. Once again the difference is that the mass no longer 
corresponds to a combination of brane and constant bulk terms. Perhaps the 
origin of such terms will be better motivated in the case that the geometry 
is driven by a non-trivial vacuum configuration of a scalar field with a 
profile in the bulk such that coupling of the gauge field to it generates 
the required mass term. At present we have no indication that this should be 
the case. 
 
 
\section{Multi-Localization of spin $\frac{3}{2}$ field} 
 
In this section we consider the (multi-) localization of a spin $\frac{3}{2}$ 
particle. The starting point will be the Lagrangian for a $\frac{3}{2}$ 
particle propagating in curved background is:  
\begin{equation} 
S = - \int d^4 x \int dy \sqrt{G} \bar{\Psi}_{M} \Gamma^{MNP} \left( D_{N} +  
\frac{m}{2} \Gamma_{N} \right) \Psi_{P} 
\end{equation} 
where the covariant derivative is  
\begin{equation} 
D_{M} \Psi_{N} = \partial_{M} \Psi_{N} - \Gamma^{P}_{MN} \Psi_{P} + \frac{1}{%
2} \omega^{AB}_{M} \gamma_{AB} 
\end{equation} 
with $\gamma_{AB} = \frac{1}{4} [ \gamma_{A},\gamma_{B} ]$ and $%
\Gamma^{MNP}=\Gamma^{[M} \Gamma^{N} \Gamma^{P]}$.  The connection is given 
by  
\begin{eqnarray} 
\omega^{AB}_{M}=\frac{1}{2}~g^{PN}~{e^{[A}}_{P} \partial_{[M} {e^{B]}}_{N]} 
+ \frac{1}{4}~g^{PN}~g^{T\Sigma}~{e^{[A}}_{P} {e^{B]}}_{T} 
\partial_{[\Sigma} ~ {e^{\Gamma}}_{N]}~ e^{\Delta}_{M} ~\eta_{\Gamma \Delta} 
\end{eqnarray} 
where $\Gamma^{M}={e^{M}}_{n}\gamma^{n}$ with ${e^{M}}_{n}=diag(e^{%
\sigma(y)},e^{\sigma(y)},e^{\sigma(y)},e^{\sigma(y)},1)$. As in the case of 
the Abelian gauge boson we will assume that we generate the mass term for 
this field in a gauge invariant way. Exploiting the gauge invariance we can 
fix the gauge setting $\Psi_{5}=0$, something that simplifies considerably 
the calculations. In this case, the above action becomes  
\begin{equation} 
S = - \int d^4 x \int dy \sqrt{G} \bar{\Psi}_{\mu} \Gamma^{\mu \nu \rho} 
\left( D_{\nu} + \frac{m}{2} \Gamma_{\nu} \right) \Psi_{\rho} - \sqrt{G}  
\bar{\Psi}_{\mu} \Gamma^{\mu 5 \rho} \left( D_{5} + \frac{m}{2} \Gamma_{5} 
\right) \Psi_{\rho} 
\end{equation} 
We can simplify the above further taking in account the following 
identities:  
\begin{eqnarray} 
\Gamma^{\mu \nu \rho}&=&e^{3\sigma(y)}\gamma^{\mu \nu \rho}  \nonumber \\ 
\Gamma^{\mu 5 \rho}&=&e^{2\sigma(y)}\gamma^{\mu 5 \rho}  \nonumber \\ 
\gamma^{\mu \nu \rho} \gamma_{\nu}&=&-2\gamma^{\mu \rho}  \nonumber \\ 
\gamma^{\mu 5 \rho}&=&-\gamma^{5} \gamma^{\mu \rho}  \nonumber \\ 
\gamma^{\mu \rho} \gamma_{\mu}&=&-2 \gamma^{\rho} 
\end{eqnarray} 
Using the above we find the following simple forms for the covariant 
derivatives  
\begin{eqnarray} 
D_{\nu}&=&\partial_{\nu}-\frac{1}{2} \sigma^{\prime}(y)e^{-\sigma(y)} 
\gamma_{\nu}\gamma^{5}  \nonumber \\ 
D_{5}&=&\partial_{5} 
\end{eqnarray} 
Using the previous relations we write the action in the form  
\begin{eqnarray} 
S = -\int d^4 x \int dy e^{-\sigma(y)} \bar{\Psi}_{\mu} \gamma^{\mu \nu 
\rho} \left( \partial_{\nu} -\frac{1}{2} \sigma^{\prime}(y) e^{-\sigma(y)} 
\gamma_{\nu} \gamma^{5} + \frac{m}{2} e^{-\sigma(y)} \gamma_{\nu} \right) 
\Psi_{\rho}  \nonumber \\ 
- e^{-2\sigma(y)} \bar{\Psi}_{\mu} \gamma^{\mu 5 \rho} \left( \partial_{5} +  
\frac{m}{2} \gamma_{5} \right) \Psi_{\rho} 
\end{eqnarray} 
The above can be brought in the form  
\begin{equation} 
S= - \int d^4 x \int dy e^{-\sigma(y)} \bar{\Psi}_{\mu} \gamma^{\mu \nu 
\rho} \partial_{\nu}\Psi_{\rho} + e^{-2\sigma(y)} \bar{\Psi}_{\mu} 
\gamma^{\mu \rho} \left[ \frac{3m}{2} + \gamma ^{5} \left( \partial_{5} - 
\sigma^{\prime}(y) \right) \right] \Psi_{\rho} 
\end{equation} 
At this stage it turns out, like in the spin $\frac{1}{2}$ case, that it is 
convenient to write $\Psi_{\mu}$ in terms of $\Psi^{R}_{\mu}$ and $%
\Psi^{L}_{\mu}$ ($\Psi_{\mu}=\Psi^{R}_{\mu}+\Psi^{L}_{\mu}$) which have 
different KK decomposition:  
\begin{equation} 
\Psi_{\mu}^{R,L}(x,y)=\sum_{n}\psi^{R,L}_{\mu ~ 
n}(x)e^{\sigma(y)}f_{n}^{R,L}(y) 
\end{equation} 
Substituting the above decompositions in the action we get  
\begin{eqnarray} 
S = - \int d^4 x \int dy e^{\sigma(y)} ( \bar{\Psi}^{R}_{\mu} \gamma^{\mu 
\nu \rho} \partial_{\nu}\Psi^{R}_{\rho}+ \bar{\Psi}^{L}_{\mu} \gamma^{\mu 
\nu \rho} \partial_{\nu}\Psi^{L}_{\rho})  \nonumber \\ 
+ \bar{\Psi}^{R}_{\mu} \gamma^{\mu \rho} \left[ \frac{3m}{2} + \gamma ^{5} 
\partial_{5} \right] \Psi^{L}_{\rho}+ \bar{\Psi}^{L}_{\mu} \gamma^{\mu \rho} %
\left[ \frac{3m}{2} + \gamma ^{5} \partial_{5} \right] \Psi^{R}_{\rho} 
\end{eqnarray} 
this can be brought to the familiar action form for massive spin $\frac{3}{2} 
$ particle in flat background  
\begin{equation} 
S = \int d^4 x \{ - \bar{\Psi}_{\mu} \gamma^{\mu \nu \rho} 
\partial_{\nu}\Psi_{\rho} + m_{n} \bar{\Psi}_{\mu} \gamma^{\mu \rho} 
\Psi_{\rho} \} 
\end{equation} 
provided that $f^{R}_{n}$ and $\Psi^{L}_{n}$ satisfy that following coupled 
differential equations  
\begin{eqnarray} 
\left( -\partial_{y} +\frac{3m}{2} \frac{\sigma^{\prime}(y)}{k}%
\right)f^{L}_{n}(y)=m_{n}e^{\sigma(y)}f^{R}_{n}(y)  \nonumber \\ 
\left( \partial_{y} +\frac{3m}{2} \frac{\sigma^{\prime}(y)}{k}%
\right)f^{R}_{n}(y)=m_{n}e^{\sigma(y)}f^{L}_{n}(y) 
\end{eqnarray} 
supplied with the orthogonality relations (taking account of the $Z_{2}$ 
symmetry):  
\begin{equation} 
\int_{-L}^{L} dy e^{\sigma(y)} {f^{L}}^{*}_{m}(y) f^{L}_{n}(y)=\int_{-L}^{L} 
dy e^{\sigma(y)} {f^{R}}^{*}_{m}(y) f^{R}_{n}(y)=\delta_{mn} 
\end{equation} 
Note that the form of the above system of differential equations is 
identical to the one of spin $\frac{1}{2}$ particle provided we substitute $%
m \rightarrow \frac{3 m}{2}$. Accordingly the corresponding Schr\"odinger 
equation and thus the mass spectrum in this case is going to be the same as 
the spin $\frac{1}{2}$ case up to the previous rescaling of the mass 
parameter. 
 
 
\section{Multi-Localization of the graviton field} 
 
In this section, for completeness, we review the multi-localization scenario 
for the graviton. The gravitational field has the characteristic that it 
creates itself the background geometry in which it and the rest of the 
fields propagate. Thus, one has first to find the appropriate vacuum 
solution and then consider perturbations around this solution. In order to 
exhibit how the multi-localization appears in this case, we will again work 
with the $^{\prime\prime}+-+^{\prime\prime}$ configuration. The starting 
point is the Lagrangian  
\begin{equation} 
S=\int d^4 x \int_{-L_2}^{L_2} dy \sqrt{-G} \{-\Lambda + 2 M^3 
R\}-\sum_{i}\int_{y=L_i}d^4xV_i\sqrt{-\hat{G}^{(i)}} 
\end{equation} 
The Einstein equations that arise from this action are:  
\begin{equation} 
R_{MN}-\frac{1}{2}G_{MN}R=-\frac{1}{4M^3} \left(\Lambda G_{MN}+ \sum_{i}V_i%
\frac{\sqrt{-\hat{G}^{(i)}}}{\sqrt{-G}} \hat{G}^{(i)}_{\mu\nu}\delta_M^{\mu}%
\delta_N^{\nu}\delta(y-L_i)\right) 
\end{equation} 
using the metric ansatz of eq.(\ref{multi2}) we find that the above equations imply 
that the function $\sigma(y)$ satisfies:  
\begin{eqnarray} 
\left(\sigma ^{\prime}\right)^2&=&k^2 \\ 
\sigma ^{\prime\prime}&=&\sum_{i}\frac{V_i}{12M^3}\delta(y-L_i) 
\end{eqnarray} 
where $k=\sqrt{\frac{-\Lambda}{24M^3}}$ is a measure of the curvature of the 
bulk. The exact form of $\sigma(y)$ depends on the brane configuration that 
we consider. For example, in the case of $^{\prime\prime}+-+^{\prime\prime}$ 
model we have  
\begin{equation} 
\sigma(y)=k\left\{L_1-\left||y|-L_1\right|\right\} 
\end{equation} 
where $L_{1}$ is the position of the intermediate brane. with the 
requirement that the brane tensions are tuned to $V_0=-\Lambda/k>0$, $%
V_1=\Lambda/k<0$, \mbox{$V_2=-\Lambda/k>0$}. In order to examine the 
localization properties of the graviton, the next step is to consider 
fluctuations around the vacuum of eq.(\ref{multi1}). Thus, we expand the field $%
h_{\mu\nu}(x,y)$ in graviton and KK states plane waves:  
\begin{equation} 
h_{\mu\nu}(x,y)=\sum_{n=0}^{\infty}h_{\mu\nu}^{(n)}(x)f_{n}(y) 
\end{equation} 
where $\left(\partial_\kappa\partial^\kappa-m_n^2\right)h_{\mu\nu}^{(n)}=0$ 
and fix the gauge as $\partial^{\alpha}h_{\alpha\beta}^{(n)}=h_{\phantom{-}%
\alpha}^{(n) \alpha}=0$ \footnote{Note that we have ignored the presence of dilaton/radion fields associated 
with the size of the extra dimension or the positions of the branes. For 
more details see Ref.\cite{Charmousis:2000rg,Pilo:2000et,Kogan:2001qx}.}. The function $f_{n}(y) 
$ will obey a second order differential equation which after a change of 
variables ($\frac{dz}{dy}=e^{\sigma(y)}$) reduces to an ordinary 
Schr\"{o}dinger equation:  
\begin{equation} 
\left\{- \frac{1}{2}\partial_z^2+V(z)\right\}\hat{f}_{n}(z)=\frac{m_{n}^{2}}{%
2}\hat{f }^{n}(z) 
\end{equation} 
with potential  
\begin{equation} 
\hspace*{0.5cm} V(z)=\frac{\frac{15}{4}(\sigma^{\prime}(y))^2}{2[g(z)]^2}-  
\frac{\frac{3}{2}}{2[g(z)]^{2}} \sigma^{\prime\prime}(y) 
\label{multi66}
\end{equation} 
where  
\begin{equation} 
\hat{f}_{n}(z)\equiv f_{n}(y)e^{\sigma/2} 
\end{equation} 
and the function $g(z)$ as $g(z)\equiv 
k\left\{z_1-\left||z|-z_1\right|\right\}+1$, where $z_1=z(L_1)$. The study 
of the mass spectrum of reveals the following structure for the mass 
spectrum (for the symmetric configuration):  
\begin{eqnarray} 
m_1&=&2\sqrt{2}ke^{-2x} \\ 
m_{n+1}&=& \xi_n k e^{-x} ~~~~~~n=1,2,3, \ldots 
\end{eqnarray} 
where $\xi_{2i+1}$ is the $(i+1)$-th root of $J_1(x)$ ($i=0,1,2, \ldots$) 
and $\xi_{2i}$ is the $i$-th root of $J_2(x)$ ($i=1,2,3, \ldots$). Again the 
first KK state is singled out from the rest of the KK tower as its mass has 
an extra exponential suppression . 
 
Let us see now how the separability argument works in the case of the 
graviton. Starting with the familiar configuration $^{\prime\prime}+-+^{%
\prime\prime}$, the mass spectrum consists of the massless graviton, the 
ultra-light first KK state and the rest of the KK tower which are massive 
spin two particles. In the limit of infinite separation the first special KK 
mode becomes the second massless mode, according to our previous general 
discussions. However, at first sight the counting of degrees of freedom 
doesn't work: we start with a massive spin $2$ state (first KK state) which 
has five degrees of freedom and we end up with a massless mode which has 
two. It has been shown that in the case of flat spacetime the extra 
polarizations of the massive gravitons do not decouple giving rise to the 
celebrated van Dam-Veltman-Zakharov discontinuity in the propagator of a 
massive spin-2 field in the massless limit \cite{vanDam:1970vg,Zakharov}\footnote{However, in Ref.\cite{Vain,Deffayet:2001uk} was shown that in 
the presence of a source with a characteristic mass scale, there is no 
discontinuity for distances smaller that a critical one. This argument is 
also supported by the results of Refs.\cite{Higuchi:1987py,Kogan:2001uy,Porrati:2001cp} where it was shown that the 
limit is smooth in $dS_{4}$ or $AdS_{4}$ background.}. However our 
separability argument is still valid: Up to this point we have ignored the 
presence of a massless scalar mode, the radion, which is related to the 
motion of the freely moving negative tension brane. It turns out that this 
scalar field is a ghost field, that is, it enters the Lagrangian with the 
wrong kinetic term sign. It can be shown that the effect of the presence of 
this field is to exactly cancel the contribution of the extra polarizations 
of the graviton making the limit of infinite brane separation smooth. Note 
that apart from the radion there is another scalar field in the spectrum, 
the dilaton, which parameterizes the overall size of the extra dimension 
which also decouples in the above limit. 
 
As we have mentioned the problems associated with the presence of the ghost 
radion can be avoided by allowing for $AdS_{4}$ spacetime on the 3-branes. 
In this case there is no need for the negative tension brane (thus there is 
no radion field) and moreover the presence of curvature on the branes makes 
the massless limit of the massive graviton propagator smooth \cite{Kogan:2001uy,Porrati:2001cp}, meaning that 
in the the $AdS_{4}$ curved background the extra polarizations of the 
massive graviton decouple in the massless limit, in agreement with our 
separability argument. 
 
 
\section{Multi-Localization and supersymmetry} 
 
It is interesting to investigate the multi-localization in the 
supersymmetric versions of the previous models. The inclusion of 
supersymmetry is interesting in the sense that it restricts the possible 
mass terms by relating the mass parameters of fermion and boson fields. It 
is well known that $AdS$ spacetime is compatible with supersymmetry \cite{Townsend:1977qa,Deser:1977uq}. In contrast to the case of flat spacetime, 
supersymmetry in $AdS$ requires that fields belonging in the same multiplet 
have different masses. In the previous discussions on the localization of 
the fields, the mass term parameters which control the localization of the 
bulk states, are generally unconstrained. Let us now examine in more detail 
the cases of supergravity, vector supermultiplets and the hypermultiplet. 
 
\paragraph{Supergravity supermultiplet} 
 
The on-shell supergravity multiplet consists of the vielbein $e^{\alpha}_{M}$ 
, the graviphoton $B_{M}$ and the two symplectic-Majorana gravitinos $%
\Psi^{i}_{M}$ ($i=1,2$). The index $i$ labels the fundamental representation 
of the SU(2) automorphism group of the $N=1$ supersymmetry algebra in five 
dimensions. The supergravity Lagrangian in $AdS_{5}$ has the form \cite 
{Gherghetta:2000qt} (in $AdS_{5}$ background 
we can set $B_{M}=0$):  
\begin{eqnarray} 
S_{5}=-\frac{1}{2} \int d^{4}x \int dy \sqrt{-g} \Bigl[ M_{5}^{3} \Big\{ R+ 
i \bar{\Psi}^{i}_{M} \gamma^{MNR} D_{N} \Psi_{R}^{i} -i \frac{3}{2} 
\sigma^{\prime}(y) \bar{\Psi}^{i}_{M}\sigma^{MN}(\sigma_{3})^{ij} 
\Psi^{j}_{N} \Big\}  \nonumber \\ 
+2\Lambda -\frac{\Lambda}{k^2} \sigma^{\prime\prime}(y) \Bigr] 
\end{eqnarray} 
where $\gamma^{MNR}\equiv \sum_{perm} \frac{(-1)^{p}}{3!} \gamma^{M} 
\gamma^{N} \gamma^{R} $ and $\sigma^{MN}=\frac{1}{2}[\gamma^{M},\gamma^{N}]$%
. From the above expression, that is invariant under the supersymmetry 
transformations \cite{Gherghetta:2000qt}, we see that the 
symplectic-Majorana gravitino mass term $m=\frac{3}{2} \sigma^{\prime}(y)$ 
is such that its mass spectrum is identical to the mass spectrum of the 
graviton. This becomes clear by comparing eq.(\ref{multi26}) for $\nu=\frac{3}{2}$ and 
eq.(\ref{multi66}). The latter implies that, in the presence of supersymmetry, 
multi-localization of the graviton field implies multi-localization of the 
gravitinos and thus the mass spectrum of these fields will contain 
ultralight KK state(s). 
 
\paragraph{Vector supermultiplet} 
 
The on-shell field content of the vector supermultiplet $V=(V_{M},%
\lambda^{i},\Sigma)$ consists from the gauge field $V_{M}$, a 
symplectic-Majorana spinor $\lambda^{i}$, and the real scalar field $\Sigma$ 
in the adjoint representation.  
\begin{eqnarray} 
S_{5}=-\frac{1}{2} \int d^{4}x \int dy \sqrt{-g} \Bigl[ \frac{1}{2g_{5}^{2}}%
F_{MN}^{2}+(\partial_{M}\Sigma)^{2}+ i\bar{\lambda}^{i} \gamma^{M} D_{M} 
\lambda^{i} + m_{\Sigma}^{2} \Sigma^{2}+i m_{\lambda}\bar{\lambda}^{i} 
(\sigma_{3})^{ij} \lambda^{j}\Bigr] 
\end{eqnarray} 
The above Lagrangian is invariant under the supersymmetry transformations if 
the mass terms of the various fields are of the form (for more details see 
Ref.\cite{Gherghetta:2000qt}):  
\begin{eqnarray} 
m_{\Sigma}^{2}&=&-4(\sigma^{\prime}(y))^{2}+2\sigma^{\prime\prime}(y)  
\nonumber \\ 
m_{\lambda}&=& \frac{1}{2} \sigma^{\prime}(y) 
\end{eqnarray} 
Assuming that $V_{\mu}$ and $\lambda_{L}^{1}$ are even while $\Sigma$ and $%
\lambda^{2}_{L}$ odd then the mass spectrum of all the fields is identical. 
This can be easily seen if we note that for the spinors we have $\nu=\frac{1%
}{2}$, for the scalar $\alpha=-4$, $\beta=2$ and for the gauge boson $%
\alpha=0$, $\beta=0$. The even fields, for the above values of the mass 
parameters, they obey the eq.(\ref{multi83}) of Appendix  \ref{multilocapp} with $\nu=\frac{1}{2}$ 
whereas the odd fields obey eq.(\ref{multi84}) for the same value of the $\nu$ 
parameter. The mass spectrum of the two potentials is identical, apart from 
the zero modes, since they are SUSY-partner quantum mechanical potentials. 
The even fields have zero modes that are not localized (which is expected 
since the massless gauge field is not localized) in contrast to the odd 
fields that have no zero modes (they are projected out due to the boundary 
conditions). 
 
\paragraph{Hypermultiplet} 
 
The hypermultiplet $H=(H^{i},\Psi)$ consists of two complex scalar fields $%
H^{i}$ ($i=1,2$) and a Dirac fermion $\Psi$. In this case the action setup 
is:  
\begin{eqnarray} 
S_{5}=- \int d^{4}x \int dy \sqrt{-g} \Bigl[ |\partial_{M}H^{i}|^{2}+ i\bar{%
\Psi} \gamma^{M} D_{M} \Psi + m_{H^{i}}^{2} |H^{i}|^{2}+i m_{\Psi} \bar{\Psi} 
\Psi \Bigr] 
\end{eqnarray} 
Invariance under the supersymmetric transformations (see Ref.\cite 
{Gherghetta:2000qt}) demand that the mass term of the scalar and fermion 
fields has the form:  
\begin{eqnarray} 
m_{H^{1,2}}^{2}&=&(c^{2} \pm c - \frac{15}{4})(\sigma^{\prime}(y))^{2}+(%
\frac{3}{2}\mp c)\sigma^{\prime\prime}(y)  \nonumber \\ 
m_{\lambda}&=&c \sigma^{\prime}(y) 
\end{eqnarray} 
from the above we identify that $\alpha=c^2 \pm c - \frac{15}{4}$ and $\beta=%
\frac{3}{2} \mp c$ for the scalar fields. Note that $\alpha=\beta^{2}-4 \beta 
$ which implies the existence of zero mode for the symmetric scalar fields. 
Moreover, for the scalar fields we find that $\nu \equiv \frac{3}{2}- 
\beta=\pm c$ which implies that the wavefunctions (in z-coordinates) and the 
mass spectrum are identical to the Dirac fermion's. Note that we are 
assuming that $H^{1}$ and $\Psi_{L}$ are even, while $H^{2}$ and $\Psi_{R}$ 
are odd. As expected if supersymmetry is realized, multi-localization of 
scalar fields implies multi-localization of Dirac fermions and the opposite. 
 
In the five dimensional $AdS$ background,  the mass terms compatible with 
supersymmetry are not the ones that correspond to degenerate supermultiplet 
partners, but are such that all members of the supermultiplets have the same 
wavefunction behaviour and the same mass spectrum. However, in the four 
dimensional effective field theory description, the states lie in degenerate 
SUSY multiplets, as is expected since the 4D theory is flat. 
 
\section{Discussion and conclusions}

In this Chapter we studied the localization behaviour and the mass spectrum of 
bulk fields in various multi-brane models with localized gravity. We showed 
that the addition of appropriate mass terms controls the strength or/and the 
location of localization of the fields and moreover can induce localization 
to Abelian spin $1$ fields. The localization of all the above fields can 
resemble that of the graviton, at least in a region of the parameter space. 
This means that it is possible for fields of all spins ($\leq 2$) can be 
localized on positive tension branes. The latter implies that in the context 
of multi-brane models emerges the possibility of multi-localization for all 
the previous fields with appropriate mass terms. We have shown, giving 
explicit examples, that when multi-localization is realized the above fields 
apart from the massless zero mode support ultra-light localized KK mode(s). 
 
In the simplest constructions with two positive branes, that we considered 
here, there is only one special KK state. However by adding more positive 
tension branes one can achieve more special light states. In the extreme 
example of a infinite sequence of positive branes instead of discrete 
spectrum of KK states we have continuum bands. In the previous case the 
special character of the zeroth band appears as the fact that it is well 
separated from the next. 
 
Summarizing, in this paper we pointed out some new characteristics of 
multi-brane scenarios in the case that multi-localization is realized. The 
new phenomenology reveals itself through special light and localized KK 
states. The idea of multi-localization and its relation to new interesting 
phenomenology is of course general and it should not be necessarily related 
to RS type models\footnote{In the case of gravity though, such a construction (or similar) with curved 
background is essential.}, although it finds a natural application in the 
context of these models.

\chapter{Summary and Conclusions}
\label{summ}

In this Thesis we have presented an account of new phenomena that can be realized in wraped brane-world models in five and six dimensions.

In the context of these models we can realize the possibility of having massive gravitons contributing significantly to intermediate distance gravitational interactions. This scenario is not excluded by current observations and furthermore it offers phenomenological signatures which are in principle testable. Taking into account that massive gravitons necessarily arise as KK states from dimensional reduction of a higher dimensional theory, these modifications of gravity can be very unusual window to extra dimensions since Multigravity models which involve brane configurations with interbrane distances of the order of the Planck length, modify gravity not only at short distances as usually do ordinary KK theories, but also at ultra large distances.

 The $''+-+''$ Bigravity model and the quasi-localized GRS model, although they had interesting phenomenology, they all shared the characteristic of moving negative tension branes and thus the moduli corresponding to the fluctuations of these branes are ghost fields and therefore unacceptable.  However, the general characteristics of these models persist in more involved constructions that have less or no problems. Foe example, the appearance of light states whenever we have more than one positive branes that localize gravitons is a generic characteristic which persists in the $AdS$ brane models or the six dimensional ones.

The $AdS$ Bigravity model, although does not have the ghost radion problem, does not give interesting phenomenology. This is due to the fact that the remnant\
cosmological constant is of negative sign, in contradiction with  observations. Moreover it does not lead to any observable modifications of gravity at any observable scale since the modifications of gravity at ultra-large scales are hidden behind the $AdS$ horizon, while the modifications at all scales due to the different propagator structure of the massive gravitons are enormously suppressed. It is, however, interesting how the $AdS$  brane models escape the van Dam-Veltman-Zakharov no-go theorem about the non-decoupling of the additional polarization states of the massive graviton. This in addition to the observation of Vainstein that the latter no-go theorem may not be valid even for flat background, suggests that the massive graviton proposal does not face any fundamental obstructions.

However, one can construct a theoretically and phenomenologically viable multigravity model in six dimensions. Due to the non-trivial curvature of the internal space one can have bounces of the warp factor with only positive tension flat branes.

There are a lot of open questions regarding multigravity  at the moment. The most important one is the construction of a cosmological model of multigravity.    So far the models we considered involved either flat or $AdS$ branes which are not valid descriptions of our Universe at large scales where modifications of gravity due to the massive gravitons are expected to appear.

Apart from applications in the gravitational sector, multi-localization can have interesting applications to other fields: Fields of all spins can be localized  in the context of braneworld models with non-factorizable geometry. Their localization properties are controlled by the form and the details of the mass term (for some fields in order to achieve the desired localization, specific mass terms must be added). Given the latter, in the context of multi-brane models emerges the possibility of multi-localization for fields of all spins provided that   appropriate mass terms are added. When multi-localization is realized the above fields apart from the massless zero mode support ultra-light localized KK mode(s).

\appendix
\chapter{The Myth of Sisyphus} 
\label{myth}


   The gods had condemned Sisyphus to ceaselessly rolling a rock to the top of a mountain, whence the
   stone would fall back of its own weight. They had thought with some reason that there is no more
   dreadful punishment than futile and hopeless labor.

   If one believes Homer, Sisyphus was the wisest and most prudent of mortals. According to another
   tradition, however, he was disposed to practice the profession of highwayman. I see no contradiction
   in this. Opinions differ as to the reasons why he became the futile laborer of the underworld. To
   begin with, he is accused of a certain levity in regard to the gods. He stole their secrets. Egina, the
   daughter of Esopus, was carried off by Jupiter. The father was shocked by that disappearance and
   complained to Sisyphus. He, who knew of the abduction, offered to tell about it on condition that
   Esopus would give water to the citadel of Corinth. To the celestial thunderbolts he preferred the
   benediction of water. He was punished for this in the underworld. Homer tells us also that Sisyphus
   had put Death in chains. Pluto could not endure the sight of his deserted, silent empire. He
   dispatched the god of war, who liberated Death from the hands of her conqueror. 

   It is said that Sisyphus, being near to death, rashly wanted to test his wife's love. He ordered her to
   cast his unburied body into the middle of the public square. Sisyphus woke up in the underworld.
   And there, annoyed by an obedience so contrary to human love, he obtained from Pluto permission to
   return to earth in order to chastise his wife. But when he had seen again the face of this world,
   enjoyed water and sun, warm stones and the sea, he no longer wanted to go back to the infernal
   darkness. Recalls, signs of anger, warnings were of no avail. Many years more he lived facing the
   curve of the gulf, the sparkling sea, and the smiles of earth. A decree of the gods was necessary.
   Mercury came and seized the impudent man by the collar and, snatching him from his joys, lead him
   forcibly back to the underworld, where his rock was ready for him.

   You have already grasped that Sisyphus is the absurd hero. He is, as much through his passions as
   through his torture. His scorn of the gods, his hatred of death, and his passion for life won him that
   unspeakable penalty in which the whole being is exerted toward accomplishing nothing. This is the
   price that must be paid for the passions of this earth. Nothing is told us about Sisyphus in the
   underworld. Myths are made for the imagination to breathe life into them. As for this myth, one sees
   merely the whole effort of a body straining to raise the huge stone, to roll it, and push it up a slope a
   hundred times over; one sees the face screwed up, the cheek tight against the stone, the shoulder
   bracing the clay-covered mass, the foot wedging it, the fresh start with arms outstretched, the
   wholly human security of two earth-clotted hands. At the very end of his long effort measured by
   skyless space and time without depth, the purpose is achieved. Then Sisyphus watches the stone rush
   down in a few moments toward that lower world whence he will have to push it up again toward the
   summit. He goes back down to the plain.

   It is during that return, that pause, that Sisyphus interests me. A face that toils so close to stones is
   already stone itself! I see that man going back down with a heavy yet measured step toward the
   torment of which he will never know the end. That hour like a breathing-space which returns as
   surely as his suffering, that is the hour of consciousness. At each of those moments when he leaves
   the heights and gradually sinks toward the lairs of the gods, he is superior to his fate. He is stronger
   than his rock.

   If this myth is tragic, that is because its hero is conscious. Where would his torture be, indeed, if at
   every step the hope of succeeding upheld him? The workman of today works everyday in his life at
   the same tasks, and his fate is no less absurd. But it is tragic only at the rare moments when it
   becomes conscious. Sisyphus, proletarian of the gods, powerless and rebellious, knows the whole
   extent of his wretched condition: it is what he thinks of during his descent. The lucidity that was to
   constitute his torture at the same time crowns his victory. There is no fate that can not be surmounted
   by scorn.

   If the descent is thus sometimes performed in sorrow, it can also take place in joy. This word is not
   too much. Again I fancy Sisyphus returning toward his rock, and the sorrow was in the beginning.
   When the images of earth cling too tightly to memory, when the call of happiness becomes too
   insistent, it happens that melancholy arises in man's heart: this is the rock's victory, this is the rock
   itself. The boundless grief is too heavy to bear. These are our nights of Gethsemane. But crushing
   truths perish from being acknowledged. Thus, Edipus at the outset obeys fate without knowing it.
   But from the moment he knows, his tragedy begins. Yet at the same moment, blind and desperate, he
   realizes that the only bond linking him to the world is the cool hand of a girl. Then a tremendous
   remark rings out: "Despite so many ordeals, my advanced age and the nobility of my soul make me
   conclude that all is well." Sophocles' Edipus, like Dostoevsky's Kirilov, thus gives the recipe for the
   absurd victory. Ancient wisdom confirms modern heroism.

   One does not discover the absurd without being tempted to write a manual of happiness.
   "What!---by such narrow ways--?" There is but one world, however. Happiness and the absurd
   are two sons of the same earth. They are inseparable. It would be a mistake to say that happiness
   necessarily springs from the absurd. discovery. It happens as well that the felling of the absurd
   springs from happiness. "I conclude that all is well," says Edipus, and that remark is sacred. It
   echoes in the wild and limited universe of man. It teaches that all is not, has not been, exhausted. It
   drives out of this world a god who had come into it with dissatisfaction and a preference for futile
   suffering. It makes of fate a human matter, which must be settled among men.

   All Sisyphus' silent joy is contained therein. His fate belongs to him. His rock is a thing Likewise,
   the absurd man, when he contemplates his torment, silences all the idols. In the universe suddenly
   restored to its silence, the myriad wondering little voices of the earth rise up. Unconscious, secret
   calls, invitations from all the faces, they are the necessary reverse and price of victory. There is no
   sun without shadow, and it is essential to know the night. The absurd man says yes and his efforts
   will henceforth be unceasing. If there is a personal fate, there is no higher destiny, or at least there is,
   but one which he concludes is inevitable and despicable. For the rest, he knows himself to be the
   master of his days. At that subtle moment when man glances backward over his life, Sisyphus
   returning toward his rock, in that slight pivoting he contemplates that series of unrelated actions
   which become his fate, created by him, combined under his memory's eye and soon sealed by his
   death. Thus, convinced of the wholly human origin of all that is human, a blind man eager to see who
   knows that the night has no end, he is still on the go. The rock is still rolling. 

   I leave Sisyphus at the foot of the mountain! One always finds one's burden again. But Sisyphus
   teaches the higher fidelity that negates the gods and raises rocks. He too concludes that all is well.
   This universe henceforth without a master seems to him neither sterile nor futile. Each atom of that
   stone, each mineral flake of that night filled mountain, in itself forms a world. The struggle itself
   toward the heights is enough to fill a man's heart. One must imagine Sisyphus happy.

\vspace*{2cm}

   ~~~~~~~~~~~~~~~~~~~~~~~~~~~~~~~~~~~~~~~~~~~~~~~~~~~~~~~~~~~~~~~~~~~~~~~~~~~~ Albert Camus

\vspace*{2cm}

\chapter{Radion in Multibrane World}
\label{radion}

In the treatment of gravitational perturbations, in the context of
of the braneworld models with flat branes, that is presented in the 
 Chapter \ref{RS} of this Thesis, we have ignored the presence
of the scalar perturbations, the dilaton and the radion(s)\footnote{The 
radion(s) appear in configurations with freely moving branes.},
that are associated with the size of the compact system   and the
fluctuations of the  freely moving 
branes respectively. In this Appendix we examine these the behaviour
of these fields and their relevance to the  phenomenology of these models. 
  
\section{The general three three-Brane system}   
 
\begin{figure}[t]  
\begin{center}  
\begin{picture}(300,200)(0,50)  
\SetOffset(0,-20)  
  
\SetWidth{2}  
\SetColor{Red}  
\Line(150,80)(150,250)  
  
\SetColor{Blue}  
\Line(240,80)(240,250)  
\Line(60,80)(60,250)  
  
\SetColor{Green}  
\Line(-30,80)(-30,250)  
\Line(330,80)(330,250)  
  
\SetWidth{0.5}  
\SetColor{Black}  
\Line(-30,160)(330,160)

\Text(150,60)[c]{$y=0$}  
\Text(60,60)[c]{$y=-r$}  
\Text(240,60)[c]{$y=r$}  
\Text(-30,60)[c]{$y=-L$}  
\Text(330,60)[c]{$y=L$}  
\Text(195,180)[c]{$\Lambda_1$}  
\Text(285,180)[c]{$\Lambda_2$}  
\Text(105,180)[c]{$\Lambda_1$}  
\Text(15,180)[c]{$\Lambda_2$}

\end{picture}  
\end{center}  
  
\caption{General three 3-brane model on an orbifold.}  
\label{model}  
  
\end{figure}
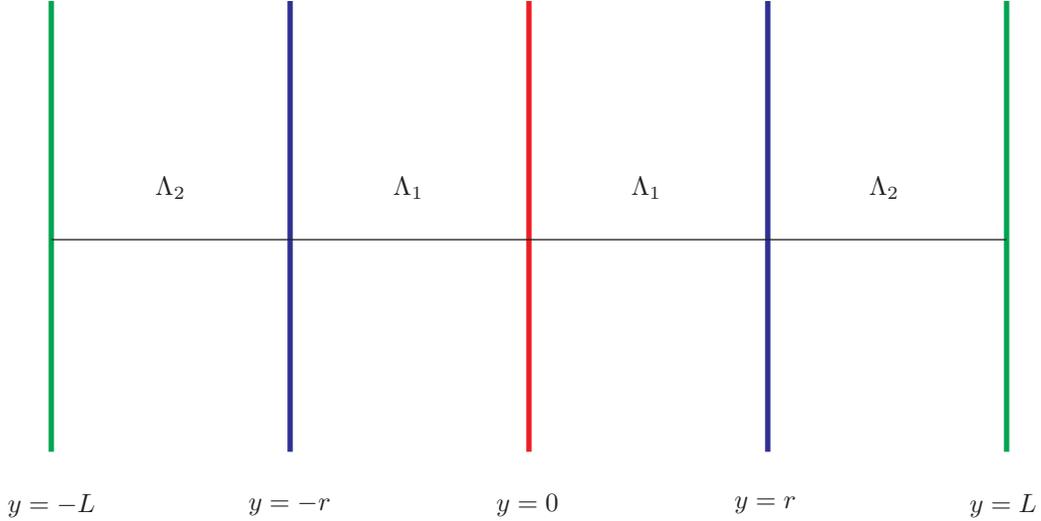

We will consider a general three three-brane model on an $S^1/Z_2$ 
orbifold\footnote{By taking various limits we will be able to
reproduce the physics of RS1, RS2, $''+-+''$ and GRS models.}.  
Two of the branes sit on the orbifold fixed points $y=y_0=0$, 
$y=y_2=L$ respectively. A  third brane is sandwiched in between at 
position $y=y_1=r$ as in Fig.(\ref{model}). In each region between the 
branes the space is $AdS_5$ and  in general the various $AdS_5$ regions  
have  different cosmological constants $\Lambda_1$, $\Lambda_2$. The 
action describing the above system is:  
 
\begin{equation} 
S = \int d^4xdy \sqrt{-G} \left[ 2 M_5^3 \, R \,- \, \Lambda(y)  
\, - \, \sum_i V_i \, \delta(y - y_i)  \,   
\frac{\sqrt{-\hat{g}}}{ \sqrt{-G}} \right] \quad.  
\label{act}  
\end{equation} 
where $M_5$ is the five dimensional Planck mass, $V_i$ the tensions of 
the gravitating branes and $\hat{g}_{\mu\nu}$ the induced metric on the  
branes. The orbifold symmetry $y \to -y $ is imposed.

We seek for a background static solution of Einstein equations for the 
following 4D Poincar\'{e} invariant metric ansatz: 
\begin{equation} 
ds^2 = a^2(y) \, \eta_{\mu \nu} dx^\mu dx^\nu \, + \, dy^2  \quad ;  
\label{back}  
\end{equation} 
 
The solution for the warp factor has the usual exponential form:   
\ba 
a(y) =  e^{-k_1 y } &  ,0 < y < r  \\[0.3cm]  
e^{-k_2 y + r(k_2 - k_1)} & ,r < y < L   
\label{sol}  
\ea 
where $k_1$ and $k_2$ are the curvatures of the bulk in the two 
regions and are related to the bulk cosmological constants as:  
\be  
k_1^2 = -\frac{\Lambda_1}{24 M_5^3} ~ , ~~   
k_2^2= -\frac{\Lambda_2}{24 M_5^3}   
\ee 
 
Moreover, the Einstein equations impose the following fine tuning 
between the brane tensions and the bulk cosmological constants: 
 
\begin{equation} 
V_0=24 M_5^3 \, k_1 \, , \qquad V_1=24 M_5^3 \frac{(k_2-k_1)}{ 2} \, , 
\qquad V_2=-24 M_5^3 \, k_2 \; . 
\end{equation}  
 
It is straightforward to recover some special models 
that have been considered in Chapter \ref{RS}. The RS1 model is obtained 
for $k_1=k_2$ where the intermediate brane is absent (zero tension).  
For $k_1>0$ and $k_2=-k_1$ we get the $''+-+''$ multigravity model 
. For $k_1>0$ and $k_2>k_1$ we 
get the $''++-''$ brane model. In 
the decompactification limit where $L \rightarrow \infty$ we get also 
two interesting models: For  $k_1>0$ and  $k_2=0$ we obtain the 
Gregory - Rubakov - Sybiriakov (GRS) model  and 
for  $k_1>0$ and $k_2=0>k_1$ the non-zero tension version 
\cite{Pilo:2000et} of the model considered in \cite{Lykken:2000nb}.

\section{Effective action}

Our purpose is to study fluctuations of the background (\ref{back}).  
The first important observation is that there exists a generalization of   
Gaussian normal coordinates such that in the perturbed geometry the embedding  
of branes is still  described by $y=0, \, y=r$ and $y=L$ (see for instance  
the  appendix of \cite{Pilo:2000et}).   
  
When analyzed from a 4D point view, in each region,  perturbations are of  
of three types.  
\begin{itemize}  
\item  
{\bf Spin two}:  
 
Tensor-like perturbation $h_{\mu\nu}(x,y)$ corresponding to massive   
(massless) 4D gravitons   
\begin{equation}  
ds^2 = a^2(y) \left[\eta_{\mu \nu} + h_{\mu \nu}(x,y) \right] \, dx^\mu dx^\nu  
+ dy^2 \quad .  
\end{equation}  
\item  
{\bf Spin zero: Dilaton}   
 
Scalar perturbation $f_1(x)$ corresponding to an overall rescaling of 
distances \cite{Charmousis:2000rg} 
\begin{equation}  
ds^2 = a^2(y) \left[1+ Q(y) f_1(x) \right] \eta_{\mu \nu} \, dx^\mu dx^\nu   
+ \left[1+ q(z) f_1(x) \right] dy^2 \quad .  
\end{equation}   
\item  
{\bf Spin zero: Radion}  
 
Scalar perturbation $f_2(x)$ corresponding to a fluctuating distance   
of the branes \cite{Pilo:2000et}  
\be 
ds^2 = a^2(y) \left[ \left(1+ B(y) f_2(x) \right) \eta_{\mu \nu} +    
2 \epsilon(y) ~ \de_\mu \de_\nu f_2(x) \right]  
dx^\mu dx^\nu   
+ \left[1+ 2 A(y) ~  f_2(x) \right] dy^2   
\ee   
\end{itemize}   
The generic perturbation can be written as   
\ba    
ds^2 =& a^2(y)& \left\{ \left[1+ \varphi_1(x,y) \right] \eta_{\mu \nu} + 2   
\epsilon(y) \, \de_\mu \de_\nu f_2(x) +   
h_{\mu \nu}(x,y) \right \} dx^\mu dx^\nu \,   \\  
& +& \left[1+  \varphi_2(x,y) \right] dy^2 \quad ;  
\label{pert}  
\ea  
where   
\ba    
\varphi_1(x,y)  =  Q(y) f_1(x) + B(y) f_2(x)\\  
\varphi_2(x,y)  = q(y) f_1(x) + 2A(y) f_2(x)       
\label{ans}  
\ea 
Given the expression (\ref{sol}) for $a$, Israel junctions conditions at $y=0,  
\, L$, simply require that $A, \, B, \,  \de_y \epsilon, \, Q, \, q$ are  
continuous there \cite{Pilo:2000et}.   
The 4D effective action $S_{eff}$ for the various modes is obtained inserting   
the ansatz (\ref{ans}) in the action (\ref{act}) and integrating out $y$.   
So far the functions $A, \, B, \, \epsilon, \, Q, \,   
q$ haven't been specified, however imposing that $S_{eff}$ contains no   
mixing terms among $h$ and  $f_i$ one determines $Q, \, q$ and $A$ is  
expressed in terms of $B$ (for more details, see the last Section of this Appendix ) which satisfies  
\ba  
\frac{d}{dy} \left(B a^2 \right) + 2 a^{-1} ~ \frac{da}{dy} ~ \frac{d}{dy}   
\left(a^4 \de_y \epsilon \right) = 0 \quad ; \label{bdiff} \\  
 \int_{- L}^{L} dy ~ a \left(\frac{da}{dy} \right)^{-1} ~   
\frac{dB}{dy} = 0 
 \label{nomix}  
\ea  
As a consequence of the no-mixing conditions the linearized Einstein equations  
for (\ref{ans}) will consist in a set independent equations for the graviton  
and the scalars.  
  
The effective Lagrangian reads  
\ba    
S_{eff} &= \int d^4 x  \, {\cal L}_{eff} =  \int d^4 x  \, \left(  
{\cal L}_{Grav} + {\cal L}_{Scal} \right) \\  
{\cal L}_{eff} &=  2 M_5^3 \int_{- L}^{L} dy  \, \Big \{a^2    
{\cal L}_{PF}(h) +   
\frac{a^4}{4} \left[(\de_y h)^2 - \de_y h_{\mu \nu} \, \de_y h^{\mu \nu}   
\right] + {\cal L}_{Scal}\Big \} \quad    
\label{eff1}  
\ea
with  
\ba  
{\cal L}_{Scal} &= {\cal K}_1 ~ f_1 \Box f_1 +  {\cal K}_2 ~ f_2 \Box f_2 \\  
 {\cal K}_1 =  2 M^3_5 \frac{3}{2} c^2\int_{- L}^{L} a^{-2}   
~ dy ~ ,& \quad   
{\cal K}_2 = - 2 M^3_5 ~ \frac{3}{4} \int_{- L}^{L} a   
\left(\frac{da}{dy} \right)^{-1} ~ \frac{d}{dy} \left( B^2 a^2 \right) ~ dy  
\label{kt}  
\ea  
In (\ref{eff1}), the spin-2 part,  as expected,  contains the 4D   
Pauli-Fierz Lagrangian ${\cal L}_{PF}(h)$ for the graviton  plus a mass term  
coming from the dimensional reduction. In the scalar part ${\cal L}_{Scal}$  
the mass terms are zero since $f_i$ are moduli fields. Notice that after the  
no-mix conditions are enforced, the effective Lagrangian  
contains the undetermined function $\epsilon$.  
 
The metric ansatz $G_{{}_{MN}}$ in (\ref{ans}) is related to a special  
coordinate choice (generalized Gaussian normal), nevertheless a residual  
gauge (coordinate) invariance is still present. Consider the class of  
infinitesimal coordinate transformations  $X^M \to {X^\prime}^M= X^M +  
\xi(X)^M$ such that  
the transformed metric ${G_{{}_{MN}}}^\prime = G_{{}_{MN}} + \delta  
G_{{}_{MN}}$ retains the original form (\ref{pert}) up to a  
redefinition of the functions $q,Q,A,B, \epsilon$ and the dilaton 
and the radion field. Consistency with the orbifold geometry and the  
requirement 
that the brane in $y=r$ is kept fixed by the diffeomorphism lead to $ 
\xi^5(x,0) = \xi^5(x,r) = \xi^5(x,L) =0 $. From  
\be 
\delta G_{{}_{MN}} = - \xi^A \, \de_A  G_{{}_{MN}} - \de_{{}_M} \xi^A \,  
 G_{{}_{AN}} - \de_{{}_N} \xi^A \, G_{{}_{MA}} \;, 
\ee 
one can show that $\xi^M$ has to be of the form 
\ba 
\xi^\mu(x,y) = \hat{\xi}^\mu(x)  - a^{-2} W(y) \, \eta^{\mu \nu} \,  
\de_\nu f_2(x) 
\, , \qquad \xi^5(x,y) = W^\prime(y) ~ f_2(x)  
\ea 
with $ W^\prime(0) =  W^\prime(r) =  W^\prime(L) = 0$. 
The case $W = 0$ corresponds the familiar pure 4D diffeormorphisms, under which $h_{\mu \nu}$  
transforms as spin two field, $f_i$ as scalars and $q,Q,A,B, \epsilon$ are 
left unchanged. On the contrary the case $\hat{\xi}^\mu = 0, \; W \neq 0 $ is 
relic of 5D diffeormorphisms and one can check that $q,Q,A,B, \epsilon$ are not 
invariant and in particular $\delta \epsilon = W$.   
As a result, the values of $\de_y \epsilon$ in  
$0$, $r$ and $L$ are gauge independent and this renders   
${\cal L}_{eff}$ free from any gauge ambiguity.

\section{Scalars Kinetic Energy}  
  
\subsection{The compact case}  
In this Section we will focus on the part of the effective Lagrangian 
involving the scalars and concentrate on the dilaton and radion kinetic 
term coefficients  ${\cal K}_1$ and ${\cal K}_2$. In particular we are  
interested in the cases when the radion becomes a ghost field, 
\textit{i.e.} ${\cal K}_2<0$. Firstly, it is trivial to obtain the dilaton kinetic term ${\cal K}_1$ by integrating (\ref{kt}):  
  
\begin{equation}  
{\cal 
K}_1=3c^2M_5^3\left[\frac{a^{-2}(r)-1}{k_1}+\frac{a^{-2}(L)-a^{-2}(r)}{k_2}\right]  
\label{K1} 
\end{equation}

\begin{figure}[t]  
\begin{center}  
\begin{picture}(200,200)(0,0)  
\LongArrow(0,100)(200,100)   
\LongArrow(100,0)(100,200)  
\SetColor{Blue}  
\Text(110,190)[lb]{$k_2$}  
\Text(195,80)[lb]{$k_1$}  
\DashLine(0,0)(200,200){10}  
\Text(190,185)[lt]{\Blue{$k_1=k_2$}}  
\Text(50,140)[rb]{\color{red}${\cal K}_2 >0$}  
\Text(150,40)[lt]{\color{red}${\cal K}_2 <0$}  
\SetColor{Red}  
\Text(155,80)[lt]{\color{green} GRS}  
\SetColor{Green} 
\Line(100,100)(195,100) 
\LongArrow(150,80)(140,95)  
\SetColor{Black}  
\Vertex(100,100){2} 
\Text(115,140)[lb]{$A$}  
\Text(85,60)[rt]{$A'$} 
\Text(140,110)[lb]{$B$} 
\Text(60,90)[rt]{$B'$} 
\Text(125,80)[lt]{$C$} 
\Text(75,120)[rb]{$C'$} 
\end{picture}  
\end{center}  
\caption{Sign of the radion kinetic term in the $(k_1,k_2)$ plane. In 
the regions $A$, $B'$, $C'$ the moving brane has positive tension 
and the radion positive kinetic energy. In the regions $B$, $C$, $A'$ 
the moving brane has negative tension and the radion  negative kinetic 
energy. We show the GRS line for the non-compact case. The dashed line  
corresponds to $k_1=k_2$, \textit{i.e.} a \textit{tensionless} moving brane.}  
\label{phase}  
\end{figure}
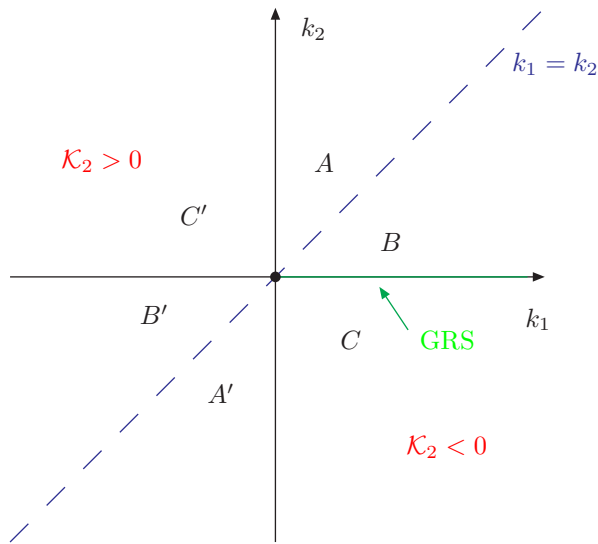

It turns out that for any possible values of $k_1$, $k_2$ and $r$, $L$  
the above quantity is positive definite. The radion kinetic term on 
the other hand is more involved.  
Integrating (\ref{bdiff}) we  get the  radion wavefunction for the  
regions ($y>0$):  
\ba  
B(y)= c_1 \, a^{-2} + 2 k_1 ~ \de_y \epsilon ~ a^2  &  0 < y < r \\[0.3cm]   
c_2 ~ a^{-2} + 2 k_2 ~ \de_y \epsilon ~ a^2  &  y > r      
\label{B}   
\ea  
where $c_1$ and $c_2$ are integration constants. The orbifold boundary   
conditions demand that $\de_y \epsilon(0) =\de_y \epsilon(L)= 0$ since  
$\epsilon$ is an even function of $y$. From the non-mixing conditions  
for radion and dilaton (\ref{nomix}) and the  
continuity of $B$ we are able to determine $c_2$ and $\de_y  
\epsilon(r)$ as the following:  
\ba   
c_2=c_1 \frac{k_2}{k_1}~\frac{a^2(r)-1}{\left(\frac{a(r)}{a(L)}\right)^2-1}\\   
\epsilon^\prime(r)=\frac{c_1k_2}{2k_1(k_2-k_1)a^4(r)}\left[\frac{k_1}{k_2}- 
\frac{a^2(r)-1}{\left(\frac{a(r)}{ a(L)}\right)^2-1}\right]   
\ea   
Therefore, the values of the radion  wavefunction $B$ at the branes positions 
are given by the following expressions:  
\ba 
B(0)=c_1 ~ \\  
B(r)=\frac{c_1k_2}{(k_2-k_1)} \frac{1-a^2(L)}{a^2(L)\left[\left( 
\frac{a(r)}{ a(L)}\right)^2-1\right]} ~\\  
B(L)=\frac{c_1k_2}{k_1}\frac{a^2(r)-1}{a^2(L)\left[\left(\frac{a(r)}{  
a(L)}\right)^2-1\right]}   
\ea  
Thus we can carry out the integral in   
(\ref{kt}) to find  the radion kinetic term coefficient:  
\ba   
{\cal K}_2&= 3 M_5^3  \left[\left(\frac{1}{  k_1} - \frac{1}{ k_2} \right)  
 B^2(r) \, a^2(r) + \frac{1}{   
k_2} \, B^2(L) \, a^2(L) - \frac{1}{ k_1} \, B^2(0)\right] \nonumber \\  
&= \frac{3 M_5^3c_1^2}{ k_1}  \left\{ \frac{k_2}{  
(k_2-k_1)}\frac{a^2(r)(a^2(L)-1)^2}{a^4(L)\left[\left(\frac{a(r)}{ 
a(L)}\right)^2-1\right]^2}+\frac{k_2}{  
k_1}\frac{(a^2(r)-1)^2}{a^2(L)\left[\left(\frac{a(r)}{  
a(L)}\right)^2-1\right]^2}-1 \right\} \, .  
\label{K2} 
\ea  
The above quantity is not positive definite. In particular, it turns out 
that it is positive whenever the intermediate brane has positive 
tension and  negative whenever the intermediate brane has negative 
tension. This result is graphically represented in Fig.(\ref{phase}) 
where the $(k_1,k_2)$ plane is divided in two regions.

\subsection{The non-compact limit} 
It is instructive to discuss the decompactification limit in which the third brane is  sent  to infinity, \textit{i.e.}  
$L \to + \infty$. To examine this limit we distinguish two cases:  
\vskip 0.5truecm  
\centerline{\bf The case $\mathbf{k_2 > 0}$}  
\vskip 0.5truecm  
\noindent  
In this case we have $\displaystyle{a(\infty) \propto 
\lim_{L\rightarrow \infty} e^{-k_2L} =0}$. The dilaton 
kinetic term is trivial since from (\ref{K1}) we obtain ${\cal K}_1 \to \infty$. In other words 
the dilaton decouples  from the 4D effective theory and the condition 
of absence of kinetic mixing between the scalars (\ref{nomix}) plays 
no role. The radion kinetic term coefficient can be read off from (\ref{K2}):   
\begin{equation}  
{\cal K}_2= \frac{3 M_5^3  c_1^2}{k_1} \left[e^{2k_1r} \frac{k_2}  
{(k_2-k_1)}-1  
\right]  
\end{equation}  
This result agrees with the computation of \cite{Pilo:2000et}. It is 
easy to see that the radion has positive kinetic energy when the moving 
brane has positive tension and negative kinetic energy when the 
tension is negative. Indeed, for $0 < k_1 < k_2$, or for $k_2 > 0$ and 
$k_1 < 0$ we have a 
positive brane and positive kinetic energy. On the other hand for $0 < 
k_2 < k_1$  we have a negative tension brane 
and negative kinetic energy. In the limit $k_2 \rightarrow 0$ we get the GRS  
model with  negative kinetic energy as in \cite{Pilo:2000et}. 
\vskip 0.5truecm  
\centerline{\bf The case $\mathbf{k_2 < 0}$}  
\vskip 0.5truecm  
\noindent  
In this case $\displaystyle{a(\infty) \propto 
\lim_{L\rightarrow \infty} e^{-k_2L} \rightarrow \infty}$. This time 
the dilaton plays in the game since ${\cal K}_1$ is finite as seen 
from (\ref{K1}) and has the value: 
\begin{equation}  
{\cal K}_1 = \frac{3 M_5^3 \, c^2 }{2 k_1 |k_2|}  \,   
\left[ e^{2k_1r} (|k_2|+k_1)  - |k_2| \right]  
 \quad   
\end{equation}  
which is manifestly positive definite. The presence of dilaton mode  
is a bit surprising since it describes the fluctuations of the overall size of  
the system which is infinite. Something similar happens in the  
model of \cite{Karch:2001ct} that has a dilaton mode although it is 
non-compact. The dilaton is a remnant of the decompactification 
process of the $''++''$ model \cite{Kogan:2001vb}  and enters in the game 
because the inverse of the warp factor is normalizable. 
 
The radion kinetic term 
coefficient can be obtained from  taking the limit $L \to  
+ \infty$ in (\ref{K2}). We get 
\begin{equation}  
{\cal K}_2 = \frac{3 M_5^3 \,  c^2_1}{k_1} \left[e^{-2 k_1 r} \, \frac{k_2}  
{(k_2-k_1)} - 1 \right]  
\end{equation}  
The same considerations for the compact case applies here. When $k_1 < k_2 < 0$ we have a 
positive tension brane and a positive kinetic energy. On the other hand when  
$k_2 < k_1 < 0$, or for $k_2 < 0$ and $k_1 > 0$ we have a negative tension  
brane and a negative kinetic energy. In the GRS limit $k_2 \rightarrow 0$,  
the radion has negative kinetic energy. 
 
\subsection{$''+-+''$ Model}

From the above it is clear that in the case of the $''+-+''$ model,
\textit{i.e.} for $k_{2}=-k_{1}$ with $k_{1}>0$, the dilaton has
positive kinetic term (${\cal K}_1$) whereas the radion negative
(${\cal K}_2$). The latter means that there is a physical ghost in the 
spectrum something that makes this model unacceptable. The appearance
of this ghost state in the spectrum is not a characteristic only of the
present model but it is a generic characteristic of models that include
moving negative tension branes.

\vskip0.8cm   
\noindent   
\centerline{\Large \bf Usefull Formulae}   
\vskip0.4cm   
\noindent   
It is convenient to define a new variable $z$ defined by   
\begin{equation}   
\frac{1}{a(y)} = \frac{dz}{dy} \quad .   
\end{equation}   
In the coordinates $(x,z)$ the metric (\ref{pert}) is conformal to a flat    
perturbed metric $\bar{G}_{AB}=   
\eta_{AB} + H_{AB}$   
\ba      
ds^2 &= a^2 \left[\bar{G}_{\mu \nu}  \, dx^\mu dx^\nu + \bar{G}_{zz} \, dz^2 \right] =    
a^2 \left[\eta_{MN} + H_{MN}  \, dx^M dx^N \right] \nonumber \\   
H_{\mu \nu} &= \varphi_1(x,z)  \,  \eta_{\mu \nu} + 2 \epsilon(z) \,    
\de_\mu \de_\nu f_2(x) +    
h_{\mu \nu}(x,z) \nonumber \\   
H_{zz} &= \varphi_2(x,z)     
\label{pert1}   
\ea   
Inserting (\ref{pert1}) in (\ref{act}) and taking into account the equation   
of motion satisfied by $a$ one gets   
\ba      
S_{eff}& = \int d^4x \; {\cal L}_{eff} =  \int d^4x  \; 2 M_5^3\int dz     
\left[a^3  {\cal L}_{PF}(h) + \frac{a^3}{4}    
\left[(\de_z h)^2 - \de_z h_{\mu \nu} \, \de_z h^{\mu \nu}\right]+ {\cal L}_{\varphi} +    
{\cal L}_{h \varphi}    
\right] \quad . \nonumber \\   
 {\cal L}_{\varphi} &={\cal L}_1  + {\cal L}_2 +  {\cal L}_{12} \quad ;      
\ea   
Where   
\ba       
{\cal L}_{h \varphi} &= \left[a^3 \left(\varphi_1 + \frac{1}{2} \varphi_2    
\right)+ f_2 \frac{d}{dz}    
\left(\epsilon^\prime a^3 \right)    
\right] \partial_\mu \partial_\nu h^{\mu \nu} \nonumber \\   
&- \left[a^3 \left(\varphi_1 +   
 \frac{1}{2}\varphi_2 \right) + f_2     
\frac{d}{dz} \left(\epsilon^\prime a^3 \right) \right] \Box h    
+\frac{3}{2} \frac{d}{dz} \left(a^2 a^\prime \varphi_2 - a^3    
\varphi_1^\prime \right) h \quad ;       
\ea   
\begin{equation}   
{\cal L}_1 = \de_\mu f_1 \de^\mu f_1 \, \frac{3}{2} a^3 \left(Q^2 + Q q \right)   
+ f_1^2 \left(3 a^3 {Q^\prime}^2 + 3 {a^\prime}^2 a q^2 - 6 a^2 a^\prime    
Q^\prime q \right) \; ; 
\end{equation}   
\ba      
{\cal L}_2 &= \de_\mu f_2 \de^\mu f_2 \left(\frac{3}{2} a^3 B^2 + 3 a^3 AB   
-3 a^3 B^\prime \epsilon^\prime + 6 a^2 a^\prime \epsilon^\prime A
\right) \nonumber \\   
&+ f_2^2 \left(3 a^3 { B^\prime}^2 + 12 {a^\prime}^2 a A^2 - 12 a^2 a^\prime   
A B^\prime \right) \; ;     
\ea   
\ba      
{\cal L}_{12} &= \de_\mu f_1 \de^\mu f_2 \left(\frac{3}{2} a^3 B q + 3a^3 A Q   
+3a^3B Q - 3 a^3 \epsilon^\prime Q^\prime + 3 a^2 a^\prime
\epsilon^\prime q \right) \nonumber \\   
&+ f_1 f_2 \left[6 a^3  B^\prime Q^\prime + 12 {a^\prime}^2 a A q - 6 a^2   
a^\prime \left(2 A Q^\prime + q  B^\prime \right) \right] \; ;      
\ea   
and ${\cal L}_{PF}(h)$ is the 4D Pauli-Fierz Lagrangian for $h$   
\begin{equation}   
{\cal L}_{PF}(h) = \frac{1}{2} \de_\nu h_{\mu \alpha} \, \de^\alpha    
h^{\mu \nu} -    
\frac{1}{4} \de_\mu h_{\alpha \beta} \, \de^\mu h^{\alpha \beta}    
- \frac{1}{2} \de_\alpha h \, \de_\beta h^{\alpha \beta} +\frac{1}{4}    
\de_\alpha h \, \de^\alpha h \, .   
\end{equation}   
Differentiation with respect of $z$ is denoted with a prime. The absence of mixing   
terms in ${\cal L}_{eff}$ yields the following constraints   
\ba   
&&A(z) = \frac{a B^\prime}{2 a^\prime} \, ,  \frac{d}{dz}    
\left(B a^2 \right) + \frac{2 a^\prime}{a^2} \, \frac{d}{dz} \left(a^3    
\epsilon^\prime \right) = 0 \;  \label{cc1}\\   
&& Q(z) = c \, a^{-2} \, , q(z) = -2 c \, a^{-2} 
 \; ; \label{cc2}\\   
&& \int dz \, a A(z) = 0  . \label{cc3}   
\ea   
(c  is a  constant)Eqns. (\ref{cc1})-(\ref{cc3}) give   
\ba      
{\cal L}_{eff} &=  2 M_5^3 \int dz  \Big \{a^3  {\cal L}_{PF}(h) +    
\frac{a^3}{4} \left[(\de_z h)^2 - \de_zh_{\mu \nu} \, \de_zh^{\mu \nu}    
\right] \nonumber \\   
&- \frac{3}{2}c^2  a^{-1} \de_\mu f_1 \de^\mu f_1 + \frac{3}{4}    
\frac{a^2}{a^\prime} \, \frac{d}{dz} \left(B^2 a^2 \right)  \, \de_\mu f_2   
\de^\mu f_2 \Big \} \quad .      
\label{eff}   
\ea   
In particular the effective Lagrangian ${\cal L}_{Scal}$ for the dilaton    
$f_1$ and the radion   
$f_2$  is   
\ba      
{\cal L}_{Scal} &= {\cal K}_1 \, f_1 \Box f_1 +  {\cal K}_2 \, f_2
\Box f_2 \nonumber \\   
& {\cal K}_1 =  2 M^3_5 \, \frac{3}{2}c^2 \int_{- L}^{L} a^{-2}    
\, dy \nonumber \\   
& {\cal K}_2 = - 2 M^3_5 \, \frac{3}{4} \int_{- L}^{L} a    
\left(\frac{da}{dy} \right)^{-1} \, \frac{d}{dy} \left( B^2 a^2 \right) \, dy   
\quad ,      
\ea   
with   
\ba      
&\frac{d}{dy} \left(B a^2 \right) + 2 a^{-1} \, \frac{da}{dy} \, \frac{d}{dy}    
\left(a^4 \de_y \epsilon \right) = 0 \quad ; \nonumber \\   
& \int_{- L}^{L} dy \; a \left(\frac{da}{dy} \right)^{-1} \,    
\frac{dB}{dy} = 0 \quad .      
\ea   
As a result only $\de_y\epsilon(0)$, $\de_y\epsilon(r)$ and 
$\de_y\epsilon(L)$ enter the radion effective action.


\vspace*{3cm}

\chapter{Dynamical Generation of Branes} 
\label{dynamical}

In the context of the RS type models the 3-branes are placed in the
Lagrangian by hand, as $\delta$-function source terms. However it is
interesting to ask if it possible to generate all brane configurations
in a dynamical way. It turns out that it is possible to reproduce
all the models that do not contain negative moving branes by
considering gravity coupled to a scalar field that acquires a (multi-)
kink type vev. The fact that no negative tension branes can be
constructed in this dynamical way is related to the violation of the weaker
energy condition by them. 

In this Appendix we will try to reproduce the familiar 
RS2  construction as a limiting of  a smooth configuration (no
$\delta$-function sources).
We start with the usual ansatz for the metric:  
\be
ds^{2}=e^{-2 \sigma(y)} g_{\mu \nu}(x^{kappa}) dx^{\mu} dx^{\nu}+dy^2
\ee
where $g_{\mu \nu}$ is the induced metric on the brane.
The usual brane term associated with the above metric has been replaced by
a scalar field $\phi$ residing in this space coupled though the
standard action:
\be
S=\int d^4 x \int_{-L_2}^{L_2} dy \sqrt{-G} 
\{2 M^3 R-\frac{1}{2}(\partial \phi)^{2}-V(\phi) \}
\ee
The corresponding equations of motion are: 
\ba
R_{MN}-\frac{1}{2}G_{MN}R=\frac{1}{4M^3} \{ \partial_{M} \phi
\partial_{N} \phi - G_{MN} \left(\frac{1}{2}(\partial \phi)^2+V(\phi)
\right) \} \\
\frac{1}{\sqrt{-G}}\partial_{M}\{\sqrt{-G} G^{MN} \partial_{N} \phi
\}=\frac{\partial V}{\partial \phi}
\ea
Our purpose is to find a vacuum solution for the scalar field (by
choosing appropriate potential $V(\phi)$) that at a certain limit
gives the RS solution \textit{i.e.} $\sigma(y) \rightarrow ky$.

\section{Flat Branes}

In the case that $g_{\mu \nu}= \eta_{\mu \nu}$, \textit{i.e.} when the
induced spacetime on the branes is flat, the equations of motion 
result to:
\ba
\frac{1}{2} (\phi')^{2}-V(\phi)=24M^3 ~(\sigma')^{2} \\
\frac{1}{2} (\phi')^{2}+V(\phi)=- 12M^3 \left( - (\sigma'')+
2(\sigma')^{2} \right)
\ea
by  adding the previous equations we have:
\be
 \sigma''=\frac{1}{12 M^3 } (\phi')^2
\ee
the previous implies that $\sigma'' \ge 0 $ - something that reminds us
the constraints that the weaker energy condition imposes to the
behaviour of the warp factor $\sigma$ in the case of the original RS
model. From the previous condition it is clear that one cannot
generate a warp factor with a bounce form in the case that the branes
are kept flat. Moreover it implies that moving negative tension branes
(not on an orbifold fixed point) cannot be generated by this mechanism 
(by the use of a scalar field). The latter agrees with the constraints
coming from the weaker energy conditions, since moving negative
tension branes violates them. 

However it is straitforward to generate positive tension branes. Following Ref.\cite{Kehagias:2000au}
it turns out that this can be achieved by  demanding that $V(\phi)$ is such that the vacuum solution of
the scalar field is: 
\be
\phi_{B}(y)=\upsilon \tanh (\alpha y)
\ee
in this case one can find that
\be
\sigma(y)=\beta \ln \cosh^{2}(\alpha y) + \frac{\beta}{2} \tanh^{2}(\alpha y)
\ee
where $\beta \equiv \frac{ \upsilon^2 }{36 M^3}$
which gives the RS limit when $\alpha \rightarrow \infty$. This can 
be seen be finding that
\be
\lim_{a \rightarrow \infty} \{e^{-2 \sigma(y)} \}= e^{-4\alpha
\beta y}+ \dots 
\ee
idendifing $k \equiv 2 \alpha \beta$ (keeping the product $\alpha
\beta$ finite)

Due to the fact that negative tension branes cannot be generated by
the previous mechanism, it is not possible to construct the $''+-+''$
model by the previous method (naively one would expect that a
multi-kink profile for the field $\phi$ would result to the that
configuration - but this is not the case)\footnote{However, one can
recover the RS1 limit (by compactification) since in this case the negative tension brane is 
on a fixed point.}. 
However, one can reproduce the $''++''$ configuration
(where the spacetime on the branes is $AdS_{4}$).

\section{$AdS_{4}$ Branes}

In the case that we allow for $AdS_{4}$ spacetime on the branes \textit{i.e.}: 
\be
g_{\mu \nu}= \frac{\eta_{\mu \nu}}{\left(1-\frac{H^2 x^2}{4} \right)^{2}}
\ee
where $x^{2}=\eta_{\mu \nu} x^{\mu} x^{\nu}$
the equations of motion give:
\ba
\frac{1}{2} (\phi')^{2}-V(\phi)=24M^3~\left( (\sigma')^{2} + H^2 e^{2 \sigma} \right) \\
\frac{1}{2} (\phi')^{2}+V(\phi)=-12M^3~\left(
-(\sigma'')+2~(\sigma')^{2}+H^2~e^{2 \sigma} \right)
\ea
The addition of the two previous equations give:
\be
\sigma''=\frac{1}{12M^3}(\phi')^{2}-H^{2} e^{2 \sigma}
\ee 
from the previous equation it is clear that in the case that we allow
for $AdS_{4}$ spacetime on the branes $\sigma''$ can take both signs,
giving the possibility of a warp  factor of a bounce form. 

In more detail if one demands to generate the $''++''$ configuration
with 
\be
\sigma(y)=-\ln (A(y))=-\ln (\frac{\cosh(k(y_{0}-|y|))}{\cosh(ky_{0})})
\ee
then the scalar field should acquire a vev of the form
\be
\phi(y) \propto \arctan(\tanh (\frac{k(y_{0}-|y|)}{2}))
\ee
which means in the smooth form it should have a multi-kink profile.
The latter can justify two things:

First in the case of bulk fermions it justifies the choice of the
profile of the scalar field: $\phi(y)=-\frac{A'(y)}{A(y)}$, since this 
approximates reasonably the exact form and at the same time allows us
to have analytic results for the mass for the wavefunctions and the
mass spectrum

Second it justifies the choice of the mass term in the case of the
$''+-+''$ model. In this case although the model cannot be assumed to
be generated by a scalar field, it has been proved (see Chapter \ref{neutrino}) that its
phenomenology resembles the one of the $''++''$ model which can be
generated by a scalar field with a multi kink profile. The latter
implies that the phenomenology of the bulk fermion with a multi-kink
mass term of the form $m(y) \bar{\Psi} \Psi= \frac{\sigma'(y)}{k}$
will resemble the phenomenology of the fermion in the $''++''$
model.


\vspace*{3cm}

\chapter{Multi-Localization}  
\label{multilocapp}

\section{Wavefunction Solutions} 
 
In this Appendix we briefly discuss the general form of the solutions of the Schr\"{o}ndiger equation of the $^{\prime \prime }+-+^{\prime \prime }$ 
configuration - a configuration that exhibits multi-localization. As we have 
seen in Chapter \ref{multiloc} sections the general form of the potential (for field 
of any spin) of the corresponding quantum mechanical problem is of the type  
\footnote{In the following expression we have assumed that boundary mass term 
contribution is universal (its absolute value) for all the branes, \textit{i.e.} $\lambda $ is the weight of all $\delta $-functions. One might consider a 
more general case, where each $\delta $-function to have its own weight. 
However, generally the requirement of the existence of zero mode implies 
that their absolute values are equal.}:  
\begin{equation} 
\hspace*{0.5cm}V(z)=\frac{\kappa }{2[g(z)]^{2}}(\sigma ^{\prime }(y))^{2}-%
\frac{\lambda }{2[g(z)]^{2}}\sigma ^{\prime \prime }(y) 
\label{multi75}
\end{equation} 
where $\kappa $, $\lambda $ are constant parameters\footnote{%
The particular values of the parameters $\kappa $, $\lambda $ depend on the 
spin of the particle. Here we are interested in the general forms of the 
solutions. We also assume that we have already performed the appropriate 
redefinition of the wavefunction ($f(y)\rightarrow \hat{f}(z)$), which also 
depends on the spin of the particle.}. For the case of $^{\prime \prime 
}+-+^{\prime \prime }$ model we have   
\begin{eqnarray} 
(\sigma ^{\prime }(y))^{2} &=&k^{2}  \nonumber \\ 
\sigma ^{\prime \prime }(y) &=&2kg(z)\left[ \delta (z)+\delta 
(z-z_{2})-\delta (z-z_{1})\right]  
\label{multi76}
\end{eqnarray} 
$z_{1}$ and $z_{2}$ are the position of the second (negative) and the third 
(positive) brane respectively in the new coordinates ($z_{1}=z(L_{1})$ and $%
z_{2}=z(L_{2})$). The convenient choice of variables, which is universal for 
fields of all spins, is:  
\begin{equation} 
\renewcommand{\arraystretch}{1.5}z\equiv \left\{  
\begin{array}{cl} 
\frac{2e^{kL_{1}}-e^{2kL_{1}-ky}-1}{k} & y\in \lbrack L_{1},L_{2}] \\  
\frac{e^{ky}-1}{k} & y\in \lbrack 0,L_{1}] 
\end{array} 
\right. \  
\end{equation} 
(the new variable is chosen to satisfy $\frac{dz}{dy}=e^{\sigma (y)}$) and 
the function $g(z)$ is defined for convenience as $g(z)\equiv e^{\sigma 
(y)}\equiv k\left\{ z_{1}-\left| |z|-z_{1}\right| \right\} +1$. Note that in 
principle $\kappa $ and $\lambda $ are not related since the first gets 
contributions from the five dimensional bulk mass whereas the second from 
the boundary mass term. 
 
\paragraph{Even fields} 
 
Let us consider first the case that the field under consideration is even 
under the reflections $y \rightarrow -y$. In this case, it is easy to show 
that the zero mode exists only in the case that $\kappa=\nu(\nu+1)$ and $%
\lambda=\nu$ or $\lambda=-(\nu+1)$ (this is derived by imposing the boundary 
conditions coming from the $\delta$-function potentials on the massless 
solution, see below). The zero mode wavefunctions in this case have the form:%
\newline 
In the case that $\lambda=\nu$,  
\begin{equation} 
\hat{f}_{0}(z)=\frac{A}{[g(z)]^{\nu}} 
\end{equation} 
and in the case that $\lambda=-(\nu+1)$,  
\begin{equation} 
\hat{f}_{0}(z)=A^{\prime}~ [g(z)]^{\nu+1} 
\end{equation} 
Where $A$,$A^{\prime}$ are normalization constants. Note that the first is 
localized on positive tension branes whereas the second on negative tension 
branes. However, only the first choice gives the possibility of multi- 
localization on positive tension branes. Thus the existence of zero mode and 
light KK state requires $\kappa=\nu(\nu+1)$ and $\lambda=\nu$. Since we are 
interested in configurations that give rise to light KK states, the 
potential of interest is:  
\begin{equation} 
\hspace*{0.5cm} V(z)=\frac{\nu(\nu+1) k^2}{2[g(z)]^2}- \frac{\nu}{2g(z)} 2k %
\left[\delta(z)+\delta(z-z_2)-\delta(z-z_1)\right] 
\label{multi80}
\end{equation} 
For the KK modes ($m_{n} \ne 0$) the solution is given in terms of Bessel 
functions. For $y$ lying in the regions $\mathbf{A}\equiv\left[0,L_1\right]$ 
and $\mathbf{B}\equiv\left[L_1,L_2\right]$, we have:  
\begin{equation} 
{\hat{f}}_{n}\left\{ 
\begin{array}{cc} 
\mathbf{A} &  \\  
\mathbf{B} &  
\end{array} 
\right\}=\sqrt{\frac{g(z)}{k}}\left[\left\{ 
\begin{array}{cc} 
A_1 &  \\  
B_ 1 &  
\end{array} 
\right\}J_{\frac{1}{2}+\nu}\left(\frac{m_n}{k}g(z)\right)+\left\{ 
\begin{array}{cc} 
A_ 2 &  \\  
B_2 &  
\end{array} 
\right\}J_{-\frac{1}{2}-\nu}\left(\frac{m_n}{k}g(z)\right)\right] 
\end{equation} 
The boundary conditions that the wavefunctions must obey are:  
\begin{eqnarray} 
{{{\hat{f}}}_{n}}~^{\prime}({0}^{+})+\frac{k \nu}{g(0)}{{\hat{f}}}_{n}(0)=0  
\nonumber \\ 
{{\hat f}_{n}}({z_{1}}^{+})-{{\hat f}_{n}}({z_{1}}^{-})=0  \nonumber \\ 
{{\hat f}_{n}}~^{\prime}({z_{1}}^{+})-{{\hat f}_{n}}~^{\prime}({z_{1}}^{-})-%
\frac{2 k \nu}{g(z_{1})}{\hat f}_{n}(z_{1})=0  \nonumber \\ 
{{\hat f}_{n}}~^{\prime}({z_{2}}^{-})-\frac{k \nu}{g(z_{2})}{\hat f}%
_{n}(z_{2})=0 
\end{eqnarray} 
The above boundary conditions give a 4~x~4 linear homogeneous system for $%
A_{1}$, $B_{1}$, $A_{2}$ and $B_{2}$, which, in order to have a nontrivial 
solution should have vanishing determinant. This imposes a quantization 
condition from which we are able to extract the mass spectrum of the bulk 
field. 
 
\paragraph{Odd fields} 
 
In the case that the field is odd under the reflections $y \rightarrow -y$, 
there is no zero mode solution since it is not possible to make it's 
wavefunction to vanish at both boundaries. In the absence of zero mode, the 
previous restrictions between $\lambda$ and $\nu$ do not apply - and thus 
they are in principle independent. However one can ask if in this case, 
despite the absence of zero mode, a light state can exist. Indeed we can 
easily find that this can be realized for special choice of parameters: It 
can be shown that the potential  
\begin{equation} 
\hspace*{0.5cm} V_{1}(z)=\frac{\nu(\nu-1)}{2[g(z)]^2}(\sigma^{%
\prime}(y))^{2}+ \frac{\nu}{2[g(z)]^2}\sigma^{\prime\prime}(y) 
\label{multi83} 
\end{equation} 
considering odd parity for the fields, gives the same spectrum (apart from 
the zero mode) with the familiar potential for fields of even parity:  
\begin{equation} 
\hspace*{0.5cm} V_{2}(z)=\frac{\nu(\nu+1)}{2[g(z)]^2}(\sigma^{%
\prime}(y))^{2}- \frac{\nu}{2[g(z)]^2}\sigma^{\prime\prime}(y) 
\label{multi84} 
\end{equation} 
which according to the previous discussions supports an ultra-light special 
KK state. This is because the previous potentials are SUSY partners and as 
expected have the same spectrum apart from the zero mode. 
 
\section{Life without negative tension branes} 
 
In Chapter \ref{5dads} it was shown  that the properties of the $^{\prime\prime}+-+^{\prime%
\prime}$ model (the bounce form of the ``warp'' factor), which contains a 
moving negative tension brane can be mimicked by the $^{\prime\prime}++^{%
\prime\prime}$ model, where the negative brane is absent provided that we 
allow for $AdS_{4}$ on the branes. Since the corresponding potential has two  
$\delta$-function wells that support bound states the multi-localization 
scenario appears also here. The previous results related to the localization 
properties of the various fields are valid also in this case. However, the 
presence of $AdS_{4}$ geometry on the branes, modifies the form of the 
potential of the corresponding Schr\"ondiger equation and thus the details 
of the form of wavefunctions of the KK states. In this section we briefly 
discuss these modifications. 
 
As previously mentioned, the spacetime on the 3-branes must be $AdS_{4}$ (in 
contrast to the $^{\prime \prime }+-+^{\prime \prime }$ models where the 
spacetime is flat). Thus in this case the background geometry is described 
by:  
\begin{equation} 
ds^{2}=\frac{e^{-2\sigma (y)}}{(1-\frac{H^{2}x^{2}}{4})^{2}}\eta _{\mu \nu 
}dx^{\mu }dx^{\nu }+dy^{2} 
\end{equation} 
By following exactly the same steps as in the case of flat branes, again the 
whole problem is reduced to the solution of a second order differential 
equation for the profile of the KK states. The differential equation is such 
that after the dimensional reduction the five dimensional physics is 
described by a infinite tower of KK states that propagate in the $AdS_{4}$ 
background of the 3-brane. It is always possible to make the coordinate 
transformation from y coordinates to z coordinates related through: $\frac{dz%
}{dy}=A^{-1}(y)$, where $A(y)=e^{-\sigma (y)}$, and a redefinition of the 
wavefunction\footnote{The form of this redefinition depends on the spin of the field.} and bring 
the differential equation in the familiar Schr\"{o}dinger-like form:  
\begin{equation} 
\left\{ -\frac{1}{2}{\partial _{z}}^{2}+V(z)\right\} {\hat{f}}_{n}(z)=\frac{%
m_{n}^{2}}{2}{\hat{f}}_{n}(z) 
\end{equation} 
where $\hat{f}_{n}(z)$ is the appropriate redefinition of the wavefunction. 
 
For the $^{\prime\prime}++^{\prime\prime}$ model the form of the potential 
for fields of different spin is different. However, in the case that it 
admits a massless mode and an anomalously light mode it has the generic form 
given in eq.(\ref{multi80}) that applied to the $^{\prime\prime}+-+^{\prime\prime}$ 
case. However the warp factor has a different form from the case with 
negative tension branes being given by  
\begin{equation} 
g(z)\equiv e^{\sigma (y)}=\frac{1}{\cosh (k(\left| z\right| -z_{0})} 
\end{equation} 
 
Note that in this case $\sigma ^{\prime }(y)$ is not constant in the bulk 
and $\sigma ^{\prime \prime }(y)$ is not confined to the branes. The 
massless modes, corresponding to the Schrodinger equation with this 
potential,are given by eqs.(\ref{multi75}) and (\ref{multi76}) as in the $^{\prime\prime}+-+^{%
\prime\prime}$ case. Note however that the constraint on the relative 
magnitude of the two terms in the potential is now required when 
solving for the propagation in the bulk whereas in the case of a negative 
tension brane it came when solving for the boundary conditions. 
 
 
The zero mode wavefunction is given by:  
\begin{equation} 
\hat{f}_{0}(z)=\frac{C}{[\cos(\tilde{k}(z_{0}-|z|))]^{\nu}} 
\end{equation} 
where $C$ is the normalization factor. By considering cases with $m_{n}\neq0$%
, we find the wavefunctions for the KK tower :  
\begin{equation} 
\renewcommand{\arraystretch}{1.5}  
\begin{array}{c} 
{\hat{f}}_{n}(z)=\cos^{\nu+1}(\tilde{k}(|z|-z_{0}))\left[C_{1}~F(\tilde{a}%
_{n},\tilde{b}_{n},\frac{1}{2};\sin^{2}(\tilde{k}(|z|-z_{0})))~~~~~~~~\right. 
\\  
\left. ~~~~~~~~~~~~~~~~~~~+C_{2}~|\sin(\tilde{k}(|z|-z_{0}))|~F(\tilde{a}%
_{n}+\frac{1}{2},\tilde{b}_{n}+\frac{1}{2},\frac{3}{2};\sin^{2}(\tilde{k}%
(|z|-z_{0})))\right] 
\end{array} 
\end{equation} 
where  
\begin{eqnarray} 
\tilde{a}_{n}=\frac{\nu+1}{2}+\frac{1}{2}\sqrt{\left(\frac{m_{n}}{\tilde{k}}%
\right)^2+{\nu}^2} \cr \tilde{b}_{n}=\frac{\nu+1}{2}-\frac{1}{2}\sqrt{\left(%
\frac{m_{n}}{\tilde{k}}\right)^2+{\nu}^2 } 
\end{eqnarray} 
The boundary conditions are given by:  
\begin{eqnarray} 
{\hat{f}}_{n}~^{\prime}({0}^{+})+k \nu \tanh(k y_{0}){\hat{f}}_{n}(0)=0  
\nonumber \\ 
{\hat{f}}_{n}~^{\prime}({z_{L}}^{-})-k \nu \frac{\sinh(k(L-y_{0}))}{\cosh(k 
y_{0})}{\hat{f}}_{n}(z_{L})=0 
\end{eqnarray} 
the above conditions determine the mass spectrum of the KK states. By 
studying the mass spectrum of the KK states it turns out that it has a 
special first mode similar to the one of the $^{\prime\prime}+-+^{\prime%
\prime}$ model as expected.

\section{Light states without multi-localization}
 
\subsection{Light states from SUSY-partner configuration} 
 
\begin{figure} 
\begin{center} 
\begin{picture}(200,200)(40,80) 
 
\SetWidth{1} 
\LongArrow(20,80)(20,240) 
\LongArrow(210,80)(210,240) 
 
\SetWidth{1} 
\LongArrow(-70,100)(110,100) 
\LongArrow(120,100)(300,100) 
 
 
\Text(15,270)[c]{Original Model} 
\Text(215,270)[c]{SUSY-partner} 
\Text(40,235)[c]{$V(z)$} 
\Text(230,235)[c]{$V_{s}(z)$} 
\Text(110,90)[c]{z} 
\Text(300,90)[c]{z} 
 
 
 
{\SetColor{Green} 
\Curve{(20,90)(30,92)(35,96)(38,100)(43,105)(48,106)(55,108)(60,110)(65,115)(77,170)(80,200)(83,230)} 
\Curve{(20,90)(10,92)(5,96)(3,100)(-2,105)(-7,106)(-14,108)(-19,110)(-24,115)(-36,170)(-39,200)(-42,230)} 
} 
 
 
{\SetColor{Red} 
\Curve{(210,136)(220,136)(225,138)(228,142)(233,140)(241,120)(245,100)(255,60)(258,65)(267,120)(270,150)(273,180)} 
\Curve{(210,136)(200,136)(195,138)(192,142)(187,140)(179,120)(175,100)(165,60)(162,65)(153,120)(150,150)(147,180)} 
} 
 
\end{picture} 
\end{center} 
 
\caption{On the left the potential $V(z)$ for the corresponding 
quantum-mechanical problem. On the right, the 
 SUSY-partner potential  $V_{s}(z)$. The two potentials have the same 
massive spectrum. The first does not support massless zero mode but the 
SUSY-partner does.} 
\label{shap1} 
\end{figure}
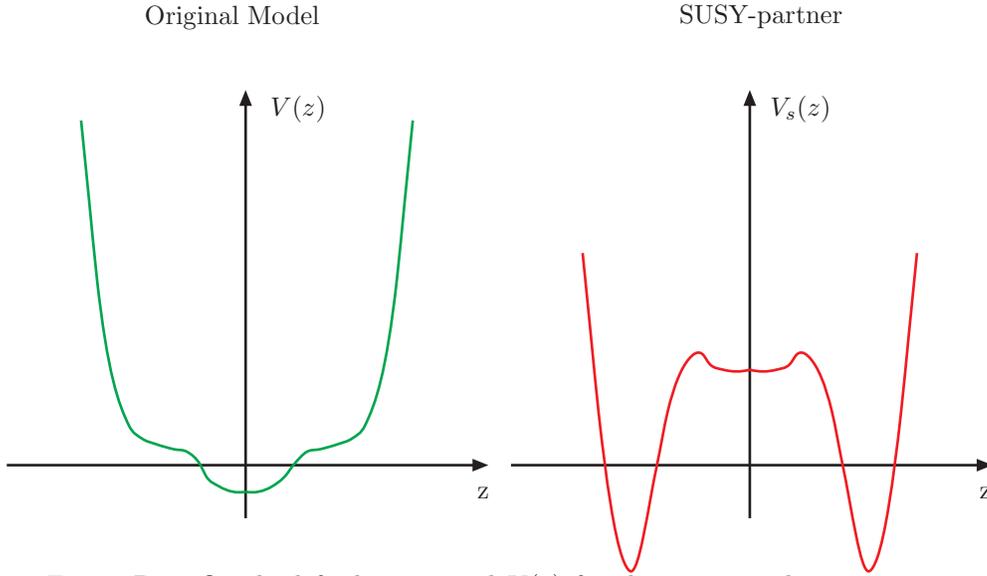

In Chapter \ref{multiloc} we focused in the relation between 
multi-localization and multi-brane configurations. Here we would like 
to point out another possibility based on the same ideas though: The 
appearance of light KK states without multi-localization \cite{Shaposhnikov:2001nz}. In order to 
demonstrate how this can occur by a simple example, 
let us  assume that we 
have a brane configuration such that the  potential of the 
corresponding Schr\"odinger equation is given by 
$V(z)=W^{2}(z)-W'(z)$  and 
has a form, for example, like the one appearing in Fig.(\ref{shap1}) 
($W(z)$ is an appropriate function giving the previous potential).     
Noting that the previous potential has the familiar form of supersymmetric 
quantum mechanics , we can easily construct the SUSY-partner 
potential $V_{s}(z)=W^{2}(z)+W'(z)$. The previous potential  has the same discrete levels 
and additionally a zero mode. The form of the potential in 
our example is given in Fig.(\ref{shap1}).  
 
Although the first potential has a  form that one would not expect to 
give rise to light KK states, by looking the SUSY-partner potential 
with the familiar double well form we can immediately realize that this 
scenario can be realized by choosing an appropriate form for the 
$W(z)$ function.     
Thus using the above, one can construct a model with light (compared to the 
rest) and localized KK mode. The difference with the case of 
the multi-localized case is the absence of zero mode\footnote{The zero 
mode can be absent also in multi-localized models if we allow for 
infinite compactification volume}. The above configuration although is  
giving a light special KK state does no provide the distinct  two scale 
KK spectrum since the zero mode is absent. All KK state mass 
differences are of the same order. However the above provides a 
mechanism of generating a light localized KK mode.  
 
 
\subsection{Light states from twisted boundary conditions} 
 
In order to give a consistent description of bulk physics one has to 
specify the boundary conditions associated with the orbifold 
symmetries. Thus one should specify if the fields are even or odd 
under the orbifold $Z_{2}$ symmetry (reflections under $y \rightarrow 
-y$). The choice of the parity properties 
of the various fields depends on the form of their 
couplings/interactions. Up to this point we have used, depending on 
the field, even or odd parities.  
 
However, apart from considering odd and even boundary conditions it turns out 
that we can also use twisted boundary conditions, that is one can 
demand  that the wavefunctions should be symmetric 
in respect to the one orbifold fixed point but antisymmetric in respect to 
the other. 
This turns out  to be  a convenient way to break SUSY in brane world models. 
But if we demand  in the case of  RS  
model  that the wavefunctions should be symmetric 
in respect to the first positive brane but antisymmetric with respect to 
the second, this results to  a model with a mass spectrum 
identical with the  spectrum of the antisymmetric states with respect to 
the central brane in the symmetric $''+-+''$ model. This implies that 
the zero mode of the spectrum will be the ultra-light KK state. In this 
case the zero mode (and all the rest states which are symmetric with 
respect to the central brane) is projected out due to the boundary 
conditions. The above gives the possibility of  ultralight states 
without having brane configurations with double well potentials. 
However, like in the previous case the KK spectrum is no longer 
characterized by two scales due to the absence of the zero mode.  


\addcontentsline{toc}{chapter} 
		 {\protect\numberline{Bibliography\hspace{-96pt}}}

\end{document}